*G. M. Garibian, C. Yang*

# X-RAY TRANSITION RADIATION

ՀԱՅԿԱԿԱՆ ՍՍՀ ԳԻՏՈՒԹՅՈՒՆՆԵՐԻ ԱԿԱԴԵՄԻԱ
ՍԱՀՄ ԱՐՊՎ ԵՐԵՎԱՆԻ ՖԻԶԻԿԱՅԻ ԻՆՍՏԻՏՈՒՏ

*Ղարիբյան Գ. Մ., Յան Շի*

# **ՌԵՆՏԳԵՆՅԱՆ**
# ԱՆՑՈՒՄԱՅԻՆ ՃԱՌԱԳԱՅԹՈՒՄ

ACADEMY OF SCIENCES OF THE ARMENIAN SSR
YEREVAN PHYSICS INSTITUTE OF THE
STATE COMMITTEE FOR THE USE OF ATOMIC ENERGY
OF THE UNION OF SOVIET SOCIALIST REPUBLICS

*G. M. Garibian, C. Yang*

# **X-RAY** TRANSITION RADIATION

The monograph is devoted to theoretical and experimental studies of transition radiation produced by fast charged particles traversing media with interfaces. Particular attention is paid to X-ray transition radiation (XTR) generated at a single interface as well as in a single plate or in a stack of plates. The foundations of the general theory of transition radiation are presented, and the results of experimental studies of XTR properties are given together with a comparison with theoretical predictions. TRDs of high-energy particles used at modern accelerators and in cosmic-ray physics are described.

The book is intended for researchers working in the field of high-energy charged-particle interactions with media, as well as for graduate students and advanced undergraduates in related disciplines.

# CONTENTS

## PREFACE

The phenomenon of transition radiation, first theoretically predicted in the Soviet Union, has attracted considerable attention over the past decade. This is particularly true of X-ray transition radiation, which has already found practical application in high-energy particle physics as the basis for a new method of detecting and identifying ultrarelativistic particles.

At present, an extensive body of original literature (over 700 scientific publications) is devoted to the theoretical and experimental study of transition radiation. These results remain insufficiently covered in the existing monographs. Studies on transition radiation carried out up to the end of the 1960s were reviewed in the well-known monograph by M. L. Ter-Mikaelian, Influence of the Medium on Electromagnetic Processes at High Energies (Yerevan: Publishing House of the Academy of Sciences of the Armenian SSR, 1969). The book by V. L. Ginzburg and V. N. Tsytovich, *Transition Radiation and Transition Scattering (Some Aspects of the Theory)*, has not yet appeared in print. Thus, at present there is a need for monographs that present and analyze the principal theoretical and experimental results on transition radiation and, especially, on X-ray transition radiation obtained to date. The present book aims, to some extent, to meet this need.

The book deals primarily with X-ray transition radiation. This is motivated both by the practical importance of X-ray transition radiation and by the authors' longstanding scientific interest in these problems. At the same time, we have aimed to provide a sufficiently complete and systematic exposition of the theoretical foundations. The fundamental equations given in Chapters I and II are general and applicable to radiation of all frequencies. For this reason, we also consider various aspects of transition radiation outside the X-ray frequency range. The material is presented from a unified standpoint and often differs from the original published papers.

The monograph contains a reasonably complete bibliography on the subjects considered. The references are arranged chronologically and numbered using the last two digits of the year of publication and a sequential number within each year in the order of citation. For example, [**72**.5] refers to a work published in 1972 and numbered 5 in that year's bibliography.

During the preparation of this book, we greatly benefited from the support of A. Ts. Amatuni and B. M. Bolotovsky, who offered a number of valuable comments on the manuscript. We also benefited from helpful discussions with M. P. Lorikyan and A. G. Oganesyan concerning Chapter V and made partial use of A. G. Oganesyan's review article prior to its publication. We gratefully acknowledge the assistance of L. A. Vardanyan, particularly with the bibliography, and of S. P. Avetisyan. We would like to express our sincere gratitude to all those mentioned above.

We would be grateful to all readers who may send us their comments. We hope that, despite possible shortcomings, the book will prove useful primarily to those beginning work in this interesting field of physics.

Yerevan
December 1982

G. M. GARIBIAN
CHI YANG

# INTRODUCTION

## REVIEW OF THE MAIN RESEARCH AREAS IN TRANSITION RADIATION

It is well known that a charged particle in uniform straight-line motion in vacuum does not emit electromagnetic radiation. In a material medium, the particle's field excites and polarizes the atoms of the medium, causing them to act as sources of secondary electromagnetic waves that propagate in all directions (Huygens, Fresnel, Laue). These secondary waves cancel each other if the medium is stationary and homogeneous and the particle's speed is less than the phase velocity of the wave in the medium. If even one of these conditions is not satisfied, the secondary waves fail to cancel completely, giving rise to a net radiation field. This radiation is fundamentally different from bremsstrahlung: it occurs even when the particle moves with constant velocity[1]. In a homogeneous medium, this is the well-known *Vavilov–Cherenkov radiation* [**37**.1] (see also Refs. [**60**.1, **68**.1, **72**.1]). When the medium is inhomogeneous—for example, when it has an interface with another medium (in particular, with vacuum)—radiation arises due to the presence of the boundary, in addition to possible Vavilov–Cherenkov radiation; namely, *transition radiation*.

This phenomenon was theoretically predicted by V. L. Ginzburg and I. M. Frank as early as 1945 [**45**.1]. This work laid the foundation for a new direction in the study of charged-particle interactions in a medium.

Since transition radiation results from the incomplete cancellation of secondary waves emitted by the atoms of a medium (or of different media) under the influence of a passing charged particle, it is generated whenever inhomogeneities in the electrodynamic properties of the medium are present. In this sense, it differs from Vavilov–Cherenkov radiation, which occurs only when the particle's velocity exceeds the phase velocity of light in the medium. At the same time, it follows that transition radiation has the same physical nature as Vavilov–Cherenkov radiation: both are forms of *collective radiation of the medium's electrons initiated by an external charge*, and under certain conditions they may be inseparable.

[1] Since the radiation energy is supplied by the particle's kinetic energy, its velocity is, strictly speaking, not constant. However, if the particle's energy is large compared with its losses, its motion can, to a very good approximation, be treated as uniform straight-line motion.

## A. EARLY STUDIES

In their pioneering work [**45**.1], Ginzburg and Frank analyzed the case of a charged particle normally crossing a plane interface between two homogeneous media. They showed that radiation—termed transition radiation—is generated, and that the component emitted into the backward hemisphere relative to the particle's motion is concentrated predominantly in the *optical frequency range*. In this case, the spectral intensity of the radiation increases logarithmically with the particle's Lorentz factor $\gamma$ (the ratio of its total energy to its rest energy). Consequently, measurement of the transition radiation intensity, in principle, offers a new method for determining the Lorentz factor of high-energy particles.

It should be noted, however, that the intensity (the number of quanta) of radiation produced at a single interface is small, averaging approximately 1/137 quantum per charged particle. Therefore, it became necessary to seek ways of increasing the intensity of transition radiation (see, for example, Frank's Nobel lecture [**59**.1]). In this context, various authors studied the formation of transition radiation at multiple interfaces during 1957–1959: Feinberg and Khizhnyak [**57**.1], and Bliokh [**59**.2] in an infinite periodically stratified medium; Pafomov [**57**.2] and Garibian and Chalikyan [**58**.1, **59**.3] in a plate; and Garibian [**58**.2] in a stack of plates.

At this stage of research, it appeared that the practical use of transition radiation for particle detection and identification faced the following difficulties: (1) the dependence of the radiation intensity on the particle's Lorentz factor $\gamma$ is weak (logarithmic), and therefore requires high measurement accuracy; (2) in order for the radiation intensities from the individual interfaces of the plates in a stack to add coherently, the plates must be placed far apart, at distances exceeding the so-called formation length of transition radiation in vacuum, which is (to order of magnitude) equal to the wavelength multiplied by $\gamma^2$ [**59**.1] (for example, for optical frequencies and $\gamma \geq 10^3$, the formation length is on the order of 10 cm or more).

In 1959, Garibian [**59**.4] showed that the radiative energy losses of an ultrarelativistic particle ($\gamma \gg 1$) in a material that has an interface with vacuum increase in direct proportion to the particle's Lorentz factor $\gamma$. As Barsukov [**59**.5] and Garibian [**59**.4] showed, this is due to the fact that the spectrum of transition radiation emitted by a particle from a semi-infinite medium under consideration into the forward hemisphere relative to its direction of motion extends into the *X-ray frequency range*, up to frequencies of order $\omega_0\gamma$ (where $\omega_0$ is the plasma frequency of the material). The total radiation intensity is proportional to the particle's Lorentz factor.

As is well known, the total energy losses of a particle consist of radiation losses and ionization losses. At large values of the Lorentz factor $\gamma$, the ionization energy loss per unit length in an infinite (or semi-infinite) medium becomes independent of $\gamma$ (the Fermi density effect). In the same paper [**59**.4], Garibian showed that when the plate thickness is small compared with a certain critical value, of the order of the formation length of optical transition radiation in medium, the density effect disappears in the ionization energy loss (see also Ref. [**81**.1]).

The theoretical discovery of *X-ray transition radiation* (XTR) [**59**.4, **59**.5] to a large extent resolved the aforementioned difficulties associated with the use of optical transition radiation. Owing to the linear $\gamma$-dependence of XTR generated at the interface between a semi-infinite medium and vacuum, and to its much shorter wavelength, real possibilities emerged for the use of this phenomenon in high-energy particle physics. To date, XTR provides virtually the only non-destructive method for identifying ultrarelativistic particles. This has motivated the intensive and comprehensive theoretical and experimental studies of transition radiation in general, and of X-ray transition radiation in particular, since 1959.

To date, numerous publications on transition radiation have appeared; a fairly comprehensive bibliography (up to 1978) can be found in Ref. [**79**.1]. There are several reviews on this subject [**61**.1, **65**.1, **65**.2, **69**.1, **72**.2, **72**.3, **73**.1, **73**.2, **78**.1]. XTR and its practical applications have been discussed repeatedly at international conferences (Versailles, 1968 [**68**.2]; Dubna, 1970 [**71**.1]; Hobart, 1971 [**71**.2]; Frascati, 1973 [**73**.2]; Munich, 1975 [**75**.1]; etc.). In 1977, an international symposium devoted specifically to transition radiation was held in Yerevan, Soviet Union [**77**.1]. In what follows, we briefly outline the principal research directions in this field and highlight some unresolved problems.

## B. DEVELOPMENT OF XTR THEORY

Following the prediction of transition radiation arising at a single interface between media, it was natural to study the phenomenon in more complex configurations. Of particular interest was the formation of transition radiation in a stack of plates and, more generally, in a periodic inhomogeneous medium, since such media were expected to yield higher radiation intensities.

In 1960, Ter-Mikaelian, building on the methods and ideas he had previously developed for coherent bremsstrahlung of charged particles in single crystals [**69**.1, **72**.2], showed that transition radiation in a periodically inhomogeneous medium is characterized by the requirement that the frequency and emission angle satisfy a certain relation ("*resonance condition*") [**60**.2], arising from the interference of waves generated in different regions of the medium (see also Ref. [**57**.1]). He referred to such coherent transition radiation as "resonant." In the same year, on the basis of this relation, Ter-Mikaelian and Gazazyan [**60**.3, **61**.2] derived relatively simple expressions for the spectral–angular and spectral (angle-integrated) intensity distributions of resonant XTR in an infinite periodically stratified medium. At the same time, Amatuni and Korkhmazyan [**60**.4], building on their earlier work on a diffuse boundary (see below), considered a medium with periodically varying density and obtained the first two terms in the expansion of the radiation field. At the same time, Garibian and Goldman [**60**.5], using the previously derived general equation for the radiation fields in a regular stack of an arbitrary number of plates [**58**.2], derived the same expressions as in Ref. [**60**.3] for the spectral–angular and spectral intensity distributions of XTR in a stratified medium.[2]

Subsequent work [**71**.3, **74**.1] established criteria for the applicability of the equations derived in Ref. [**60**.3] and [**60**.5] (see Secs. 3.4 and 3.5).

---

[2] The works [**60**.2–**60**.5] were carried out in the same laboratory of the Physics Institute of the Academy of Sciences of the Armenian SSR, and their authors were familiar with the course and results of the other works.

By the mid-1970s, theoretical studies of X-ray transition radiation had also begun in the United States. Müller et al. [**74**.2] and Durand [**75**.2] analyzed the corresponding equations, with some generalizations, for the XTR intensity at a single interface, in a plate, and in a stack of plates. Yodh et al. [**75**.3] derived analogous equations in dimensionless variables and carried out their analysis. Cherry [**78**.2] provided a fairly complete summary of frequently used theoretical results.

In all the works mentioned above, it was assumed that the interfaces between the media are "sharp," i.e., that the electrodynamic properties undergo a discontinuous change across the boundary. In reality, this assumption does not hold.

Amatuni and Korkhmazyan [**60**.6, 65.3], and Tsytovich [**61**.3] (see also Refs. [**64**.1, **76**.1, **80**.1]), studied the influence of *boundary diffuseness* on the formation of XTR. It was shown that this effect is negligible if the thickness of the diffuse layer is much smaller than the formation length of the radiation, which for XTR can reach macroscopic dimensions. If, however, the thickness of the diffuse layer is much larger than the formation length of transition radiation, then the radiation intensity becomes exponentially small.

The phenomenological *quantum* theory of X-ray transition radiation, which includes the recoil of a fast charged particle, was formulated by Garibian [**60**.7]. This theory was further developed in subsequent works by Zaretsky, Lomonosov, and others [**77**.2, **80**.2, **81**.2], which examined "transition" pion production in collisions of fast nucleons with nuclear matter, as well as stimulated transition radiation and absorption. Stimulated transition radiation and absorption were also considered in Refs. [**77**.19, **80**.16]. The quantitative relationship between these phenomena and the onset of modulation in an initially unmodulated beam of charged particles was discussed in the review [**82**.10].

The theory of transition radiation in a medium with *random inhomogeneities* was developed by S. Kapitza [**60**.8] and Ter-Mikaelian [**60**.2, **61**.2]. In Refs. [**60**.2, **61**.2], an irregularly stratified medium was also considered in the "quasiclassical" approximation, and an equation valid in the absence of radiation absorption was obtained (see Sec. 6.1). The fields of transition radiation in an arbitrary stratified medium were derived by Garibian and Muradian [**68**.3], Pafomov [**69**.2], and Magomedov [**72**.4, **74**.3]. In order to provide a theoretical justification for the use of porous media as XTR radiators, Garibian, Gevorgyan, and Yang Chi [**73**.3, **74**.4] were the first to derive and analyze a general equation for the spectral–angular distribution of the average XTR intensity produced in an *arbitrary irregular stack* (which, for sufficiently large irregularity, loses its resonant character), and they also calculated the corresponding spectral distributions.

When a fast charged particle traverses a crystal, the secondary waves emitted by the atoms are expected to form an interference pattern with maxima determined by the Bragg condition. Ter-Mikaelian analyzed this phenomenon within the framework of perturbation theory, describing it as "resonant radiation" in a three-dimensional periodic medium (Ref. [**61**.13]; see also Ref. [**69**.1]). This issue was subsequently examined by Rivlin [**65**.4], and by Kudryavtsev and Ryazanov [**70**.1]. This phenomenon exhibits a close similarity to Vavilov–Cherenkov radiation in an amorphous medium and is therefore often termed "quasi-Cherenkov radiation" (see, for example, Ref. [**82**.7]).

Based on a quantum microscopic approach and taking into account the interaction of the field of a charged particle with individual atoms of the crystal, Garibian and Yang Chi developed a quantitative theory of quasi-Cherenkov radiation of relativistic particles, first for a thin crystalline plate [**71**.4], then for a plate of arbitrary thickness [**72**.5, **73**.4], and finally for a stack of crystalline plates [**75**.5, **81**.3]. The role of "strongly diffracted waves" was properly accounted for by analogy with the dynamical theory of X-ray diffraction (see, for example, Refs. [**67**.1, **72**.6, **72**.7]). The authors [**74**.5, **75**.6–**75**.9] carried out a detailed study of the properties of the sharp maxima ("dynamical maxima") of the peculiar "*Laue pattern*" of X-ray quasi-Cherenkov radiation. A similar problem was analyzed by Baryshevsky and Feranchuk [**71**.5, **73**.5, **74**.6] from a phenomenological perspective; however, their initial estimate of the radiation intensity exceeded the correct value by many orders of magnitude. Subsequently, the estimate was corrected, and it was found that their results were consistent with those of other authors.

Belyakov and Orlov [**72**.30] developed the theory of quasi-Cherenkov radiation of a particle in cholesteric liquid crystals, and Afanasyev and Aginyan [**77**.3] extended it to the case of a mosaic crystal. The latter work also provides a simpler equation for the total radiation intensity in individual spots of the Laue pattern. Dialetis [**78**.3] also analyzed the generation of coherent X-ray radiation by a uniformly moving relativistic charged particle in a crystal. However, no experiment aimed at detecting and studying the Laue pattern of quasi-Cherenkov radiation has yet been carried out.

Ryazanov [**82**.8] analyzed resonant transition radiation of relativistic charged particles from ordered defects in crystals, such as chains and lattices of vacancies.

At high radiation frequencies, the imaginary part of the medium's polarizability, associated with absorption, can become comparable to—or even exceed—the real part (see, for example, Ref. [**74**.7]). Garibian and Yang Chi showed [**76**.2] that in this case the *absorptive properties* must play a significant role in the formation of XTR. This results in a stronger, specifically quadratic, dependence of the total XTR energy on the particle's Lorentz factor at sufficiently large $\gamma$. The same effect also occurs for a plate and a stack of plates (see the work of Avakyan et al. [**77**.4, **77**.1, p. 592]).

A particle traversing medium is inevitably scattered, to some degree, by the atoms of the medium. The problem of the influence of *multiple particle scattering* on the formation of XTR was first formulated by Garibian and Pomeranchuk as early as 1959 [**59**.6]. This question was subsequently discussed by Pafomov [**60**.9, **64**.2], Goldman [**60**.10], Garibian [**60**.11] (for a semi-infinite medium–vacuum interface), Ternovskii [**60**.12], and again by Pafomov [**65**.5] (for a plate). At the same time, Goldman, Garibian, and Ternovskii, on the one hand, and Pafomov, on the other, adopted different approaches to separating the boundary-induced component from the total radiation. In studies of the influence of multiple scattering, absorption of the medium was taken into account by Pafomov [**67**.2] in the optical frequency range and by Varfolomeev et al. [**72**.8, **74**.8] at higher frequencies (see also Refs. [**74**.7, **74**.9, **77**.5]). The problem of multiple scattering is also examined in the review [**74**.10].

Ter-Mikaelian [**68**.4, **69**.1, **72**.2] analyzed the influence of multiple scattering in a stratified medium, assuming that the dielectric permittivity varies periodically in space and that the particle's scattering probability is uniform throughout the medium.

The equation for the spectral distribution of the total radiation generated in a plate by a relativistic charged particle, taking multiple scattering into account, was derived and analyzed by Garibian and Yang Chi [**76**.3]. They proposed a simple and physically transparent method for isolating the edge contribution—the radiation component associated with the presence of boundaries—from the total radiation. These authors and Vardanyan showed that, in addition to bremsstrahlung, multiple scattering of the particle leads to a smoothing and enrichment of the spectral distribution [**76**.3, **76**.4, **77**.7, **79**.2, **80**.3, **82**.14].

The possibility of using XTR produced in a periodically inhomogeneous medium as an *intense X-ray source* was noted by Zhevago [**77**.6]. Pantell and co-workers in the United States [**80**.8, **80**.9, **81**.8, **82**.3, **82**.4] subsequently proposed a similar concept, noting that the source is readily tunable. The authors of these works carried out not only theoretical but also experimental studies.

## C. EXPERIMENTAL STUDIES OF XTR AND PARTICLE DETECTOR DEVELOPMENT

As early as 1961, *methods* for the experimental detection of resonant transition radiation in a stratified medium in the X-ray range were discussed by Alikhanian, Arutyunyan, Ispiryan, and Ter-Mikaelian [**61**.4], as well as by Garibian [**61**.14]. Somewhat later, one of the ideas proposed in Ref. [**61**.4] was implemented, and resonant XTR in a layered medium generated by cosmic-ray muons was observed [64.3, 65.6].

Beginning in 1969, Yuan and collaborators in the United States carried out a series of experimental studies of transition radiation [**69**.3, **69**.4, **70**.2–**70**.4, **71**.2, **71**.6, **72**.9, **72**.10, **75**.10–**75**.13] and obtained good agreement with the theoretical prediction [**59**.4] of the linear increase in total radiation intensity with the Lorentz factor (see, for example, Yuan's review lecture [**73**.2, p. 334]). In these experiments, transition radiation was detected using a germanium detector, a CsI scintillation crystal, and a multiwire proportional chamber.

Following the commissioning of the electron synchrotron in 1967 at the Yerevan Physics Institute, systematic studies of the properties of XTR were carried out by Alikhanian, Ispiryan, Lorikyan, Oganesyan, and others [**70**.5–**70**.7, **71**.7, **72**.11–**72**.13, **73**.6–**73**.12, **74**.11, **74**.12, **75**.14] (see also Ref. [**77**.1]). The possibility of studying XTR using a *streamer chamber*, in which both the primary charged particle and the XTR photons are detected simultaneously, was proposed by Lorikyan and implemented in Refs. [**70**.6, 70.7]. There, for the first time, transition radiation produced in an irregular medium—*plastic foam*—was observed. Subsequently, XTR produced in an irregular medium (plastic foam) was also reported in Ref. [**71**.6, **71**.8]. The properties of XTR in irregular (porous) media were studied in detail in Refs. [**73**.10, **73**.11], and good agreement with the theory [**74**.4, **75**.4] was subsequently reported in Ref. [**75**.27]. A streamer chamber was also used for XTR studies at the Lebedev Physical Institute of the USSR Academy of Sciences by Slavatinsky et al. [**72**.14].

The first transition radiation detector (*TRD*) for the identification of high-energy particles was developed at the Yerevan Physics Institute [**72**.15, **73**.14, **74**.16]. A large installation employing a TRD for the detection and identification of cosmic-ray hadrons (the Pion facility) was developed and constructed at the same institute on Mount Aragats [**74**.17, **76**.7, **76**.8, **78**.5].

A TRD was also installed on the Intercosmos-17 Earth satellite for the study of cosmic electrons [**78**.6, **79**.4, **81**.13], and on high-altitude balloons for studying the spectrum and composition of primary cosmic-ray nuclei in the USSR [**82**.9, **82**.13].

From 1973 onward, several other groups of physicists in the United States and Europe began experimental studies of XTR. Yodh's group in the United States and England [**73**.13, **73**.34, **75**.3] studied the applicability of XTR for particle identification at $\gamma \geq 10^3$ and employed a TRD in cosmic-ray experiments. Müller et al. in the United States [**74**.2, **74**.13, **75**.15, **77**.8, **78**.2] carried out a detailed study of the frequency spectra of XTR produced in regular stacks of plates and of their dependence on the particle Lorentz factor. The results were in good agreement with the theory [**60**.3, **60**.5, **71**.3]. An irregular medium such as plastic foam was also used as a XTR radiator by Yodh's group [**73**.13] and by Müller [**75**.15]. The latter group applied a TRD to measure the spectrum of primary cosmic-ray electrons on balloons [**77**.9]. Mack and Osborne in the United States [**74**.14] separated XTR from the primary particle using the Compton effect (see also Ref. [**71**.7]). In Austria, Fabian experimentally studied the frequency spectrum of XTR generated in plastic foam [**77**.10] and obtained good agreement with the theoretical predictions.

Willis et al. in the United States [**74**.15, **75**.16, **75**.17, **76**.5] used stacks of lithium plates as an XTR radiator. Based on such a radiator, an installation [**77**.1, **77**.11] was developed and used in experiments at CERN (Geneva, Switzerland), in particular for studies of the properties of the newly discovered J/ψ and ϒ particles [**77**.12–**77**.14, **78**.4, **79**.3, **80**.4, **80**.5]. A new TRD capable of separating protons from pions was developed by a Soviet–American collaboration (Oganesyan, Atac, et al.) [**77**.1, **77**.15] at Fermi National Accelerator Laboratory (Fermilab), Batavia, USA.

A novel concept of a TRD employing superheated superconducting granules was proposed and tested by Drukier, Yuan, et al. [**75**.18, **76**.6, **77**.1]. However, a quantitative theory of this method is still lacking (see Vardanyan et al. [**77**.1, p. 374, **76**.16] on this subject).

At CERN, a Soviet–European–American group (Dolgoshin, Willis, et al.) achieved for the first time the experimental separation of pions from kaons using a TRD [**81**.5–**81**.7]. This group also used a medium composed of thin *fibers* as an XTR radiator [**81**.7]. Müller et al. [**82**.1] conducted an experimental study of similar, apparently promising radiators made of polymer fibers.

TRDs were also used for electron identification in an experiment on hyperon leptonic decays at CERN's 400-GeV proton accelerator (SPS) [**79**.5, **79**.6, **80**.6, **82**.2], and for pion identification at Fermilab in studies of high-transverse-momentum *jet* production in hadron–hadron interactions [**80**.7].

## D. GENERAL ISSUES OF THE THEORY OF TRANSITION RADIATION. EXPERIMENTAL STUDIES OF OPTICAL TRANSITION RADIATION

Between 1958 and 1970, the theory of transition radiation was generalized by Korkhmazyan, Garibian, Pafomov, and others to the case of *oblique* incidence of a charged particle on the interface between media [**58**.3, **58**.4, **60**.13, **60**.14, **62**.1, **62**.2, **69**.5, **70**.8], on a plate [**66**.1, **67**.3], and on a stack of plates [**70**.9]. Equations were derived for the spectral–angular distributions of radiation intensity for different polarizations. An analysis of these equations shows, in particular, that transition radiation in the optical frequency range has maxima in directions determined by geometrical optics (to within an angle of order $\gamma^{-1}$). In the X-ray and higher frequency range, the radiation is concentrated primarily along the trajectory of the charged particle.

The generation of transition radiation under *arbitrary* particle motion was studied by Garibian and Magomedov [**63**.1, **63**.2, **67**.4, **69**.6].

Pafomov [**67**.5, **69**.2] developed a general method for treating radiation problems in the presence of interfaces.

At the same time, along with transition radiation, other types of radiation are produced, such as bremsstrahlung.

The theory of transition radiation generated by electric and magnetic dipole *moments*, as well as by *bunches* of charged particles, was developed by Amatuni, Tsytovich, and others [**60**.15, **61**.5, **62**.3, **75**.19]. It was shown, in particular, that if the longitudinal size of the bunch is smaller than the radiation wavelength and the transverse size is smaller than γ times the wavelength, all particles in the bunch radiate coherently.

In connection with experimental searches for the Dirac monopole, the transition radiation of a magnetic charge (*monopole*) was studied by Mergelyan [**63**.3] and examined in greater detail by Dooher [**71**.9] and Frank [**79**.7]. The possibility of using transition radiation of relativistic particles to study the optical properties of matter was considered by Frank [**65**.1], and later by Korkhmazyan and Elbakyan [**67**.6, **71**.10].

Taking into account the thermal motion of the electrons in the medium leads to spatial dispersion of the dielectric permittivity. This issue, in the context of transition radiation formation, was studied by Zhelnov, Yakovenko, Amatuni, Sekhposyan, and others [**61**.6, **61**.7, **62**.4, **62**.5, **64**.4, **70**.10, **70**.11, **74**.18]. It was shown that in many cases the effect of spatial dispersion is negligible, but it becomes significant in a relativistic plasma.

The theory of transition radiation in moving media, as well as in optically active (gyrotropic) media, was developed by Barsukov, Bolotovsky, Mergelyan, and others [**60**.20, **63**.4, **64**.5, **64**.6, **65**.7, **68**.5, **68**.6, **69**.7, **73**.15, **81**.4].

Optical transition radiation was first observed experimentally by Goldsmith and Jelley (United Kingdom) [**59**.7] in 1959.[3] Somewhat later, radiation was observed in the United States [**60**.16, **60**.17, **61**.10] which, according to Ferrell's theory [**58**.5], could arise from plasma waves excited by external electrons. It was subsequently established [**61**.11, **62**.6, **63**.5] that Ferrell's theory is a special case of transition radiation theory. Detailed experimental studies of optical transition radiation of non-relativistic electrons were carried out in the Federal Republic of Germany by Bersch, Dobbershein, et al. [**61**.8, **61**.9, **64**.7, **65**.8, **70**.12–**70**.14]; in the USSR by Mikhaliak (Moscow) [**66**.2–**66**.4, **73**.16] and by Arutyunyan et al. (Yerevan) [**66**.5–**66**.8, **72**.16–**72**.18, **73**.17–**73**.19, **74**.19, **79**.11]; and in the USA by Arakawa et al. [**63**.6, **64**.8–**64**.10, **65**.9, **66**.9, **67**.7, **67**.8].

---

[3] Apparently, this radiation was inadvertently observed by Lilienfeld in Germany as early as 1919 (see Refs. [**61**.8, **61**.9, **65**.1]). In addition, it was studied in the USSR in the 1950s by Chudakov, Mikhaliak, et al. (see Ref. [**61**.1]).

In these works, emission in the visible and infrared regions of the spectrum was studied as a function of the electron energy and angle of incidence for a wide range of metals and dielectrics (a review of early experiments is given in Frank's paper [**65**.1]). It turned out that, for the most part, the experimental results were in excellent agreement with the theory of transition radiation. However, at grazing incidence a significant increase—by as much as one to two orders of magnitude—in the radiation intensity was observed. This "anomalous" behavior was studied in detail by a group of physicists in Yerevan. It was established that the "anomalous" component of the radiation was unpolarized and was due to surface *roughness*.

Korkhmazyan and Gevorgyan [**77**.1, p. 434, **81**.9, **81**.10] advanced a plausible interpretation of the aforementioned "anomaly." However, no quantitative theory yet fully accounts for the observed "anomaly" at grazing particle incidence. The theory of transition radiation on rough surfaces was also presented in Refs. [**72**.19–**72**.21, **76**.9, **79**.8, **79**.9].

Transition radiation from non-relativistic protons was experimentally studied by Zrelov and Ruzicka [75.20]. Similar studies for relativistic particles were carried out in the USA by Yuan's group [**67**.9, **68**.7]. In France, Wartski's group [**73**.20, **73**.21, **75**.21, **75**.22] experimentally studied optical transition radiation from relativistic electrons and applied it to the diagnostics of accelerated particle beams.

In addition to the aforementioned paper by Frank [**65**.1], a review of experimental results on optical transition radiation can also be found in the books [**69**.1, **72**.2] and in Yuan's report [**73**.2, p. 334], discussed in the previous section in connection with X-ray transition radiation. This report provides a list of theoretical propositions confirmed experimentally. In particular, this list includes the logarithmic dependence of the total intensity of optical transition radiation on the particle energy.

## E. TRANSITION RADIATION AT MICROWAVE FREQUENCIES

The earliest theoretical studies of transition radiation in a *waveguide* were conducted by Feinberg and Khizhnyak [**57**.1], Barsukov [**59**.5, **60**.18], and Lomize [**60**.19, **61**.12]. The latter also carried out the appropriate experiment and compared its findings with the theory.

The development of electron accelerator technology that produces dense electron bunches has stimulated further theoretical and experimental studies of microwave generation using Cherenkov and transition radiation. Such studies were conducted beginning in 1965 by Barsukov, Laziev, and collaborators [**65**.10, **71**.11, **72**.22–**72**.24, **73**.22, **73**.23, **76**.10–**76**.13]. In particular, experiments performed with a linear accelerator electron beam of about 50 MeV by Laziev, Oksuzyan, and Serov [**72**.23] confirmed that the spectral intensity of microwave transition radiation generated in a stack of plates placed in a waveguide exhibits maxima at certain frequencies (the so-called *parametric Cherenkov radiation* [**57**.1]). It was found experimentally that the spot intensity exceeds that of Cherenkov radiation at the same frequency by about two orders of magnitude. The experimental results were in good agreement with the theory [**57**.1, **65**.10, **71**.11, **72**.22, **76**.10].

In addition, microwave transition radiation provides an effective method for beam diagnostics [**77**.16], for example, in studies of the beam phase structure.

An interesting case is that of a waveguide filled with a moving and non-stationary medium [**76**.11–**76**.13], in which energy transfer from the moving medium to the radiation, Doppler frequency conversion, and related effects may occur.

## F. TRANSITION RADIATION IN ASTROPHYSICS

In 1971–1974, several papers were published [**71**.12, **72**.25, **72**.26, **73**.24–**73**.26, **74**.20, **75**.34] addressing the possibility of X-ray transition radiation occurring under astrophysical conditions. The question concerned the extent to which the isotropic (or diffuse) cosmic X-ray radiation observed in near-Earth space can be explained by XTR produced by cosmic-ray particles interacting with interstellar dust. Later, this issue was clarified, and it was found that the contribution of XTR to diffuse radiation of both galactic and intergalactic origin is negligible [**81**.11]. In addition, the possibility of XTR generation by fast electrons in molecular clouds was discussed [**73**.26, **74**.20]. Using the latest astronomical data obtained with the dedicated Einstein satellite, Aharonyan and Ambartsumyan showed that the observed soft X-ray luminosity of the molecular clouds in the Orion Nebula can be accounted for by XTR [**81**.11].

The possibility of optical transition radiation under astrophysical conditions, in particular in peculiar nebulae, was considered by Gurzadyan [**72**.27, **73**.27, **75**.23].

## G. TRANSITION SCATTERING

If the dielectric permittivity of a medium traversed by a charged particle varies in time (for instance, owing to density variations), the secondary waves emitted by the atoms under the influence of the particle no longer undergo complete cancellation, and radiation is produced. The possibility of radiation generation by a charge was first considered by Ginzburg and Tsytovich [**73**.28, **73**.29]. This process may be interpreted as the scattering of a dielectric-permittivity wave by a charge, accompanied by its partial conversion into electromagnetic waves; accordingly, it was termed *transition scattering*. It is noteworthy that radiation in a non-stationary medium can arise even when the external charge is at rest.

Ginzburg and Tsytovich also considered an analog of transition scattering—the conversion of gravitational waves into electromagnetic waves, occurring in vacuum in the presence of a charge or, more generally, any external electromagnetic field source [**73**.30, **75**.24]. In particular, the transformation coefficient for a charge and for a magnetic moment was calculated. Later, the same authors studied yet another possibility of transition radiation and transition scattering—the conversion of waves of one kind into those of another, caused by the nonlinearity of the vacuum in the presence of strong electromagnetic fields [**78**.7].

A comprehensive exposition of the theory of transition scattering is presented in the review [**78**.1] (see also Ref. [**79**.10]).

# CHAPTER I

## MACROSCOPIC THEORY OF TRANSITION RADIATION

In the present chapter, the basic equations of classical macroscopic electrodynamics are presented, and, by rigorously solving these equations, the formation of transition radiation and Vavilov–Cherenkov radiation is investigated for the perpendicular passage of a single charged particle through a plane interface between two homogeneous media, a plane-parallel plate, and a regular stack of plates. Particular attention is given to the case where the radiation frequency greatly exceeds the characteristic atomic frequencies (so that the dielectric permittivity is close to unity), i.e., to the case of X-ray or even harder radiation.

## § 1 BASIC EQUATIONS. PASSAGE OF A PARTICLE THROUGH AN INTERFACE BETWEEN MEDIA. X-RAY TRANSITION RADIATION (XTR)

To describe the formation of transition radiation by a fast charged particle in amorphous media, we will use classical macroscopic electrodynamics. At wavelengths much larger than atomic dimensions, i.e., in the optical and radio-frequency ranges, the use of classical macroscopic electrodynamics is natural and justified. In the X-ray and higher-frequency ranges, the applicability of macroscopic electrodynamics follows from the fact that, for forward radiation, the formation length (see Sec. 1.5) is macroscopic in scale. Furthermore, so long as one does not consider extremely high frequencies, quantum processes associated with the transformation of radiation into particles are negligible, and the electromagnetic fields may be regarded as classical. In addition, the particle loses only a negligible fraction of its energy to transition radiation; therefore, its motion may also be regarded as classical and prescribed.

## 1.1. Maxwell's Equations

In classical macroscopic electrodynamics, the electromagnetic fields ***E****(r, t)*, ***H****(r, t)*, and the displacement vectors ***D****(r, t)*, ***B****(r, t)* in a medium are governed by Maxwell's equations:

$$\begin{aligned}
\operatorname{rot} \boldsymbol{H}(r,t) &= \frac{1}{c}\,\frac{\partial \boldsymbol{D}(r,t)}{\partial t} + \frac{4\pi}{c}\,\boldsymbol{j}_{\text{ext}}(r,t), \\
\operatorname{rot} \boldsymbol{E}(r,t) &= -\frac{1}{c}\,\frac{\partial \boldsymbol{B}(r,t)}{\partial t}, \\
\operatorname{div} \boldsymbol{B}(r,t) &= 0, \\
\operatorname{div} \boldsymbol{D}(r,t) &= 4\pi\rho_{\text{ext}}(r,t).
\end{aligned} \tag{1.1}$$

where $\rho_{ext}(r,\ t)$ and $j_{ext}(r,\ t)$ are the external charge and current densities.

If the spatial dispersion of the medium is neglected and the medium is assumed to be time-independent, then the quantities ***D*** and ***E***, and ***B*** and ***H*** are related by constitutive relations following from the principle of causality:

$$\begin{aligned}
\boldsymbol{D}(r,t) &= \int_{-\infty}^{t} \hat{\varepsilon}(r,t')\,\boldsymbol{E}(r,t')\,dt', \\
\boldsymbol{B}(r,t) &= \int_{-\infty}^{t} \hat{\mu}(r,t')\,\boldsymbol{H}(r,t')\,dt'.
\end{aligned} \tag{1.2}$$

Performing the Fourier transform of *t* with respect to ***E(r****, t)*,

$$\begin{aligned}
\boldsymbol{E}(r,t) &= \int_{-\infty}^{\infty} \boldsymbol{E}(r,\omega)\,\exp(-i\omega t)\,d\omega, \\
\boldsymbol{E}(r,\omega) &= \frac{1}{2\pi}\int_{-\infty}^{\infty} \boldsymbol{E}(r,t)\,\exp(i\omega t)\,dt.
\end{aligned} \tag{1.3}$$

and, similarly for the other quantities, we obtain from relations (1.2):

$$\begin{aligned}
\boldsymbol{D}(r,\omega) &= \varepsilon(r,\omega)\,\boldsymbol{E}(r,\omega),\\
\boldsymbol{B}(r,\omega) &= \mu(r,\omega)\,\boldsymbol{H}(r,\omega),\\
\varepsilon(r,\omega) &= \int_0^\infty \hat{\varepsilon}(r,t)\,\exp(i\omega t)\,dt,\\
\mu(r,\omega) &= \int_0^\infty \hat{\mu}(r,t)\,\exp(i\omega t)\,dt.
\end{aligned} \tag{1.4}$$

The dielectric permittivity and magnetic permeability ε(***r***, ω) and μ(***r***, ω) are, in general, complex quantities whose imaginary parts are determined by the absorption of the medium.

Equations (1.1) for the Fourier components of the electromagnetic fields and displacement vectors take the following form:

$$\begin{aligned}
\operatorname{rot}\boldsymbol{H}(r,\omega) &= -\frac{i\omega\varepsilon}{c}\,\boldsymbol{E}(r,\omega) + \frac{4\pi}{c}\,\boldsymbol{j}_{\text{ext}}(r,\omega),\\
\operatorname{rot}\boldsymbol{E}(r,\omega) &= \frac{i\omega\mu}{c}\,\boldsymbol{H}(r,\omega),\\
\operatorname{div}\big(\mu\boldsymbol{H}(r,\omega)\big) &= 0,\\
\operatorname{div}\big(\varepsilon\boldsymbol{E}(r,\omega)\big) &= 4\pi\rho_{\text{ext}}(r,\omega).
\end{aligned} \tag{1.5}$$

where ε = ε(***r***, ω), μ = μ(***r***, ω).

Let us introduce the Fourier components of the scalar potential φ(**r**, ω) and the vector potential ***A***(***r***, ω) in the form:

$$\begin{aligned}
\boldsymbol{H}(r,\omega) &= \frac{1}{\mu}\operatorname{rot}\boldsymbol{A}(r,\omega),\\
\boldsymbol{E}(r,\omega) &= \frac{i\omega}{c}\,\boldsymbol{A}(r,\omega) - \operatorname{grad}\varphi(r,\omega).
\end{aligned} \tag{1.6}$$

Upon imposing the following additional condition on ***A***(***r***, ω) and φ(***r***, ω),

$$\operatorname{div}\boldsymbol{A}(r,\omega) - \frac{i\varepsilon\mu\omega}{c}\,\varphi(r,\omega) = 0. \tag{1.7}$$

then Eq. (1.5) leads to the equations for $\boldsymbol{A}(\boldsymbol{r}, \omega)$ and $\varphi(\boldsymbol{r}, \omega)$:

$$\begin{aligned} &\Delta \boldsymbol{A}(r,\omega) + \frac{\varepsilon\mu\omega^2}{c^2}\,\boldsymbol{A}(r,\omega) + \varepsilon\mu\ \mathrm{grad}\left(\frac{1}{\varepsilon\mu}\,\mathrm{div}\,\boldsymbol{A}(r,\omega)\right) \\ &\qquad -\mu\ \mathrm{grad}\left(\frac{1}{\mu}\right)\times \mathrm{rot}\,\boldsymbol{A}(r,\omega) = -\frac{4\pi\mu}{c}\,\boldsymbol{j}_{\text{вн}}(r,\omega), \\ &\Delta\varphi(r,\omega) + \frac{\varepsilon\mu\omega^2}{c^2}\,\varphi(r,\omega) \\ &\qquad -\frac{1}{\varepsilon}\,\mathrm{grad}\,\varepsilon\cdot\left(\frac{i\omega}{c}\,\boldsymbol{A}(r,\omega) - \mathrm{grad}\,\varphi(r,\omega)\right) = -\frac{4\pi}{\varepsilon}\,\rho_{\text{ext}}(r,\omega). \end{aligned} \tag{1.8}$$

If the medium consists of several homogeneous regions in which $\varepsilon(\boldsymbol{r}, \omega)$ and $\mu(\boldsymbol{r}, \omega)$ are independent of $\boldsymbol{r}$, Eq. (1.5) (or (1.8)) may be solved separately in each region, subject to the boundary conditions requiring continuity of tangential $\boldsymbol{E}_t$ and $\boldsymbol{H}_t$ and normal $D_n$ and $B_n$ across the interfaces.

## 1.2. Passage of a Charged Particle Through a Plane Interface Between Two Homogeneous Media

Let a charged particle move with constant velocity $\boldsymbol{v}$ along the z-axis and cross the plane interface $z$=0 perpendicularly between two homogeneous isotropic media. The quantities ε and μ take the values $\varepsilon_1$, $\mu_1$ and $\varepsilon_2$, $\mu_2$ in the first ($z$<0) and in the second ($z$>0) media, respectively.

For each of these homogeneous media it is convenient to perform a Fourier transform also with respect to $\boldsymbol{r}$:

$$\boldsymbol{E}(\boldsymbol{r},\omega) = \int_{-\infty}^{\infty} \boldsymbol{E}(\chi, z, \omega)\ \exp(i\boldsymbol{\kappa}\cdot\boldsymbol{\rho})\,d^2\chi = \int_{-\infty}^{\infty} \boldsymbol{E}(\boldsymbol{k},\omega)\ \exp(i\boldsymbol{k}\cdot\boldsymbol{r})\,d^3k. \tag{1.9}$$

$$\boldsymbol{E}(\boldsymbol{k},\omega) = \frac{1}{2\pi}\int_{-\infty}^{\infty} \boldsymbol{E}(\chi, z, \omega)\, \exp(-ik_z z)\, dz$$
$$= \frac{1}{(2\pi)^3}\int \boldsymbol{E}(\boldsymbol{r},\omega)\, \exp(-i\boldsymbol{k}\cdot\boldsymbol{r})\, d^3\boldsymbol{r}.$$

and similarly for the remaining quantities. Here □={$k_x$, $k_y$, 0}, **ρ**={$x$, $y$, 0}.

Then, for example, Eqs. (1.8) take the following form:

$$\begin{aligned} \left(-k^2 + \frac{\varepsilon\mu\omega^2}{c^2}\right)\boldsymbol{A}(\boldsymbol{k},\omega) &= -\frac{4\pi\mu}{c}\, \boldsymbol{j}_{\rm ext}(\boldsymbol{k},\omega), \\ \left(-k^2 + \frac{\varepsilon\mu\omega^2}{c^2}\right)\varphi(\boldsymbol{k},\omega) &= -\frac{4\pi}{\varepsilon}\, \rho_{\rm ext}(\boldsymbol{k},\omega). \end{aligned} \tag{1.10}$$

It is straightforward to write down analogous equations corresponding to Eq. (1.5):

$$\begin{aligned} i\boldsymbol{k}\times\boldsymbol{H}(\boldsymbol{k},\omega) &= -\frac{i\omega\varepsilon}{c}\,\boldsymbol{E}(\boldsymbol{k},\omega) + \frac{4\pi}{c}\,\boldsymbol{j}_{\rm ext}(\boldsymbol{k},\omega), \\ i\boldsymbol{k}\times\boldsymbol{E}(\boldsymbol{k},\omega) &= \frac{i\omega\mu}{c}\,\boldsymbol{H}(\boldsymbol{k},\omega), \\ \mu\,\boldsymbol{k}\cdot\boldsymbol{H}(\boldsymbol{k},\omega) &= 0, \\ i\varepsilon\,\boldsymbol{k}\cdot\boldsymbol{E}(\boldsymbol{k},\omega) &= 4\pi\rho_{\rm ext}(\boldsymbol{k},\omega). \end{aligned} \tag{1.11}$$

The external charge and current densities $\rho_{ext}(r,\ t)$ and $j_{ext}(r,\ t)$, and their Fourier components, are given by:

$$\begin{aligned} \rho_{\rm ext}(\boldsymbol{r},t) &= e\,\delta(\boldsymbol{r}-\boldsymbol{v}t), \\ \boldsymbol{j}_{\rm ext}(\boldsymbol{r},t) &= \boldsymbol{v}\,\rho_{\rm ext}(\boldsymbol{r},t), \\ \rho_{\rm ext}(\boldsymbol{k},\omega) &= \frac{e}{(2\pi)^3 v}\,\delta\!\left(k_z - \frac{\omega}{v}\right). \end{aligned} \tag{1.12}$$

$$\boldsymbol{j}_{\mathrm{ext}}(\boldsymbol{k},\omega) = \boldsymbol{v}\,\rho_{\mathrm{ext}}(\boldsymbol{k},\omega).$$

where *e* is the charge of the particle, *δ(x)* is the Dirac delta function.

Thus, the task is to find a solution to Eqs. (1.11) (or (1.10)) taking into account the right hand part of Eq. (1.12) that satisfies the border conditions.

### 1.3. Fields of a Charged Particle

Let us first find the fields of a charged particle moving with constant velocity in a homogeneous isotropic medium. From Eq. (1.10) for ***A*(*k*,** ω) and using Eq. (1.12), we obtain the Fourier component of the vector potential of the field of the charge in the medium:

$$\boldsymbol{A}_{\mathrm{ch}}(\boldsymbol{k},\omega) = \frac{e\mu \boldsymbol{n}_z\,\delta(k_z - \omega/v)}{2\pi^2 c\,(k^2 - \varepsilon\mu\omega^2/c^2)}. \qquad (1.13)$$

where $\boldsymbol{n}_2 = \boldsymbol{v}/v$ is the unit vector directed along the -axis.

According to Eqs. (1.6) and (1.7), the Fourier components of the field of the charge are

$$\begin{aligned} \boldsymbol{E}_{\mathrm{ch}}(\boldsymbol{k},\omega) &= -\frac{eiv}{2\pi^2\varepsilon\omega^2}\,\frac{\boldsymbol{\chi} + \boldsymbol{n}_z(\omega/v)\left(1-\varepsilon\mu\beta^2\right)}{k^2v^2/\omega^2 - \varepsilon\mu\beta^2}\,\delta(k_z - \omega/v), \\ \boldsymbol{H}_{\mathrm{ch}}(\boldsymbol{k},\omega) &= \frac{c}{\omega\mu}\,\boldsymbol{k}\times\boldsymbol{E}_{\mathrm{ch}}(\boldsymbol{k},\omega). \end{aligned} \qquad (1.14)$$

where $\beta = v/c$. Recall that the quantities ε and μ are also functions of the frequency ω.

In the two-medium case under consideration, the potential and the fields of the charge in each medium are given by Eqs. (1.13) and (1.14) with $\varepsilon = \varepsilon_S$ and $\mu = \mu_S$ (*s*=1, 2).

Substituting Eqs. (1.13) and (1.14) into Eqs. (1.9) and (1.3), and then integrating over $k_z$, we obtain triple Fourier integrals over **χ** and ω for the potential and the fields of the charge, in which the Fourier components are products of the expressions multiplying $\delta(k_z - \omega/\upsilon)$ in Eqs. (1.13) or (1.14) and $exp(i\omega z/\upsilon)$. The conditions at the interface *z*=const require that the tangential components of the fields and the normal components of the displacement vectors be continuous at $z$ and for arbitrary $\boldsymbol{\rho}$ and $t$. This means that the corresponding Fourier components in $\boldsymbol{\chi}$ and $\omega$ for both media must coincide at the specified value of $z$. Obviously, this is not the case if one relies only on Eqs. (1.13) and (1.14). In other words, the field of the charge *does not satisfy* the required boundary conditions. To satisfy these requirements, it is necessary to add to the particular solution of the inhomogeneous Eq. (1.10) for $\boldsymbol{A}(\boldsymbol{r}, \omega)$ a corresponding solution of the homogeneous equation, which, as will be shown below, represents transition radiation and Vavilov–Cherenkov radiation arising in the present problem.

### 1.4. Radiation Fields

It is convenient to write the required solution of the homogeneous equation obtained from Eq. (1.10) in such a form that, upon substitution into Eqs. (1.9) and (1.3), we can readily perform the integration over $k_z$ and obtain the corresponding triple Fourier integral (see Sec. 1.3). To this end, let us write the above homogeneous equation in the following form [**57**.3]:

$$(-k_z^2+\lambda_s^2)\boldsymbol{A}(\boldsymbol{k}, \omega) = 0, \tag{1.15}$$

where

$$\lambda_s^2 = \frac{\omega^2}{c^2}\varepsilon_s\mu_s - \chi^2 \quad (s=1,\ 2). \tag{1.16}$$

A solution of Eq. (1.15) is a function $\boldsymbol{A}(\boldsymbol{k}, \omega)$ such that it vanishes for $k_z^2 \neq \lambda_s^2$, but upon integration over $k_z$ yields a finite and, in general, nonzero result. Since the particular solution of the inhomogeneous equation (1.13) has only a $z$-component, the required solution of the homogeneous equation (1.15) must also have only a $z$-component. Therefore, the general solution of the aforesaid equation can be written in the following form [**59**.9]:

$$\boldsymbol{A}_{\rm rad}(\boldsymbol{k},\omega) = \left[\boldsymbol{A}^{(+)}(\chi,\omega)\,\delta(k_z-\lambda_s) + \boldsymbol{A}^{(-)}(\chi,\omega)\,\delta(k_z+\lambda_s)\right]n_z. \tag{1.17}$$

where $A^{(+)}(\chi, \omega)$ and $A^{(--)}(\chi, \omega)$ are at present arbitrary functions. These will be determined from the requirement that the sum of the solutions of Eqs. (1.13) and (1.17) satisfies the boundary conditions.

The two terms in Eq. (1.17) correspond to free waves propagating along the positive ($A^{(+)}$) and negative ($A^{(--)}$) $z$-directions [**57**.3]. By $\lambda_S$ in Eq. (1.17) we mean the root in Eq. (1.16) such that $Re\lambda_S \equiv \lambda'_S > 0\ (< 0)$ for $\omega > 0\ (< 0)$. Since we consider only equilibrium media, $Im\lambda_S \equiv \lambda''_S > 0$. In this case, the requirement that the fields be real implies the following relation:

$$\lambda_s(-\omega) = -\lambda_s^*(\omega). \qquad (1.18)$$

According to the Sommerfeld radiation condition, i.e., the requirement that the radiation fields be outgoing at infinity, in the first medium ($z<0$, $s=1$) we retain the term proportional to $A^{(--)}$, and in the second medium ($z>0$, $s=2$) the term proportional to $A^{(+)}$.

The total vector potential $\boldsymbol{A}$ is the sum of the vector potential of the charge field (Eq. (1.13)) and that of the radiation field (Eq. (1.17)):

$$\boldsymbol{A}=\boldsymbol{A}_{\mathbf{ch}} + \boldsymbol{A}_{\mathbf{rad}}.$$

It is readily verified that, owing to the axial symmetry of the problem, the continuity conditions for the total fields imply, in accordance with Eqs. (1.6) and (1.17), the continuity of the quantities $\mu^{-1}A_z(\chi, z, \omega)$ and $(\mu\varepsilon)^{-1}\delta A_z(\chi, z, \omega)/\delta z$ at the interface $z=0$. This leads to the following two equations:

$$\frac{1}{2} e\beta v\pi^{-2}\omega^{-2}(1-u_1^2)^{-1}+\mu_1^{-1}A^{(-)}(\varkappa, \omega) =$$

$$= \frac{1}{2} e\beta v\pi^{-2}\omega^{-2}(1-u_2^2)^{-1}+\mu_2^{-1}A^{(+)}(\varkappa, \omega), \qquad (1.19)$$

$$\frac{1}{2} e\beta v\pi^{-2}\omega^{-2}\varepsilon_1^{-1}(1-u_1^2)^{-1}-u_1\varepsilon_1^{-1}\mu_1^{-1}A^{(-)}(\varkappa, \omega)=$$

$$=\frac{1}{2} e\beta v\pi^{-2}\omega^{-2}\varepsilon_2^{-1}(1-u_2^2)^{-1}+u_2\varepsilon_2^{-1}\mu_2^{-1}A^{(+)}(\varkappa, \omega),$$

where

$$u_s = \lambda_s v/\omega = \beta\sqrt{\varepsilon_s\mu_s - (\varkappa c/\omega)^2}. \tag{1.20}$$

Solving Eqs. (1.19), we obtain the Fourier components of the vector potential, with the help of which it is easy to find the Fourier components of the radiation fields. For example, the tangential component of the radiation field propagating forward (i.e., in the positive **z**-direction) in the second medium is given by

$$\boldsymbol{E}_t^{(+)}(\boldsymbol{\varkappa}, \omega) = \frac{ei v \boldsymbol{\varkappa}}{4\pi^2\omega^2}\left(\frac{1}{\varepsilon_2(1-u_2)} - \frac{R_{21}}{\varepsilon_2(1+u_2)} - \frac{P_{12}}{\varepsilon_1(1+u_1)}\right), \tag{1.21}$$

where

$$R_{21} = \frac{\varepsilon_2 u_1 - \varepsilon_1 u_2}{\varepsilon_1 u_2 + \varepsilon_2 u_1}, \qquad P_{12} = \frac{2\varepsilon_1 u_2}{\varepsilon_1 u_2 + \varepsilon_2 u_1}. \tag{1.22}$$

The quantities $R_{21}$ and $P_{12}$ are the Fresnel coefficients. The coefficient $R_{21}$ represents the reflection coefficient for a wave incident from the second medium onto the interface with the first, whereas $P_{12}$ is the transmission coefficient for a wave passing from the first medium into the second (for polarization in the plane of incidence). Thus, the three terms in Eq. (1.21) correspond to three types of waves (Fig. 1.1): (1) a wave emitted forward in the second medium; (2) a wave emitted initially backward in the second medium and subsequently reflected from the interface; (3) a wave emitted forward in the first medium and transmitted through the interface.

The Fourier component $E_t^{(-)}(\chi, \omega)$ of the backward-propagating radiation field can be obtained from Eq. (1.21) by interchanging the indices 1 and 2 and reversing the sign of $u_S$. In this case, three waves are also obtained, which are shown in Fig. 1.2.

From Eq. (1.21) it follows that the tangential component of the electric field of the radiation is proportional to $\chi$. This means that the radiation (both forward and backward) is linearly polarized in the plane of observation, i.e., in the plane defined by the direction of propagation and the direction of motion of the charged particle.

The normal component of the radiation electric field can be obtained from the transversality condition:

$$\boldsymbol{\varkappa} \cdot \boldsymbol{E}_t^{(\pm)}(\boldsymbol{\varkappa}, \omega) \mp \lambda_s E_z^{(\pm)}(\boldsymbol{\varkappa}, \omega) = 0. \tag{1.23}$$

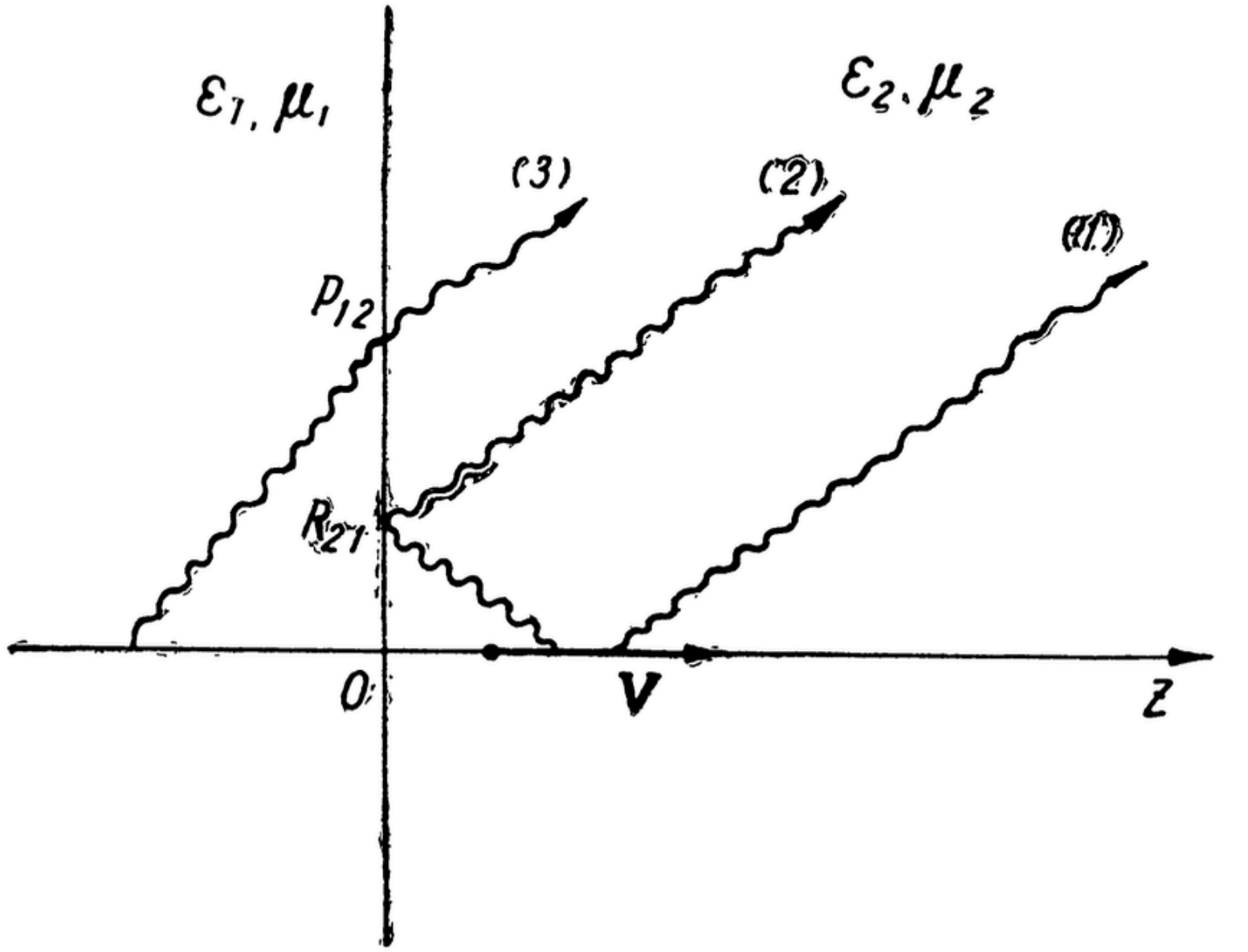

Fig. 1.1 Schematic representation of the three waves constituting the forward transition radiation emitted by a particle crossing the interface between two media. The three waves correspond to the three terms in Eq. (1.21)

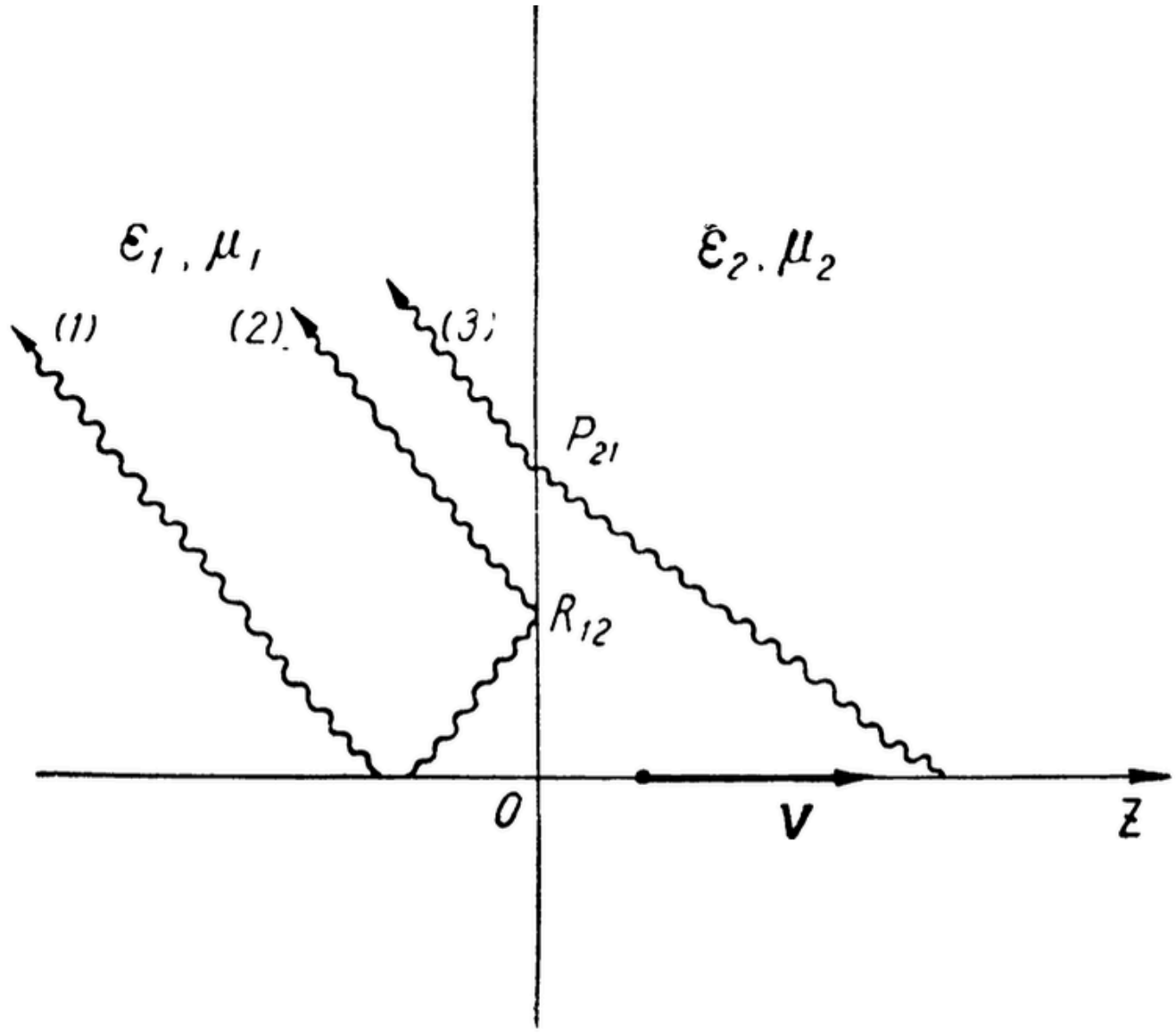

Fig. 1.2. Same as Fig. 1.1, but for backward radiation

The Fourier components of the magnetic field can be obtained from those of the electric field using the following equation:

$$\boldsymbol{H}^{(\mp)}(\boldsymbol{\varkappa},\ \omega)=c(\omega\mu_s)^{-1}[(\boldsymbol{\varkappa}\mp\boldsymbol{n}_z\lambda_s)\times\boldsymbol{E}^{(\mp)}(\boldsymbol{\varkappa},\ \omega)]. \qquad (1.24)$$

The fields in the coordinate representation are calculated according to Eqs. (1.9) and (1.3). The asymptotic expressions for these fields far from the interface are conveniently evaluated using the method of steepest descent (see [**57**.3]).

## 1.5. Electromagnetic Energy Flux. Radiation Formation Length

Let us calculate the flux of the Poynting vector through an arbitrary transverse (i.e., perpendicular to the *z*-axis) plane over the entire duration of the particle's passage:

$$S=c(4\pi)^{-1}\int\limits_{-\infty}^{\infty}dt\int\limits_{-\infty}^{\infty}[\boldsymbol{E}(\boldsymbol{r},\ t)\times\boldsymbol{H}(\boldsymbol{r},t)]_z\,dx\,dy. \qquad (1.25)$$

Here $\boldsymbol{E}(\boldsymbol{r}, t)$ and $\boldsymbol{H}(\boldsymbol{r}, t)$ are the total fields taken at the transverse plane $z=z_{obs}$.

Let us express $\boldsymbol{E}(\boldsymbol{r}, t)$ and $\boldsymbol{H}(\boldsymbol{r}, t)$ in terms of their Fourier components and integrate over $x$, $y$, and $t$ from $-\infty$ to $\infty$. As a result, we obtain a product of three delta functions: $(2\pi)^3\delta(k_x + k'_x)$. After straightforward integration over $k'_x$, $k'_y$, and $\omega'$ we obtain

$$\begin{aligned} S^{(\mp)} &= 2\pi^2 c\int\Big[E_{зap}(\chi,\omega)\exp(i\omega z_{obs}/v)+E^{(\mp)}(\chi,\omega)\exp(\mp i\lambda_s(\omega)z_{obs})\Big] \\ &\quad\times\Big[H_{ch}(-\chi,-\omega)\exp(-i\omega z_{obs}/v)+H^{(\mp)}(-\chi,-\omega)\exp(\pm i\lambda_s(-\omega)z_{obs})\Big]_z d^2\chi d\omega \\ &= S_{ch}+S_{rad}^{\mp}+S_{int}^{(\mp)}. \end{aligned} \qquad (1.26)$$

where

$$S_{ch}=2\pi^2 c\int\Big[E_{ch}(\chi,\omega)\times H_{зap}(-\chi,-\omega)\Big]_z d^2\chi d\omega. \qquad (1.27)$$

$$S_{\text{ch}}^{(\mp)} = 2\pi^2 c \int \left[ E^{(\mp)}(\chi, \omega) \times H^{(\mp)}(-\chi, -\omega) \right]_z \times$$

$$\times \exp[\pm i\,(\lambda_s(\omega)+\lambda_s(-\omega))z_{\rm obs}]d^2\chi\, d\omega, \qquad (1.28)$$

$$S_{\rm int}^{(\mp)} = 2\pi^2 c \int \Big\{ \Big[E_{\rm ch}(\chi,\omega)\times H^{(\mp)}(-\chi,-\omega)\Big]_z \exp[i(\omega/v \mp \lambda_s(-\omega))z_{\rm obs}] + \Big[E^{(\mp)}(\chi,\omega)\times H_{\rm ch}(-\chi,-\omega)\Big]_z \exp[i(-\omega/v \pm \lambda_s(\omega))z_{\rm obs}]\Big\} d^2\chi d\omega. \qquad (1.29)$$

In the equations above, the "−" or "+" sign attached to the fields $\boldsymbol{E}^{(\mp)}$ and $\boldsymbol{H}^{(\pm)}$, as noted earlier, corresponds to the position of the plane $z=z_{\rm obs}$ in either the first ($z_{\rm obs}<0$) or the second ($z_{\rm obs}>0$) medium. Accordingly, the index $s$ in $\lambda_S$ takes the value $s=1$ or $s=2$. The three terms in Eq. (1.26) have a clear physical interpretation: the quantity $S_{\rm ch}$ represents the energy flux associated with the field of the charged particle; the quantity $S_{rad}^{(\pm)}$ represents the energy flux of the emitted radiation; and the quantity $S_{int}^{(\pm)}$ arises from the interference between the fields of the charge and the radiation.

The three terms $S_{\rm ch}$, $S_{rad}^{(\pm)}$, and $S_{int}^{(\pm)}$ depend differently on $z_{\rm obs}$. If $S_{\rm ch}$ depends only on whether $z_{\rm obs}$ is positive or negative, then $S_{rad}^{(\pm)}$ and $S_{int}^{(\pm)}$ depend explicitly on $z_{\rm obs}$. At the same time, owing to the presence of an exponential factor with a purely real argument in $S_{rad}^{(\pm)}$ (in view of Eq. (1.18))

$$exp[\mp\, i(\lambda_S(\omega)\, +\, \lambda_S(-\,\omega))z_{obs}] \,=\, exp(\pm\, 2\lambda''_S z_{obs}),$$

the magnitude $|S_{rad}^{(\pm)}|$ decreases monotonically as the observation plane moves away in either direction from the interface between the media (i.e., with increasing $|z_{\rm obs}|$) for equilibrium absorbing ($lm\varepsilon_s \equiv \varepsilon_s > 0$) media. This is entirely natural and is associated with the absorption of the generated radiation in the media.

The quantity $S_{int}^{(\pm)}$ contains complex-conjugate exponential factors of the following form:

$$exp(-\, i\pi z_{obs}/z_S^{(\mp)}),$$

where

$$z_s^{(\mp)} = \frac{\pi}{\omega/v \pm \lambda_s} = \frac{\pi v}{\omega(1 \pm u_s)}. \qquad (1.30)$$

When

$$z_{\rm obs} \gg \left|\lambda_s^{(\mp)}\right|. \qquad (1.31)$$

the above exponential factors in $S_{int}^{(\pm)}$ oscillate rapidly and vanish upon averaging over a small frequency range, or decay due to the imaginary part $\lambda''_S$. Physically, this means that far from the interface the field of the generated radiation separates from the field of the charged particle and becomes a free field (in a sufficiently transparent medium).

The quantities $z_S^{(\pm)}$, defined by Eq. (1.30), play an important role in the formation of radiation by a charged particle (we shall return to this issue repeatedly). They are commonly referred to as the *coherence lengths* or *formation lengths* for backward radiation (upper sign) and forward radiation (lower sign). In the general case of complex dielectric permittivity and magnetic permeability ($\varepsilon(\omega) = \varepsilon' + i\varepsilon''$, $\mu(\omega) = \mu' + i\mu''$), the coherence lengths $z_S^{(\pm)}$ are likewise complex. Their imaginary part is due to absorption, whereas the real part (if absorption is weak), as follows from Eq. (1.30), represents the path length over which the phase difference between waves emitted by the charge at different points along this path does not exceed π (see [**42**.1]; see also [**82**.5]).

Let us calculate the intensity of the radiation propagating forward relative to the direction of motion of the charge. To this end, let us substitute expression (1.24) into Eq. (1.28) and take into account Eqs. (1.23) and (1.18):

$$S_{\rm rad}^{(+)} = 4\pi^2 \int_0^\infty \omega d\omega \int \left|E_t^{(+)}(\chi,\omega)\right|^2 {\rm Re}\left(\frac{\varepsilon_2}{\lambda_2}\right) \exp\left(-2\lambda_2'' z_{\rm obs}\right) d^2\chi. \qquad (1.32)$$

Let us note that at large distances the quantities $\lambda_2$ and $\boldsymbol{\chi}$ in Eq. (1.32) may be regarded as the longitudinal and transverse components of the wave vector, related by Eq. (1.16). If we introduce the polar and azimuthal angles $\vartheta$ and $\phi$ of the wave vector with respect to the $z$-axis, then

$$\varkappa = (\omega/c)(\varepsilon_2\mu_2)^{1/2}\sin\vartheta,\ \lambda_2 = (\omega/c)(\varepsilon_2\mu_2)^{1/2}\cos\vartheta,$$
$$d^2\varkappa = (\omega/c)^2|\varepsilon_2\mu_2|\cos\vartheta d^2\Omega, \tag{1.33}$$

where $d^2\Omega = sin\vartheta d\vartheta d\phi$ is the element of solid angle. Substituting Eqs. (1.21) and (1.33) into Eq. (1.32), we obtain

$$W^{(+)} = S_{\rm rad}^{(+)} = \int_0^{\infty} d\omega \int_0^{\pi/2} d\vartheta\, W^{(+)}(\omega, \vartheta). \tag{1.34}$$

where the spectral–angular distribution of the radiation intensity $W^{(+)}(\omega, \vartheta)$ is given by the following equation:

$$W^{(+)}(\omega,\vartheta) = \frac{2e^2\beta^3}{\pi c}\left|\frac{\left[(\varepsilon_1/\varepsilon_2 - 1)(1-u_1) - \beta^2(\varepsilon_1\mu_1 - \varepsilon_2\mu_2)\right]u_2}{(1-u_2^2)(1-u_1)(\varepsilon_1 u_2 + \varepsilon_2 u_1)}\right|^2$$
$$\times|\varepsilon_2\mu_2|^2\,\mathrm{Re}\left(\frac{\varepsilon_2}{u_2}\right)\exp\left(-2u_2'' z_{\rm obs}\,\omega/v\right)\sin^3\vartheta\cos\vartheta, \tag{1.35}$$
$$u_s = v\lambda_s/\omega = \beta\left(\varepsilon_s\mu_s - \varepsilon_2\mu_2\sin^2\vartheta\right)^{1/2} = u_s' + iu_s''.$$

The intensity of the backward radiation $W^{(-)}$ is given by the following equation:

$$W^{(-)} = -S_{\rm rad}^{(-)} = \int_0^{\infty} d\omega \int_0^{\pi/2} d\vartheta\, W^{(-)}(\omega, \vartheta). \tag{1.36}$$

where $S_{rad}^{(-)}$ is obtained from $S_{rad}^{(+)}$ by interchanging the indices 1 and 2 and changing the signs of y $u_s$ and β. In this case, the polar angle $\vartheta$ is measured from the negative $z$-direction.

## 1.6. Forward Radiation

Let us analyze Eq. (1.35) for the intensity of radiation emitted in the forward direction. For definiteness, let the particle enter from a medium ($\varepsilon_1 = \varepsilon = \varepsilon' + i\varepsilon''$, $\mu_1 = 1$) into vacuum ($\varepsilon_2 = \mu_2 = 1$). In this case, we obtain [57.3]

$$W^{(+)}(\omega, \vartheta) = \frac{2e^2\beta^2|\varepsilon-1|^2\,|1-\beta(\varepsilon-\sin^2\vartheta)^{1/2}-\beta^2|^2}{\pi c(1-\beta^2\cos^2\vartheta)^2|1-\beta(\varepsilon-\sin^2\vartheta)^{1/2}|^2}\times$$

$$\times\frac{\sin^3\vartheta\cos^2\vartheta}{|\varepsilon\cos\vartheta+(\varepsilon-\sin^2\vartheta)^{1/2}|^2}. \qquad (1.37)$$

## *A. Vavilov–Cherenkov Radiation*

First of all, let us note that if the medium is sufficiently transparent ($\varepsilon''\ll 1$), then under certain conditions (see below) Eq. (1.37), considered as a function of the angle $\vartheta$ (at fixed frequency $\omega$), contains a narrow and pronounced maximum, which may be interpreted as the well-known *Vavilov–Cherenkov radiation* [**37**.1] (see also Refs. [**60**.1, **68**.1, **72**.1]). Indeed, in Eq. (1.37) the denominator contains the factor $|D|^2$, where

$$D = 1-\beta(\varepsilon-\sin^2\vartheta)^{1/2} = D' + iD'', \qquad (1.38)$$

$$D' = 1-\frac{\beta}{\sqrt{2}}\{\varepsilon'-\sin^2\vartheta+[(\varepsilon'-\sin^2\vartheta)^2+\varepsilon''^2]^{1/2}\}^{1/2},$$

$$D'' = -\frac{\beta\varepsilon''}{\sqrt{2}}\{\varepsilon'-\sin^2\vartheta+[(\varepsilon'-\sin^2\vartheta)^2+\varepsilon''^2]^{1/2}\}^{-1/2}. \qquad (1.39)$$

Let us determine the value of $\vartheta_c$ for which the quantity $|D|^2$ attains its minimum. For sufficiently small values of $\varepsilon''$, we obtain for $\vartheta_c$ the following:

$$\sin^2\vartheta_C = \varepsilon' - \beta^{-2} - \frac{\beta^2\varepsilon''}{4}. \qquad (1.40)$$

This equation has a solution for real values of $\vartheta_c$ if

$$0 \leqslant \varepsilon' - \beta^{-2} - \beta^2\varepsilon''^2/4 \leqslant 1. \qquad (1.41)$$

The conditions in Eq. (1.41) have a clear physical meaning. They simply mean that Vavilov–Cherenkov radiation is generated in the first medium (the right inequality) and that this radiation can leave the medium and enter the vacuum without undergoing total internal reflection at the interface (the left inequality). The angle $\vartheta_c$, defined by Eq. (1.40), is nothing other than the well-known Cherenkov angle arc $\cos(\beta\varepsilon'^{1/2})^{-1}$, taking into account refraction at the medium–vacuum interface.

In the vicinity of $\vartheta_c$, we have the following:

$$|D|^2 = \beta^4 \left[ (\vartheta - \vartheta_C)^2 \sin^2 \vartheta_C \cos^2 \vartheta_C + \frac{\varepsilon''}{4} \right]. \qquad (1.42)$$

The quantity in Eq. (1.37) then takes the following form:

$$W^{(+)}(\omega, \vartheta) = \frac{2e^2}{\pi c} \frac{\sin^3 \vartheta_C \cos^2 \vartheta_C}{\left[(\vartheta - \vartheta_ч)^2 \sin^2 \vartheta_C \cos^2 \vartheta_C + \varepsilon''^2/4\right] \left|\beta\varepsilon \cos \vartheta_C + 1\right|^2}. \qquad (1.43)$$

It follows immediately that the radiation intensity exhibits a maximum at $\vartheta = \vartheta_C$. The height of this maximum is proportional to $(\varepsilon'')^{-2}$, whereas its width is proportional to ε". After integrating Eq. (1.43) over the angle ϑ, we obtain the following:

$$W_{\mathrm{VC}}^{(+)}(\omega) = \frac{e^2\omega}{c^2} \left(1 - \frac{1}{\beta^2 \varepsilon'}\right) \left\{ \frac{4\varepsilon' \beta \cos \vartheta_C}{(\varepsilon' \beta \cos \vartheta_C + 1)^2} \right\} l_{\mathrm{abs}}, \qquad (1.44)$$

where $l_{abs}=(2\lambda''_1)^{-1}$ is the absorption length, which at the angle $\vartheta_C$ is expressed in terms of ε" by the following equation:

$$l_{\mathrm{abs}} = \frac{c}{\beta\omega\varepsilon''}. \qquad (1.45)$$

The quantity in the curly braces in Eq. (1.44) represents the transmission coefficient (in intensity) for radiation passing from the medium into the vacuum through the medium–vacuum interface. Thus, Eq. (1.44) determines the intensity of Vavilov–Cherenkov radiation at a given ω, emitted from a length equal to $l_{\mathrm{abs}}$ and transmitted into the vacuum.

The quantity $W_{VC}^{(+)}(\omega)$ vanishes at $\vartheta_C = 0$ and $\vartheta_C = \pi/2$. In the first case this is due to the threshold of radiation formation (see Eq. (1.40)), and in the second, to its total internal reflection.

It should be noted that the intensity of Vavilov–Cherenkov radiation given by Eq. (1.44) at high particle energies, when $\beta \to 1$, is practically independent of the particle energy[4].

### *B. Transition Radiation*

The difference between the total radiation given by Eq. (1.37) and the Vavilov–Cherenkov radiation given by Eq. (1.44) (and, in the case where the conditions of Eq. (1.41) are not satisfied, the entire quantity given by Eq. (1.37)) is determined solely by the presence of the interface ("transition"). Therefore, this difference may naturally be called *transition radiation*.

The fact that Vavilov–Cherenkov radiation and transition radiation are described by the same equation (1.37) (or, in the more general case, Eq. (1.35)) is not accidental. The reason is that both types of radiation occur under uniform rectilinear motion of a charged particle (or, in other words, in the limit of infinite rest mass of the particle), and therefore both represent radiation of the medium electrons that acquire recoil momentum when interacting with the passing external charge. In this sense, these two types of radiation cannot be regarded as fundamentally different, and their separation is to some extent conventional.

In addition to the Cherenkov maximum at $\vartheta = \vartheta_C$ (if the conditions in Eq. (1.41) are satisfied), the quantity given by Eq. (1.37), as a function of the angle $\vartheta$, possesses another maximum, which is likewise very sharp at high particle energies. This transition-radiation maximum is associated with the factor $1 - \beta^2 \cos^2\vartheta$ in the denominator, which for small emission angles reduces to $1 - \beta^2 + \vartheta^2$. For ultrarelativistic particles, i.e., when the Lorentz factor

$$\gamma = (1 - \beta^2)^{-1/2} \gg 1, \tag{1.46}$$

---

[4] The above statement applies to the case where the quantity ε' is not close to unity and therefore the Cherenkov angle is not small. When the Cherenkov angle is small and approaches the angle at which the maximum of transition radiation occurs (Sec. 1.6B), the Cherenkov and transition maxima overlap, and the corresponding radiation intensity becomes dependent on the particle energy.

Eq. (1.37) takes the following form at small angles (assuming that the quantity ε is not close to zero):

$$W^{(+)}(\omega, \vartheta) = \frac{2e^2|\varepsilon^{1/2} - 1|^2 \vartheta^3}{\pi c(\gamma^{-2} + \vartheta^2)^2 \left|1 - \beta(\varepsilon - \vartheta^2)^{1/2}\right|^2}. \tag{1.47}$$

It follows that for

$$\vartheta \sim \gamma^{-1} \tag{1.48}$$

the spectral–angular distribution given by Eq. (1.47), as a function of ϑ, indeed exhibits a maximum whose height is proportional to γ. In the next two sections, forward transition radiation will be considered separately for the cases where ε is not close to unity (Sec. 1.7) and where ε is very close to unity (Sec. 1.8).

## 1.7. Optical Transition Radiation

For condensed media in the optical and near-optical frequency ranges, the characteristic values of ε are not close to unity:

$$|\varepsilon - 1| \gtrsim 1. \tag{1.49}$$

The transition radiation emitted in this case will, for brevity, sometimes be referred to as “optical.” The spectral–angular distribution of the intensity for $\gamma \gg 1$ and $\vartheta \ll 0$ (Eq. (1.47)), taking into account Eq. (1.49), takes a simple form:

$$W_{\text{op}}^{(+)}(\omega, \vartheta) \simeq \frac{2e^2}{\pi c} \frac{\vartheta^3}{(\gamma^{-2} + \vartheta^2)^2}. \tag{1.50}$$

It follows from this equation that, in the case under consideration, the spectral–angular distribution of the intensity does not depend on the dielectric permittivity ε(ω) of the medium.

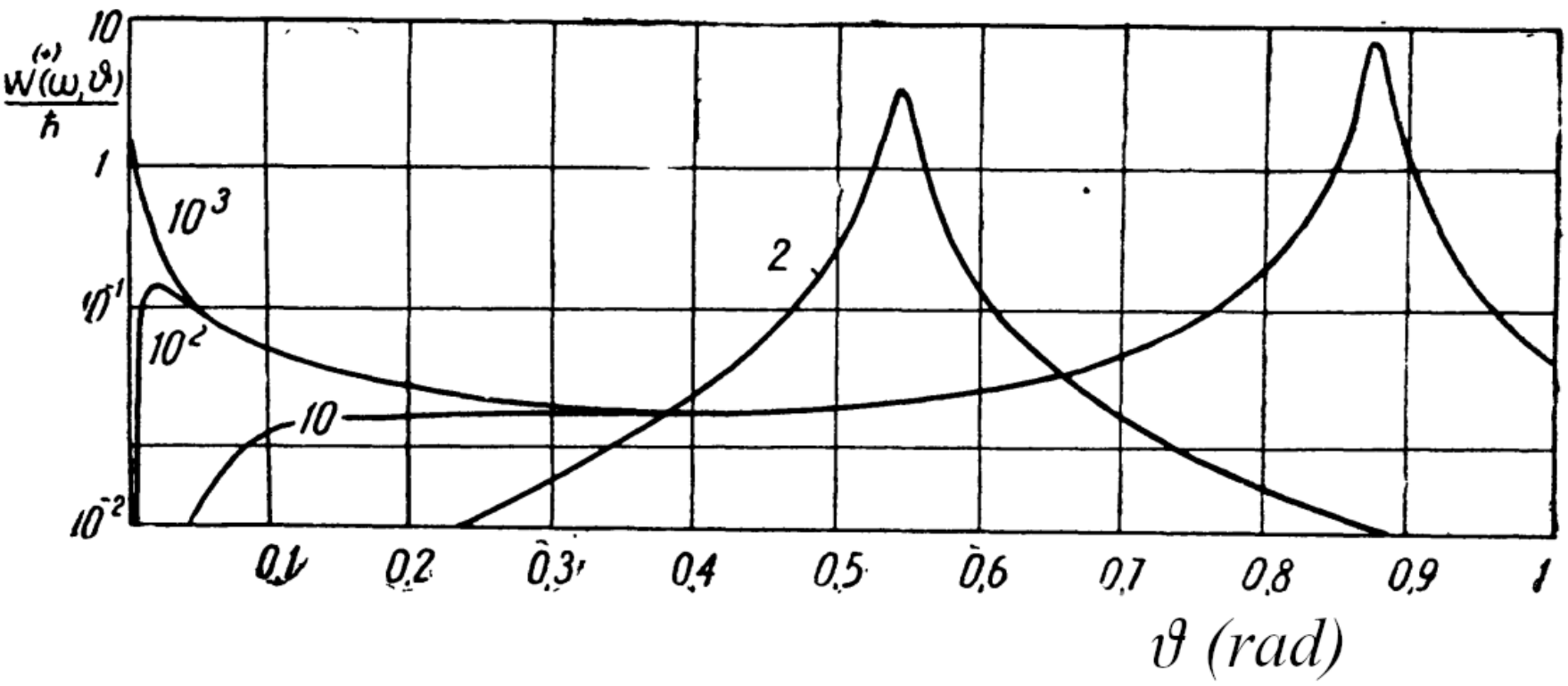

Fig. 1.3. Angular dependence of the intensity of optical transition and Cherenkov radiation (at a fixed frequency) emitted in the forward direction by a particle crossing a medium–vacuum interface ($\varepsilon' = 1.6$; $\varepsilon'' = 0.01$). The maxima on the right (at $\vartheta \approx 0.545$ and 0.875 rad) correspond to Cherenkov radiation (for $\gamma = 2$ and $\gamma > 10$), whereas the maxima at $\vartheta < 0.05$ rad correspond to transition radiation. The numbers indicated next to the curves correspond to the values of the particle Lorentz factor

In Fig. 1.3, the curves of the angular dependence of the spectral–angular distribution of the radiation intensity $W_{op}^{(+)}(\omega, \vartheta)$, calculated from Eq. (1.37), are shown for different values of $\gamma$. The quantity $W_{op}^{(+)}(\omega, \vartheta)$, which has the dimension of action, is expressed in units of the Planck constant ħ.[5] The figure clearly shows that, for sufficiently large values of γ (greater than 100), transition-radiation maxima appear at small angles of order . At angles $\vartheta=\vartheta_C$ (see Eq. (1.40)), Cherenkov maxima arise, which cease to depend on γ for γ > 10.

After integrating Eq. (1.50) with respect to ϑ from zero to $\vartheta_1$, one obtains the contribution of small emission angles to the spectral distribution of transition radiation:
It follows that, for , the contribution of small angles is equal to , i.e., it depends very strongly on the Lorentz factor of the traversing charged particle. However, as $\vartheta_1\gamma$ increases, the dependence of the small-angle contribution on $\vartheta_1\gamma \gg 1$ becomes progressively weaker and, for , becomes logarithmic.
It should be noted that the contribution from "large" emission angles[6] (from $\vartheta_1$ to π/2) may, in general, be comparable to the contribution (1.51) from the maximum of transition radiation. To this end, let us consider the simplest case, for example that of an ideal conductor, in which
Then
It follows that, if , the contribution of "large" angles in Eq. (1.53) is much smaller than that of small angles in Eq. (1.51) only under the rather stringent condition .
In the case of an arbitrary value of ε (see Eq. (1.49)), the contribution from angles between $\vartheta_1$ and π/2 depends on ε, but the estimate (without taking into account the contribution of a possible Cherenkov maximum) remains the same.

[5] To obtain the number of radiation quanta in the frequency range *d*ω, one should multiply *W(*ω, ϑ)/*ħ* by *d*ω/ω.
[6] Here we do not mean the possible contribution (Eq. (1.44)) associated with the Cherenkov maximum, which can always be taken into account separately.

Thus, the spectral intensity of optical transition radiation

$$W_{\rm op}^{(+)}(\omega) = \int_0^{\pi/2} W_{\rm op}^{(+)}(\omega, \vartheta)\, d\vartheta. \qquad (1.54)$$

is of the order of

$$W_{\rm op}^{(+)}(\omega) \sim \frac{e^2}{c} \ln \gamma. \qquad (1.55)$$

although, generally speaking, this intensity is determined by the specific dependence of ε on ω.

The number of quanta in a frequency range Δω~ω is, to order of magnitude, equal to the fine-structure constant [**45**.1]:

$$\frac{W_{\rm op}^{(+)}(\omega)\,\Delta\omega}{\hbar\omega} \sim \frac{e^2}{\hbar c} = \frac{1}{137}. \qquad (1.56)$$

i.e., *comparatively small.*

## 1.8. X-Ray Transition Radiation

Let the quantity ε now be very close to unity:

$$|\varepsilon - 1| \ll 1. \qquad (1.57)$$

Such values of ε are characteristic of the frequency range far exceeding atomic frequencies, i.e., the X-ray region (with photon energies from several keV to several hundred keV) and the harder region (with photon energies of the order of MeV and higher). In this frequency range, ε may be written in the following form (if nuclear dispersion is neglected [**70**.15, **75**.25, **76**.14]):

$$\varepsilon = 1 - \frac{\omega_0^2}{\omega^2} + \frac{i\eta c}{\omega}, \qquad (1.58)$$

where $\omega_0 = (4\pi Ne^2/m)^{1/2}$ is the plasma frequency of the medium, *N* is the number of electrons per unit volume, ***e*** and *m* are the charge and mass of the electron, and η = η(ω) is the linear absorption coefficient of radiation in the medium.

Then, in the small-angle region, from Eq. (1.47) we obtain the spectral–angular distribution of XTR radiation emitted from a single boundary:

$$W_{\rm b}^{(+)}(\omega,\vartheta)=\frac{2e^2}{\pi c}\frac{|1-\varepsilon|^2\vartheta^3}{(\gamma^{-2}+\vartheta^2)^2|\gamma^{-2}+\vartheta^2+1-\varepsilon|^2}. \qquad (1.59)$$

This quantity exhibits different dependences on ϑ, ω, and γ in different ranges of ϑ:

$$W_{\rm b}^{(+)}(\omega,\vartheta)=\begin{cases}\dfrac{2e^2\vartheta^3|1-\varepsilon|^2\gamma^4}{\pi c|\gamma^{-2}+1-\varepsilon|^2} & \text{for } \vartheta\ll\gamma^{-1},\\ \dfrac{2e^2|1-\varepsilon|^2}{\pi c\,\vartheta\,|\gamma^{-2}+1-\varepsilon|^2} & \text{for } \gamma^{-1}\ll\vartheta\ll|\gamma^{-2}+1-\varepsilon|^{1/2}\\ \dfrac{2e^2|1-\varepsilon|^2}{\pi c\,\vartheta^5} & \text{for } |\gamma^{-2}+1-\varepsilon|^{1/2}\ll\vartheta.\end{cases} \qquad (1.60)$$

The maximum of the spectral–angular distribution (see Eq. (1.59)), regarded as a function of ϑ, occurs at $\vartheta\approx(3/5)^{1/2}\gamma^{-2}\gg|1-\varepsilon|$, and at $\vartheta\approx3^{1/2}\gamma^{-1}$ when $\gamma^{-2}\ll|1-\varepsilon|$.

Figure 1.4 shows the angular dependence of $W_b^{(+)}(\omega,\vartheta)$ for X-ray transition radiation at several values of γ. It is clearly seen that, as γ increases, the maximum shifts toward smaller values of ϑ and becomes more pronounced.

It follows from Eq. (1.60) that, as in the case of optical transition radiation (see Sec. 1.7), the contribution of small angles ($\vartheta\ll\gamma^{-1}$) to the XTR intensity exhibits a very strong dependence on γ. This contribution is proportional to $\gamma^4$ when $\gamma^{-2}\ll|1-\varepsilon|$ and to $\gamma^8$ when $\gamma^{-2}\gg|1-\varepsilon|$. With increasing ϑγ, the dependence of $W_b^{(+)}(\omega,\vartheta)$ on γ becomes significantly weaker.

However, small angles $\vartheta\ll\gamma^{-1}$ contribute only very little to the total radiation intensity (integrated over all angles). The effective emission angles of XTR, which contribute more than 90% of the total intensity, are determined by the region (see also Ref. [**74**.22]):

$$\vartheta_{\min}\leqslant\vartheta\leqslant\vartheta_{\max}, \qquad (1.61)$$

where

$$\vartheta_{\min}\sim\gamma^{-1}/4,\quad \vartheta_{\max}\sim4|\gamma^{-2}+1-\varepsilon|^{1/2}. \qquad (1.62)$$

In particular, when ε is given by Eq. (1.58), we obtain

$$\vartheta_{max} \sim 4\left[\left(\gamma^{-2}+\frac{\omega_0^2}{\omega^2}\right)^2+\left(\frac{\eta c}{\omega}\right)^2\right]^{1/4}. \qquad (1.63)$$

Integrating Eq. (1.59) over the angle $\vartheta$, we obtain the spectral distribution of XTR generated at the medium–vacuum interface and propagating forward relative to the direction of motion of the charge:

$$W_b^{(+)}(\omega) = \int_0^{\vartheta_{max}} W_b^{(+)}(\omega,\vartheta)\,d\vartheta = \frac{e^2}{\pi c}\,\mathrm{Re}\left[\frac{g+\gamma^{-2}}{g-\gamma^{-2}}\ln\left(g\gamma^2\right)+\frac{g^*+g}{g^*-g}\ln g-1\right]. \qquad (1.64)$$

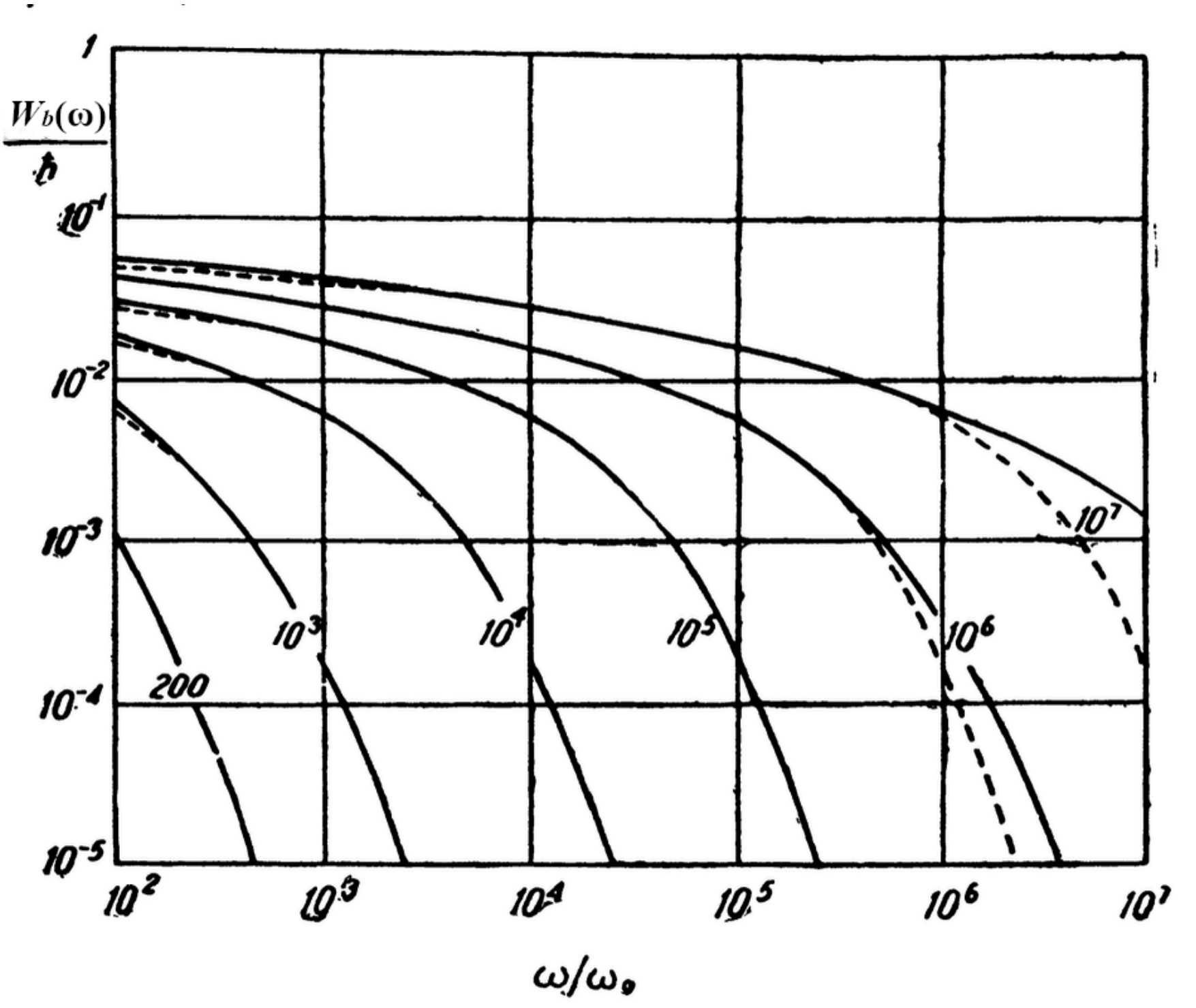

Fig. 1.4. Angular dependence of the intensity of X-ray transition radiation in the forward direction (at a frequency $\omega=10^3 \cdot \omega_0$) generated at the medium–vacuum interface. The numbers next to the curves denote the particle Lorentz factor

where

$$g = \gamma^{-2} + 1 - \varepsilon = g' + ig''. \tag{1.65}$$

The * sign means complex conjugation, as always.

If

$$\frac{\eta c}{\omega} \ll \frac{\omega_0^2}{\omega^2}, \tag{1.66}$$

then Eq. (1.64) for frequency spectrum takes the following form [**60.11**]:

$$W_{\mathrm{b}}^{(+)}(\omega) = \frac{2e^2}{\pi c^2}\left[\left(\frac{1}{2} + \frac{\omega^2}{\omega_0^2\gamma^2}\right)\ln\left(1 + \frac{\omega_0^2\gamma^2}{\omega^2}\right) - 1\right]. \tag{1.67}$$

(see Fig. 1.5).

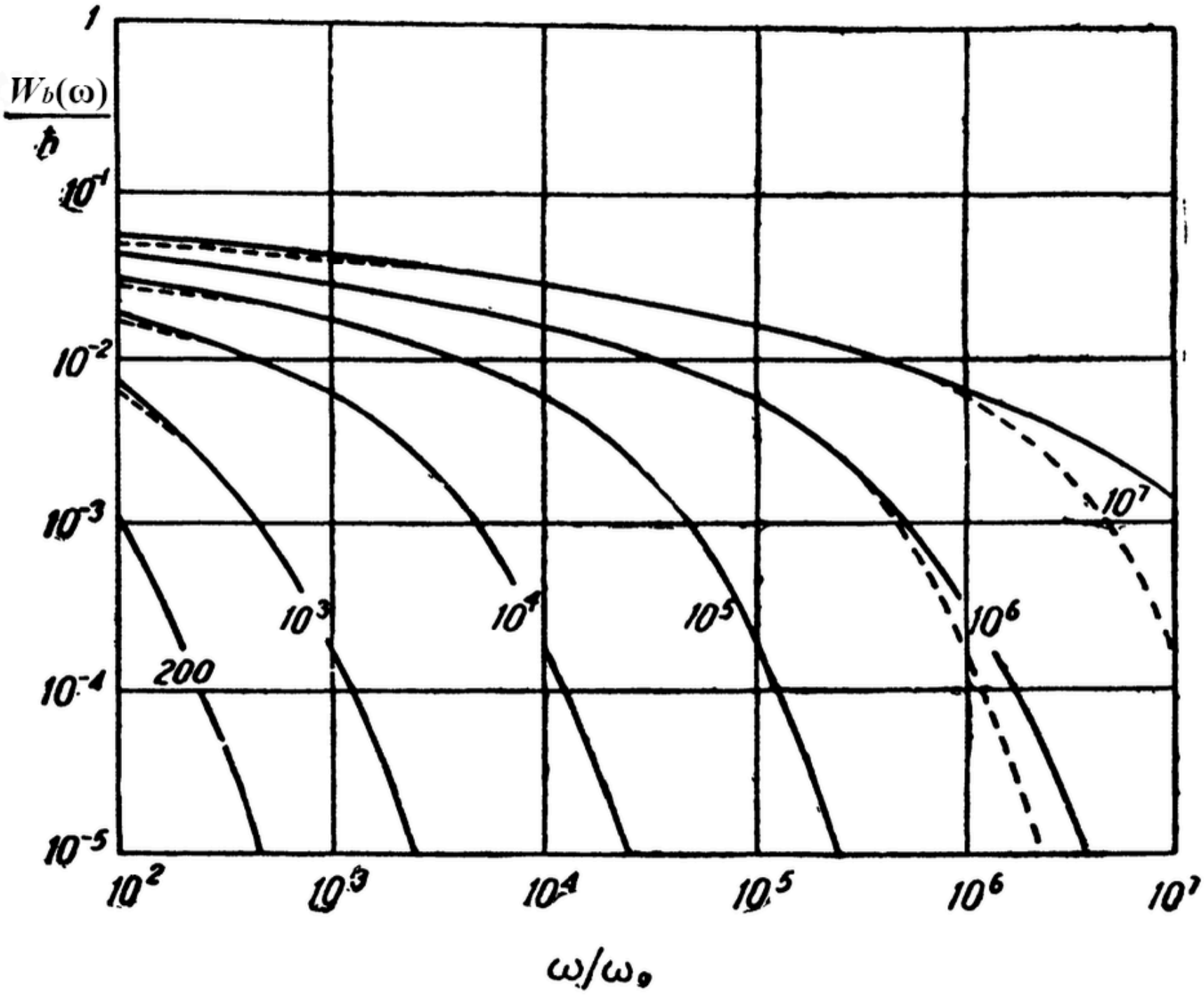

Fig. 1.5. Frequency dependence of the angle-integrated intensity (frequency spectrum) of forward XTR generated at the medium–vacuum interface. The dashed curves correspond to a non-absorbing medium, whereas the solid curves correspond to a medium (Pb) with finite absorption. The numbers next to the curves denote the particle Lorentz factor.

It follows that when $\omega_0^2/\omega^2$ is greater than or of order $\gamma^{-2}$, the radiation intensity (as well as the number of quanta in a frequency range of order ω) depends approximately logarithmically on the frequency and the particle energy and is of the same order of magnitude as in the optical or near-optical region. At high frequencies, when

$$\frac{\omega_0^2}{\omega^2} \ll \gamma^{-2}, \qquad (1.68)$$

from Eq. (1.67) we obtain

$$W_{\mathrm{b}}^{(+)}(\omega) \simeq \frac{e^2}{6\pi c}\frac{\omega_0^4\gamma^4}{\omega^4}. \qquad (1.69)$$

i.e., the intensity decreases rapidly (as $\omega^{-4}$), but exhibits a strong dependence on γ (as $\gamma^4$).

Thus, for ultrarelativistic particles, the frequency spectrum of forward transition radiation from a single medium–vacuum boundary extends into the *X-ray region* up to frequencies of the order of the cutoff frequency [**59**.4, **59**.5]

$$\omega_{\mathrm{cutoff}} = \omega_0\gamma. \qquad (1.70)$$

The total energy of XTR can be calculated by integrating Eq. (1.67) over the frequency. In doing so, let us extrapolate this equation into the low-frequency region as well, bearing in mind that the low-frequency region contributes only slightly to the total energy of transition radiation and therefore the error introduced by such an extrapolation is small. As a result, we obtain [**59**.4, **59**.5]

$$W_{\mathrm{b}}^{(+)} = \int W_{\mathrm{b}}^{(+)}(\omega)\, d\omega = \frac{e^2}{3c}\,\omega_0\gamma. \qquad (1.71)$$

It is important that both the spectral–angular distribution (see Eq. (1.59)) and the spectral distribution (see Eq. (1.67)) of the XTR intensity depend on the energy (Lorentz factor γ) of the charged particle, and not on its velocity, as in the case of Vavilov–Cherenkov radiation (see Sec. 1.6A). Moreover, the *total energy* of forward transition radiation (see Eq. (1.71)) *increases in proportion to the particle energy*. These two circumstances are particularly important for high-energy particles whose velocity is close to the speed of light.

To calculate the total number of transition-radiation quanta, one must divide the intensity (see Eq. (1.67)) by $\hbar\omega$ and then integrate the resulting expression over $\omega$. In doing so, we can no longer extrapolate Eq. (1.67) into the low-frequency region, since this region makes a considerable contribution to the number of quanta. For this region one must use the more accurate Eq. (1.54) taking into account Eqs. (1.37) or (1.47). However, if we are interested in the number of quanta in a given frequency range ($\omega_1$, $\omega_2$) in the X-ray region, this number can be obtained from the following equation (for weak absorption) [**61**.14]:

$$N(\omega_1, \omega_2) = \frac{2}{137\pi}\int_{\omega_1}^{\omega_2}\left[\left(\frac{1}{2}+\frac{\omega^2}{\omega_0^2\gamma^2}\right)\ln\left(1-\frac{\omega_0^2\gamma^2}{\omega^2}\right)-1\right]\frac{d\omega}{\omega} =$$

$$= \Phi\left[\left(\frac{\omega_2}{\omega_0\gamma}\right)^2\right] - \Phi\left[\left(\frac{\omega_1}{\omega_0\gamma}\right)^2\right], \qquad (1.72)$$

where

$$\Phi(x) = \frac{1}{2}L(x) - \frac{1}{4}\ln^2 x + (1+x)\ln\frac{1+x}{x}. \qquad (1.73)$$

Here we introduce the function

$$L(x) = \int_0^x \frac{\ln(1+t)dt}{t}. \qquad (1.74)$$

For small and large values of $x$ we have the following approximate expressions:

$$\Phi(x) = \begin{cases} -\frac{1}{4}\ln^2 x - \ln x + \left(\frac{3}{2}-\ln x\right)x + \frac{3}{8}x^2 - \frac{x^3}{9} + \ldots \\ 1 + \frac{\pi^2}{12} - \frac{1}{24x^2} + \frac{1}{36x^3} - \ldots \quad \text{for } x \ll 1, \quad \text{for } x \gg 1. \end{cases} \qquad (1.75)$$

For intermediate $x$, the values of the function $\Phi(x)$ are given in Table 1.1.

*Tablet 1.1*

*Values of the function $\Phi(x)$ (calculated from Eqs. (1.73) and (1.74))*

| $x$ | $\Phi(x)$ |
|---|---|
| 0.1 | 1.3610 |
| 0.2 | 1.5979 |
| 0.3 | 1.6839 |
| 0.4 | 1.7269 |
| 0.5 | 1.7520 |
| 0.6 | 1.7681 |
| 0.8 | 1.7871 |
| 1.0 | 1.7975 |
| 1.5 | 1.8097 |
| 2.0 | 1.8147 |
| 3.0 | 1.8187 |
| 4.0 | 1.8202 |
| 5.0 | 1.8210 |

A simple result for $N(\omega_1, \omega_2)$ is obtained in two limiting cases: when both frequencies are much smaller than the cutoff ($\omega_1 \ll \omega_2 \ll \omega_0\gamma$), and when both frequencies are much larger than the cutoff ($\omega_0\gamma \ll \omega_1 \ll \omega_2$)). In the first case, we have

$$N_1 \approx \frac{1}{137\pi} \ln \frac{\omega_0^2\gamma^2}{\omega_1\omega_2} \ln \frac{\omega_2}{\omega_1}, \tag{1.76}$$

i.e., the number of quanta depends logarithmically on $\gamma$. In the second case the number of quanta

$$N_2 \approx \frac{\omega_0^4\gamma^4}{137 \cdot 24\pi} \left(\frac{1}{\omega_1^4} - \frac{1}{\omega_2^4}\right) \tag{1.77}$$

is proportional to $\gamma^4$. In the intermediate case, when one of the frequencies $\omega_1$, $\omega_2$, or both of them, are close to the cutoff, one should expect an intermediate $\gamma$-dependence, steeper than $\ln\gamma$ and weaker than $\gamma^4$.

## 1.9. Effect of Absorption

The linear dependence of the cutoff frequency (Eq. (1.70)) and the total energy (Eq. (1.71)) of transition radiation on the particle's Lorentz factor γ was obtained under the assumption (Eq. (1.66)) that the absorption of the medium is small. However, there may be cases when the imaginary part of the polarizability $\chi'' = \eta c/4\pi\omega$ is of the same order as, or even larger than, its real part $|\chi'| = |1 - \varepsilon'|/4\pi$, i.e., the condition (1.66) is not satisfied. Such cases occur, for example, in the vicinity of absorption edges and in the high-frequency region with energies of order MeV or higher. The frequency range near absorption edges in which condition (1.66) is violated are small; therefore, they do not qualitatively affect the linear dependence of $\omega_{cutoff}$ and $W^{(+)}_{(b)}$. This is automatically taken into account by using the actual values of ε" at different frequencies (see Ref. [**74**.21]). As for the effect of absorption in the high-frequency region, it leads to a stronger—namely, almost quadratic—dependence of the cutoff frequency and the total energy of transition radiation on γ [**76**.2, **77**.17].

Indeed, Eq. (1.59) shows that the XTR intensity drops sharply when

$$\gamma^{-2} \gg |1-\varepsilon|. \qquad (1.78)$$

For this, two inequalities must be satisfied: (1.68) and

$$\gamma^{-2} \gg \frac{\eta c}{\omega}. \qquad (1.79)$$

In other words, the radiation is small when

$$\omega \gg \max\{\omega_{\text{cutoff}}, \omega'_{\text{cutoff}}\}. \qquad (1.80)$$

where $\omega_{cutoff}$ is determined by Eq. (1.70), and

$$\omega'_{\text{cutoff}} = \eta c \gamma^2. \qquad (1.81)$$

When inequality (1.74) is satisfied, from Eq. (1.64) we obtain

$$W_{\text{b}}^{(+)}(\omega) = \frac{e^2}{6\pi c}\left\{\left(\frac{\omega_{\text{cutoff}}}{\omega}\right)^4 + \left(\frac{\omega'_{\text{cutoff}}}{\omega}\right)^2\right\} \ll \frac{e^2}{\pi c}. \qquad (1.82)$$

The absorption coefficient $\eta(\omega)$ in the frequency range $\omega$ of the order of several MeV (for heavy elements) and larger (for light elements) has a minimum $\gamma_{min}$, then increases slowly with increasing $\omega$ up to the region of hundreds of MeV (see, for example, Ref. [**73**.31]).

From Eqs. (1.70) and (1.75) for $\omega_{cutoff}$ and $\omega'_{cutoff}$ it follows that if

$$\gamma > \gamma_0 = \frac{\omega_0}{c\,\eta_{\min}}. \qquad (1.83)$$

then $\omega'_{cutoff} > \omega_{cutoff}$, and the frequency spectrum of transition radiation extends further, up to the new cutoff frequency $\omega_{cutoff}$, which depends quadratically on $\gamma$ (see Fig. 1.5). This, in turn, means that the total energy of transition radiation should also depend approximately quadratically on $\gamma$ (see curve 1 in Fig.[7] 2.7). Numerical estimates give for $\gamma_0$ values of the order of $4 \cdot 10$ for light substances such as carbon and $3 \cdot 10^6$ for lead.

## 1.10. XTR in the Medium–Medium Case

Let us now consider XTR emitted forward, in the general case of two different media. Starting from the general equation (1.35), instead of Eqs. (1.59) and (1.64) we obtain the expressions:

$$W^{(+)}(\omega,\vartheta) = \frac{2e^2}{\pi c}\frac{|g_1-g_2|^2\vartheta^3\exp(-2\lambda_2'' z_{\rm obs})}{|\vartheta^2+g_1|^2\,|\vartheta^2+g_2|^2}. \qquad (1.84)$$

$$W^{(+)}(\omega) = \frac{2e^2}{\pi c}|g_1-g_2|^2\,\mathrm{Re}\left\{\frac{g_1\ln g_1}{(g_1^*-g_1)(g_2-g_1)(g_2^*-g_1)} + \frac{g_2\ln g_2}{(g_2^*-g_2)(g_1-g_2)(g_1^*-g_2)}\right\}\exp(-2\lambda_2'' z_{\rm obs}). \qquad (1.85)$$

where $g_s = \gamma^{-2} + 1 - \varepsilon_s$ (s = 1, 2) (in the case of transparent media, the equation corresponding to Eq. (1.85) was obtained in Ref. [61.14]).

First of all, let us note that, apart from the obvious attenuation factor $exp(-2\lambda_2'' z_{obs})$, the intensity of XTR given by Eqs. (1.84) and (1.85) does not change upon interchanging the indices 1 and 2. For simplicity, the attenuation factor will be omitted in what follows. For definiteness, let us assume that

$$|1-\varepsilon_2|<|1-\varepsilon_1|. \tag{1.86}$$

In addition, for simplicity we assume that the imaginary parts of the quantities $g_1$ and $g_2$ are small.

An analysis of the spectral–angular distribution given by Eq. (1.84) shows that the maximum is determined by $g_2$, i.e., by the less dense medium. It occurs at angles

$$\vartheta = \begin{cases} \sqrt{3}\,|g_2|^{1/2}, & \text{if } |g_1| \gg |g_2|, \\ \sqrt{\dfrac{3}{5}}\,|g_2|^{1/2}, & \text{if } |g_1 - g_2| \ll |g_2|. \end{cases} \tag{1.87}$$

At

$$\gamma^{-2} \ll |1-\varepsilon_2| \tag{1.88}$$

the spectral–angular distribution given by Eq. (1.84) does not depend on $\gamma$ at all. In particular, in this case the position and the height of the transition maximum are also independent of $\gamma$.

Let us now consider the spectral distribution given by Eq. (1.85). It can be seen that at

$$\left|\frac{g_1-g_2}{g_2}\right| \ll 1 \tag{1.89}$$

the XTR intensity decreases sharply:

$$W^{(+)}(\omega) = \frac{e^2}{6\pi c}\left|\frac{g_1-g_2}{g_2}\right|^2. \tag{1.90}$$

If the imaginary parts of $g_s$ ($s = 1, 2$) are neglected, then Eqs. (1.89) and (1.90) take the following form:

$$\omega_{01}^2 - \omega_{02}^2 \ll \omega_{02}^2 + \omega^2\gamma^{-2} \tag{1.91}$$

and

$$W^{(+)}(\omega) = \frac{e^2}{6\pi c}\left(\frac{\omega_{01}^2-\omega_{02}^2}{\omega_{02}^2+\omega^2\gamma^{-2}}\right)^2, \tag{1.92}$$

where $\omega_{0s}$ are the plasma frequencies of the two media.

If the densities of the two media differ significantly, inequality (1.91) can be satisfied only under the condition

$$\omega \gg (\omega_{01}^2 - \omega_{02}^2)^{1/2}\gamma \approx \omega_{01}\gamma. \qquad \textbf{(1.93)}$$

If the densities of the two media are close, i.e., $\omega_{01} - \omega_{02} \ll \omega_{02}$, then inequality (1.91), and hence Eq. (1.92), hold for all $\omega$ and $\gamma$. In this case, if the inequality $\omega \gg \omega_{02}\gamma$ is satisfied, the intensity given by Eq. (1.92) is proportional to $\gamma^4\omega^{-4}$.

Naturally, when condition given by condition (1.88) holds, the spectral distribution is also independent of $\gamma$.

The total energy of transition radiation (neglecting the imaginary parts of $g_s$) is obtained by integrating the spectral distribution:

$$W^{(+)} = \int W^{(+)}(\omega)\,d\omega = \frac{e^2}{3c}\gamma\frac{(\omega_{01} - \omega_{02})^2}{\omega_{01} + \omega_{02}}, \qquad (1.94)$$

the linear dependence of the total radiation energy on $\gamma$, analogous to Eq. (1.71), also holds in the general case of two media.

## 1.11. Backward Radiation

Let us now consider the backward radiation produced when a particle enters a medium from vacuum. Replacing $\beta$ with $-\beta$ in Eq. (1.37), we obtain the Ginzburg–Frank formula [**45**.1]*[7]:

$$W^{(-)}(\omega, \vartheta) = \frac{2e^2\beta^2}{\pi c}\frac{|\varepsilon - 1|^2|1 + \beta(\varepsilon - \sin^2\vartheta)^{1/2} - \beta^2|^2}{(1 - \beta^2\cos^2\vartheta)^2|1 + \beta(\varepsilon - \sin^2\vartheta)^{1/2}|^2} \times$$
$$\times \frac{\sin^3\vartheta\cos^2\vartheta}{|\varepsilon\cos\vartheta + (\varepsilon - \sin^2\vartheta)^{1/2}|^2}. \qquad (1.95)$$

It should be noted at once that the Cherenkov maximum is absent in this formula. This is quite natural, since Vavilov–Cherenkov radiation does not occur in vacuum, whereas in a medium it can be produced only in the forward direction.

---

[7] There is an inaccuracy in the corresponding equation of Ref. [**45**.1] (see Refs. [**57**.2, **57**.8]).

For ultrarelativistic particles, when condition in Eq. (1.46) is satisfied, at small angles $\vartheta \ll 1$ Eq. (1.95) yields

$$W^{(-)}(\omega, \vartheta) = \frac{2e^2}{\pi c}\left|\frac{\varepsilon - 1}{\varepsilon + 1}\right|^2 \frac{\vartheta^3}{(\gamma^{-2} + \vartheta^2)^2}. \qquad (1.96)$$

Eq. (1.96) differs from Eq. (1.50) only by the presence of the reflection coefficient. Therefore, in the case when ε is not close to unity (see Eq. (1.49)), the intensities given by Eqs. (1.96) and (1.50) are of approximately the same order. In this case, backward transition radiation also has a maximum at $\vartheta \sim \gamma^{-1}$, and the contribution of small angles dominates provided that $ln\gamma \gg 1$.

If, however, one considers frequencies much higher than atomic ones (i.e., if condition in Eq. (1.57) holds), then because of the small reflection coefficient the intensity given by Eq. (1.95) (or Eq. (1.96)) decreases sharply. Thus we conclude that *backward* transition radiation occurs mainly in the optical and lower-frequency ranges, whereas in the *forward* direction XTR is also emitted.

## § 2. TRANSITION RADIATION GENERATED IN A PLATE

Let us now consider the more general problem in which a charged particle passes through a plane-parallel plate of thickness $a$ and dielectric permittivity ε. For simplicity, we assume that the plate is in vacuum and that the particle enters the plate perpendicularly.

### 2.1. General Equation for the Spectral–Angular Distribution of Radiation Intensity

We shall find the intensities of radiation propagating backward in the vacuum region before the plate ($z < 0$) and forward beyond the plate ($z > a$). As in the previous section, the total field is the sum of the particle field and the radiation field. At the same time, according to the Sommerfeld radiation condition, in the region $z < 0$ there exists only the backward-propagating radiation field $E^{(-)}$, whereas in the region $z < a$ there exists only the forward-propagating radiation field $E^{(+)}$. Inside the plate ($0 < z < a$) radiation fields of both types are present. The continuity conditions at the interfaces $z = 0$ and $z = a$ yield four algebraic equations for the Fourier components of the radiation field. Solving these equations yields the following expressions for the Fourier components of the tangential components of the electric field of the radiation outside the plate (for brevity, the arguments $\varkappa$ and ω of the Fourier components are omitted):

$$E_t^{(-)} = E_{t,\text{V-M}}^{(+)} P_a + E_{t,\text{M-V}}^{(+)} \exp\left[i\left(\frac{\omega}{v} - \lambda_0\right)a\right] + E_{t,\text{M-V}}^{(-)} R P_a \exp\left[i\left(\frac{\omega}{v} + \lambda\right)a\right]. \quad (2.1)$$

$$E_t^{(-)} = E_{t,\text{V-M}}^{(-)} + E_{t,\text{V-M}}^{(+)} R_a P + E_{t,\text{M-V}}^{(-)} P \exp\left[i\left(\frac{\omega}{v} + \lambda\right)a\right]. \quad (2.2)$$

where $E_{t,v-m}^{(\pm)}$ and $E_{t,m-v}^{(\pm)}$ are the Fourier components of the tangential components of the electric fields of the radiation generated at the vacuum-medium and medium-vacuum interfaces and propagating forward (the “+” sign) or backward (the “−” sign). These quantities are determined by Eq. (1.21) and by an analogous equation obtained by interchanging the indices 1 and 2 and changing the sign of $u_s$ (for waves propagating backward). The quantity $R$ is the radiation reflection coefficient:

$$R = \frac{\varepsilon u_0 - u}{\varepsilon u_0 + u}, \quad (2.3)$$

and $P$ is the transmission coefficient taking into account multiple reflections of the radiation within the plate:

$$P = \frac{2\varepsilon u_0}{\varepsilon u_0 + u} \frac{1}{1 - R^2 \exp(2i\lambda a)} \quad (2.4)$$

($u$ and $u_0$ are defined by Eq. (1.20)). The quantities $R_a$ and $P_a$ are the corresponding coefficients with the phase shift taken into account (when the interface plane is located at $z = a$):

$$R_a = R\exp(2i\lambda a), \quad P_a = P\exp\{i(\lambda - \lambda_0)a\}. \tag{2.5}$$

Thus, Eqs. (2.1) and (2.2), obtained by formally solving the equations, have the following transparent physical interpretation. Specifically, the terms in Eq. (2.1) describe three waves: (1) a wave generated at the first boundary of the plate ($z = 0$) and emerging from the plate in the forward direction; (2) a wave generated at the second boundary of the plate ($z = a$) and propagating forward; (3) a wave generated at the second boundary of the plate and initially propagating backward, but then reflected from the first boundary and ultimately emerging from the plate in the forward direction (Fig. 2.1). The factors $exp\{i(\omega/v - \lambda_0)a\}$ and $exp\{i(\omega/v + \lambda)a\}$ in Eq. (2.1) account for the phase shifts for waves generated at the interface $z = a$. The three terms in Eq. (2.2) have an equally simple physical meaning. Moreover, as indicated in Sec. 1.2, each of these three waves generated at the interface in turn consists of three waves arising on the corresponding segments of the charged particle trajectory (Figs. 1.1 and 1.2).

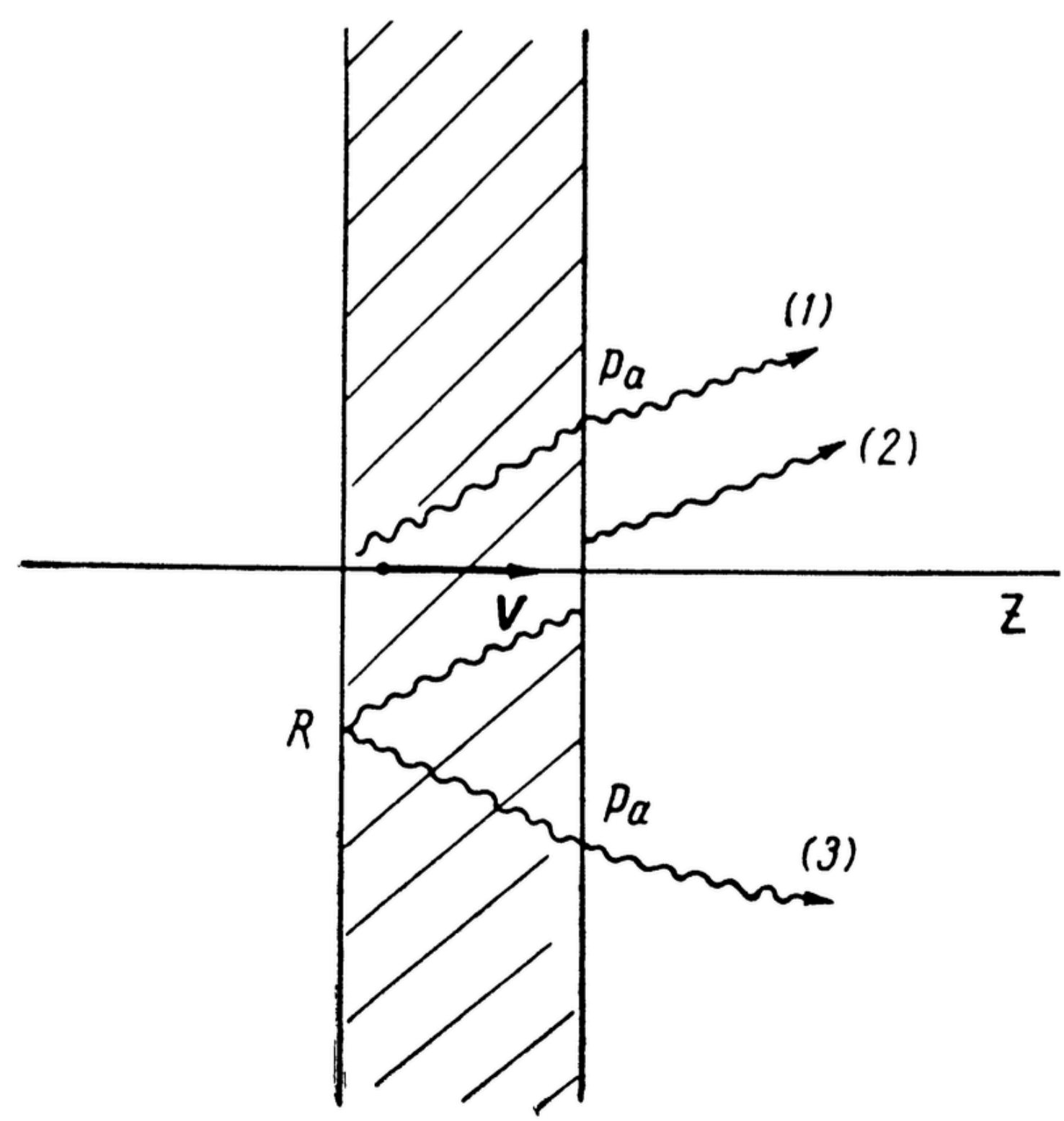

Fig. 2.1. Schematic representation of the three waves that constitute the transition radiation emitted by a particle in the forward direction as it passes through a plate. These three waves correspond to the three terms in Eq. (2.1)

By calculating the flux of the Poynting vector through the transverse plane $z = z_{obs}$ in the same way as in Sec. 1.3, we obtain the spectral–angular distribution of the radiation intensity at large distances from the plate. For forward radiation we obtain

$$W_{\mathrm{pl}}^{(+)}(\omega,\vartheta) = \frac{2e^2\beta^2}{\pi c}\frac{\sin^3\vartheta\, u_2^2}{(1-u_2^2)^2}\left|\frac{\varepsilon-1}{1-u^2}\right|^2\left|\frac{B}{Q_1}\right|^2. \qquad (2.6)$$

where

$$B = 2u(1+u_0)(1-u_0-\beta^2\varepsilon)\exp[-ia_0(1-u)] -$$

$$-(\varepsilon u_0+u)(1+u)(\gamma^{-2}-u)+(\varepsilon u_0-u)(1-u)(\gamma^{-2}+u)\exp(2ia_0u), \quad (2.7)$$

$$Q_1 = (\varepsilon u_0+u)^2-(\varepsilon u_0-u)^2\exp(2ia_0u),$$

$$u = \beta(\varepsilon-\sin^2\vartheta)^{1/2}, \quad u_0 = \beta\cos\vartheta, \quad a_0 = \frac{\omega}{v}a. \quad (2.8)$$

It follows that, when the plate thickness $a$ is measured in units of $v/\omega$, the quantity $W_{pl}^{(+)}(\omega, \vartheta)$ depends on the frequency only through $\varepsilon(\omega)$.

The spectral–angular distribution of the radiation intensity propagating backward can be obtained from Eqs. (2.6)–(2.8) by replacing $v$ to $-v$ and $\beta$ to $-\beta$. In this case, the angle $\vartheta$ is measured from the negative $z$-direction.

Let us analyze the quantity in Eq. (2.6) separately in the optical frequency range (when $\varepsilon$ is not close to unity) and in the X-ray frequency range (when $|\varepsilon - 1| \ll 1$).

## 2.2. Optical Frequency Range

Let the dielectric permittivity of the plate $\varepsilon$ be not close to unity, i.e., assume that condition (1.49) is satisfied. As shown in Sec. 1.4, in the case of a semi-infinite medium a maximum corresponding to Vavilov–Cherenkov radiation appears in the angular distribution of the radiation intensity at the angle $\vartheta_C$ (Eq. (1.40)). It is natural in the case of a plate to first analyze the quantity in Eq. (2.6) as a function of $\vartheta$ in the vicinity of $\vartheta_C$.

Note that Eq. (2.6) contains in the denominator the same factor $1 - u$ (cf. Eq. (1.38)), which attains a minimum at $\vartheta = \vartheta_C$ if conditions in Eq. (1.41) are satisfied. Let us consider the quantity $B/Q_1$ at $\vartheta \approx \vartheta_C$. It is easy to verify that, when $|1 - u| \ll 1$, we obtain

$$\frac{B}{Q_1}=\frac{4i\beta^2(\varepsilon u_0+1)\exp[-ia_0(1-u)/2]\sin[a_0(1-u)/2]}{(\varepsilon u_0+1)^2-(\varepsilon u_0-1)^2\exp(2ia_0u)}+O(1-u), \tag{2.9}$$

where by $O(1-u)$ we mean terms of order $(1-u)$ or higher, and the quantity $u_0$ is evaluated at $\vartheta=\vartheta_C$.

For not very large values of $a_0$, the first term in Eq. (2.9) is also a small quantity proportional to $(1-u)$. In this case the factor $(1-u)$ in the denominator in Eq. (2.6) is cancelled by a quantity of the same order in the numerator $B$. As a result, no radiation maximum appears at $\vartheta=\vartheta_C$. In other words, for the occurrence of the Cherenkov maximum in the case of a *plate*, in addition to the conditions of Eq. (1.41) the following conditions must also be satisfied [**57**.2, **58**.1, **59**.3]:

$$a_0=\frac{a\omega}{v}\gg 1. \tag{2.10}$$

For fast particles ($v\approx c$), condition (2.10) essentially means that the plate thickness must be much greater than the wavelength.

If condition (2.10) holds (as well as Eq. (1.41)), the terms $O(1-u)$ in the numerator in Eq. (2.9) can be neglected. Substituting Eq. (2.9) into Eq. (2.6) (at $\vartheta\approx\vartheta_C$), we obtain (cf. Refs. [**54**.1, **81**.14])

$$W^{(+)}_{\text{V-M}}(\omega,\vartheta)=\frac{2e^2}{\pi c}\frac{\sin^3\vartheta_{\text{ч}}\cos^2\vartheta_{\text{ч}}\beta^4\exp\left(-a_0\beta^2\varepsilon''\right)}{|\varepsilon u_0+1|^2|1-R^2_{\text{ч}}\exp(2ia_0u)|^2}\left|\frac{\sin^{a_0(1-u)/2}}{(1-u)/2}\right|^2. \tag{2.11}$$

where $R_C$ is determined by Eq. (2.3) with $u_0$ and $u$ taken at $\vartheta=\vartheta_C$.

From Eq. (2.11) it follows that as long as

$$a \ll \frac{c}{\omega\beta^2\varepsilon''} = l_{\rm abs}. \qquad (2.12)$$

(this condition does not contradict Eq. (2.10), since $\varepsilon'' \ll 1$), the maximum at $\vartheta=\vartheta_C$ becomes higher and narrower as $a$ increases. Integrating Eq. (2.11) in this case, we obtain

$$W^{(+)}_{\text{V-M}}(\omega) = \frac{e^2\omega}{c^2}\left(1-\frac{1}{\beta^2\varepsilon'}\right)a\,\frac{4\varepsilon'\beta\cos\vartheta_{\rm ч}}{(\varepsilon'\beta\cos\vartheta_{\rm ч}+1)^2}\cdot\frac{1}{|1-R_{\rm C}^2\exp(2ia_0)|^2}. \qquad (2.13)$$

This equation differs from the corresponding Eq. (1.44) for a semi-infinite medium by the presence of the plate thickness $a$ (instead of $l_{abs}$ and the factor $|1-R^2_C\exp(2ia_0)|^{-2}$, which describes multiple reflections of the radiation inside the plate.

With a further increase of $a$ to values of the order of $l_{abs}$, the growth of the maximum in Eq. (2.11) and its narrowing slow down due to the complex nature of $\varepsilon$ (and hence of $u$). When

$$a \gg l_{\rm abs}. \qquad (2.14)$$

Eq. (2.11) ceases to depend on $a$ altogether and exactly reduces to Eq. (1.43) for a semi-infinite medium. After integration over $\vartheta$, we obtain quantity (1.44)—the intensity of Vavilov–Cherenkov radiation emitted over a length equal to $l_{abs}$ and emerging into vacuum.

A completely analogous situation takes place for backward radiation. It is easy to verify that at $\vartheta=\vartheta_C$

$$W^{(-)}_{\rm VC}(\omega,\vartheta) = W^{(+)}_{\rm VC}(\omega,\vartheta)|R_{\rm C}|^2\exp(-2a/l_{\rm abs}). \qquad (2.15)$$

where $W^{(+)}_{VC}(\omega, \vartheta)$ is given by Eq. (2.11). The physical meaning of Eq. (2.15) is clear: the radiation emitted backward at an angle $\vartheta_C$ can be interpreted as Vavilov–Cherenkov radiation reflected from the second boundary ($z=a$) of the plate and emerging through the first boundary ($z=0$), with allowance for absorption in the plate.

In addition to the possible radiation maximum at the Cherenkov angle $\vartheta_C$ considered above, the angular distribution in Eq. (2.6), as in Eq. (1.37) in the case of a semi-infinite medium, has a second maximum at an angle of order $\gamma^{-1}$(for $\gamma \gg 1$). This maximum, due to the denominator $(1 - u_0^2)^2$, will be referred to as the transition-radiation maximum (see Sec. 1.6.B).

From Eq. (2.6) it is seen that when the plate is sufficiently thin, i.e., the conditions are satisfied:

$$|a_0(1-u)| \ll 1 \text{ and } |a_0 u| \ll 1, \qquad (2.16)$$

the fraction $B/Q_1$ can be expanded in powers of the small quantities defined in Eq. (2.16), yielding

$$\frac{B}{Q_1} \approx \frac{ia_0}{2\varepsilon u_0}(1-u^2)(\varepsilon u_0 - \gamma^{-2}). \qquad (2.17)$$

Substituting Eq. (2.17) into Eq. (2.6), in the case under consideration we have

$$W_{\text{pl}}^{(+)}(\omega,\vartheta) = \frac{e^2\beta^2 a_0^2}{2\pi c}\frac{|\varepsilon-1|^2}{|\varepsilon|^2}\frac{|\varepsilon\beta\cos\vartheta - \gamma^{-2}|^2}{(1-\beta^2\cos^2\vartheta)^2}\sin^3\vartheta. \qquad (2.18)$$

For backward radiation the conditions under which the plate can be regarded as thin have the following form:

$$|a_0(1+u)| \ll 1 \text{ and } |a_0 u| \ll 1, \qquad (2.19)$$

and instead of Eq. (2.18) we have a analogous equation with $\beta$ replaced by $-\beta$. It is easy to see that under this substitution the quantity in Eq. (2.18) remains practically unchanged if the particle is ultrarelativistic ($\gamma \gg 1$). In other words, for ultrarelativistic particles the forward and backward radiation intensities in the case of a thin plate are in fact equal at the same values of $\vartheta$ and $\omega$.

The conditions in Eqs. (2.16) and (2.19), for values of $\varepsilon$ not close to 1 (when Eq. (1.49) holds), essentially mean

$$|a_0| \ll 1, \qquad (2.20)$$

i.e., the plate thickness being small compared with the wavelength (for fast particles $v \approx c$). It is known (see Ref. [**73**.32], Sec. 67) that in this case the radiation is dipole, i.e., all parts of the plate along the trajectory of the traversing charged particle radiate coherently. As follows from Eq. (2.18) and an analogous equation for $W_{(-)}(\omega, \vartheta)$, the intensities of dipole radiation both in the forward and backward directions are proportional to $a^2$.

When the dielectric permittivity ε is close to unity (for example, in a dilute medium or in the X-ray frequency range), the first of the conditions in Eq. (2.16) implies that the plate thickness is small compared with the radiation formation length, which may greatly exceed the wavelength for $v \approx c$, $|\varepsilon| \approx 1$, $\vartheta \ll 1$ (see Sec. 2.3 for details).

The factor $1 - u$ does not appear in the denominator in Eq. (2.18). This means that for thin plates no radiation maximum occurs at the Cherenkov angle $\vartheta_C$ as already noted above in the discussion of Eq. (2.9).

In the general case of a plate of arbitrary thickness one must use Eq. (2.6) (or the analogous equation with the replacement β → −β for backward radiation).

For illustration, Fig. 2.2 shows the angular dependences of $W_{(\pm)}^{(\pm)}(\omega, \vartheta)$ for different plate thicknesses at a fixed frequency ω. As seen from Fig. 2.2, for $a_0 = 5$ the angular distribution of the radiation intensity has only one maximum at an angle of order $\sim \gamma^{-1}$ (the transition radiation maximum). As $a_0$ increases, a whole series of maxima and minima of interference origin appears, with the spacing between them decreasing as $a_0$ increases. For $a_0 > 50$, a maximum at the angle $\vartheta_C$ (the Cherenkov maximum) gradually emerges, becoming higher and narrower as $a_0$ increases, until the plate thickness becomes much greater than the absorption length. The intensity of the radiation emitted backward has a similar character, only it is significantly weaker due to the small value of the reflection coefficient.

When $|\varepsilon| \approx 1$ ($\varepsilon' > 1$), the angle $\vartheta_C$ is small (see Eq. (1.40)); thus the maxima of transition radiation and Cherenkov radiation may lie very close to each other and merge. As seen in Fig. 2.3, that in this case, in the angular distribution of the intensity $W_{pl}^{(\pm)}(\omega, \vartheta)$ at a given frequency ω, there is either a single distorted maximum (the curves on the left and those corresponding to Lorentz-factor values 50 and 100 on the right) or two separate but broadened maxima (the curve with $\gamma = 500$ on the right). As noted in Sec. 1.6, the separation of radiation into transition and Cherenkov components is, in general, somewhat conventional, and in the case considered here such a separation becomes quite difficult.

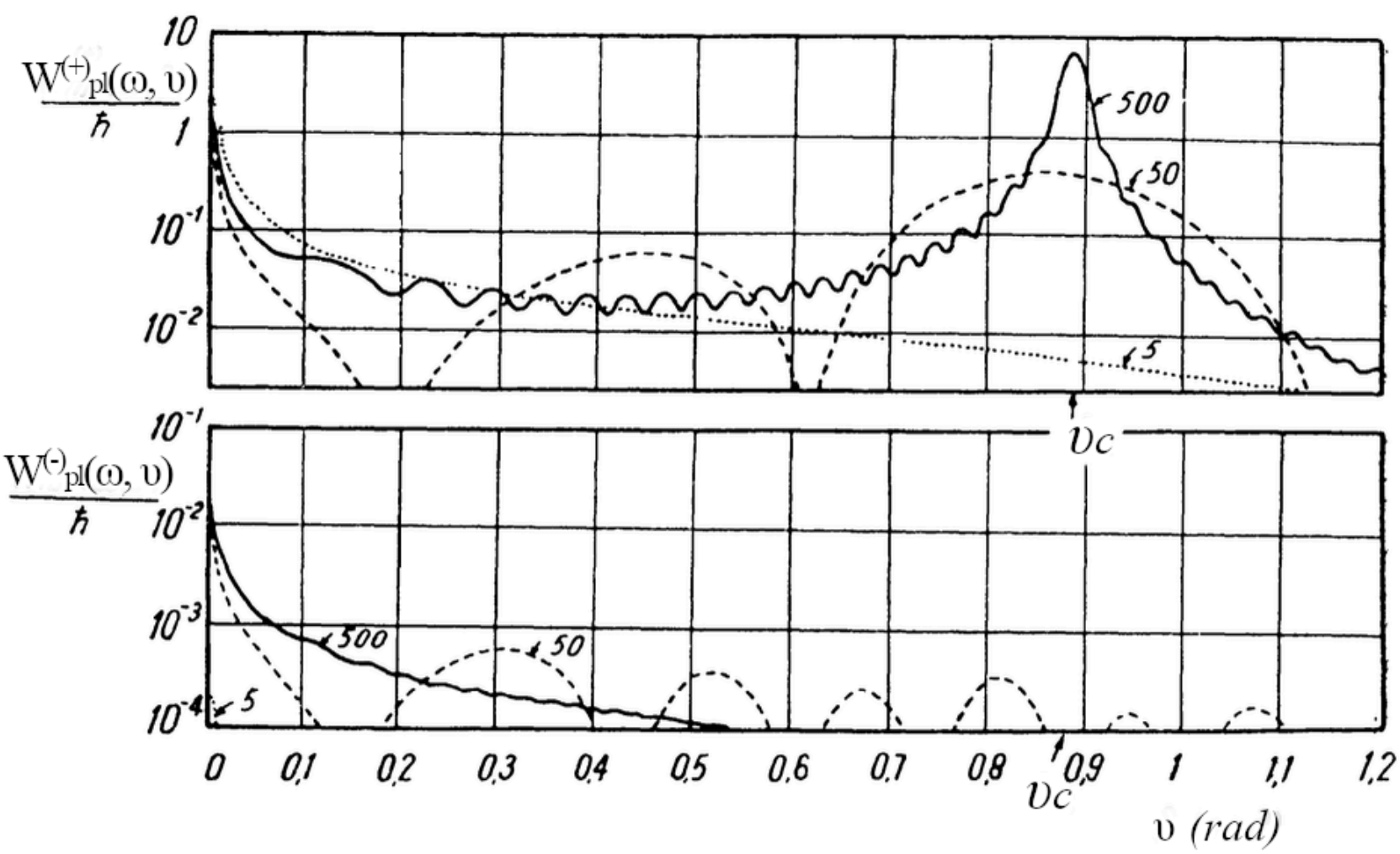

Fig. 2.2. Angular dependence of the spectral–angular intensity distribution of transition and Cherenkov radiation produced in a plate in the forward (upper plot) and backward (lower plot) directions. The numbers next to the curves indicate the value $a_0 = a\omega/v$; $\varepsilon' = 1.6$; $\varepsilon'' = 0.001$; the Lorentz factor of the particle is $10^3$

Let us now consider the spectral distribution of the radiation intensity $W_{pl}^{(\pm)}(\omega)$. The dependence of this quantity on the particle Lorentz factor is primarily *logarithmic* because the intensity of optical transition radiation depends logarithmically on γ, whereas the intensity of Vavilov–Cherenkov radiation is practically independent of γ (for $\gamma \gg 1$).

The dependence of $W^{(\pm)}_{pl}(\omega)$ on $a$ is somewhat more complicated (Fig. 2.4). For sufficiently small $a_0 = a\omega/v$ (of order unity or less), the dependence of $W^{(+)}_{pl}(\omega)$ on $a_0$ is quadratic owing to the dipole character of the radiation. In the range of $a_0$ values of order 10–100, this dependence becomes oscillatory due to interference. With a further increase in $a_0$, the dependence becomes linear because the main contribution to $W^{(+)}_{pl}(\omega)$ comes from the Cherenkov maximum. Finally, for sufficiently large $a_0$, the quantity $W^{(+)}_{pl}(\omega)$ becomes independent of $a_0$ because of absorption. All this, naturally, is in full agreement with the general analysis carried out above.

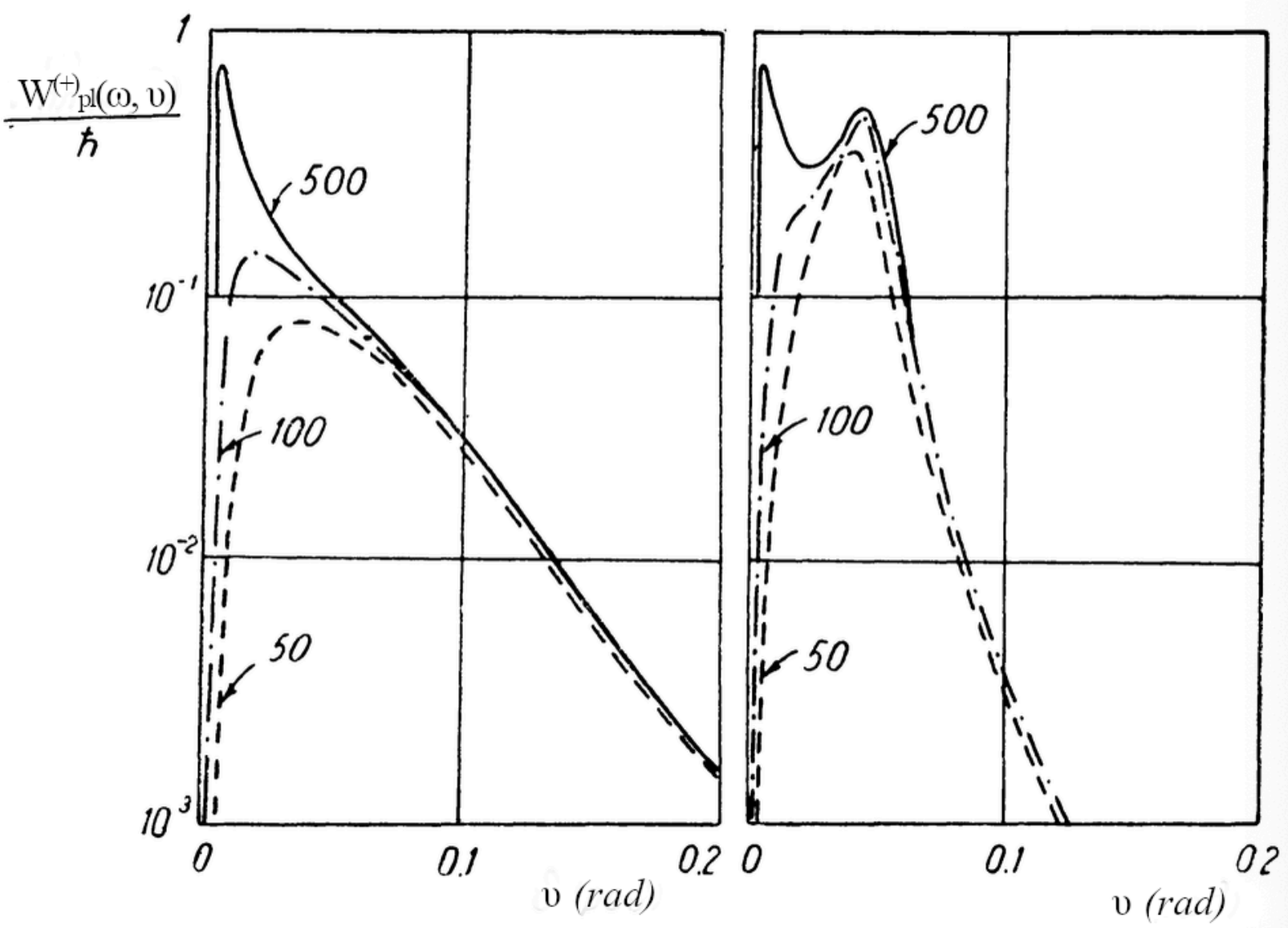

Fig. 2.3. Angular dependence of the spectral–angular distribution of radiation intensity $W^{(+)}_{pl}(\omega, \vartheta)$. The transition and Cherenkov maxima are mixed. The dimensionless thickness of the plate $a_0 = a\omega/v$ equals $10^5$; $\varepsilon' = 1.002$. The curves on the left correspond to the case $\varepsilon'' = 0.01$; the curves on the right correspond to $\varepsilon'' = 0.001$. The numbers indicated next to the curves correspond to the values of the particle Lorentz factor.

The dependence of the spectral intensity $W_{pl}^{(-)}(\omega)$ of the backward radiation on $a_0$ has a similar character. However, because of the reflection coefficient, the contribution of the Cherenkov maximum to $W_{pl}^{(-)}(\omega)$ is considerably smaller; therefore, the linear region in the dependence on $a_0$ appears at larger values of $a_0$. If absorption is already significant at such values of $a_0$, the linear region may not be observable at all.

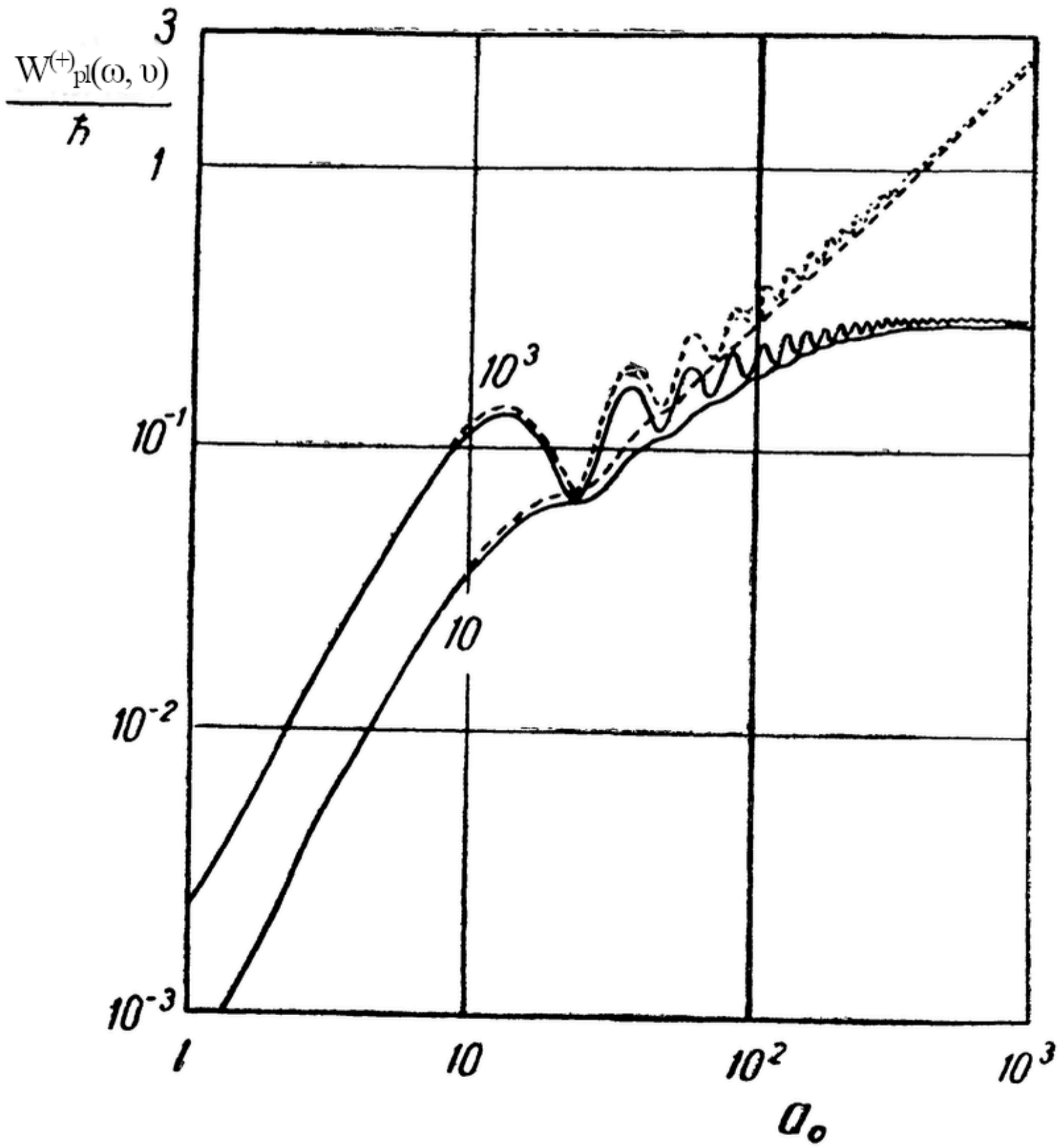

Fig. 2.4. Dependence of the angle-integrated forward radiation intensity $W^{(+)}_{pl}(\omega)$ on the dimensionless thickness $a_0 = a\omega/v$; $\varepsilon' = 1.6$; $\varepsilon'' = 0.01$ (solid curves) and $\varepsilon'' = 0$ (dashed curves); the numbers by the curves denote the particle Lorentz factor

## 2.3. X-ray Frequency Range

Let now the dielectric permittivity of the plate ε be close to unity, i.e., condition in Eq. (1.57) holds. Then for ultrarelativistic particles ($\gamma \gg 1$) and small radiation angles ($\vartheta \ll 1$) from Eq. (2.6) we obtain

$$W^{(+)}_{\rm pl}(\omega,\vartheta) = W^{(+)}_{\rm b}(\omega,\vartheta)\,F_{\rm pl}, \qquad (2.21)$$

where $W^{(+)}_{gr}(\omega,\vartheta)$ is given by Eq. (1.59), and $F_{pl}$ is the factor describing the interference of the radiation produced at both boundaries of the plate:

$$F_{\rm pl} = \left(1 - Q^{1/2}\right)^2 + 4Q^{1/2}\sin^2 Y, \qquad (2.22)$$

$$Q = \exp(-\varepsilon''\omega a/c) = \exp(-\eta a), \qquad (2.23)$$

$$Y = \frac{\omega a}{4c}\left(\vartheta^2 + \gamma^{-2} + 1 - \varepsilon'\right) = \mathrm{Re}\,\frac{\pi a}{2z_{\rm med}(\vartheta)}. \qquad (2.24)$$

In the latter equation, the quantity $z_{med}(\vartheta)$ is the *formation length* of forward XTR *in the medium*, defined as follows (see Eq. (1.30) for the medium–vacuum case):

$$z_{\rm ext}(\vartheta) = \frac{2\pi v}{\omega\left(\gamma^{-2} + \vartheta^2 + 1 - \varepsilon\right)}. \qquad (2.25)$$

Similarly, let us also define the *formation length* of XTR emitted forward *in vacuum*:

$$z_{\rm vac}(\vartheta) = \frac{2\pi v}{\omega\left(\gamma^{-2} + \vartheta^2\right)}. \qquad (2.26)$$

For backward radiation, from Eqs. (2.6)–(2.8), after replacing $v$ by $-v$ and taking into account the conditions $|1-\varepsilon| \ll 1$, $\gamma \gg 1$ and $\vartheta \ll 1$, we obtain

$$W_{\mathrm{pl}}^{(-)}(\omega,\vartheta) = \frac{e^2}{8\pi c}\frac{|1-\varepsilon|^2\vartheta^3}{(\gamma^{-2}+\vartheta^2)^2\,|\gamma^{-2}+\vartheta^2+1-\varepsilon|^2}\times$$
$$\times\left|-(\gamma^{-2}+\vartheta^2)\exp[ia_0(1+u)]+(\gamma^{-2}+\vartheta^2+1-\varepsilon)+(\varepsilon-1)\exp(2ia_0u)\right|^2. \tag{2.27}$$

Let us note at once that, as in the case of a single boundary (see Sec. 1.11), the intensity of XTR emitted forward (Eq. (2.21)) is much greater than that of XTR emitted backward (Eq. (2.27)).

The angular distribution of the XTR intensity has a sharp maximum at $\vartheta \sim \gamma^{-1}$. At the same time, as a result of the interference of radiation produced at the two boundaries of the plate, a whole series of additional maxima and minima may appear in the angular distribution, and the maximum at $\vartheta \sim \gamma^{-1}$ may be strongly suppressed. From Eq. (2.21) it follows that for forward XTR these interference maxima appear at odd integer values of $Re(a/z_{med}(\vartheta))$, i.e., at angles

$$\vartheta_n^2 = \frac{4\pi c}{\omega a}\left(n + \frac{1}{2}\right) - (\gamma^{-2} + 1 - \varepsilon'), \qquad (2.28)$$

where $n$ is a positive integer such that the right-hand side of Eq. (2.28) is greater than zero. Similarly, the minima appear at even integer values of $Re(a/z_{med}(\vartheta))$, i.e., at angles

$$\vartheta_m^2 = \frac{4\pi c}{\omega a} m - (\gamma^{-2} + 1 - \varepsilon'), \qquad (2.29)$$

where $m$ is a positive integer such that the right-hand side of Eq. (2.29) is greater than or equal to zero.

From Eq. (2.22) for $F_{pl}$ It follows that the amplitude of the interference oscillations depends on the transparency $Q$ of the plate. When the plate is sufficiently transparent, i.e.,

$$Q \approx 1, \text{ i.e. } \eta a \ll 1, \qquad (2.30)$$

destructive interference at the angles (Eq. (2.29)) reduces the radiation intensity almost to zero. In this case, if

$$\mathrm{Re}\left|\frac{a}{z_{\mathrm{med}}(\vartheta)}\right| \gg 1, \qquad (2.31)$$

then the factor $F_{pl}$ oscillates strongly and its average value is $(1 + Q)$. In this case, the two boundaries of the plate radiate *independently*.

As the transparency $Q$ decreases, the amplitude of oscillations of the factor $F_{pl}$ decreases. In other words, the spectral–angular distribution of the radiation intensity is smoothed out. In the limit where the plate is almost opaque, i.e.,

$$Q \ll 1, \text{ i.e. } \eta a \gg 1, \qquad (2.32)$$

the oscillations disappear, and the factor $F_{pl}$ becomes practically equal to unity. This is quite clear, since in this case the radiation produced at the first interface ($z = 0$) is almost completely absorbed in the plate and does not reach the second interface ($z = a$); i.e., only the radiation produced at the second interface is emitted forward.

From Eqs. (2.28) and (2.29) it follows that, as the plate thickness $a$ decreases, the interference maxima and minima in the angular distribution shift to larger angles. In the case of a sufficiently thin plate, as noted in Sec. 2.1, the radiation becomes dipole, and the interference extrema disappear altogether. In the case of backward XTR, as seen from Eq. (2.27), the dipole character occurs under the conditions in Eq. (2.19). However, for forward XTR the corresponding condition is much weaker. Indeed, as seen from Eq. (2.21), the radiation becomes dipole upon fulfillment of the following inequality:

$$a \ll |z_{\text{med}}(\vartheta)|, \qquad (2.33)$$

which corresponds to the first of the conditions in Eq. (2.16). The second of the conditions in Eq. (2.16) does not play a role in the XTR case under consideration because the third term in Eq. (2.7) is small; this term corresponds to the wave emitted backward at the second interface ($z = a$) and then reflected from the first interface ($z = 0$).

As already noted in Sec. 1.6 and as follows from Eq. (2.25), the quantity $|z_{med}(\vartheta)|$ is much larger than the wavelength and is macroscopic for $\gamma \gg 1, \vartheta \ll 1$, and $|1 - \varepsilon| \ll 1$.

When condition (2.33) is satisfied, from Eq. (2.22) we have

$$F_{\text{pl}} = \frac{(\pi a)^2}{|z_{\text{med}}(\vartheta)|^2}. \qquad (2.34)$$

In this case, according to Eq. (2.33), the quantity in Eq. (2.34) is much less than unity, i.e., the intensity

$$W^{(+)}_{\rm pl}(\omega,\vartheta) = \frac{e^2}{2\pi c}\frac{|1-\varepsilon|^2\vartheta^3}{(\gamma^{-2}+\vartheta^2)^2}\left(\frac{\omega a}{c}\right)^2. \qquad (2.35)$$

is small compared to the intensity (2.21) of XTR at the same values of ω, θ, and γ, but larger *a*, such that

$$a \gtrsim |z_{med}(\vartheta)|.$$

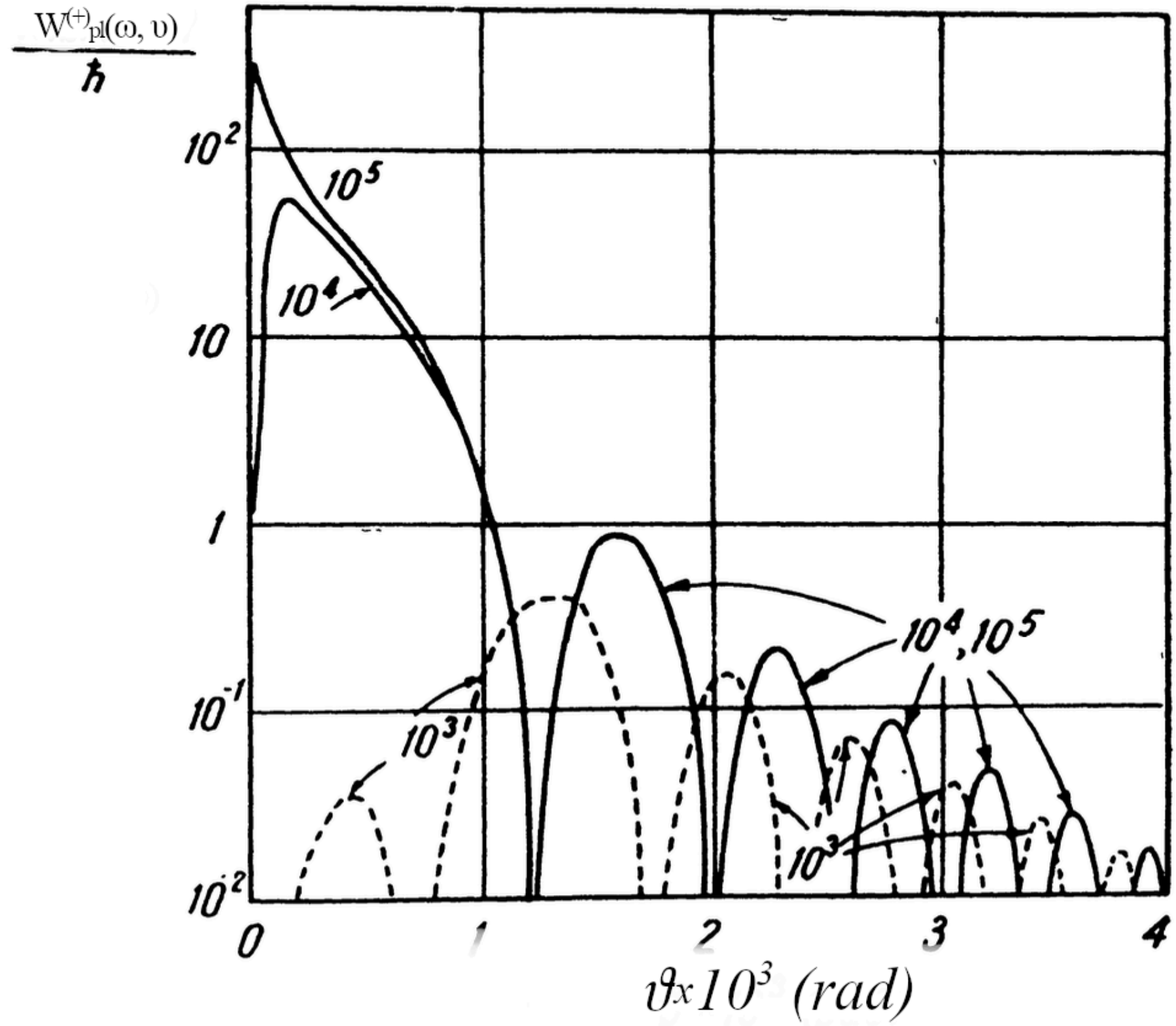

Fig. 2.5. Angular dependence of the spectral–angular distribution of the intensity of XTR+ (forward), produced on a plate, for different values of the particle Lorentz factor γ. The curves are calculated for the frequency $\omega=10^3\ \omega_0$. The numbers near the curves denote γ, $a\omega_0/v = 5\times10$^3, absorption is small

Figure 2.5 shows the angular dependence curves of $W^{(+)}_{pl}(\omega,\vartheta)$ for XTR at several values of γ. Similarly to the case of a single interface (Fig. 1.4), with increasing γ the transition maximum shifts to the left and becomes higher. Interference maxima and minima occur at angles determined by Eqs. (2.28) and (2.29). The positions of these maxima and minima do not depend on γ if $\gamma^{-2} \ll 1 - \varepsilon'$ (solid curves).

## 2.4. Frequency Spectrum of XTR

As a result of integrating Eq. (2.21) over the angle ϑ we obtain the frequency spectrum of forward XTR for a plate in the following form:

$$W_{\mathrm{pl}}^{(+)}(\omega) = (1 + Q)\, W_{\mathrm{b}}^{(+)}(\omega) + W_{\mathrm{int}}^{(+)}(\omega), \qquad (2.36)$$

where $Q = exp(-\eta a)$ is the plate transparency, and $W_b^{(+)}(\omega)$ is the spectral distribution (1.64) of forward XTR for a single interface,

$$W_{\mathrm{int}}^{(+)}(\omega) = \frac{2e^2}{\pi c}\mathrm{Re}\left\{ \exp(iX_g^*)\left[1 + \left(\frac{|g|^2 - \gamma^{-4}}{|g-\gamma^{-2}|^2} - iX_\gamma\right)G(iX_\gamma)\right.\right.$$
$$\left.\left. + \left(\frac{g^*}{g-g^*} - \frac{\gamma^{-2}}{g-\gamma^{-2}}\right)G(iX_g) + \left(\frac{g}{g^*-g} - \frac{\gamma^{-2}}{g^*-\gamma^{-2}}\right)G(iX_g^*)\right]\right\}. \qquad (2.37)$$

$$G(z) = \exp(-z)\mathrm{Ei}(z). \qquad (2.38)$$

Here $g = \gamma^{-2} + 1 - \varepsilon$ (see Eq. (1.65)), and Ei (z) is the exponential integral of a complex argument *z* (see, for example, Ref. [**71**.13]),

$$X_\gamma = \frac{\pi a}{2\bar{z}_{\mathrm{vac}}}, \qquad X_g = \frac{\pi a}{2\bar{z}_{\mathrm{med}}}. \qquad (2.39)$$

$$\bar{z}_{\mathrm{vac}} = \frac{\pi c}{\omega}\gamma^2, \qquad \bar{z}_{\mathrm{med}} = \frac{\pi c}{\omega g}. \qquad (2.40)$$

The quantity $\bar{z}_{vac}$ represents the formation length (2.26) of XTR in vacuum, $z_{vac}(\vartheta)$, taken at $\vartheta = \gamma^{-1}$. Similarly, the quantity $\bar{z}_{med}$ may be regarded as $z_{med}(\vartheta)$ (Eq. (2.25)) taken (formally) at $\vartheta = g^{1/2}$.

The first term in Eq. (2.36) represents the XTR intensity from two separate independent interfaces, taking into account absorption of the radiation from the first interface. The second term $W_{int}^{(+)}(\omega)$ is due to the interference of radiation produced at both interfaces of the plate.

For large plate thickness $a$,

$$a \gg |\bar{z}_{\text{med}}|, \qquad (2.41)$$

the interference terms in Eq. (2.37) oscillate rapidly and make no contribution on average (or simply vanish if $a \gg \eta^{-1}$), so that only the first term remains in Eq. (2.36) (the plate boundaries radiate independently).

When the condition opposite to Eq. (2.41) is satisfied, i.e.

$$a \ll |\bar{z}_{\text{med}}| \ll \bar{z}_{\text{vac}}, \qquad (2.42)$$

after expanding Eqs. (2.36) and (2.37) in powers of the small parameter (Eq. (2.42)), we obtain

$$W_{\text{pl}}^{(+)}(\omega) = \frac{e^2}{8\pi c}\left(\frac{\omega a}{c}\right)^2 |1-\varepsilon|^2 \left(1 - 2C - 2\ln X_\gamma\right), \qquad (2.43)$$

where $C = 0.5772$ is Euler's constant.

In the case, however, when the thickness of the plate is on the order of the formation length,

$$a \sim |\bar{z}_{\text{med}}|, \qquad (2.44)$$

it is necessary to consider the following two cases separately.

(1) The difference between the formation lengths in vacuum and in a medium is large,

$$\bar{z}_{\text{vac}} - |\bar{z}_{\text{med}}| \gtrsim |\bar{z}_{\text{med}}|, \qquad (2.45)$$

(the so-called subcutoff frequency range, see Sec. 1.8).

If the plate is also sufficiently transparent ($\eta a \ll 1$), the spectral distribution (Eq. (2.36)) exhibits interference maxima and minima.

(2) The formation lengths in vacuum and in a medium are almost equal,

$$\bar{z}_{\text{vac}} - |\bar{z}_{\text{med}}| \ll \bar{z}_{\text{vac}}, \qquad (2.46)$$

(the frequency range above the cutoff frequency, see Eq. (1.80)).

Then Eqs. (2.36) and (2.37) can be expanded in powers of the small difference (Eq. (2.46)). As a result, we obtain (for $\eta a \ll 1$)

$$W_{\rm pl}^{(+)}(\omega) = \frac{e^2}{3\pi c}\,|1-\varepsilon|^2\Big\{1-\cos X_\gamma - X_\gamma\Big[2\sin X_\gamma - X_\gamma\big(\cos X_\gamma + 3\,{\rm ci}\,X_\gamma + X_\gamma\,{\rm si}\,X_\gamma\big)\Big]\Big\} \ll \frac{e^2}{\pi c}, \tag{2.47}$$

where ci$X$ and si$X$ are the cosine and sine integrals (see, e.g., Ref. [**71**.13]). For $X_\gamma \ll 1$, i.e., $a \ll \bar{z}_{vac}$, the above equation coincides with Eq. (2.43), which is quite natural, since in this case, by virtue of Eq. (2.46), condition (2.42) is also satisfied.

In the opposite case $X_\gamma \gg 1$, the quantity inside the curly braces in Eq. (2.47) is practically equal to unity. As a result, in the frequency region greater than the cutoff frequency, we obtain twice the spectral intensity of XTR for a single interface (see Eq. (1.82)).

Thus, the spectral distribution (Eq. (2.36)) of XTR is essentially determined by the ratio of the plate thickness $a$ to the formation length $|\bar{z}_{med}|$ in the medium, as well as by the ratio of the formation lengths $|\bar{z}_{med}|$ and $\bar{z}_{vac}$.

In order to clarify the character of the spectral distribution of XTR in various frequency ranges ω and for different values of the particle Lorentz factor γ, it is first necessary to analyze the behavior of the formation lengths $\bar{z}_{vac}$ and $|\bar{z}_{med}|$. In doing so, we will assume that the dielectric permittivity of the medium ε is determined by Eq. (1.53).

The quantity $\bar{z}_{vac}$ (Eq. (2.40)) decreases monotonically as $\omega^{-1}$ with increasing ω, and increases monotonically as $\gamma^2$ with increasing $\gamma$. As for the quantity $|z_{med}|$, in the case under consideration (Eq. (1.58)) we have

$$|\bar{z}_{\rm med}| = \frac{\pi c}{\omega\Big[\big(\gamma^{-2}+\omega_0^2/\omega^2\big)^2 + (\eta c/\omega)^2\Big]^{1/2}}. \tag{2.48}$$

For simplicity, we shall for the time being neglect the quantity $\eta c/\omega$ in comparison with $\omega_0^2/\omega^2$. Then from Eq. (2.48) it follows that the quantity $|\bar{z}_{med}|$ for $\omega \ll \omega_0\gamma$ increases approximately linearly with increasing ω, and at $\omega = \omega_0\gamma$ reaches a maximum value

$$|\bar{z}_{\mathrm{med}}|_{\max} = \frac{\pi c\gamma}{2\omega_0}. \qquad (2.49)$$

and, finally, for $\omega \gg \omega_0\gamma$ it decreases with increasing ω as $\omega^{-1}$. If, however, one takes into account $\gamma_i c/\omega$, then, as already indicated in Sec. 1.9, for sufficiently large γ (see Eq. (1.83)) a new situation arises. In this case, the linear growth of the quantity $|\bar{z}_{med}|$ occurs only in the frequency range

$$\omega \ll \omega_0\gamma.$$

For frequencies

$$\omega_0\gamma < \omega < \eta c\gamma^2 \qquad (2.50)$$

the quantity $|\bar{z}_{med}|$ depends only weakly on the frequency and is equal to $\pi/\eta$ (if the absorption coefficient η depends only weakly on the frequency in this range). Finally, for even higher frequencies

$$\max\{\omega_0\gamma, \quad \eta c\gamma^2\} \ll \omega \qquad (2.51)$$

the quantity $|\bar{z}_{med}|$ decreases inversely with ω. In case (2.51) the formation lengths $|\bar{z}_{med}|$ and $\bar{z}_{med}$ practically coincide (i.e., Eq. (2.46) holds), whereas at lower frequencies $|\bar{z}_{med}|$ is noticeably smaller than $\bar{z}_{vac}$ (i.e., Eq. (2.45) holds).

Figure 2.6 shows the corresponding curves of the frequency dependence of $\bar{z}_{vac}$ and $|\bar{z}_{med}|$, calculated for carbon and lead. The same figures also show the curves of the formation length of bremsstrahlung $\bar{z}_{brem}$, taking into account multiple scattering, which will be discussed in Chapter IV.

Having clarified the behavior of $\bar{z}_{vac}$ and $|\bar{z}_{med}|$, it is not difficult to analyze the character of the spectral distribution of XTR for various values of ω, γ, and $a$. To avoid repeating the single-boundary case, we shall assume that the plate has finite transparency (i.e., condition (2.32) is not satisfied), at least starting from some frequency.

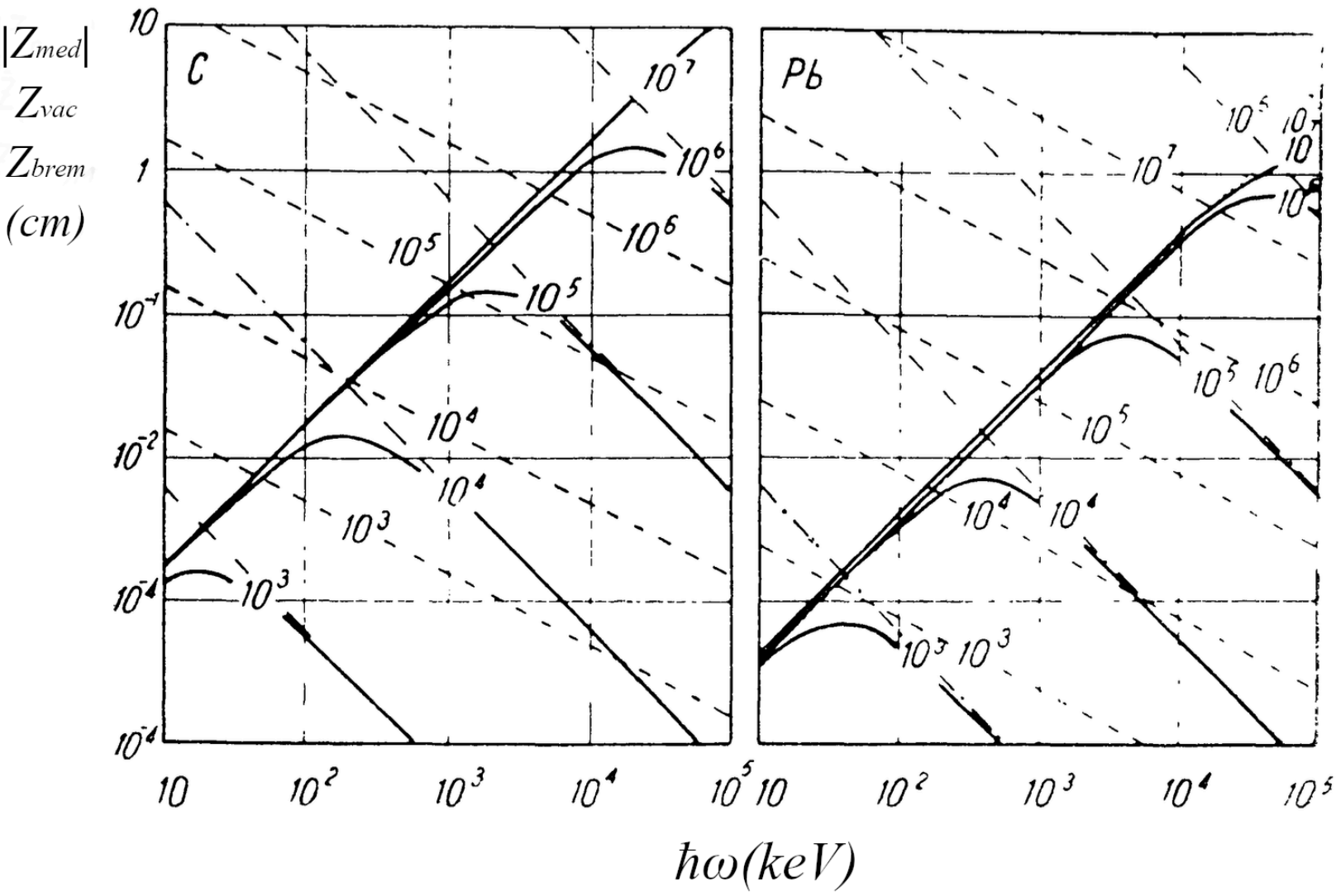

*Fig. 2.6. Curves of the formation lengths of XTR $|\bar{z}_{med}|$ (solid curves) and $\bar{z}_{vac}$ (dash-dotted curves) versus frequency. The dashed curves correspond to the formation length of bremsstrahlung $z_{brem}$.*

A. In the case when

$$\gamma \lesssim \frac{a\omega_0}{c}, \tag{2.52}$$

for all frequencies the formation length $|\bar{z}_{med}|$ is of the order of the plate thickness $a$ or smaller, therefore the XTR frequency spectrum has interference maxima and minima and extends up to frequencies of the order of $\omega_b = \omega_0\gamma$, as long as $|\bar{z}_{med}|$ differs significantly from $\bar{z}_{vac}$, after which the spectrum drops sharply (see Eq. (2.47)).

B. When, however, the following condition holds:

$$\frac{a\omega_0}{c} \ll \gamma, \tag{2.53}$$

it is necessary to consider three frequency regions separately.

(1) In the frequency range

$$\omega < \frac{a\omega_0^2}{c} \qquad (2.54)$$

the formation length $|\bar{z}_{med}|$ is smaller than the plate thickness, the XTR spectrum has interference maxima and minima.

(2) In the region of higher frequencies

$$\frac{a\omega_0^2}{c} \ll \omega < \frac{2c}{a}\gamma^2 \qquad (2.55)$$

the condition in Eq. (2.42) is satisfied, when the frequency spectrum is determined by Eq. (2.43), where

$$|1-\varepsilon|^2 = \frac{\omega_0^4}{\omega^4} + \left(\frac{\eta c}{\omega}\right)^2. \qquad (2.56)$$

It follows that in the absence of absorption in the material of the plate (i.e., for $\eta = 0$) the frequency spectrum would fall off sufficiently rapidly, namely as $\omega^{-2}$ over the entire frequency range (Eq. (2.55)). However, in the case of a finite absorption index $\eta$ such a decrease of the frequency spectrum takes place only up to frequencies of the order of $\omega_0\gamma_0$, after which the spectral intensity of the radiation is approximately equal to $e^2a^2\eta^2/(8\pi c)$. Since at such frequencies the absorption index $\eta$ depends weakly on frequency, a characteristic "plateau" appears in the frequency spectrum, which extends approximately from $\omega_0\gamma^2$ to $2c\gamma^2/a$. The height of this "plateau" is proportional to $(\gamma a)^2$ and depends logarithmically on $\gamma$.

(3) Finally, in the region of very high frequencies

$$\frac{2c\gamma^2}{a} < \omega \qquad (2.57)$$

the lengths $|\bar{z}_{med}|$ and $\bar{z}_{vac}$ are almost equal (i.e., Eq. (2.46) holds), and the frequency spectrum again drops rapidly as $e^2\eta^2c/(3\pi\omega^2)$.

In practice, it is often important to estimate the frequencies at which the XTR spectrum has maxima. Let condition (2.49) be satisfied and the absorption be small. If, in addition, $a \gtrsim \bar{z}_{vac}$, then inequality (2.41) holds; that is, the interference maxima and minima alternate very frequently, and on average the plate radiates as two independent borders. If, however, $a \ll \bar{z}_{vac}$, , it is easy to see that the dominant term in the interference contribution (2.37) is

$$\frac{2e^2}{\pi c} \ln X_\gamma \cos |X_g|.$$

Therefore, the maxima in the transition radiation spectrum correspond to the value -1 of $\cos|X_g|$, i.e., they occur at frequencies [**71**.3]

$$\omega_n^{\rm pl} \approx \frac{a\omega_0^2}{2c\pi(2n+1)}, \qquad (n = 0, 1, 2, \ldots). \qquad (2.58)$$

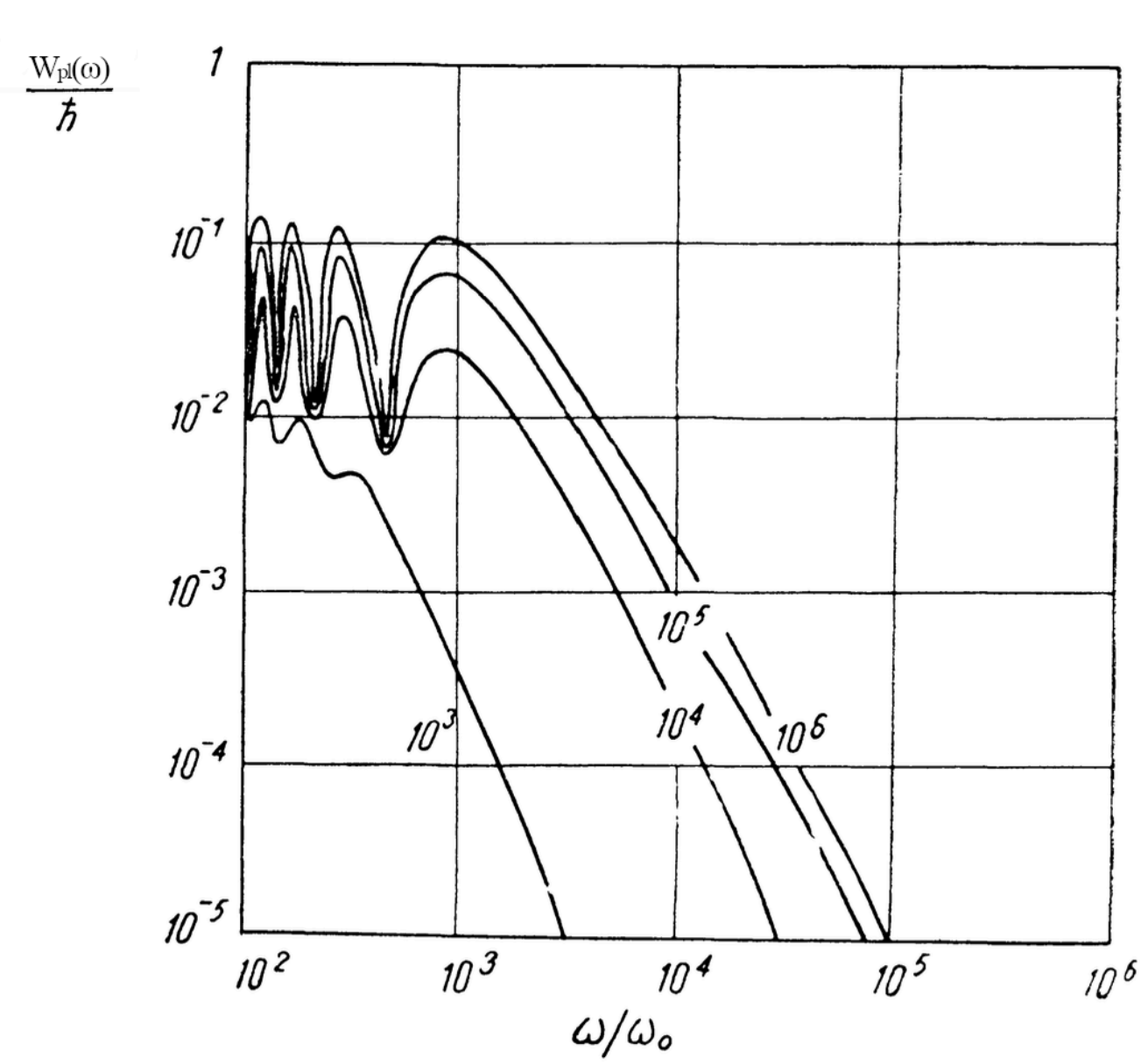

Fig. 2.7. Frequency spectrum of XTR generated in a plate; $a\omega_0/c = 5 \cdot 10^3$, absorption is small; the numbers by the curves indicate the particle Lorentz factor

Figure 2.7 shows the XTR frequency spectra in case of a plate. Comparing these spectra with analogous spectra in case of a single interface, we observe that the former have characteristic interference maxima and minima in the frequency range (Eq. 2.54), for Eq. 2.53.

Positions of the maxima in Fig. 2.7 for $\gamma \gg 10^4$ is well corresponding with the valued defined by Eq. 2.58.

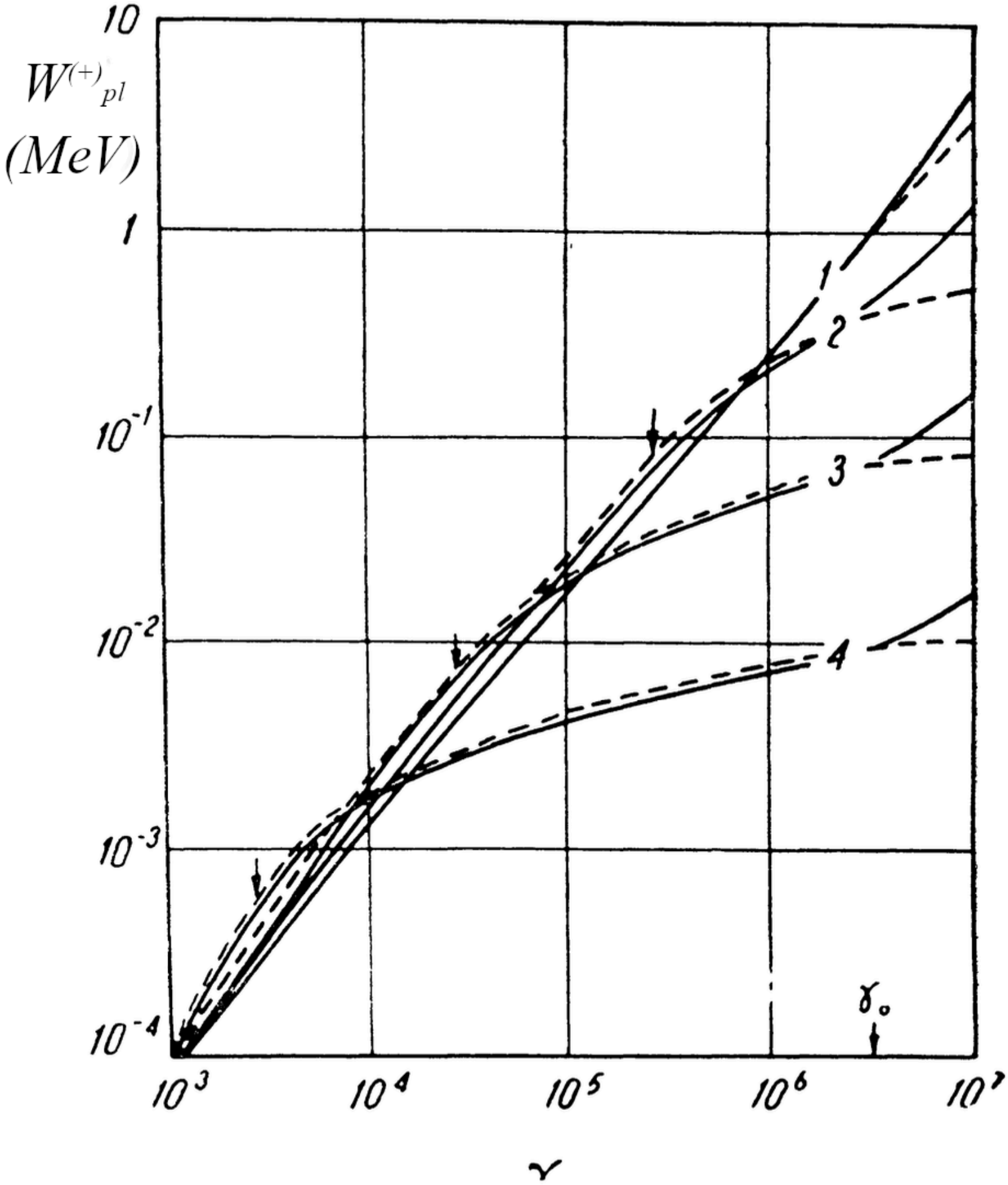

Fig. 2.8. Dependence of the total forward XTR energy for a plate (Pb) on the particle Lorentz factor $\gamma$ (the dashed curves were calculated without taking absorption in the medium into account): curve 1—infinitely large thickness (i.e., the case of a medium–vacuum interface), 2—$a = 10^3$ μm, 3—$a = 10^2$ μm, 4—$a = 10$ μm. The arrows on the curves indicate the values $\gamma = a\omega/c$.

Let us not consider the *total energy* of XTR integrated along all frequencies in the case of the plate (Eq. 2.8). In contrast with the case of a single interface for sufficiently large particle Lorentz factors $\gamma$ (when conditions (2.53) and (2.42) are satisfied), the frequency spectrum extends approximately up to a frequency of order $a\omega_0^2/(2\pi c)$ (see Eq. (2.58)), practically independent of $\gamma$. Therefore, the $\gamma$-dependence of the total energy of XTR, starting from values

$$\gamma > a\omega_0/c, \tag{2.59}$$

becomes very *weak* (approximately logarithmic). With a further increase of $\gamma$, when $\gamma > \gamma_0$ (see Sec. 1.9), a steeper increase should be observed owing to absorption.

## § 3. REGULAR STACK OF PLATES

As already noted earlier (see Secs. 1.7 and 1.8), the intensity (as well as the number of quanta) of transition radiation (within a frequency range $\Delta\omega \sim \omega$) emitted at a single interface between media is comparatively small. In the case of a plate, under the most favorable conditions, at certain frequencies and emission angles the intensity can be only four times larger. To achieve a substantial increase in the intensity (number of quanta) of transition radiation, it is necessary to turn to a system with a large number of interfaces, for example, a *stack of plates*.

In this section we shall consider transition radiation emitted by a fast charged particle upon perpendicular traversal through a stack of identical plates, regularly spaced in vacuum and parallel to one another.

### 3.1. General Equation for the Spectral–Angular Distribution of Radiation Intensity

As in Secs. 1 and 2, the total field of the problem is the sum of the particle field and the radiation field. According to the Sommerfeld condition (Sec. 1.4), up to the first boundary (the left boundary of the first plate) there exists only the backward-propagating radiation field, and beyond the last boundary (the right boundary of the *N*th plate) there exists only the forward-propagating radiation field. Inside the plates and in the gaps between the plates of the stack there are radiation fields of both types. In the case of a regular stack it is possible, from the corresponding continuity ("matching") conditions, to obtain recurrence relations expressing, for example, the fields in the *n*th gap in terms of the fields in the preceding $(n-1)$th gap. By successively applying these recurrence relations, one can express the field in the *N*th "gap" (after the last, *N*th plate) in terms of the field in the zeroth "gap" (before the first plate). After solving these equations, we obtain the radiation fields outside the stack; their exact expressions are rather cumbersome and therefore are not given here (see Refs. [**58**.2, **69**.10]). We give only the equation for the spectral–angular distribution of the radiation intensity propagating forward beyond the stack (in the *N*th "gap") [**63**.8, **69**.11]*[8]:

$$W_{\mathrm{st}}^{(+)}(\omega,\vartheta)=\frac{2e^2\beta^2}{\pi c}\frac{u_0^2\sin^3\vartheta}{(1-u_0^2)^2}\left|\frac{\varepsilon-1}{1+u^2}\right|^2|F_N(\omega,\vartheta)|^2, \tag{3.1}$$

$$F_N(\omega,\vartheta)=\frac{1}{2Q_N}\left\{2BU_{N-1}+\frac{B+B_1\exp[ib_0(1+u_0)]}{\cos(a_0+b_0)-\zeta}\times\right.$$

$$\left.\times[U_{N-2}-U_{N-1}\exp(-i(a_0+b_0))+\exp(-iN(a_0+b_0))]\right\}, \tag{3.2}$$

$$\zeta=\cos(a_0u)\cos(b_0u_0)-\frac{1}{2}\left(\frac{u}{\varepsilon u_0}+\frac{\varepsilon u_0}{u}\right)\sin(a_0u)\sin(b_0u_0),$$

$$Q_N=Q_1U_{N-1}-4U_{N-2}\varepsilon uu_0\exp[i(a_0u+b_0u_0)], \tag{3.3}$$

$$B_1=2u(1+u_0)(1-u_0-\beta^2\varepsilon)\exp[ia_0(1+u)]-$$

$$-(\varepsilon u_0+u)(1+u)(\gamma^{-2}+u)\exp(2ia_0u)+(\varepsilon u_0-u)(1-u)(\gamma^{-2}+u),$$

$$a_0=\frac{a\omega}{v},\qquad b_0=\frac{b\omega}{v},$$

*a*—the thickness of each plate, *b*—the width of the vacuum gap, ε—the dielectric permittivity of the plate material. The quantities $B$, $Q_1$, $u$ and $u_0$ are defined by Eqs. (2.7) and (2.8); $U_m\equiv U_m(\zeta)$ are Chebyshev

---

[8] In Ref. [**58**.2] the radiation fields were obtained correctly; however, the analysis of the radiation intensity was carried out incorrectly.

polynomials of the second kind with argument ζ (generally complex), which satisfy the recurrence relation

$$U_N - 2\zeta U_{N-1} + U_{N-2} = 0. \quad (3.4)$$

The equation for the spectral–angular distribution of the intensity of radiation propagating backward far from the stack can be obtained from Eq. (3.1) by replacing $v$ with $-v$. It is easy to see that for $N = 1$ the quantity contained in the square brackets in Eq. (3.2) vanishes, and Eq. (3.1) naturally reduces to Eq. (2.6) for a single plate.

## 3.2. Parametric Cherenkov Maxima

The second term of the quantity $F_N(\omega, \vartheta)$ appearing in Eq. (3.1) has the denominator

$$T_0 = \cos(a_0 + b_0) - \zeta. \quad (3.5)$$

The presence of this denominator is due to the interference of the radiation produced by a charged particle as it passes through different plates of a regular stack. When the denominator $T_0$ is small, the quantity $W_{st}^{(+)}(\omega, \vartheta)$ can become large [**74**.23].

It can be shown that the second term of the quantity $F_N(\omega, \vartheta)$, for small values of $|T_0|$, has the following form:

$$\frac{\sin[NT_0/\sin(a_0 + b_0)]}{T_0/\sin(a_0 + b_0)} \quad (3.6)$$

(it is assumed that absorption is small). It follows that when the number of plates in the stack is sufficiently large, sharp maxima may appear in the radiation intensity at those frequencies and angles that minimize $|T_0|$. The height of these maxima is proportional to $N^2$ and is also determined by the preceding factors, whereas the width of the maxima is proportional to $1/N$. Note that the equation $T_0 = 0$, or

$$\cos(a_0+b_0)=\cos(a_0u)\cos(b_0u_0)-\frac{1}{2}\left(\frac{u}{\varepsilon u_0}+\frac{\varepsilon u_0}{u}\right)\sin(a_0u)\sin(b_0u_0) \tag{3.7}$$

for ε" = 0, corresponds to the condition for the occurrence of the so-called "*parametric Cherenkov radiation*" in an infinite non-absorbing periodic layered medium [**57**.1].

From Eq. (3.5) for $T_0$, as well as from Eq. (3.1), it follows that for $\gamma \gg 1$ the positions and magnitudes of the aforementioned ("parametric Cherenkov") maxima practically do not depend on γ.

The general solution of Eq. (3.7) cannot be obtained in explicit form. In the special case when

$$|a_0u|\ll 1, \quad b_0u_0\ll 1 \tag{3.8}$$

(practically, this means that the radiation wavelength is much larger than the period of the stack). The quantity $T_0$ can therefore be expanded in powers of the small parameters given by Eq. (3.8). As a result, we find that the minimum of $|T_0|$ is attained at $\vartheta = \vartheta_{PC}$, where

$$\sin^2\vartheta_{\mathrm{PC}}=\frac{\varepsilon' a_0^2+b_0^2+(1+\varepsilon')a_0b_0-(a_0+b_0)^2\beta^{-2}}{a_0^2+b_0^2+(1+\varepsilon'/2)\,a_0b_0/\varepsilon'}, \tag{3.9}$$

$$|\varepsilon''|\ll 1.$$

In order for the solution $\vartheta_{PC}$ of Eq. (3.9) to have physical meaning, the following conditions must be satisfied

$$1+\frac{(1-\varepsilon')^2}{\varepsilon'}\cdot\frac{a_0b_0}{(a_0+b_0)^2}\geqslant\bar{\varepsilon}'-\beta^{-2}\geqslant 0, \tag{3.10}$$

where

$$\bar{\varepsilon}'=\frac{a_0\varepsilon'+b_0}{a_0+b_0}. \tag{3.11}$$

Equations (3.9) and (3.10) are, respectively, the analogs of Eq. (1.40) for the Cherenkov angle and the conditions (1.41) for the occurrence of Cherenkov radiation.

In another special case, when

$$|\varepsilon-1|\ll 1, \tag{3.12}$$

after the corresponding expansion of $T_0$ in powers of the small parameter given by Eq. (3.12), we obtain that the minimum of $|T_0|$ is attained under the condition $\vartheta = \vartheta_n$, where [**60**.2]

$$\cos\vartheta_n = \frac{1}{\sqrt{\overline{\varepsilon'}}}\left(\frac{1}{\beta} + \frac{2n\pi c}{(a+b)\omega}\right), \qquad (3.13)$$

$\overline{\varepsilon'}$ is defined by Eq. (3.11), $n = 0, \pm 1, \pm 2, \dots$. The values of $n$ are such that the right-hand side of Eq. (3.13) does not exceed unity in absolute value. From Eq. (3.13) it follows that this equation also has a solution in the case $\varepsilon' < 1$, for example in the X-ray frequency range. Moreover, if the quantity $a_0 + b_0$ is sufficiently large, Eq. (3.13) may have several solutions corresponding to different values of $n$.

Thus, the parametric Cherenkov maxima arising in a sufficiently "thick" stack (i.e., when $N \gg 1$) are indeed, in a certain sense, analogs of the Cherenkov maximum arising in a sufficiently thick plate ($a_0 \gg 1$).

However, in contrast to the latter, the parametric Cherenkov maxima at a given frequency, in general (when the conditions (3.8) are not satisfied), may simultaneously arise at different angles. In addition, they may arise both for $\varepsilon' > 1$ and for $\varepsilon' < 1$.

Equations (3.9) and (3.13) determine the positions of the parametric Cherenkov maxima in cases (3.8) and (3.12), respectively. Expressions for the intensities of these maxima are obtained from Eq. (3.1) with account of the specified conditions. In the case of Eq. (3.8), after expansion in powers of $a_0$, $b_0$, $a_0u$, and $b_0u_0$, one obtains an expression that to some extent resembles Eq. (2.6) for Cherenkov and transition radiation in a single plate. Case (3.12) will be considered in detail in the following sections.

In the general case, the positions of the parametric Cherenkov maxima must be determined numerically from the condition of minimizing $|T_0|^2$, where $T_0$ is given by Eq. (3.5), and the radiation intensity must be calculated directly from Eq. (3.1).

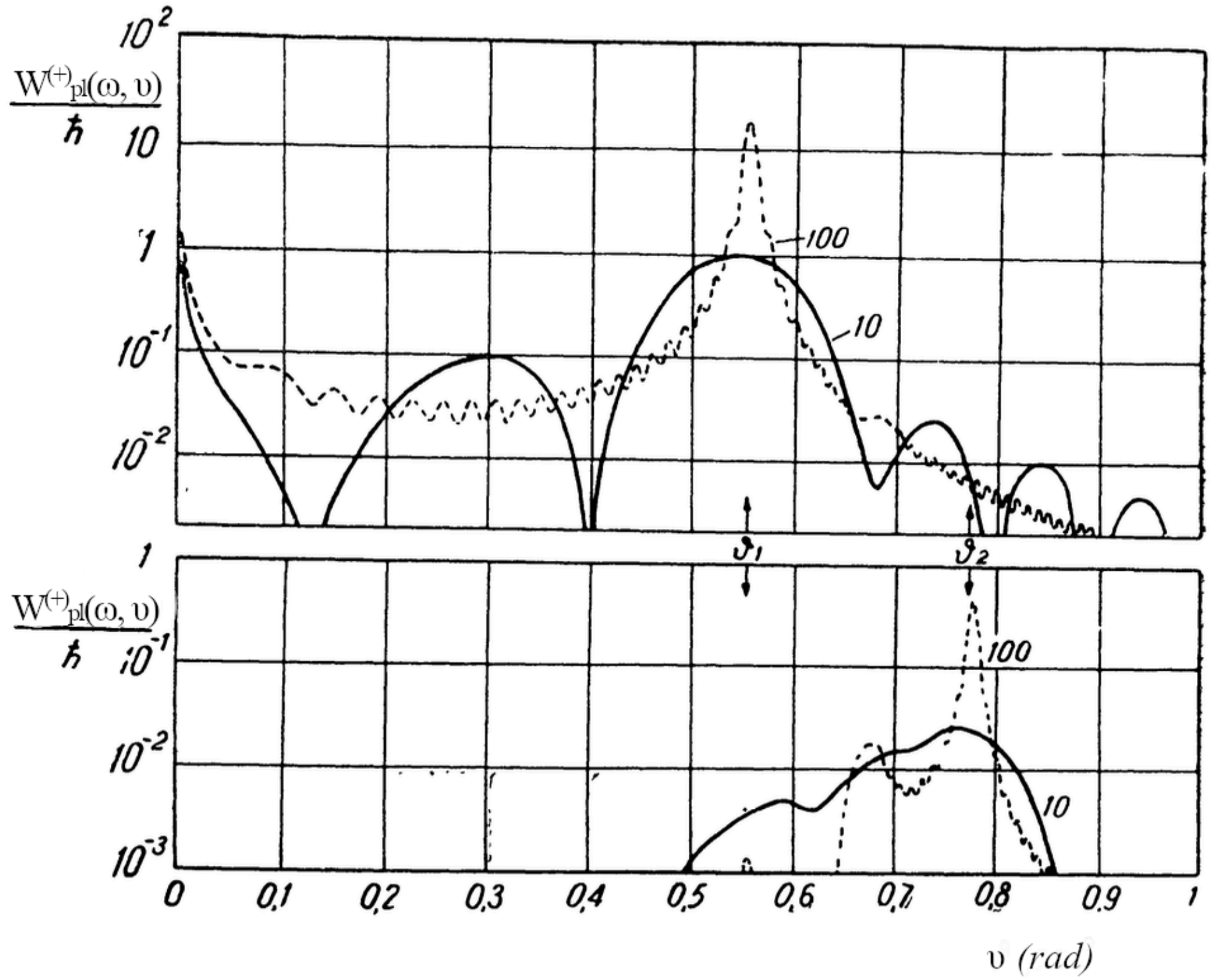

Fig. 3.1. Angular dependence of the spectral–angular distribution of the intensity of transition and Cherenkov radiation formed in a stack, forward (upper plot) and backward (lower plot). Stack parameters: $a\omega/v = b\omega/v = 5$, $\varepsilon' = 1.6$, $\varepsilon'' = 0.01$, $\gamma = 10^3$. The numbers by the curves denote the number $N$ of plates in the stack. The angles $\vartheta_1$ and $\vartheta_2$ correspond to the minimum of $|T_0|^2$.

For illustration, Fig. 3.1 shows the results of such a calculation. It is clearly seen that for a sufficiently large number of plates in the stack sharp parametric Cherenkov radiation maxima arise at the angles $\vartheta_1$ and $\vartheta_2$, both in the forward and backward directions.

### 3.3. X-Ray Frequency Range

Let conditions (3.12), $\gamma \gg 1$, and $\vartheta \ll 1$ be satisfied. Then the quantity (3.2) can be expanded in powers of the small parameters $|1-\varepsilon|$, $\gamma^{-1}$, and $\vartheta$. If, in addition, we assume that

$$N|1-\varepsilon| \ll 1, \tag{3.14}$$

which physically means the smallness of the reflected waves compared to those propagating forward, then for the forward-propagating XTR we obtain[9]

$$W_{\rm st}^{(+)}(\omega,\vartheta) = W_{\rm b}^{(+)}(\omega,\vartheta)\, F_{\rm pl}\, F_{\rm st} = W_{\rm pl}^{(+)}(\omega,\vartheta)\, F_{\rm st}, \tag{3.15}$$

where $W_b^{(+)}(\omega, \vartheta)$ is defined by Eq. (1.59), the interference factor $F_{pl}$ for a plate by Eq. (2.22), and the factor $F_{st}$, which describes the interference of radiation produced on different plates of the stack, is defined by

$$F_{\rm st} = F_{\rm st}(X) = \frac{(1-Q^{N/2})^2 + 4Q^{N/2}\sin^2 NX}{(1-Q^{1/2})^2 + 4Q^{1/2}\sin^2 X}. \tag{3.16}$$

$$X = \frac{\pi}{2}\,{\rm Re}\left(\frac{a}{z_{\rm med}(\vartheta)} + \frac{b}{z_{\rm vac}(\vartheta)}\right). \tag{3.17}$$

(the quantities $Q$, $z_{med}(\vartheta)$, $z_{vac}(\vartheta)$ are defined by Eqs. (2.23), (2.25), and (2.26), respectively).

When the plate thickness $a$ is much larger than the absorption length $\eta^{-1}$, i.e., $Q \ll 1$, the factor $F_{\rm st}$ is practically equal to unity and does not depend on $\omega$ and $\vartheta$. Then Eq. (3.15) reduces to Eq. (2.21) for a single plate, which in this case, as noted in Sec. 2.3, reduces to the equation for a single interface ($F_{pl}=F_{st}=1$).

Let now

$$1 - Q^{N/2} \ll 1 \quad \text{or} \quad N\eta a \ll 1, \tag{3.18}$$

[9] As will be shown in Sec. 6.1, Eq. (3.15) can also be obtained directly from the general equation (6.9) for an irregular stack when the irregularity of the stack is zero.

i.e., the entire stack of plates is weakly absorbing with respect to radiation at the frequency under consideration. Then the interference factor of the stack, $F_{st}$, takes the form

$$F_{st} = \frac{\sin^2 NX}{\sin^2 X},$$

and Eq. (3.15) coincides with the corresponding equation for a non-absorbing layered medium, first found in Refs. [**60**.3, **60**.5].

When the distance between plates is much larger than the XTR formation length in vacuum, i.e.,

$$b \gg z_{\text{vac}}(\vartheta), \qquad (3.19)$$

the factor $F_{st}$ is a rapidly oscillating function of $\omega$ and $\vartheta$. In this case, after averaging over a small interval of frequencies or angles we obtain

$$W_{\text{st}}^{(+)}(\omega, \vartheta) = W_{\text{pl}}^{(+)}(\omega, \vartheta)\,\overline{F}_{\text{st}}, \qquad (3.20)$$

where

$$\overline{F}_{\text{st}} = \frac{1}{\pi}\int_{-\pi/2}^{\pi/2} F_{\text{st}}(X)\,dX. \qquad (3.21)$$

If absorption in the plate is completely absent, i.e., $\eta = 0$ or $Q = 1$, then it is easy to verify that

$$\overline{F}_{\text{st}} = N \quad (\eta = 0). \qquad (3.22)$$

In other words, in the case of a stack of non-absorbing plates spaced far apart from one another (condition in Eq. (3.19)), after averaging over a small interval of emission angles or frequencies there is a simple addition of the intensities (numbers of quanta) of the radiations produced in each plate of the stack. This result naturally generalizes to the case of a stack of plates with finite absorbing ability. At the same time, however, it is necessary to take into account the absorption of the radiation as it passes through the subsequent plates. Indeed, the quantity $\overline{F}_{st}$, defined according to Eq. (3.21), coincides to high accuracy with the so-called effective number of plates in the stack, equal to the sum of the first $N$ terms of a geometric progression with common ratio $a$:

$$N_{\text{eff}} = \frac{1-Q^N}{1-Q}. \qquad (3.23)$$

It is easy to see that $N_{eff} \to N$ as $Q \to 1$ $(a\eta \ll 1)$ and $N_{eff} \to 1$ as $Q \to 0$ $(a\eta \gg 1)$.

Let us now consider another limiting case, when

$$\mathrm{Re}\left[\frac{a}{z_{\text{med}}(\vartheta)}\right] \ll 1, \quad b \ll z_{\text{vac}}(\vartheta). \qquad (3.24)$$

In addition, by virtue of Eq. (3.18), we also have $a \ll \eta^{-1}$.

Then, after expanding the quantities $Q$, $X$, and $Y$, it is easy to show that the spectral–angular distribution given by Eq. (3.15) coincides with the analogous distribution for a single *equivalent plate* of thickness $a_{eq} = N(a + b)$, with the real part of the dielectric permittivity

$$\varepsilon'_{\text{eq}} = \frac{a\varepsilon' + b}{a+b}. \qquad (3.25)$$

and the absorption coefficient

$$\eta_{\text{eq}} = \frac{a\eta}{a+b}. \qquad (3.26)$$

In the general case, due to the presence of the stack interference factor $F_{st}$, the spectral–angular distribution (Eq. (3.15)), in comparison with the analogous distribution (Eq. (2.21)) for a single plate, has a number of additional maxima, which are nothing but the “parametric Cherenkov maxima” noted in Sec. 3.2, only in the region of high (X-ray) frequencies. These maxima are determined by the condition $X = m\pi$, i.e., taking into account Eqs. (3.17), (2.25), and (2.26),

$$\frac{\omega(a+b)}{4\pi v}\left[(1-\varepsilon')\frac{a}{a+b} + \gamma^{-2} + \vartheta^2\right] = m, \qquad (3.27)$$

where $m$ is an integer such that the frequency $\omega$ and angle $\vartheta$ satisfying Eq. (3.27) have physical meaning, i.e., are real (and positive). The condition in Eq. (3.27) (for a non-absorbing stratified medium) was first obtained by Ter-Mikaelian [**60**.2] from the laws of conservation of energy and momentum and was called by him the "*resonance condition*" (see also Ref. [**69**.1]). When this condition is satisfied, the factor $F_{st}$ attains its maximum value

$$F_{\rm st}^{\rm max} = \left( \frac{1 - Q^{N/2}}{1 - Q^{1/2}} \right)^2 . \qquad (3.28)$$

In the absence of absorption

$$F_{st}^{max} = N^2,$$

and from this it is seen that when the resonance condition (3.27) is satisfied, the plates of the stack radiate coherently, so that the radiation intensity at a given frequency ω and a given angle ϑ is proportional to the square of the number of plates. When $N \gg 1$ (more precisely, $N_{eff} \gg 1$), the maxima determined by condition (3.27) are very sharp and indeed have a resonant character. The width (both angular and spectral) of these maxima, as seen from Eq. (3.16), is proportional to $1/N_{eff}$. Therefore, after integration, for example over ϑ, the intensity is proportional to $N_{eff}$ (see Sec. 3.4 for details, in particular Eq. (3.31)).

The radiation corresponding to the maxima determined by condition (3.27) may be called "resonant transition radiation" [**78**.1]. In this case different maxima can be regarded as "harmonics" with different indices *m*. From the requirement that real values ω and ϑ must satisfy condition (3.27), it follows that each harmonic arises only when the particle Lorentz factor γ exceeds a certain threshold value, which is larger the smaller the harmonic number *m*, and the harmonic itself (if it arises) can lie only within a certain frequency range determined by the stack parameters, the particle γ-factor, and the harmonic number *m*.

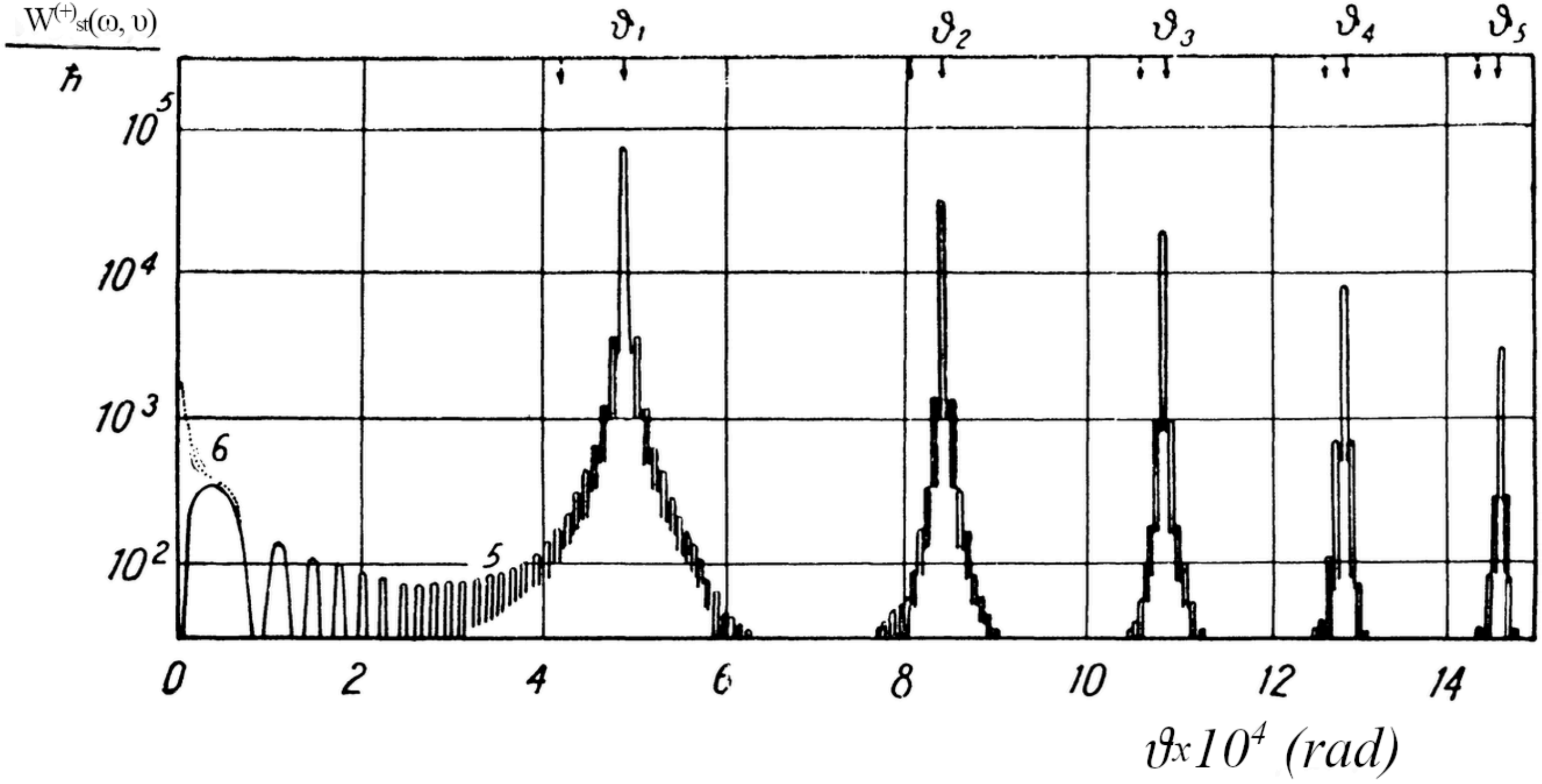

Fig. 3.2. Angular dependence of the spectral–angular distribution of the XTR intensity formed in a stack with parameters a = 30 μm, b = 500 μm, N = 50, $\hbar\omega_0 = 20$ eV. Radiation quantum energy $\hbar\omega = 10$ keV, $\gamma = 2.21$ cm$^{-1}$. The solid curve corresponds to the case $\gamma = 10^5$, the dotted

curve to $\gamma = 10^6$. In the region $\vartheta > 8 \cdot 10^{-5}$ rad the curve does not depend on $\gamma$ for $\gamma \geq 10^5$

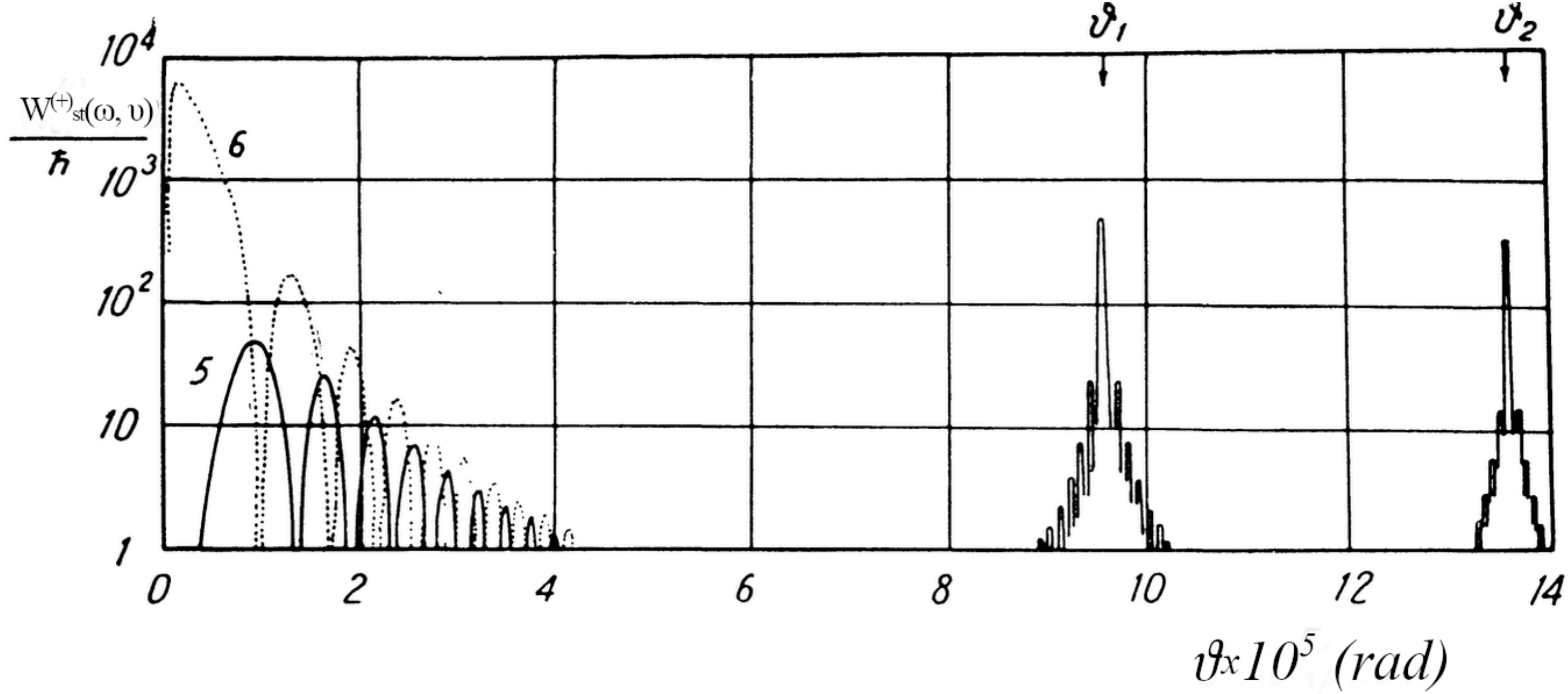

Fig. 3.3. The same as in Fig. 3.2, for $\hbar\omega = 500$ keV and $q = 8.67 \cdot 10^{-2}$ $cm^{-1}$.

Typical curves of the angular dependence of the quantity $W_{st}^{(+)}(\omega, \vartheta)$ (see Eq. (3.15)) are shown in Figs. 3.2 and 3.3. The selected frequencies ($\hbar\omega$ = 10 and 500 keV) correspond to maxima of the frequency spectrum (see Fig. 3.4), caused by interference of different types (see Sec. 3.5 for details). It can be seen from Figs. 3.2 and 3.3 that the curves of the spectral–angular radiation intensity contain a series of sharp resonance maxima that practically do not depend on $\gamma$ for all $\gamma \lesssim 10^4$ (Fig. 3.2) or $\gamma \gtrsim 10^5$ (Fig. 3.3). The positions of these maxima are determined by Eq. (3.27) for $m = 1, 2, ...$ and are indicated in the figures by solid arrows. In Fig. 3.2 the positions of the resonance maxima for $\gamma = 10^3$, obtained from Eq. (3.27) for $m = 3, 4, ...$, are marked by dashed arrows. The harmonics with indices $m = 2$ and $m = 1$ arise for $\gamma > 1190$ and $\gamma > 2040$, respectively.

Thus, when the Lorentz factor $\gamma$ is sufficiently large so that all harmonics arise (starting from $m = 1$ for the given radiation frequency), no new harmonics appear with further increase of $\gamma$; on this basis one usually speaks of *saturation* of resonant transition radiation. However, it should be noted that saturation holds approximately only in the limiting case of a stack consisting of a very large number of non-absorbing plates. In a real stack consisting of a finite number of plates with finite (though small) absorption, however, saturation does not occur, because the spectral–angular intensity given by Eq. (3.15) of the radiation at angles $\vartheta \sim \gamma^{-1}$ (for large $\gamma$ these angles lie between zero and the first resonance maximum) continues to *increase* with increasing $\gamma$ (see Figs. 3.2 and 3.3). For finite values of $N_{eff}$, the contribution of this part of the radiation is not always negligible compared with the contributions of the aforementioned resonance maxima. Under conditions in Eq. (3.24), i.e., when the entire finite stack radiates as a single equivalent plate, the contribution of the radiation at $\vartheta \sim \gamma^{-1}$ becomes quite significant, even decisive (see Fig. 3.3 and Secs. 3.4 and 3.5).

It should be emphasized once again that the aforementioned maxima ("resonances") occur in the *spectral–angular* distribution of the intensity, i.e., in the angular dependence of the intensity at a fixed frequency or, conversely, in the frequency dependence at a fixed angle.

After integrating over one of these variables, for example over the angle $\vartheta$, the intensity distribution practically ceases to exhibit a resonant character.

## 3.4. Integration over the Radiation Angle

In order to obtain the frequency spectrum of XTR emitted from a stack, it is necessary to integrate the spectral–angular distribution (Eq. (3.15)) over the radiation angle $\vartheta$. For small values of $N_{eff}$ such integration can, generally speaking, be carried out only numerically. However, it turns out that if

$$N_{\text{eff}} \gg 1, \qquad (3.29)$$

under certain conditions the integration of Eq. (3.15) over $\vartheta$ can be carried out analytically with sufficiently good accuracy. The calculation method uses the fact that, as noted in Sec. 3.3, when condition in Eq. (3.29) holds, the factor $F_{st}$ exhibits sharp maxima. If the function $W_n^{(+)}(\omega, \vartheta)$ that multiplies $F_{st}$ in Eq. (3.15) varies only slightly within the width of a single maximum, then in the integration this function may be replaced by its value at the point where the given maximum is located. Thus, instead of integrating the entire expression (3.15) over $\vartheta$, it remains only to sum over the maxima of the factor $F_{st}$ with the corresponding weights. Analysis shows that when condition in Eq. (3.29) is satisfied, the function $W_n^{(+)}(\omega, \vartheta)$ varies only slightly within the widths of all maxima of the factor $F_{st}$, except, perhaps, for the first maximum corresponding to the smallest value of the angle $\vartheta_1$. If, in addition to Eq. (3.29), the condition [**71**.3, **74**.1] is also satisfied:

$$N_{\text{eff}}(a+b) \gg \overline{z}_{\text{vac}}, \qquad (3.30)$$

then the function $W_n^{(+)}(\omega, \vartheta)$ is "smooth" even within the width of the first maximum, and the frequency spectrum of the radiation is given by the following equation [**60**.3, **60**.5, **71**.3]

$$W_{\text{st}}^{(+)}(\omega) = \int W_{\text{st}}^{(+)}(\omega, \vartheta)\, d\vartheta = \frac{4e^2 N_{\text{eff}} |1-\varepsilon|^2}{\omega(a+b)} \times$$

$$\times \sum_{n=1,2,3,\ldots} \vartheta_n^2 \left\{ \frac{(1-Q^{1/2})^2 - 4Q^{1/2}\sin^2\left|\frac{\omega a}{4c}\left(\vartheta_n^2+\gamma^{-2}+1-\varepsilon'\right)\right|}{(\vartheta_n^2+\gamma^{-2})^2\left|\vartheta_n^2+\gamma^{-2}+1-\varepsilon\right|^2} \right\}. \quad (3.31)$$

where $N_{\mathrm{eff}}$ and Q are defined by expressions (3.23) and (2.23),

$$\vartheta_n^2 = \frac{4\pi c}{\omega(a+b)}(n-d), \quad (3.32)$$

$$d = A - \mathrm{entier}\,(A),$$

$$A = \frac{\omega(a+b)}{4\pi c}\left[(1-\varepsilon')\frac{a}{a+b} + \gamma^{-2}\right], \quad (3.33)$$

entier ($A$) means the highest integer not exceeding $A$. According to Eq. (3.33), $0 \le d < 1$.

Equation (3.31) is a sum over “harmonics” (Sec. 3.3). In this case the harmonic number m corresponds to n + entier (A). From Eq. (3.31) it follows that for

$$\vartheta_n^2 \gg |\gamma^{-2} + 1 - \varepsilon| \quad (3.34)$$

the corresponding terms make a negligible contribution to the sum, and therefore the summation can be terminated at this point. This is natural, since condition (3.34) corresponds to the upper bound of the effective emission angles (see Eqs. (1.61) and (1.62)).

However, it should be noted that Eq. (3.31) does not always correctly describe the frequency spectrum of radiation in a real stack consisting of a finite number of plates. Let us consider, for example, the case when

$$a \ll |\bar{z}_{\mathrm{med}}| \quad \text{and} \quad b \ll \bar{z}_{\mathrm{med}}, \quad (3.35)$$

i.e., when the entire stack radiates as a single equivalent plate (at least at angles $\vartheta \sim \gamma^{-1}$, see Sec. 3.3). In this case d = A < 1 and $\vartheta_1^2 \approx 4\pi c/(\omega(a+b))$ (see Eq. (3.32)), i.e., if one takes Eq. (3.35) into account, then the angle $\vartheta_1$ corresponding to the first maximum of the factor $F_{st}$ is much larger than the effective emission angles of XTR from a single interface. This means that if we replace this first maximum by a δ-function, the contribution of the indicated effective angles to the spectral intensity will be taken into account incorrectly (see Fig. 3.3). Therefore, in the case under consideration one should use the following equation [**74**.1]:

$$W_{\mathrm{st}}^{(+)}(\omega) = \int_0^{\vartheta_1'} W_{\mathrm{st}}^{(+)}(\omega, \vartheta)\, d\vartheta + S_{\mathrm{st}}. \qquad (3.36)$$

$$\vartheta_1' = \sqrt{\vartheta_1^2 + \vartheta_2^2/2}.$$

where the term $S_{st}$ corresponds to Eq. (3.31) without the first term of the sum ($n \geq 2$). The integration in the first term of Eq. (3.36) must be performed numerically.

Equation (3.36) should also be used in the more general case when condition (3.30) is not satisfied; therefore, within the width of the first maximum of the factor $F_{st}$, the function $W_{pl}^{(+)}(\omega, \nu)$ cannot be regarded as "smooth."

Of course, the contributions of both terms (the integral and the sum $S_{st}$) to the spectral intensity in Eq. (3.36) are not always comparable. At the same time, it should be borne in mind that the value of the first term in Eq. (3.36) increases with increasing γ, whereas the value of the sum $S_{st}$ saturates (see Eq. (3.3)). For sufficiently large γ the first term (the integral) becomes dominant, and the spectral intensity does not saturate but continues to grow with increasing γ (see Sec. 3.5).

## 3.5. Frequency Spectrum

The spectral intensity of XTR emitted by a stack of plates generally depends in a rather complicated way both on the stack parameters $a$, $b$, $N$, and the plate material, and on ω and γ.

Let us consider the case often encountered in practice: $b \gg a$, $N_{eff} \gg 1$. As long as $b \gg z^{-}_{vac}$, the radiation intensities from the plates of the stack simply add (the plates radiate independently), and the spectrum of the radiation produced in the stack coincides with the spectrum of radiation from a single plate multiplied by $N_{eff}$. For $\gamma \ll a\omega_0/c$ such a spectrum has no maxima or minima and spans until the frequencies of order $\omega_0\gamma$. In the higher frequency range, the spectral intensity is small, but significantly depends on $\gamma$. For $\gamma > a\omega_0/c$, the cutoff spectrum frequency ceases to increase with $\gamma$ and remains at value of order $a\omega_0/c$ (see Sec. 2.4).

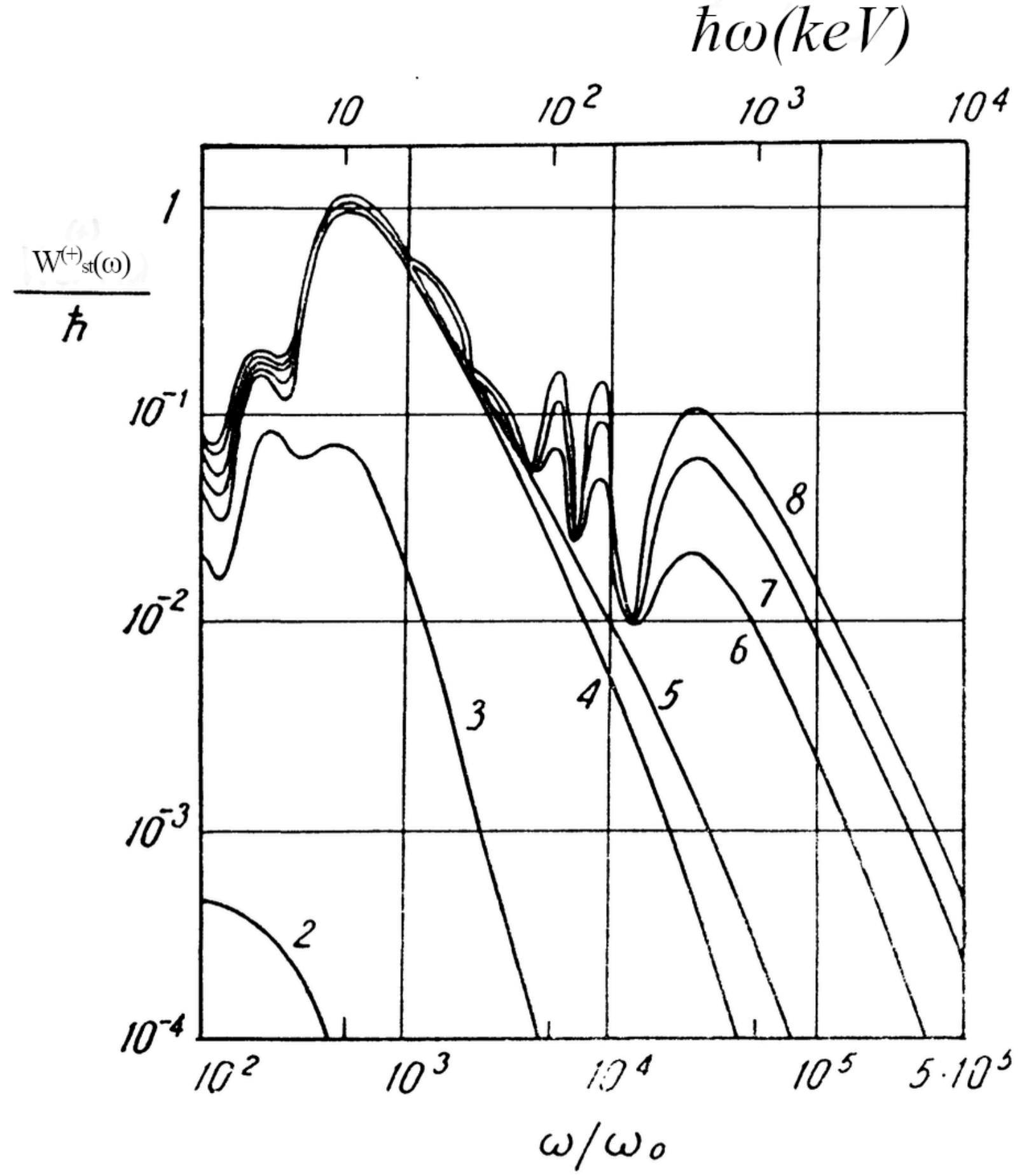

*Fig. 3.4. Spectral distribution of the XTR intensity produced in a stack with parameters $a = 30\ \mu m$, $b = 500\ \mu m$, $N = 50$. Plates of the stack are made of a light material such as polyethylene ($\hbar\omega_0 = 20$ eV), the numbers at the curves denote $\lg\gamma$*

Characteristic interference maxima appear in the spectrum, whose positions are determined by Eq. (2.58). This situation is well illustrated by the curves corresponding to $\gamma = 10^2$ and $10^3$ in Fig. 3.4. The maxima at $\hbar\omega \approx 10$ and 3.3 keV are well described by Eq. (2.58) for $n = 0$ and 1. These frequencies do not differ much from $\omega_0\gamma$ (for $\gamma = 10^3$), therefore the spectral intensity of the radiation increases substantially as $\gamma$ changes from $10^3$ to $10^4$ (in the region $\hbar\omega \approx 10$ keV—by more than a factor of 10).

When the Lorentz factor of the particle reaches a value such that the formation length $\bar{z}_{vac}$ at a frequency of order $a\omega_0^2/c$ becomes larger than the distance $b$ between the plates, i.e., when

$$\gamma > \frac{\sqrt{ab}\,\omega_0}{c}, \qquad (3.37)$$

the interference of the radiation generated in different plates of the stack is significant at all frequencies of the spectrum $\hbar\omega < a\omega_0^2/c$. In an idealized infinite non-absorbing stack, this would lead to saturation of the spectral intensity (and hence also of the total intensity) of resonant transition radiation (see the paper by Cherry [**78**.2]). In a real stack consisting of a finite number of plates with finite absorption, however, the spectrum in the indicated frequency range does not fully saturate but continues to grow, albeit much more slowly (see the curves for $\gamma \geq 10^4$ in the region $\hbar\omega < 60$ keV in Fig. 3.4).

In addition, in a real finite stack, *"edge interference"* is possible when the formation length $\bar{z}_{med}$ at large $\gamma$ becomes comparable to the total length of the stack $N(a + b)$. In this case, the conditions in Eq. (3.24) are satisfied over the entire range of effective radiation angles, i.e., the whole stack radiates as a single equivalent plate whose boundaries coincide with the ends of the stack. "Edge interference" leads to the appearance in the radiation spectrum of new, additional maxima whose intensities increase monotonically with increasing $\gamma$.

The positions of these additional maxima can be determined by analogy with the case of a single isolated plate. To this end, it is sufficient to substitute the quantities $a_{eq} = N(a + b)$ and $\omega^2_{0,eq} = a\omega^2_0/(a + b)$ (cf. (3.25)) into Eq. (2.58) in place of $a$ and $\omega^2_0$. As a result we obtain

$$\omega_n^{\mathrm{edge}} = \frac{Na\omega_0^2}{2\pi c(2n+1)}, \qquad (n = 0, 1, 2, \ldots). \qquad (3.38)$$

By an analogous substitution in inequality (2.52) we obtain the condition for the appearance of the indicated additional maxima:

$$\gamma > \frac{N\sqrt{a(a+b)}\,\omega_0}{c}. \qquad (3.39)$$

As can be seen from Fig. 3.4, for $\gamma \geq 10^6$ (in this case the right-hand side of inequality (3.39) is approximately equal to $6 \cdot 10^5$), well-pronounced non-saturating maxima indeed appear at frequencies $\hbar\omega \approx 100$, 167, and 500 keV (Eq. (3.38) for $n = 2$, 1, and 0).

The spectra shown in Fig. 3.4 were calculated using Eq. (3.36) when condition (3.29) is satisfied; otherwise (when the number $N_{eff}$ is small, which occurs mainly at $\hbar\omega < 10$ keV) they were obtained by direct numerical integration of the spectral–angular intensity given by Eq. (3.15) over the entire range of effective emission angles. If the spectra are calculated using Eq. (3.31) (in the region $\hbar\omega > 10$ keV, where $N_{eff} \gg 1$), then for $\gamma \leq 10^5$, the results agree well with those shown in Fig. 3.4. Interestingly, in this case the agreement (with relative errors not exceeding 6%) holds not only when condition (3.30) is satisfied, but also at those frequencies for which that condition is not satisfied. This means that inequality (3.30) is only a sufficient condition for the applicability of Eq. (3.31), but by no means a necessary one. Equation (3.31) is manifestly inapplicable when $N_{eff}(a + b) \leq |\bar{z}_{med}|$.

For $\gamma > 10^5$, the spectra calculated according to Eq. (3.31) for the stack parameters under consideration cease to depend on $\gamma$ and practically coincide with the spectrum at $\gamma=10^5$. In particular, no additional maxima appear in the frequency range given by Eq. (3.38). This is quite natural, since Eq. (3.31) corresponds to an infinite, non-absorbing, periodic layered medium in which there is no "edge interference," and, consequently, no additional maxima caused by such interference. In the region of these maxima, the result of the correct calculation (by Eq. (3.36), Fig. 3.4) exceeds the result of the calculation by Eq. (3.31) by approximately a factor of 4–13 for $\gamma = 10^6$ and by a factor of 8–40 for $\gamma = 10^7$.

Let us now consider the total intensity of XTR emitted by a stack of plates. From the standpoint of practical applications, it is important that the total intensity (number of quanta) in a given range of detectable frequencies be sufficiently large and depend appreciably on the particle's Lorentz factor. Unfortunately, it is often difficult to satisfy these two requirements simultaneously. Therefore, in each specific case it is necessary to choose the stack parameters, as well as ω and γ, so as to achieve an optimal result [**75**.30, **75**.31, **76**.15, **81**.12].

For illustration, Figs. 3.5–3.7 show some dependences of the number of XTR quanta on $a$, $b$, and the plate material. As can be seen from Figs. 3.5 and 3.6, the maxima in the number of quanta occur approximately at values $a \approx |\bar{z}_{med}|$ and $b \approx \bar{z}_{vac}$ (for some mean frequency within the frequency range under consideration).

For smaller values of $a$ and $b$, the XTR intensity decreases rather rapidly (especially for smaller $a$), and the γ-dependence becomes weaker. For larger $a$, the radiation intensity also decreases (Fig. 3.5) due to increasing absorption. With increasing $b$, the XTR intensity reaches a plateau (Fig. 3.6).

With increasing atomic number of the condensed matter, the plasma frequency $\omega_0$ usually increases, which leads to an increase in the XTR intensity and in the cutoff frequency of the spectrum. However, absorption also increases. From Fig. 3.7 it follows that in the considered photon-energy range of 50–60 keV, with an appropriate choice of the thickness $a$ the advantage is attributed to beryllium and aluminum. The yield of transition quanta is relatively small, but there is a pronounced γ-dependence associated with the fact that the frequency range lies in the vicinity of the cutoff frequency. The situation is quite different in the 10–20 keV photon energy range, where the yield of XTR quanta is much larger, but the γ-dependence is much weaker.

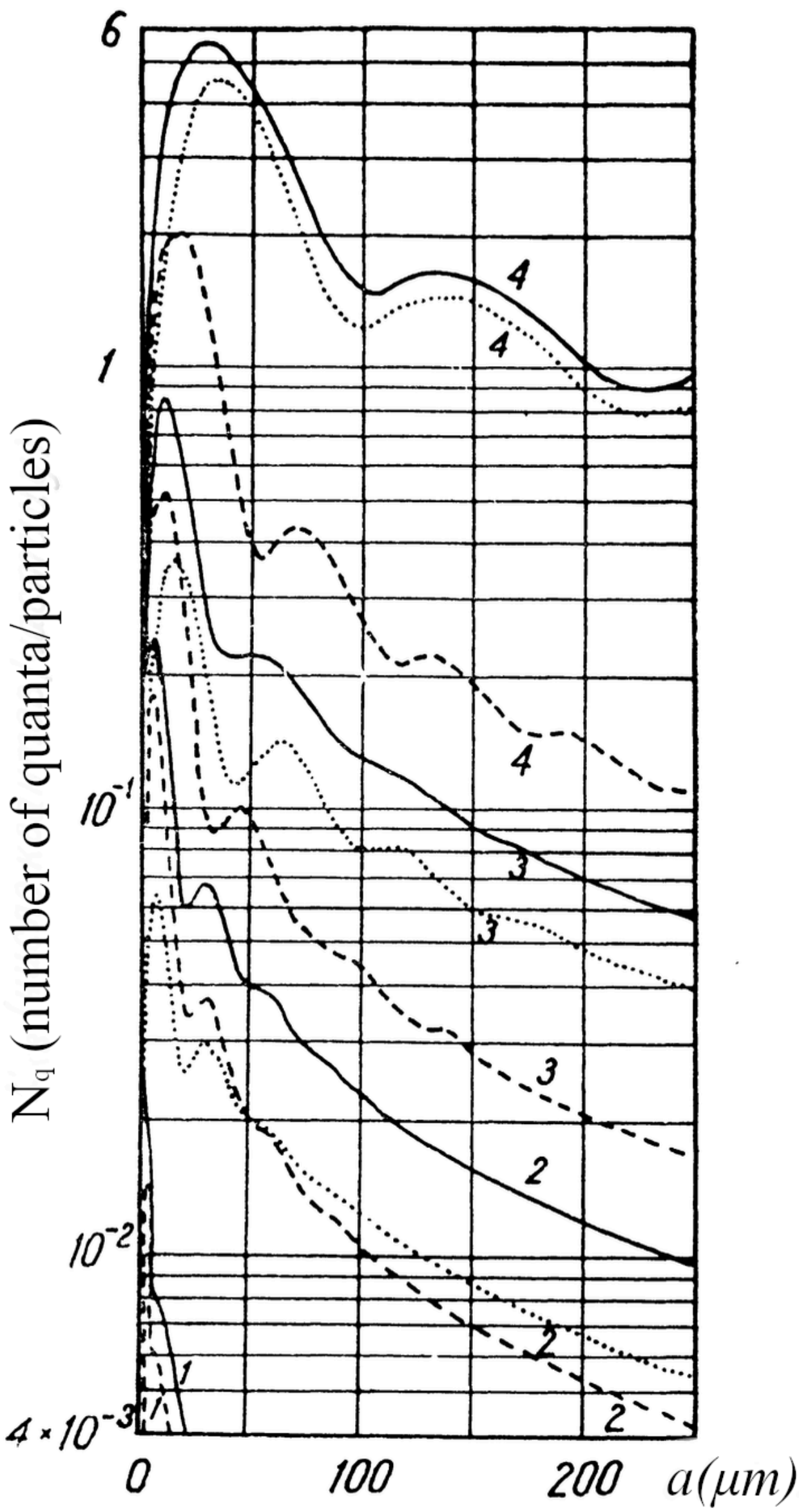

Fig. 3.5. Dependence of the number of XTR quanta in photon energy ranges from 5 to 10 keV (dashed curves), from 10 to 20 keV (dotted curves), and from 5 to 20 keV (solid curves), emitted from a stack of 500 paper plates, on the thickness $\underline{a}$. The spacing $b$ of the vacuum gap between the plates is 1 mm. Curve numbers 1–4 correspond to $\gamma$ = 200, 500, $10^3$, and $10^4$

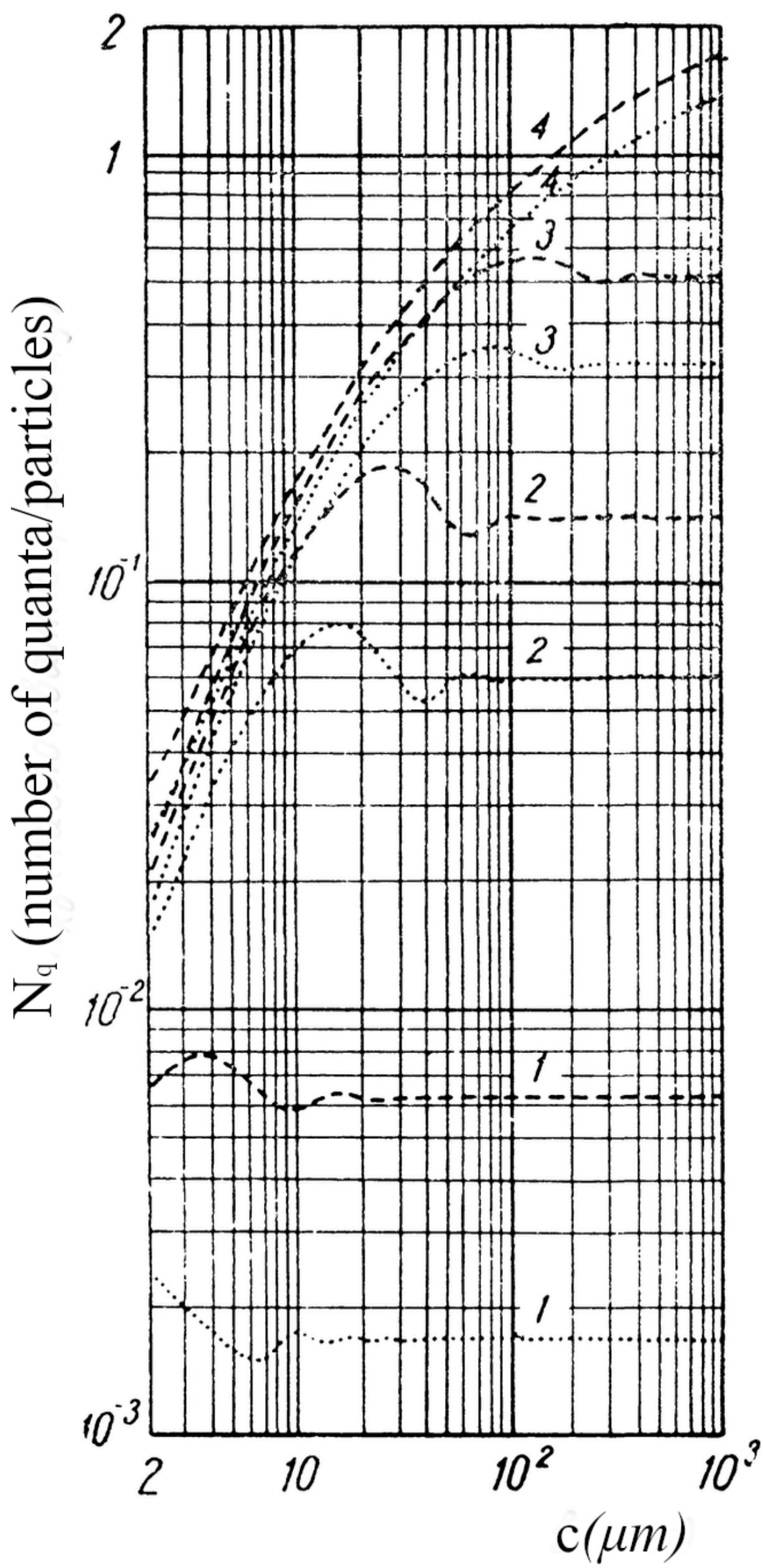

Fig. 3.6 Dependence, analogous to that shown in Fig. 3.5, on the distance b between the plates for a plate thickness a = 10 μm

If the recorded frequencies lie in the region of the maxima (Eq. (2.58)) due to interference within individual plates of the stack, then at the values in Eq. (3.37) the growth of the total intensity (number of quanta) with $\gamma$ decelerates strongly (Fig. 3.8). Usually, in experiments for generating XTR by a particle with $\gamma \sim 10^3$—$10^4$, one uses a stack designed specifically for "intra-plate interference." However, it is difficult to avoid saturation (or, more precisely, sharp slowing down) of the radiation intensity at the frequencies given by Eq. (2.58) for $\gamma > 10^5$ (see Eq. (3.37)) because there are practical limitations on the distance *b* between the plates. To generate XTR by a particle with $\gamma \sim 10^5$–$10^7$, it is apparently better to use "edge interference" and switch to higher frequencies (Eq. (3.38)).

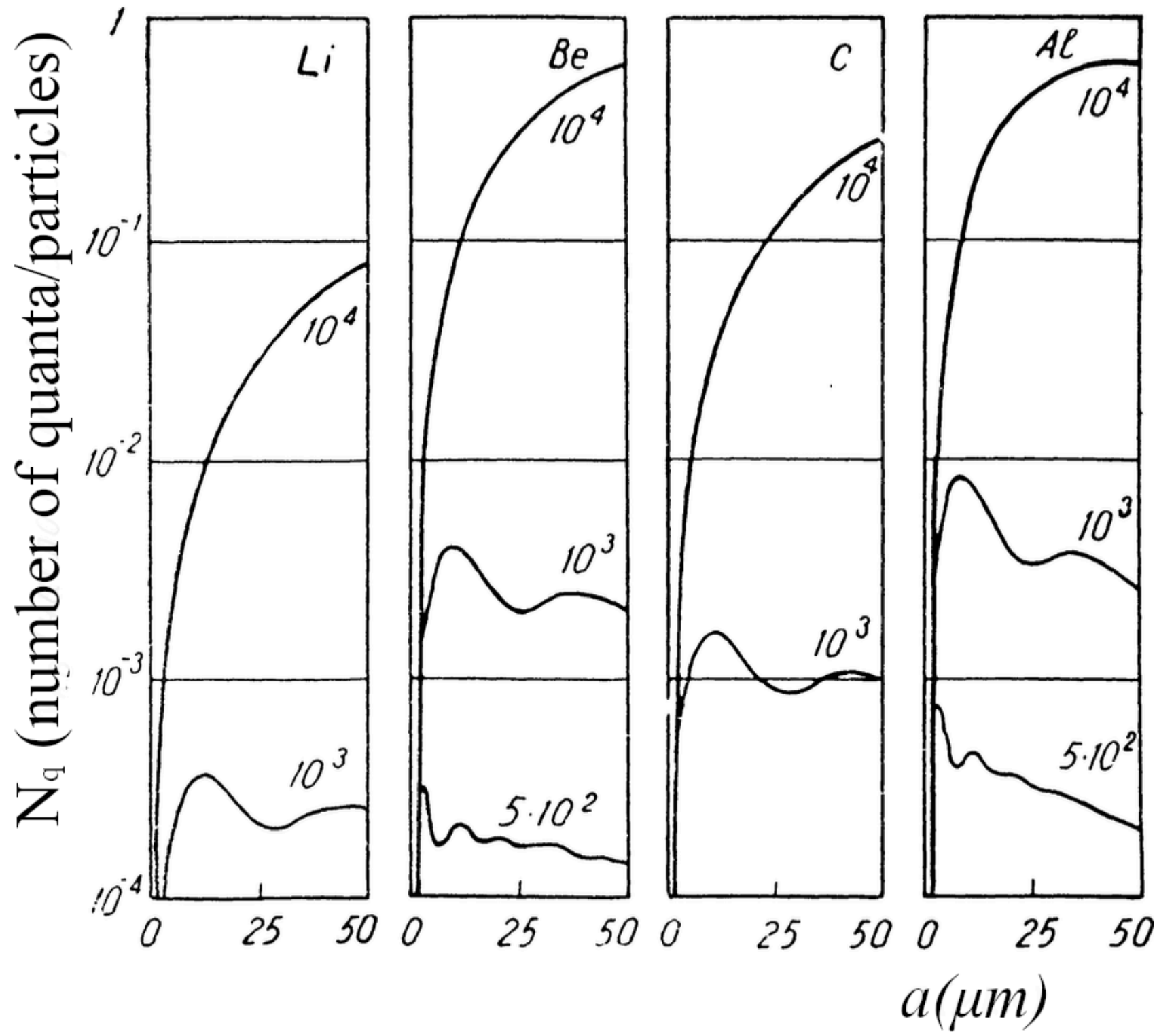

Fig. 3.7. Dependence of the number of XTR quanta emitted from the stack, in the photon energy range from 50 to 60 keV, on the plate thickness a for different materials. The stack consisted of 500 plates with a spacing b = 1 mm between them. The numbers on the curves denote the particle's Lorentz factor γ

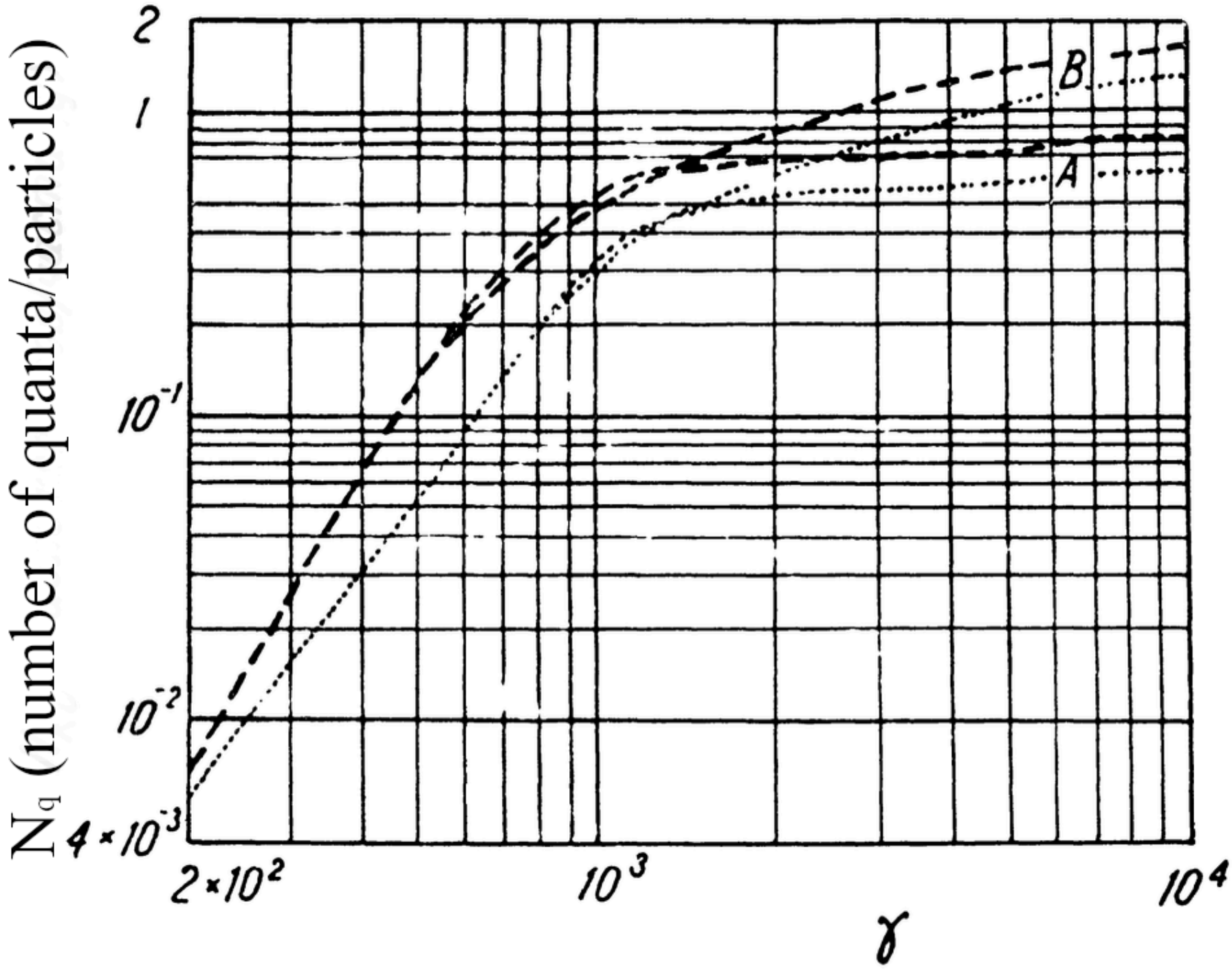

Fig. 3.8. Dependence of the number of XTR quanta in the ranges 5–10 keV (dashed curves) and 10–20 keV (dotted curves), emitted from a stack of paper plates ($\hbar\omega_0$ = 20 eV) with parameters a = 10 μm, N = 500, b = 10 μm (curves A) and b = 1000 μm (curves B), on the particle Lorentz factor γ

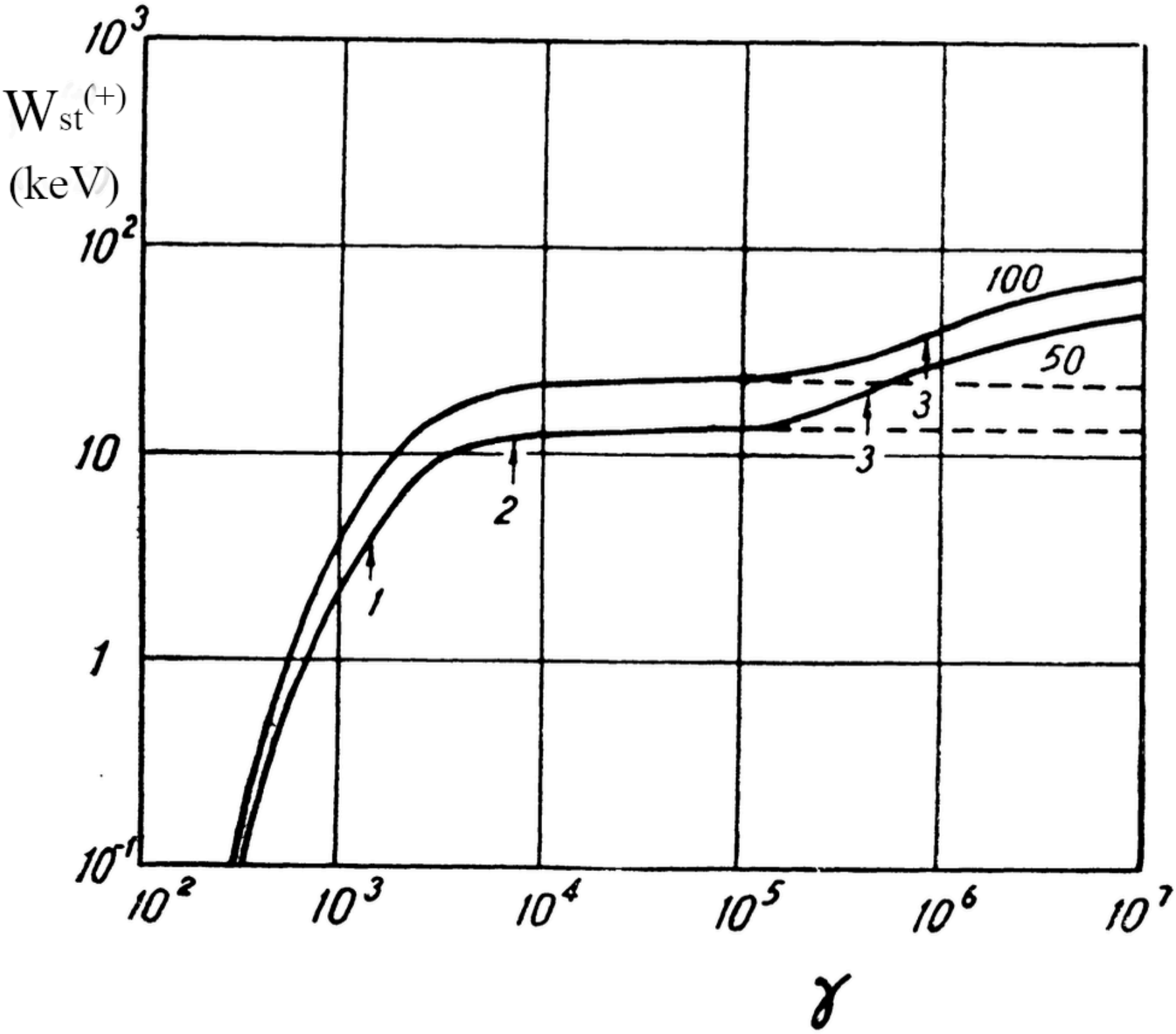

Fig. 3.9. Dependence of the total intensity ($\hbar\omega \geq 2$ keV) of transition radiation formed in a stack of plates (a = 15 μm, b = 500 μm) made of a light material (such as polyethylene, $\hbar\omega_0 = 20$ eV) on the Lorentz factor γ of the particle. The numbers by the curves indicate the number N of plates in the stack. Arrow 1 indicates the value $\gamma = a\omega_0/c$, arrow 2 — $\gamma = \sqrt{a(a+b)}\omega_0/c$, and arrow 3 — $\gamma = N\sqrt{a(a+b)}\,\omega_0\,c$. The dashed curves correspond to the results of calculations according to Eq. (3.31)

If, however, one detects all frequencies of the radiation spectrum (for example, from 2 keV and above), then the total radiation intensity (Fig. 3.9) for comparatively small γ ($\gamma < a\omega_0/c$) first increases rapidly with γ, and then, in the range $a\omega_0/c < \gamma < \sqrt{ab}\omega_0/c$, the growth slows down somewhat and, for $\sqrt{ab}\omega_0/c < \gamma < N\sqrt{a(a+b)}\omega_0/c$ becomes very slow. At large $\gamma$ (see Eq. (3.39)), the total intensity begins to increase again due to the appearance of additional maxima (Eq. (3.38)) in the frequency spectrum, caused by “edge interference.”

# CHAPTER II

## DEVELOPMENT OF THE MACROSCOPIC THEORY

In Chapter I we examined the generation of transition radiation during the normal passage of a single charged particle through a sharp interface between media, a plate, and a regular stack of plates. These cases are to some extent idealized. In reality, a particle may cross the interface at an arbitrary angle, the interface itself may be somewhat diffuse, and the plates in a stack may have unequal thicknesses and be arranged irregularly. It is often important to know what radiation is produced by a group (bunch) of charged particles. In addition, a charged particle loses its energy when passing through a medium not only in producing radiation, but also in excitation and ionization of the atoms of the medium. The issues listed are addressed in the present chapter.

## § 4. OBLIQUE INCIDENCE OF A PARTICLE

Let a charged particle now cross a plane interface between two media at an angle $\psi$ with respect to the normal, moving uniformly and rectilinearly from the first medium, characterized by the parameters $(\varepsilon_1, \mu_1)$, into the second medium $(\varepsilon_2, \mu_2)$. The electric and magnetic fields in this problem can be found from Maxwell’s equations and the boundary conditions in the same way as in the case of normal incidence (see Sec. 1). Without writing out the corresponding equations for the fields (see Ref. [58.3]), let us note only the following circumstance. In the case of normal incidence of a charge on an interface, the resulting transition radiation is always linearly polarized in the plane passing through the wave vector and the normal to the interface, i.e., in the so-called observation plane (Sec. 1.2). In the case of oblique incidence, however, the polarization of the transition radiation becomes *elliptical*.[10] In other words, in addition to the component of the radiation field lying in the observation plane (the "parallel" component), there is also a component perpendicular to it. In the case of a medium–vacuum interface, the spectral–angular distributions of the radiation intensities of the two indicated polarizations are given by the following equations [**62**.1, **62**.2]:

$$W^{\parallel}(\omega, \boldsymbol{n}) = \frac{e^2\beta_z^2 n_z^2 |1-\varepsilon|^2}{\pi^2 c\left[(1-\beta_y n_y)^2 - \beta_z^2 n_z^2\right]^2 (1-n_z^2)} \times$$

$$\times \left| \frac{(1-\beta_z\tau-\beta_z^2-\beta_y n_y)(1-n_z^2)+\beta_y\beta_z n_y\tau}{(1-\beta_y n_y-\beta_z\tau)(\varepsilon n_z+\tau)} \right|^2, \tag{4.1}$$

$$W^{\perp}(\omega, \boldsymbol{n}) = \frac{e^2\beta_y^2\beta_z^4 n_x^2 n_z^2 |1-\varepsilon|^2}{\pi^2 c\left[(1-\beta_y n_y)^2-\beta_z^2 n_z^2\right]^2(1-n_z^2)} \times$$

$$\times \left|(1-\beta_y n_y-\beta_z\tau)(n_z+\tau)\right|^{-2}, \tag{4.2}$$

where $\tau = \sqrt{\varepsilon - 1 + n_z^2}$. The $z$-axis is directed along the normal to the interface, with $v = \{0, v_y, v_z\}$, $\beta_a = v_a/c$ $(a = y, z)$, and $n = \{cos\vartheta_x, cos\vartheta_y, cos\vartheta_z\}$ being the unit vector along the direction of radiation. It is easy to see that for $\beta_y = 0$ (normal incidence) the perpendicular polarization vanishes ($W^{\perp}(\omega, n) = 0$), and Eq. (4.1), after multiplication by the solid-angle element $d^2\Omega = sin\vartheta\, d\vartheta\, d\varphi$, coincides with Eq. (1.37) multiplied by the corresponding element $d\vartheta\, d\varphi/2\pi$.

The spectral–angular distribution of the total radiation intensity is determined by the sum of the expressions given:

$$W(\omega, \boldsymbol{n}) = W^{\parallel}(\omega, \boldsymbol{n}) + W^{\perp}(\omega, \boldsymbol{n}). \tag{4.3}$$

[10] An error was made in one of the earliest papers [**58**.4], where transition radiation at oblique incidence of a particle was considered.

To obtain the corresponding equations for the vacuum–medium interface, it is sufficient to replace $\beta_z$ by $-\beta_z$ in Eqs. (4.1) and (4.2).

Let us now consider the case of relativistic particles ($1 - \beta^2 \ll 1$) and frequencies much higher than optical frequencies ($|1 - \varepsilon| \ll 1$). It is convenient to rotate the coordinate system about the $x$-axis by an angle $\psi$ so that the new $z'$-axis coincides with the direction of the particle's velocity $v$. Then in the new coordinate system ($x'$, $y'$, $z'$) we have

$$
\begin{aligned}
n_{x'} &= n_x, \\
n_{y'} &= n_y \cos\psi - n_z \sin\psi, \\
n_{z'} &= n_y \sin\psi + n_z \cos\psi.
\end{aligned} \tag{4.4}
$$

Let us analyze the following two cases.

1. *Incidence at an Angle not Close to* $\pi/2$. Let

$$
\cos\psi \gg |\gamma^{-2} + 1 - \varepsilon|^{1/2}. \tag{4.5}
$$

In this case, as can be readily verified, the radiation is emitted predominantly at small angles with respect to the particle's direction of motion (see Sec. 1.8). Introducing the polar and azimuthal radiation angles $\vartheta'$ and $\varphi'$ such that $n_{x'} = \sin\vartheta' \sin\phi'$, $n_{y'} = \sin\vartheta' \cos\phi'$, $n_{z'} = \cos\vartheta'$, and noting that $\vartheta' \ll 1$, from Eqs. (4.1) and (4.2) we obtain:

$$
\begin{aligned}
W^{\parallel}(\omega, \mathbf{n}') &= C(\vartheta') \frac{|2\vartheta' \cos\phi' \operatorname{tg}\psi - (\gamma^{-2} + \vartheta'^2)\operatorname{tg}^2\psi + 2\vartheta'^2|^2}{4(\operatorname{tg}^2\psi + \vartheta'^2)}, \\
W^{\perp}(\omega, \mathbf{n}') &= C(\vartheta') \frac{\vartheta'^2 \sin^2\phi' \operatorname{tg}'^2\psi}{\operatorname{tg}^2\psi + \vartheta'^2}, \\
C(\vartheta') &= \frac{e^2 |1-\varepsilon|^2}{\pi^2 c(\gamma^{-2} + \vartheta'^2)^2 |\gamma^{-2} + \vartheta'^2 + 1 - \varepsilon|^2}.
\end{aligned} \tag{4.6}
$$

If $\tan\psi \gg \vartheta'$ and $|\vartheta' \cos\phi'| \gg \gamma^{-2} \tan\psi$, then Eqs. (4.6) take the simple form:

$$
\begin{aligned}
W^{\parallel}(\omega, \mathbf{n}') &= C(\vartheta')\vartheta'^2 \cos^2\phi', \\
W^{\perp}(\omega, \mathbf{n}') &= C(\vartheta')\vartheta'^2 \sin^2\phi'.
\end{aligned} \tag{4.7}
$$

Thus, the spectral–angular distributions of the XTR intensities for both polarizations, in the case of incidence at angles not close to $\pi/2$ (with the exception of the region of very small emission angles $\vartheta'$), do not depend on the incidence angle of the charge $\psi$, and their sum (after multiplication by the corresponding angular elements) coincides with the analogous distribution (Eq. (1.59)) for normal incidence:

$$[W^{\parallel}(\omega, \boldsymbol{n}') + W^{\perp}(\omega, \boldsymbol{n}')]\vartheta' d\vartheta' d\phi' = W_{\text{rp}}^{(+)}(\omega, \vartheta) d\vartheta d\phi / (2\pi). \qquad (4.8)$$

Deviations in the small-angle region ($|\vartheta' \cos\varphi'| \lesssim \gamma^{-2} \tan\psi$) with respect to the particle's direction of motion are due to the refraction of the radiation. However, this angular region corresponds to a low intensity per unit solid angle and therefore makes only a very small contribution to the total radiation intensity after integration over angles.

From Eq. (4.7) for the degree of polarization we obtain

$$P(\omega, \boldsymbol{n}') = \left| \frac{W^{\parallel}(\omega, \boldsymbol{n}') - W^{\perp}(\omega, \boldsymbol{n}')}{W^{\parallel}(\omega, \boldsymbol{n}') + W^{\perp}(\omega, \boldsymbol{n}')} \right| = |\cos 2\phi'|. \qquad (4.9)$$

For the resulting intensity (after integration over the emission angles) we have

$$W^{\parallel}(\omega) = W^{\perp}(\omega) = W_{\text{rp}}^{(+)}(\omega), \qquad (4.10)$$

where $W_b^{(+)}(\omega)$ is given by Eq. (1.64) for normal incidence.

2. *Grazing Incidence*. Let now the angle $\psi$ be close to $\pi/2$, i.e.:

$$\cos\psi \sim |\gamma^{-2} + 1 - \varepsilon|^{1/2} \ll 1. \qquad (4.11)$$

Then from Eqs. (4.1) and (4.2) we obtain

$$W^{\parallel}(\omega, \boldsymbol{n}') = C_{\psi}(\vartheta')(\gamma^{-2} + \vartheta'^2 - 2\vartheta' \cos\psi \cos\phi')^2,$$
$$W^{\perp}(\omega, \boldsymbol{n}') = C_{\psi}(\vartheta') 4\vartheta'^2 \cos^2\psi \sin^2\phi', \qquad (4.12)$$
$$C_{\psi}(\vartheta') = (16e^2/\pi^2 c)\cos^2\psi(\cos\psi - \vartheta' \cos\phi')^2 |1 - \varepsilon|^2 \times$$
$$\times (\gamma^{-2} + \vartheta'^2)^{-2} (4\cos^2\psi - 4\vartheta' \cos\psi \cos\phi' + \gamma^{-2} + \vartheta'^2)^{-2} \times$$

$$\times|(\gamma^{-2}+\vartheta'^2+2\cos^2\psi-2\tau\cos\psi-2\vartheta'\cos\psi\cos\phi')\times$$
$$\times(\tau+\cos\psi-\vartheta'\cos\phi')|^{-2},$$

where

$$\tau=[\varepsilon-1+(\cos\psi-\vartheta'\cos\phi')^2]^{1/2}.$$

It follows from these equations that in the case of grazing incidence of the particle, the intensity of XTR depends appreciably on the angle of incidence, decreasing as $(\pi/2-\psi)^2$ (for parallel polarization) or $(\pi/2-\psi)^4$ (for perpendicular polarization) as $\psi\to\pi/2$.

However, in comparing with experimental results, one must bear in mind that Eqs. (4.12) were obtained under the assumption that the surface is perfectly plane and the motion is rectilinear and uniform. On the other hand, it is clear that precisely for grazing incidence of the particle, both the actual state of the surface and deviations of the particle's trajectory from a straight line can be of great importance. Then the radiation may turn out to be completely different (see in more detail Ref. [**65**.1, **73**.17, **74**.19, **79**.11, **81**.9]).

## § 5. GEOMETRICAL OPTICS APPROXIMATION. DIFFUSE BOUNDARIES

Up to now we have considered problems with sharp interfaces, i.e., we have assumed that the dielectric permittivity $\varepsilon$ and magnetic permeability $\mu$ of the medium change discontinuously at these interfaces and are constant outside them. Since in reality the changes of $\varepsilon$ and $\mu$ occur over finite distances, it is natural to pose the question of the influence of the diffuse boundary on the formation of transition radiation [**60**.6, **61**.3, **64**.1, **65**.3, **76**.1, **80**.1]. Moreover, one can pose the general question of the radiation of a charged particle in an arbitrary inhomogeneous medium [**60**.2, **61**.2].

For simplicity, let us assume the magnetic permeability $\mu$ to be equal to unity everywhere in what follows. Then, from Maxwell's equations (1.1) for the vector potential $A(r,\omega)$ of the electromagnetic field (Eq. (1.8)), we obtain the equation

$$\Delta A + \frac{\varepsilon\omega^2}{c^2} A + \varepsilon \operatorname{div}\left( A \nabla \frac{1}{\varepsilon} \right) = -\frac{4\pi}{c} j_{\text{ext}}(r, \omega), \qquad (5.1)$$

where $\varepsilon = \varepsilon(r, \omega)$ is a function of the spatial coordinates.

Consider an isotropic "one-dimensional" inhomogeneous medium in which $\varepsilon$ varies only in one spatial direction: $\varepsilon = \varepsilon(z, \omega)$. Let a charged particle move in the same $z$-direction.

Let us perform a Fourier transform over the transverse coordinates $\rho = \{x, y, 0\}$:

$$\boldsymbol{A}(\boldsymbol{r}, \omega) = \int \boldsymbol{A}(\boldsymbol{\varkappa}, z, \omega) \exp(i\boldsymbol{\varkappa} \cdot \boldsymbol{\rho}) d^2\boldsymbol{\varkappa}. \qquad (5.2)$$

Because of the symmetry of the problem, $A_x = A_y = 0$, and for the $z$-component $A_z(\varkappa, z, \omega)$ we obtain the equation. To simplify the notation, we omit the index $z$ and the arguments:

$$\varepsilon \frac{\partial}{\partial z}\left( \frac{1}{\varepsilon} \frac{\partial A}{\partial z} \right) + \left( \frac{\omega^2}{c^2} \varepsilon - \varkappa^2 \right) A = -\frac{ev}{2\pi^2 c} \int \delta(z - z(t)) e^{i\omega t} dt. \qquad (5.3)$$

To find a solution of the linear inhomogeneous Eq. (5.3) it is first necessary to find the general solution of the corresponding homogeneous equation. For an arbitrary dependence of $\varepsilon$ on $z$, an exact solution cannot be found. However, in a special case of practical importance (especially for X-ray frequencies), when the relative variation of the dielectric permittivity over distances of order the wavelength is small, i.e.

$$\left| \frac{c}{\varepsilon^{3/2}\, \omega} \frac{\partial \varepsilon}{\partial z} \right| \ll 1, \qquad (5.4)$$

the problem can be solved with good accuracy in the so-called *geometrical optics approximation* (quasiclassical approximation). Moreover, as will be seen in what follows, this approximation turns out to be convenient for solving some other problems of the theory of transition radiation [69.1].

Let us seek the solution of Eq. (5.3) in the following form:

$$A = \exp\left(\frac{i}{\hbar}\psi\right), \qquad (5.5)$$

where $\psi = \psi(\varkappa, z, \omega)$ is a new unknown function (the so-called "eikonal"), and ħ is a formal small parameter[11].

Substituting (5.5) into the left-hand side of Eq. (5.3), we obtain the equation for $\psi$:

$$\frac{i\varepsilon}{\hbar}\frac{\partial}{\partial z}\left(\frac{1}{\varepsilon}\frac{\partial\psi}{\partial z}\right) - \frac{1}{\hbar^2}\left(\frac{\partial\psi}{\partial z}\right)^2 + \left(\frac{\omega^2}{c^2}\varepsilon - \varkappa^2\right) = 0. \qquad (5.6)$$

The solution of Eq. (5.6) will be found by the method of successive approximations, expanding the function $\psi$ in a series in the small parameter ħ;

$$\psi = \psi_0 + \hbar\psi_1 + \ldots. \qquad (5.7)$$

For the first two terms of the expansion we have:

$$\frac{1}{\hbar^2}\left(\frac{\partial\psi_0}{\partial z}\right)^2 = \frac{\omega^2}{c^2}\varepsilon - \varkappa^2,$$

$$\frac{\partial\psi_1}{\partial z} = \frac{i}{2}\frac{\partial}{\partial z}\ln\frac{\lambda}{\varepsilon}, \qquad (5.8)$$

where (see Sec. 1.4)

$$\lambda = \sqrt{\frac{\omega^2}{c^2}\varepsilon - \varkappa^2}\,. \qquad (5.9)$$

From this we obtain two linearly independent solutions of the homogeneous equation for $A$:

$$A_\pm(z) = \sqrt{\frac{\varepsilon}{\lambda}}\exp\left(\pm i\int^z \lambda\, dz\right). \qquad (5.10)$$

[11] Here the problem is considered in classical electrodynamics; therefore, this parameter is, of course, not related to the Planck constant. We use this notation in order to emphasize the analogy with the quasiclassical approximation in quantum mechanics.

These solutions are meaningful provided that the corrections introduced by higher-order approximations are small. This means that, first, the semiclassical condition (5.4) must be satisfied and, second, the corrections to the phase of the vector potential due to higher-order approximations must not accumulate significantly over distances where $\varepsilon$ varies appreciably. Furthermore, the quasiclassical solutions (Eq. (5.10)) are inapplicable near the "turning points," where the quantity $\lambda$ is close to zero.

The solution of the inhomogeneous Eq. (5.3) can be expressed in terms of linearly independent solutions of the homogeneous equation. In the case of motion of a particle with constant velocity $\upsilon$ we have

$$A = A_z(\varkappa, z, \omega) = \frac{ie}{4\pi^2 c}\left\{A_+(z)\int\limits_{-\infty}^{z}\varepsilon^{-1}A_-(\xi)\exp\left(i\frac{\omega\xi}{v}\right)d\xi + \right.$$

$$\left. + A_-(z)\int\limits_{z}^{\infty}\varepsilon^{-1}A_+(\xi)\exp\left(i\frac{\omega\xi}{v}\right)d\xi\right\}. \qquad (5.11)$$

Let, for sufficiently large $|z|$, the dielectric permittivity practically not depend on z and take the values $\varepsilon_1$ and $\varepsilon_2$ for $z \to -\infty$ and $z \to +\infty$, respectively. Let us consider Eq. (5.11) for $z \to +\infty$. Let us rewrite Eq. (5.11) in the following form:

$$A = \frac{ie}{4\pi^2 c}\left\{A_+(z)\int\limits_{-\infty}^{+\infty}\varepsilon^{-1}A_-(\xi)\exp\left(\frac{i\omega\xi}{v}\right)d\xi - \right.$$

$$\left. - \int\limits_{z}^{\infty}\varepsilon^{-1}[A_+(z)A_-(\xi) - A_-(z)A_+(\xi)]\exp\left(\frac{i\omega\xi}{v}\right)d\xi\right\}. \qquad (5.12)$$

In the second integral one may assume that $\varepsilon = \varepsilon_2$ and $\lambda = \lambda_2$. Then this integral can be evaluated explicitly. Comparing the result with Eqs. (1.9) and (1.10), one can verify that this quantity corresponds to the vector potential of the field of a particle in a medium with dielectric permittivity $\varepsilon_2$. Since we are interested in the radiation field, we shall omit this quantity henceforth.

Thus, the vector potential of the radiation field emitted forward, at large distances ($z \to +\infty$), is described by the integral

$$A^{(+)} = \frac{ie}{4\pi^2 c} A_+(z) \int_{-\infty}^{+\infty} \varepsilon^{-1} A_-(\xi) \exp\left(\frac{i\omega\xi}{v}\right) d\xi. \tag{5.13}$$

Similarly, for radiation emitted backward, at large distances ($z \to -\infty$) we have

$$A^{(-)} = \frac{ie}{4\pi^2 c} A_-(z) \int_{-\infty}^{\infty} \varepsilon^{-1} A_+(\xi) \exp\left(\frac{i\omega\xi}{v}\right) d\xi. \tag{5.14}$$

In this case the quantities $\kappa$ and $\lambda$ are related to the radiation angle $\theta$ by relations analogous to Eq. (1.33).

Let us now consider a single diffuse boundary between two media, i.e., assume that the dielectric permittivity depends on z primarily within a layer of thickness $z_0$, where its value changes smoothly from $\varepsilon_1$ to $\varepsilon_2$.

Let us substitute Eq. (5.10) into Eqs. (5.13) and (5.14). After integration we obtain that $A^{(\uparrow)}$ consists of terms containing exponents of the type $\exp[i(\omega/v-\lambda_s)z_0/2]$, while $A^{(-)}$ contains $\exp[i(\omega/v+\lambda_s)z_0/2]$ ($s = 1, 2$). This means that the quantities $A^{(\pm)}$ depend strongly on the ratio of the diffuseness length $z_0$ to the radiation formation lengths, respectively forward or backward, for both media (cf. Eq. (1.30)):

$$z_s^{\text{forward}} = \frac{\pi}{\omega/v - \lambda_s}, \tag{5.15}$$

$$z_s^{\text{backward}} = \frac{\pi}{\omega/v + \lambda_s}, \tag{5.16}$$

When the diffuseness length $z_0$ of the boundary is much larger than the radiation formation length:

$$z_0 \gg \left|z_s^{\text{forward}}\right| \quad \left(\text{or} \quad z_0 \gg \left|z_s^{\text{backward}}\right|\right), \tag{5.17}$$

the integrand on the right-hand side of Eq. (5.13) (or, respectively, Eq. (5.14)) oscillates rapidly, and therefore the magnitude of A is small. However, for its evaluation we cannot substitute the functions given by Eq. (5.10) into Eqs. (5.13) or (5.14), since even a small inaccuracy in the phase of these functions leads to large relative errors in the value of the integral given by Eqs. (5.13) or (5.14). The situation here closely resembles the so-called “above-barrier reflection” of particles in quantum mechanics. A correct treatment of this issue shows that when Eq. (5.17) holds, the transition radiation emitted forward (or backward) is exponentially small [**60**.6, **61**.3, **64**.1].

The formation length of backward radiation (Eq. (5.16)) is always of the order of the wavelength or smaller. As for the formation length of forward radiation (5.15), when the particle is non-relativistic, or the dielectric permittivity ε is not close to unity, or the radiation angle is not small, the quantity $z_s^{forward}$ is also of the order of the wavelength. Therefore, in these cases Eq. (5.17) can be replaced by the condition that the diffuseness length must be much greater than the wavelength.

However, as noted in Sec. 1, in the case when

$$\gamma \gg 1, \quad |1-\varepsilon| \ll 1 \quad \text{and} \quad \vartheta \ll 1. \qquad (5.18)$$

(in particular, for XTR), the formation length (Eq. (5.15)) greatly exceeds the wavelength and becomes a macroscopic quantity. Then the first condition (5.17) may turn out to be not satisfied.

When inequalities (5.17) do not hold, the integrand in Eqs. (5.13) or (5.14) is no longer rapidly oscillating. Therefore, it is entirely legitimate to substitute the functions given by Eq. (5.10) into these equations. Substituting Eq. (5.10) into Eq. (5.13) and taking into account Eq. (5.18), we obtain that the intensity of forward radiation is given by the following equation:

$$W^{(+)}(\omega,\vartheta) = \frac{e^2\omega^2\vartheta^3}{2\pi c^3}\left|\int_{-\infty}^{\infty} \exp\left\{i\left(\frac{\omega}{v}\xi - \int_{z_{obs}}^{\xi}\lambda(z')\,dz'\right)\right\}d\xi\right|^2. \qquad (5.19)$$

($z_{obs}$ is the coordinate of the observation plane; see Sec. 1.3).

Let now

$$z_0 \ll \left|z_s^{\text{forward}}\right|, \qquad (5.20)$$

(with $z_0$ possibly being much greater than the wavelength). Evaluating the integral in Eq. (5.19) for a diffuse boundary and expanding the result in powers of the small parameter (Eq. (5.20)), we obtain

$$W^{(+)}(\omega, \vartheta) = W_{\rm b}^{(+)}(\omega, \vartheta) \left|1 - \alpha\right|^2, \qquad (5.21)$$

where $W_b^{(+)}(\omega, \vartheta)$ is the spectral–angular distribution of the forward XTR intensity for a sharp boundary (see Eq. (1.78)), and δ is a small quantity with a positive real part. When, for example, ε depends linearly on $z$ within the boundary layer, we have

$$\delta = \frac{\pi^2}{24} \frac{z_0^2}{z_1 z_2}, \qquad (5.22)$$

where $z_{1,2}$ are the forward-radiation formation lengths (5.15). For simplicity, the subscript “forward” is omitted.

It follows from Eq. (5.21) that in the case under consideration the diffuse boundary leads to a decrease in intensity for all emission angles and frequencies. To illustrate this, Fig. 5.1 shows the curves of the angular dependence of the forward XTR intensity (at a fixed frequency) for a diffuse boundary when a particle exits the medium into vacuum. The curves were calculated using Eq. (5.19) under the assumption of a linear dependence ε(z) within the boundary layer. The figure clearly shows that when conditions (5.20) are satisfied, the XTR intensities for diffuse and sharp boundaries differ little. However, when the diffuseness length $z_0$ is of the order of the corresponding formation lengths or larger, the radiation intensity decreases substantially. Since the formation lengths decrease with decreasing Lorentz factor γ, the XTR intensity for a diffuse boundary decreases with decreasing γ faster than for a sharp boundary.

Similarly, one can solve the problem of a plate with diffuse boundaries. As in the previous case, the radiation is exponentially small if the diffuse boundary lengths of the plate are much larger than the formation lengths given by Eqs. (5.15) and (5.16). When conditions (5.20) are satisfied, the forward radiation intensity becomes appreciable.

For XTR in a diffuse plate we obtain

$$W^{(+)}(\omega, \vartheta) = W_{\rm b}^{(+)}(\omega, \vartheta) \left|1 - \delta_2 - (1 - \delta_1) \exp(-i\pi a / z_1)\right|^2, \qquad (5.23)$$

where $a$ is certain average thickness of the plate, and $\delta_1$ and $\delta_2$ are small quantities determined by the smearing lengths $z_{01}$ and $z_{02}$ of the first and second (along the motion of the charged particle) boundaries of the plate, respectively. In the quantity $W_b^{(+)}(\omega, \vartheta)$ (Eq. (1.78)) $z_{obs}$ is measured from the second boundary, $\varepsilon_1$ and $\varepsilon_2$ are the values of $\varepsilon$ far from the boundary layers inside and outside the plate, respectively.

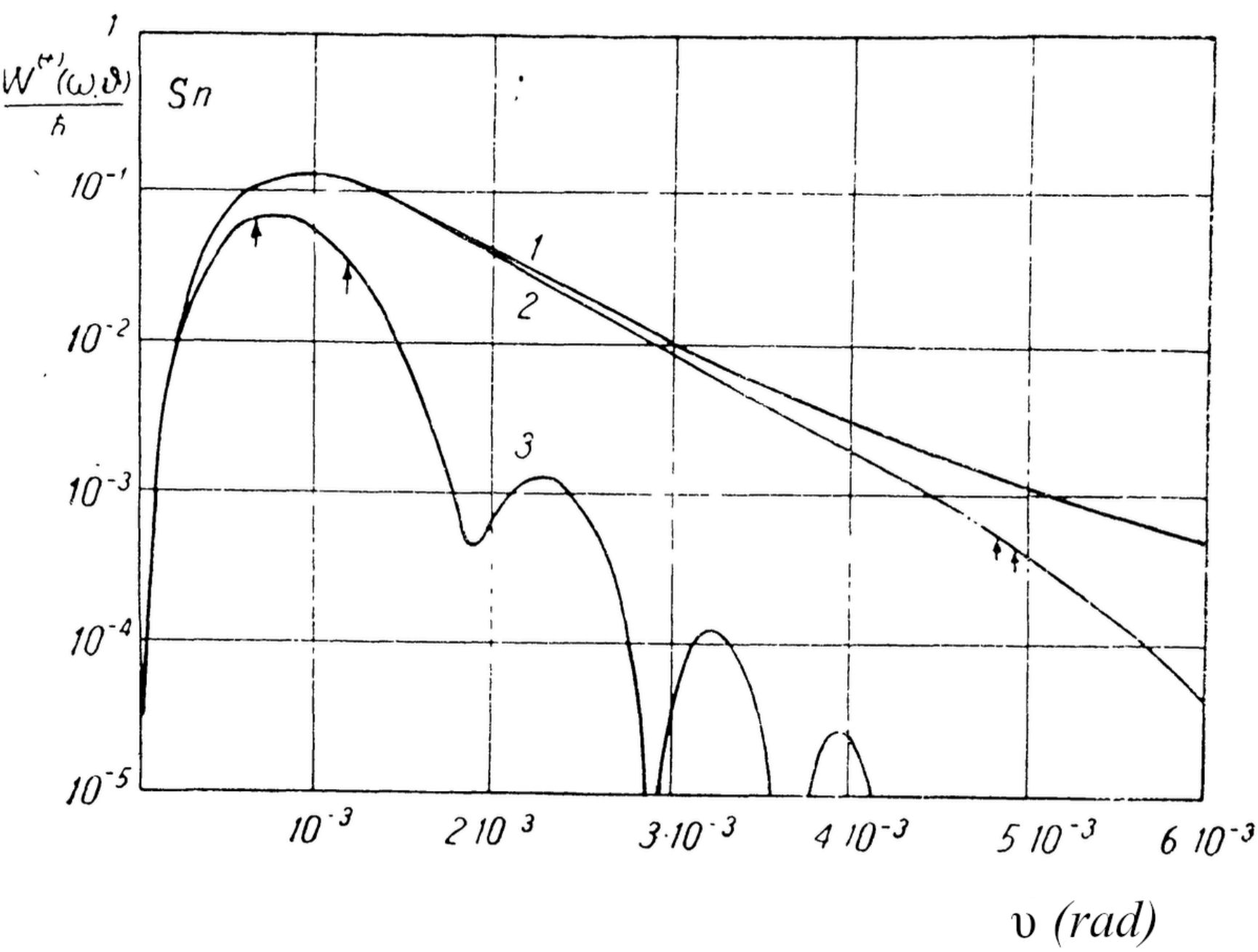

Fig. 5.1. Angular dependence of the spectral–angular distribution of the XTR intensity for a diffuse tin–vacuum boundary: curve 1—sharp boundary; 2—diffuse boundary $z_0 = 10^{-4}$ cm; 3—$z_0 = 10^{-3}$ cm; $\hbar\omega_0 = 50$ eV, $\gamma = 10^3$, $\hbar\omega = 50$ keV. The arrows indicate the angles at which the formation lengths $z_{med}(\vartheta)$ and $z_{vac}(\vartheta)$ are equal to the diffuseness length $z_0$.

In particular, when $\varepsilon$ varies linearly in the boundary layers, we have

$$\delta_s = \frac{\pi^2}{24} \frac{z_{0s}^2}{z_1 z_2} \qquad (s=1,\ 2). \tag{5.24}$$

It is easy to see that when the boundaries of the plate are sharp, i.e., δ1 = δ2 = 0, Eq. (5.23) reduces to the well-known equation for XTR emitted forward in the case of a plate with sharp boundaries (in particular, if the plate is in vacuum, one obtains Eq. (2.21)).

When the thickness of the plate (with diffuse boundaries) is much greater than the absorption length of radiation in the plate material, i.e., $a\, Im\, \lambda_1 \gg 1$, Eq. (5.23), as expected, reduces to Eq. (5.21) for a single diffuse boundary.

Let the plate and the external medium now be transparent, i.e., $Im\lambda_1 = Im\lambda_2 = 0$. Then we have

$$W^{(+)}(\omega,\vartheta) = W_{\mathrm{b}}^{(+)}(\omega,\vartheta)\left[4(1-\delta_1-\delta_2)\,\sin^2\left(\frac{\pi a}{2z_1}\right)+(\delta_1-\delta_2)^2\right]. \qquad (5.25)$$

It follows that the diffuseness of the boundaries leads to a decrease in the radiation intensity for those frequencies and angles for which the sine squared is appreciable. In particular, the maxima at $a = (2n + 1)z_1$ ($n$ is an integer) decrease. However, if the degrees of diffuseness of the two boundaries of the plate are different ($\delta_1 \neq \delta_2$), then the intensity at the minima at $a = 2nz_1$, on the contrary, increases somewhat; that is, complete destructive interference of the waves formed in such a plate does not occur. In other words, the oscillations in the spectral–angular distribution of the radiation intensity are smoothed out.

In Fig. 5.2, the curves of the frequency spectrum of forward XTR are shown, obtained by integrating Eq. (5.19) over the angle ϑ, for a plate with diffuse boundaries under the assumption of a linear dependence ε(z) in the interfacial layers. It is clearly seen that when $z_0 \gtrsim |z_s|$, the radiation intensity decreases sharply. The merging of curves 1 and 2, as well as 3 and 4, at ħω < 8 keV is explained by the fact that at such frequencies the absorption length of the radiation becomes of the order of, or smaller than, the mean thickness of the plate, and therefore the state of the first boundary does not play any appreciable role.

The generalization of the above analysis to the case of a stack is quite straightforward. However, it should be borne in mind that, as already noted after Eq. (5.8), for the validity of the geometrical optics approximation, the corrections to the solution must not “accumulate.” In the case of a stack, this means that the number of interfaces in the stack should not be too large.

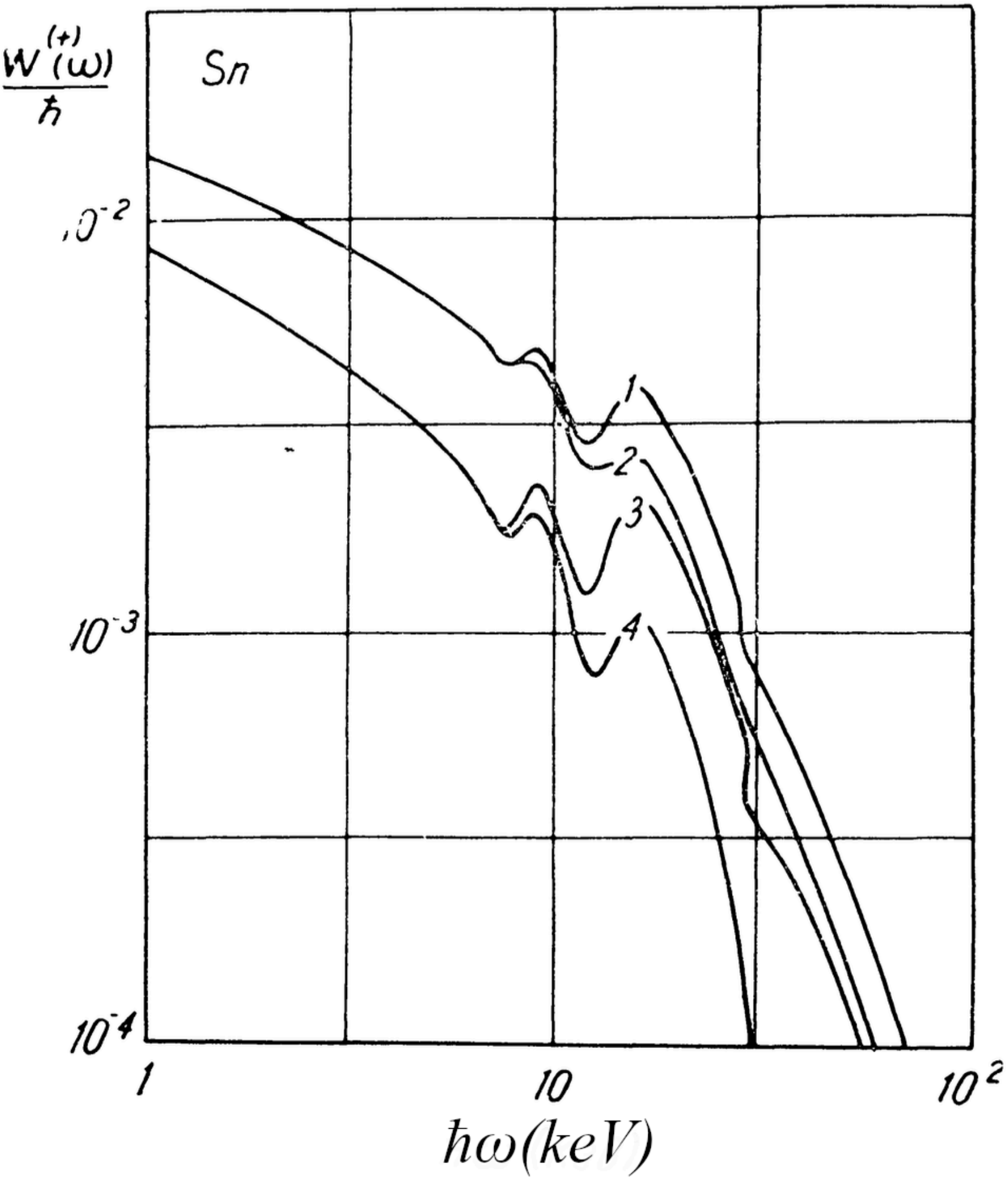

Fig. 5.2. Frequency spectrum of XTR for a plate with diffuse boundaries: curve 1—plate with sharp boundaries; 2—diffuseness of the first boundary $z_{01} = 2 \cdot 10^{-3}$ cm, second boundary sharp; 3—first boundary sharp, diffuseness of the second boundary $z_{02} = 2 \cdot 10^{-3}$ cm; 4—diffuseness of both boundaries $z_{01} = z_{02} = 2 \cdot 10^{-3}$ cm; $\gamma = 10^{3}$, $a = 2 \cdot 10^{-3}$ cm.

It is important to note that the geometrical optics approximation is in a certain sense suitable for calculating the intensity of forward XTR even in problems with sharp boundaries. Indeed, for XTR condition (5.4) means that

$$z_0 \gg |\varepsilon - 1| c/\omega. \tag{5.26}$$

And since the formation lengths of forward XTR are large:

$$|z_s| \gg \frac{c}{\omega} \gg |\varepsilon - 1| \frac{c}{\omega}, \tag{5.27}$$

then we can always, instead of the problem with sharp boundaries, consider an analogous problem with diffuse boundaries such that conditions (5.20) and (5.26) are satisfied simultaneously. As already noted, the forward XTR intensities in these two problems practically coincide, and therefore we may altogether neglect the diffuseness and calculate the spectral–angular distribution of the XTR intensity from Eq. (5.19).

In what follows, the geometric optics approximation will be used to consider the problems of an irregular stack (Sec. 6) and the effect of multiple scattering on transition radiation (Chapter IV).

## § 6. IRREGULAR STACK OF PLATES

Let us now consider a stack of parallel plates with arbitrary thicknesses irregularly arranged in some medium. Such a system is not only of independent interest but also serves as a convenient model for the theoretical description of X-ray and higher-energy transition radiation in porous and similar irregular media. Indeed, for X-ray and higher frequencies the effective transverse dimensions of the electromagnetic field of an ultrarelativistic charged particle remain, up to quite large values of the Lorentz factor, much smaller than the pore sizes, for example in plastic foam. One can then ignore the curvature of the walls and regard them as planar. In addition, the intensity of XTR is almost independent of the angle at which the charge enters the material (see Sec. 4). Therefore, the walls may be regarded as mutually parallel and perpendicular to the charge trajectory.

The problem of a charged particle passing through an irregular stack of plates can be solved exactly [**69**.2, **72**.4, **74**.3]. However, the resulting equations for the fields and the radiation intensity are quite cumbersome and difficult to analyze. On the other hand, in the previous Section we saw that the equations for the forward XTR can be obtained in the geometrical optics approximation, which is distinguished by its simplicity and clarity. Therefore, we will solve the problem of an irregular stack in this approximation, assuming that conditions in Eqs. (5.20) and (5.26) are satisfied [**73**.3, **74**.4, **75**.4].

## 6.1. Spectral–Angular Distribution of the Mean Radiation Intensity

Let a particle with charge $e$ move with velocity $v = \{0, 0, v\}$ perpendicular to $N$ plates with thicknesses $a_n$ ($n$ is the plate number) made of a material with dielectric permittivity $\varepsilon_1$. The plates are located in a medium with dielectric permittivity $\varepsilon_2$ at distances $b_n$ between the $n$-th and ($n$+1)-th plates.

We will calculate the forward XTR intensity produced in such an *irregular stack* within the geometrical optics approximation, in which reflection of the wave is neglected. This means that if the plates are weakly absorbing, the following condition must be satisfied:

$$|N R_{12}|^2 \ll 1, \tag{6.1}$$

where $R_{12}$ is the Fresnel coefficient for reflection of a wave at the interface between the plate and the medium (see Eq. (1.22)):

$$R_{12} = \frac{\varepsilon_1 \lambda_2 - \varepsilon_2 \lambda_1}{\varepsilon_1 \lambda_2 + \varepsilon_2 \lambda_1}. \tag{6.2}$$

In the X-ray and higher-frequency range, the quantity $R_{12}$ is small, and condition in Eq. (6.1) is well satisfied even for a sufficiently large number $N$. Then, by virtue of what was said at the end of Sec. 5 and in accordance with Eq. (5.19), we have

$$W^{(+)}(\omega,\vartheta) = W_{\mathrm{b}}^{(+)}(\omega,\vartheta)\left|\sum_{m=0}^{N-1}\exp(iB_m)\left[1-\exp(i\delta_m)\right]\right|^2, \qquad (6.3)$$

where $W_b^{(+)}(\omega, \vartheta)$ is defined by Eq. (1.78), in which $z_{obs}$ is the distance from the last boundary of the system to the observation plane,

$$B_m = -\varphi_1 \sum_{k=m+2}^{N} a_k - \varphi_2 \sum_{k=m+2}^{N} b_{k-1}\,,$$
$$\delta_m = -\varphi_1 a_{m+1}, \qquad (6.4)$$
$$\varphi_s = \frac{\omega}{v} - \lambda_1 \qquad (s = 1,\ 2).$$

The method of geometrical optics for solving problems in the theory of transition radiation was first used by Ter-Mikaelian [**60.2**], and he also obtained an equation (see [**69.1**], p. 314) analogous to Eq. (6.3), only essentially without accounting for absorption of radiation in the medium.

If the plate thicknesses $a_m$ and the spacings $b_m$ between them are random variables, then the XTR intensity (Eq. 6.3)) is also a random variable with a probability distribution determined by the distribution functions of the quantities $a_m$ and $b_m$.

Let $a_m$ and $b_m$ be independent random variables. We shall assume that both $a_m$ and $b_m$ are independent for different values of $m$. Obviously, the distribution functions of the quantities $a_m$ and $b_m$ are such that these quantities take only non-negative values.[12]

---

[12] Ter-Mikaelian regarded the positions of the plate boundaries $r_m$ as independent random variables (see Ref. [**69.1**]). However, by the definition of these quantities, the following inequality must always be satisfied:

$$r_m \leq r_n \text{ for } m<n.$$

The assumption of independence of the quantities $r_m$ leads to a violation of this inequality, i.e., to the appearance of "negative" plate thicknesses (or distances between plates) $r_m+1-r_m$, which has no physical meaning. However, in the case of weak irregularity, when the mean-square deviations $\langle r_m^2\rangle$ are much smaller than $\langle a^2\rangle$ and $\langle b^2\rangle$, the assumption of independence of the

Let us find the average value of the quantity in Eq. (6.3). To this end, it suffices to calculate the quantity

$$F \equiv \left\langle \left| \sum_{m=0}^{N-1} \exp(iB_m)[1-\exp(i\delta_m)] \right|^2 \right\rangle =$$
$$= \left\langle \sum_{n=0}^{N-1} \sum_{m=0}^{N-1} \exp[i(B_m - B_n^*)][1-\exp(i\delta_m)][1-\exp(-i\delta_n)] \right\rangle, \quad (6.5)$$

where angle brackets denote averaging.

Using the well-known theorem stating that the mean value of the product of independent quantities equals the product of their mean values, we obtain [**74**.4, **75**.4]

$$F = 2\mathrm{Re}\left\{ \frac{1-(Q_aQ_b)^N}{1-Q_aQ_b} \, \frac{(1+Q_a)/2 - h_a - [Q_a - h_a(1+Q_a)/2]h_b}{1-h_ah_b} + \right.$$
$$\left. + \frac{(1-h_a)(Q_a-h_a)h_b[(Q_aQ_b)^N - (h_ah_b)^N]}{(1-h_ah_b)(Q_aQ_b - h_ah_b)} \right\}, \quad (6.6)$$

where

$$h_a = \langle \exp(-i\varphi_1 a_k) \rangle, \quad h_b = \langle \exp(-i\varphi_2 b_k) \rangle$$
$$Q_a = \langle \exp(-\eta_1 a_k) \rangle, \quad Q_b = \langle \exp(-\eta_2 b_k) \rangle, \quad (6.7)$$
$$\eta_s = 2\mathrm{Im}\lambda_s \qquad (s = 1,\ 2).$$

In the case when both the plate material and the external medium are non-absorbing, i.e., $\eta_1 = \eta_2 = 0$, from Eq. (6.6) we obtain a simpler expression:

$$F = 2\,\mathrm{Re}\left\{ N \frac{(1-h_a)(1-h_b)}{1-h_ah_b} + \frac{(1-h_a)^2 h_b[1-(h_ah_b)^N]}{(1-h_ah_b)^2} \right\}. \quad (6.8)$$

The spectral–angular distribution of the mean intensity of the radiation is given by the following expression:

$$\langle W^{(+)}(\omega, \vartheta) \rangle = W_{\mathrm{b}}^{(+)}(\omega, \vartheta)\, F, \quad (6.9)$$

where $F$ is given by Eq. (6.6) or, in the special case, Eq. (6.8).

---

quantities rm does not lead to large errors, since in this case the probability of “negative” plate thicknesses (or distances between plates) is small. The introduction of independent nonnegative random variables $a_m$ and $b_m$, as described in this section, is free from this drawback.

## 6.2. Weak Irregularity

The regular stack considered in Sec. 3 is a special case of an irregular stack, when the random variables $a_m$ and $b_m$ assume the same values $a$ and $b$. In this case we have

$$|h_a|^2 = Q_a, \quad |h_b|^2 = Q_b,$$

and Eq. (6.6) takes a very simple form:

$$F = F_{pl} F_{st},$$

where

$$F_{pl} = |1 - h_a|^2, \; F_{st} = \frac{|1 \cdot h_a^N h_b^N|^2}{|1 - h_a h_b|^2}$$

It is not difficult to verify that these expressions for $F_{\mathrm{pl}}$ and $F_{\mathrm{st}}$ coincide exactly with the previously given expressions (2.22) and (3.16) for the corresponding interference factors. Thus, Eq. (3.15) for XTR in a regular stack can be obtained directly from Eq. (6.9), as expected.

Let us now consider the case of a weakly disordered medium, when the plates have almost identical thicknesses and are arranged almost regularly, so that the higher moments in the distributions of $a$ and $b$ can be neglected, i.e.

$$h_a \approx \exp(-i\varphi_1 \langle a \rangle)\left(1 - \frac{\varphi_1^2 \langle \Delta a^2 \rangle}{2}\right),$$
$$h_b \approx \exp(-i\varphi_2 \langle b \rangle)\left(1 - \frac{\varphi_2^2 \langle \Delta b^2 \rangle}{2}\right), \tag{6.10}$$

where $\langle \Delta a^2 \rangle = \langle a^2 \rangle - \langle a \rangle^2$, $\langle \Delta b^2 \rangle = \langle b^2 \rangle - \langle b \rangle^2$ are the corresponding mean-square deviations.

First, for simplicity, we neglect absorption, i.e., let $\gamma_1 = \gamma_2 = 0$. Then, expanding Eq. (6.8) in powers of the small quantities $\varphi_1^2 \langle \Delta a^2 \rangle$ and $\varphi_2^2 \langle \Delta b^2 \rangle$, we obtain

$$F = F_{\mathrm{reg}} - \left(\varphi_1^2 \langle \Delta a^2 \rangle + \varphi_2^2 \langle \Delta b^2 \rangle\right) \sin^2 \frac{\varphi_1 \langle a \rangle}{2} \cos(NX) \times$$

$$\times\frac{(N+1)\sin(N-1)X-(N-1)\sin(N+1)X}{\sin^3 X}+$$
$$+\varphi_1^2\langle\Delta a^2\rangle\left|N\cos\varphi_1\langle a\rangle-\frac{\sin\varphi_1\langle a\rangle\,(N\sin 2X-\sin 2NX)}{2\sin^2 X}\right|, \tag{6.11}$$

where

$$F_{\text{reg}}=F_{\text{pl}}F_{\text{st}}=4\sin^2\frac{\varphi_1\langle a\rangle}{2}\,\frac{\sin^2 NX}{\sin^2 X}, \tag{6.12}$$

$$X=\frac{\varphi_1\langle a\rangle+\varphi_2\langle b\rangle}{2}. \tag{6.13}$$

The interference factor $F_{\text{reg}}$ determines the spectral–angular distribution of the radiation intensity for the corresponding non-absorbing regular stack (cf. (3.15)). As was already noted in Sec. 3.3, this distribution for large $N$ ($N\gg 1$) has a number of sharp maxima under the following conditions:

$$X=n\pi \qquad (n=1,\ 2,\ \ldots) \tag{6.14}$$

and

$$N^2\sin^2\frac{\varphi_1\langle a\rangle}{2}\gg 1. \tag{6.15}$$

At these points, from Eq. (6.11) we obtain

$$F\approx 4\sin^2\frac{\varphi_1\langle a\rangle}{2}\left[N^2-\frac{N(N^2-1)}{6}\left(\varphi_1^2\langle\Delta a^2\rangle+\varphi_2^2\langle\Delta b^2\rangle\right)\right], \tag{6.16}$$

whence it follows that the irregularity leads to a certain decrease of the indicated maxima.

The minima of the spectral–angular distribution of the radiation intensity for a regular stack are determined by the zeros of the factor $F_{\text{reg}}$.

After taking the irregularity into account, as can be seen from Eq. (6.11), the factor $F$ becomes nonzero (and, naturally, positive). Thus, the irregularity also leads to some increase in the minima.

Thus, the correction due to the irregularity of the plate thicknesses and their arrangement somewhat *smooths* the radiation intensity distribution. A similar result can also be obtained when absorption is taken into account.

The physical nature of such smoothing is entirely clear. The sharp maxima and subsequent minima in the radiation intensity distribution are caused by interference of the radiation fields arising on different plates of a regular stack. Irregularity disturbs the phases of these fields, and therefore the interference pattern becomes somewhat blurred, i.e., smoothed.

Equation (6.16) also yields an estimate of the applicability criterion of Eq. (6.11), i.e., the smallness of the correction terms:

$$N(\varphi_1^2 < \Delta a^2 > + \varphi_2^2 < \Delta b^2 >) \ll 1. \qquad (6.17)$$

When this condition is satisfied, an irregular stack differs little from a regular one with respect to the spectral–angular distribution of XTR.

## 6.3. Special Cases of the General Equation

In the simplest case, when $N = 1$, i.e., when the stack contains only one plate, from Eq. (6.6) we obtain

$$F = F_{\mathrm{pl}} = 1 + Qa - 2\,\mathrm{Re}\,h_a. \qquad (6.18)$$

Substituting Eq. (6.18) into Eq. (6.9), we obtain the equation for the spectral–angular distribution of the XTR intensity generated in a plate with a random thickness.

It also follows from Eq. (6.6) that in the case $b_h = 0$, i.e.,

$$h_b = 1 \quad \text{and} \quad Q_b = 1, \qquad (6.19)$$

we have

$$F = 1 + Q_a^N - 2\,\mathrm{Re}\,h_a^N, \qquad (6.20)$$

which represents the interference factor for $N$ plates with random thicknesses placed in direct contact with each other.

The usual expression for a plate of thickness $a$ can be obtained from Eq. (6.18) (or Eq. (6.20)) by replacing $Q_a$ (or $Q_a^N$) and $h_a$ (or $h_a^N$) with $exp(-\eta_1 a)$ and $exp(-i\varphi_1 a)$, respectively (cf. Eqs. (2.21) and (2.22)). A small spread in thickness naturally again leads to some smoothing of distributions (Eqs. (6.18) or (6.20)) compared with Eq. (2.22).

Let us consider another limiting case, when

$$\langle\cos(\varphi_2 b_k)\rangle = \langle\sin(\varphi_2 b_k)\rangle = 0, \quad \text{i. e.} \quad h_b = 0. \qquad (6.21)$$

Since by their physical meaning all $b_k$ are non-negative, conditions in Eq. (6.21) mean that $|\langle\varphi_2 b_k\rangle| \gg \pi$ or

$$\frac{\langle b\rangle}{|z_2|} \gg 1 \qquad (z_2 = \pi/\varphi_2). \qquad (6.22)$$

Taking Eq. (6.21) into account, from Eq. (6.6) we obtain

$$F = N_{\text{eff}}\,(1 + Q_a - 2\,\text{Re}\,h_a) = N_{\text{eff}} F_{\text{pl}}, \qquad (6.23)$$

where $F_{pl}$ is defined according to Eq. (6.18), and

$$N_{\text{eff}} = \frac{1 - (Q_a Q_b)^N}{1 - Q_a Q_b}. \qquad (6.24)$$

Thus, Eq. (6.23) shows that when the mean distance between the plates is much greater than the radiation formation length, the mean intensity of XTR emitted forward from an irregular stack is equal to the sum of the mean radiation intensities from all plates, taking into account absorption in the plates and in the surrounding medium.

If

$$\langle a\rangle \gg z_1, \qquad (6.25)$$

then from Eq. (6.6) we obtain

$$F = N_{\text{eff}}\,(1 - Q_a - 2Q_a\,\text{Re}\,h_b) - 2Q_a^N Q_b^{N-1}\,\text{Re}\,h_b. \qquad (6.26)$$

If both inequalities (6.22) and (6.25) are satisfied, then from Eqs. (6.23) or (6.26) we have

$$F = N_{\text{eff}}(1 + Q_a), \qquad (6.27)$$

i.e., the intensity of the resulting radiation is the sum of the intensities of the radiation formed at all interfaces, taking into account absorption in the plates and in the surrounding medium.

Finally, when absorption in the plates is large, so that

$$Q_a \ll 1, \qquad |h_a| \ll 1, \tag{6.28}$$

from Eq. (6.6) we obtain

$$F \approx 1,$$

i.e., practically only that part of the radiation formed at the last interface emerges from the stack.

### 6.4. Gamma Distribution

For illustration, let us consider specific distributions for $a_k$ and $b_k$. For this purpose, it is convenient to take the so-called gamma distribution (see, for example, Ref. [**73**.33], p. 581):

$$f(y) = \beta_0^\alpha \frac{y^{\alpha-1} \exp(-\beta_0 y)}{\Gamma(\alpha)}, \tag{6.29}$$

where $\alpha$ and $\beta_0$ are real parameters, $\alpha > 0$, $\beta_0 > 0$, and $\Gamma(x)$ is the gamma function.

The random variable y in this distribution takes only non-negative values. The distribution (Eq. (6.29)) is normalized to unity:

$$\int_0^\infty f(y)\, dy = 1. \tag{6.30}$$

The mean value (mathematical expectation) $\langle y \rangle$ and the mean-square deviation (dispersion) $\langle \Delta y \rangle^2$ are expressed in terms of the parameters $\alpha$ and $\beta_0$ by the equations:

$$\langle y \rangle = \frac{\alpha}{\beta_0}, \quad \langle \Delta y^2 \rangle = \frac{\alpha}{\beta_0^2} = \frac{\langle y \rangle^2}{\alpha}. \tag{6.31}$$

It is convenient to introduce the concept of relative spread (or degree of irregularity):

$$\zeta = \frac{<\Delta y^2>^{1/2}}{<y>} = \frac{1}{\alpha^{1/2}}. \qquad (6.32)$$

Plots of the gamma distribution function (Eq. (6.29)) are given in Fig. 6.1 for the cases α−1 = 1; 5; 50; 500 <*y*> at = 10. These cases correspond to the values ζ = 70.7; 40.8; 14.0; 4.47%

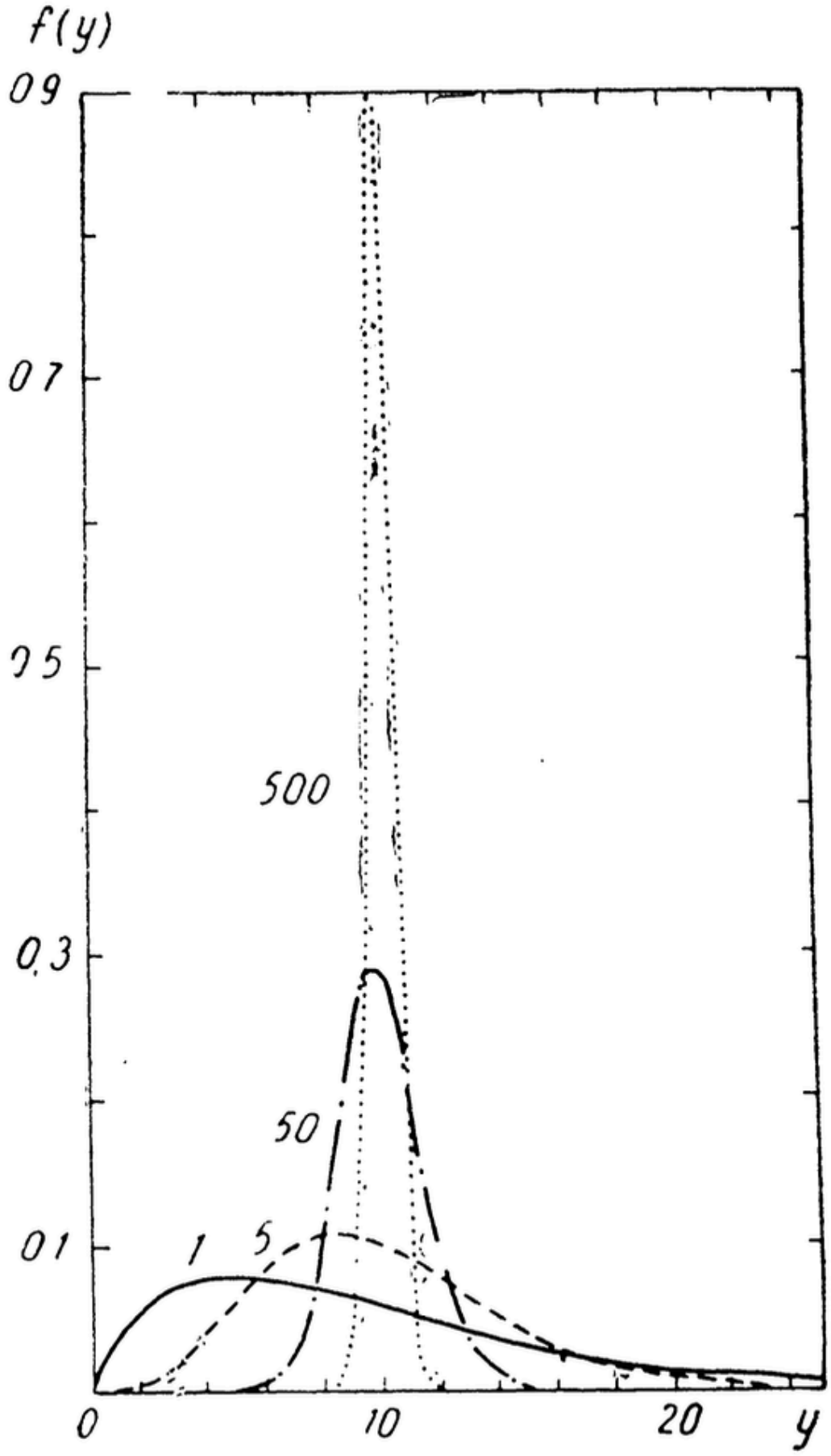

Fig. 6.1. Plots of the gamma-distribution function f(y) for the values α−1 = 1; 5; 50; 500 (indicated at the curves) at <y> = 10. The corresponding values of the degree of irregularity are ζ = 70.7; 40.8; 14.0; 4.47%.

When $\alpha = 1$, the value $f(y)$ at $y = 0$ equals $Z_0{}^0$, and when $0 < \alpha < 1$, function $f(y)$ at $y \to 0$ tends to infinity. Thus, the large relative spread of $\zeta$ can be obtained only when the random variable $y$ takes values close to zero with high probability.

Using the distribution in Eq. (6.29), it is easy to calculate the quantities $h_a$, $h_b$, $Q_a$, and $Q_b$. For example:

$$h_a = \frac{\beta^\alpha}{\Gamma(\alpha)} \int_0^\infty a^{\alpha-1} \exp\left[(-i\varphi_1 - \beta_0)\, a\right] da =$$

$$= (1 + i\varphi_1 \zeta_a^2 \langle a \rangle)^{-1/\zeta_a^2}, \tag{6.33}$$

$$Q_a = (1 + \eta_a \zeta_a^2 \langle a \rangle)^{-1/\zeta_a^2}. \tag{6.34}$$

The quantities $h_b$ and $Q_b$ are expressed by analogous equations with the replacement of index 1 by 2 and $a$ by $b$. Here $\zeta_a$ and $\zeta_b$ are the degrees of irregularity for the plate thicknesses and the spacings between them, respectively. For example, if $\zeta_a \to 0$ and $\langle a \rangle = const$, we have $h_a \to exp(-\, i\varphi_1 \langle a \rangle)$. This is the case of plates with identical thicknesses.

For $\alpha_b \to \infty$ and $\beta_{0b} = const$ we obtain $h_b = 0$ (see Eq. (6.21)); then $\langle \Delta b^2 \rangle \to \infty$ and $\langle b \rangle \to \infty$, but $(\langle \Delta b^2 \rangle)^{1/2} / \langle b \rangle = \zeta_b \to 0$. In this case we also have $Q_b = 0$.

As for the cases of infinite relative spreads of the plate thicknesses and the spacings between them, in one case ($\zeta_b \to \infty$) we obtain expressions (6.19) and (6.20), whereas in the other ($\zeta_a \to \infty$) we have $h_a = Q_a = 1$ and $F = 0$.

Both of these results are due to the fact that for large relative spreads the parameter $\alpha$ of the gamma distribution is a small quantity (see Eq. (6.32)), and the probability density (Eq. (6.29)) tends to infinity at small values of the random variable.

Using equations (6.33), (6.34), and analogous equations for $h_b$ and $Q_b$, together with equations (6.6) and (6.9), one can calculate the

spectral–angular distribution of the XTR intensity produced in an irregular stack.

In Figs. 6.2 and 6.3, the distributions of the numbers of XTR quanta computed by the indicated method are shown for stacks with given values of ⟨a⟩, ⟨b⟩ and irregularity parameters ζa and ζb, for different quantum energies and particle Lorentz factors. All curves were calculated for a light material such as carbon with $\hbar\omega_0 = 20$ eV. The spectral–angular distribution for a regular stack has a strongly oscillatory character and is therefore not shown in the figures. We only note that it resembles the curves corresponding to $\zeta_a = \zeta_b = 4.47\%$, but with maxima at the same angles, only narrower and higher.

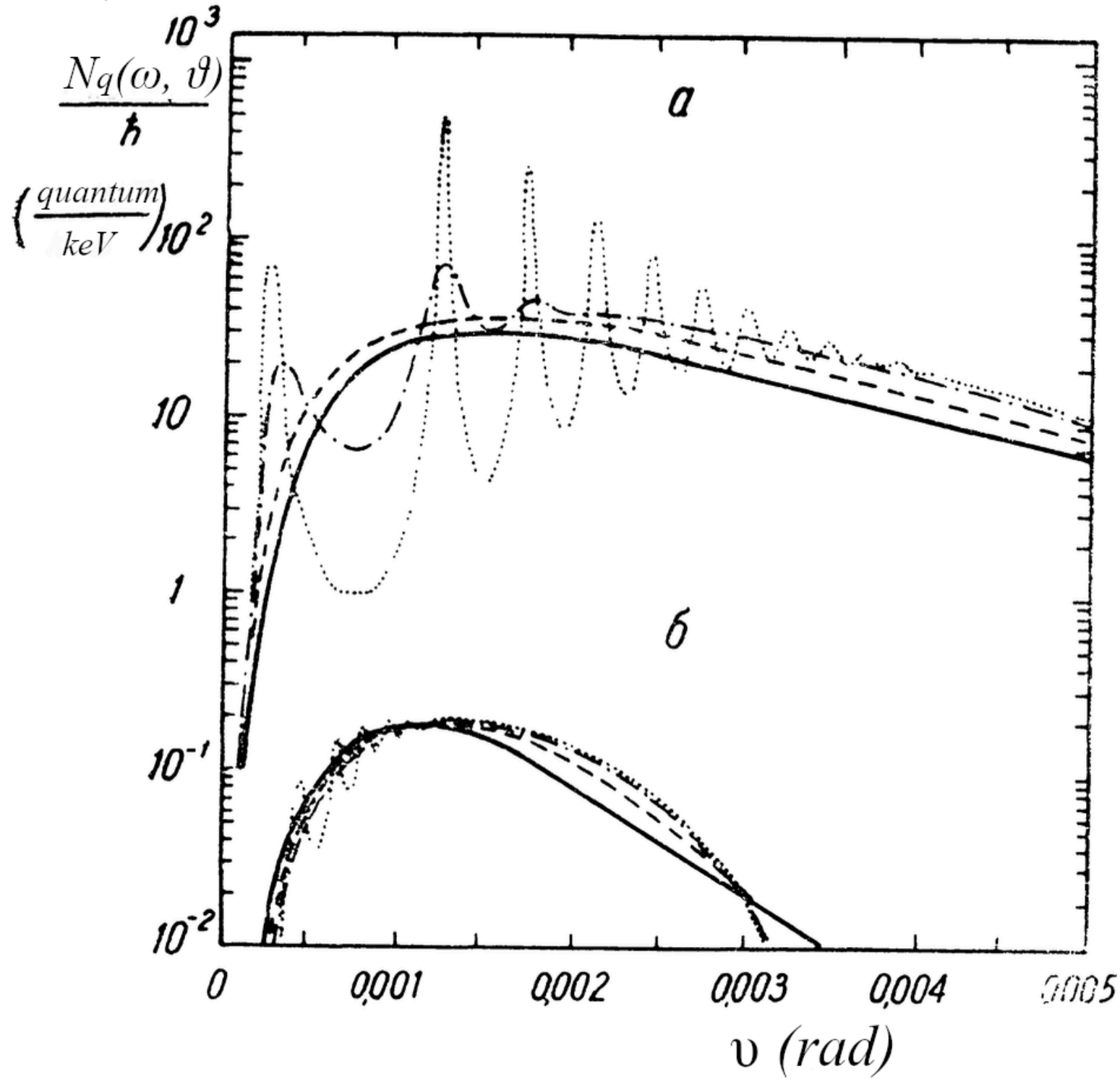

Fig. 6.2. Angular dependence of the spectral–angular distribution of the numbers of XTR quanta $N_q(\omega, \vartheta) = W(\omega, \vartheta)/\hbar\omega$ for irregular stack with <a> = 7 μm and <b> = 410 μm for γ = 10^3 )\ and ħω = keV (curves a) and 25 keV (curves b). The number of plates in the stack is N = 50, the plate material is carbon-like with $\hbar\omega_0 = 20$ eV. The degree of irregularity $\zeta_a = \zeta_b$ takes the values 4.47% (dotted curves); 14.0% (dash–dot curves); 40.8% (dashed curves); 70.7% (solid curves).

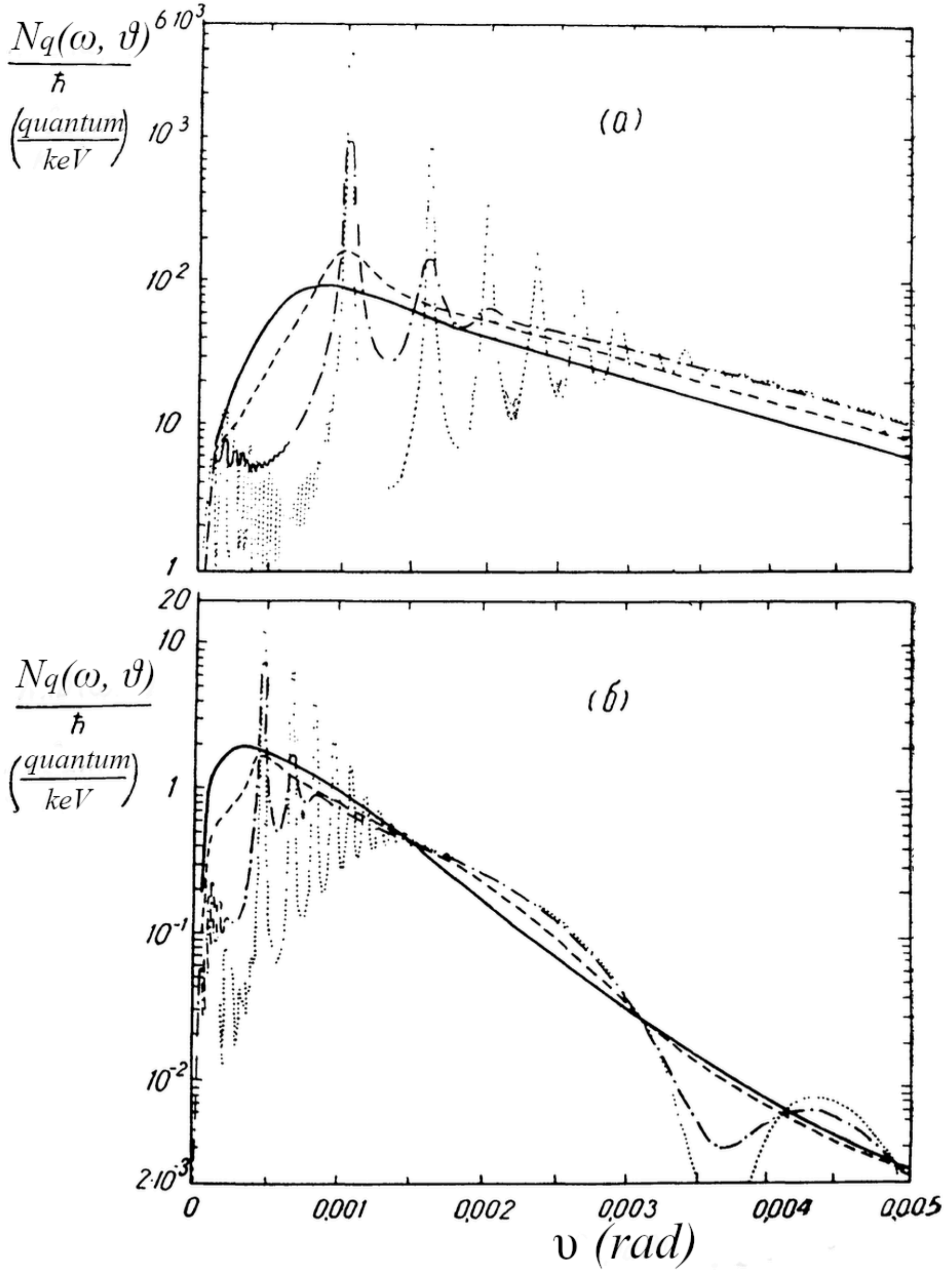

Fig. 6.3. The same as in Fig. 6.2, for $\gamma = 10^4$.

The frequency spectrum (Figs. 6.4 and 6.5) was obtained by numerically integrating the spectral–angular distribution over the emission angle.

It follows from Figs. 6.2–6.5 that irregularity mainly leads to the disappearance of the interference of radiation formed at different interfaces of the plates in an irregular stack. When the irregularity is small, i.e., when the condition in Eq. (6.17) is satisfied, the transition radiation spectrum of an irregular stack is very close to the corresponding spectrum of a regular stack. When condition in Eq. (6.17) is not satisfied, the interference disappears and the distribution (both spectral and angular) becomes monotonic.

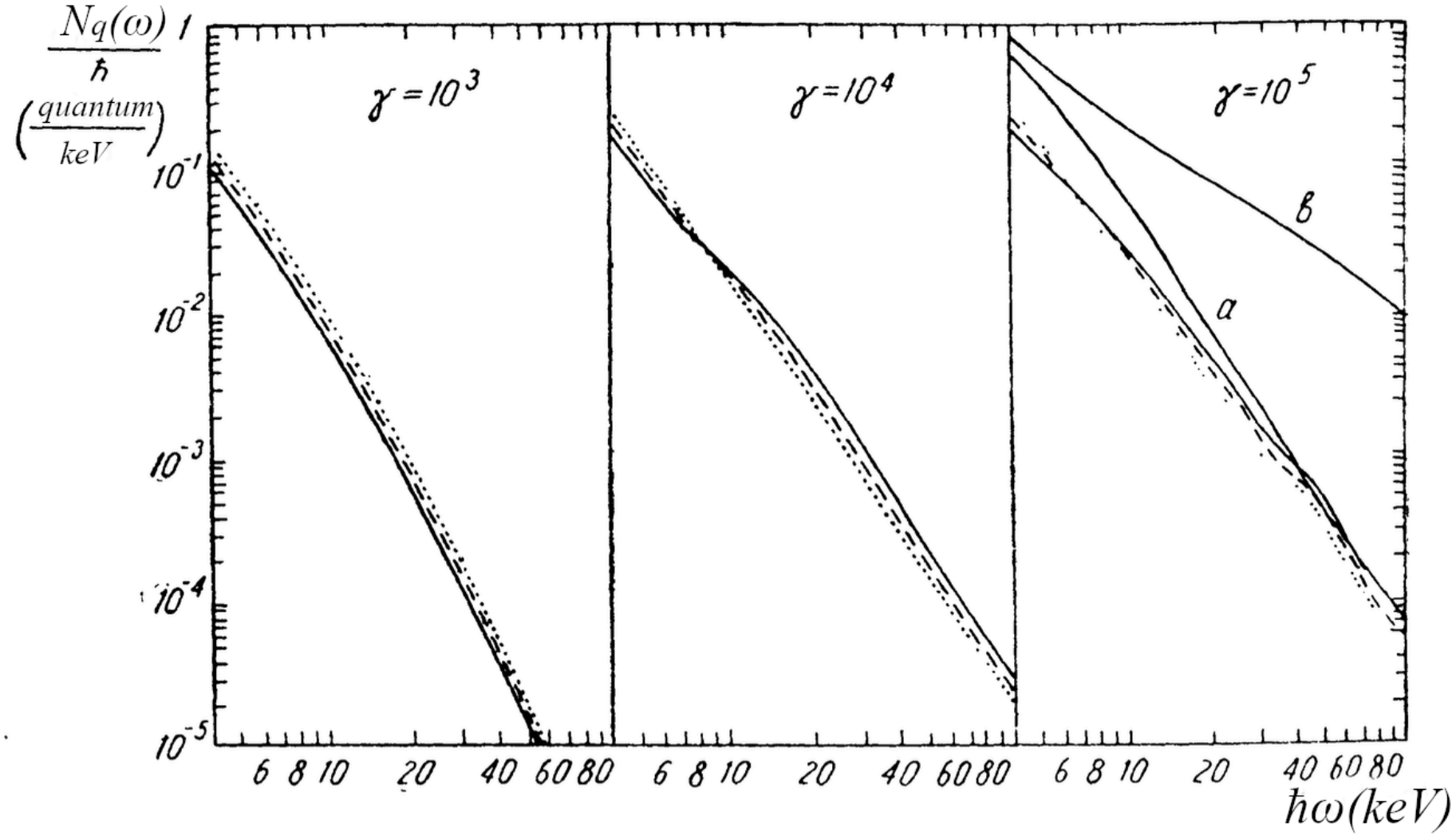

Fig. 6.4. Frequency spectrum of XTR $N_q(\omega) = W(\omega)/\hbar\omega$ for an irregular stack with $\langle a\rangle = 7$ μm, $\langle b\rangle = 410$ μm, $N = 50$, $\hbar\omega_0 = 20$ eV. $\zeta_a = \zeta_b = 4.47$ (dotted curves); 40.8% (dashed curves); 70.7% (solid curves without letters). The solid curve $a$ corresponds to the spectrum obtained by summing the numbers of quanta from $N$ plates of thickness 7 μm; the solid curve $b$ corresponds to summing the numbers of quanta from all $2N$ interfaces.

For a regular stack with $a = 7$ μm and $b = 410$ μm, the frequency spectrum of XTR from 4 to 100 keV has no maxima or minima. In this case, the frequency spectra of XTR for various irregular stacks with the same $\langle a\rangle$ and $\langle b\rangle$ differ little from the spectrum of the regular stack (Fig. 6.4). However, for $a = 250$ μm, $b = 2000$ μm, and $\gamma = 7500$, the frequency spectrum of XTR for a regular stack has a minimum in the region of 30 keV. For a stack with the same $\langle a \rangle$ and $\langle b \rangle$, but sufficiently irregular (for example, when $\zeta_a, \zeta_b > 40\%$), the frequency spectrum of XTR becomes smoother and in the region 20–50 keV differs quite significantly (by up to a factor of 2.5) from the spectrum of the regular stack (Fig. 6.5).

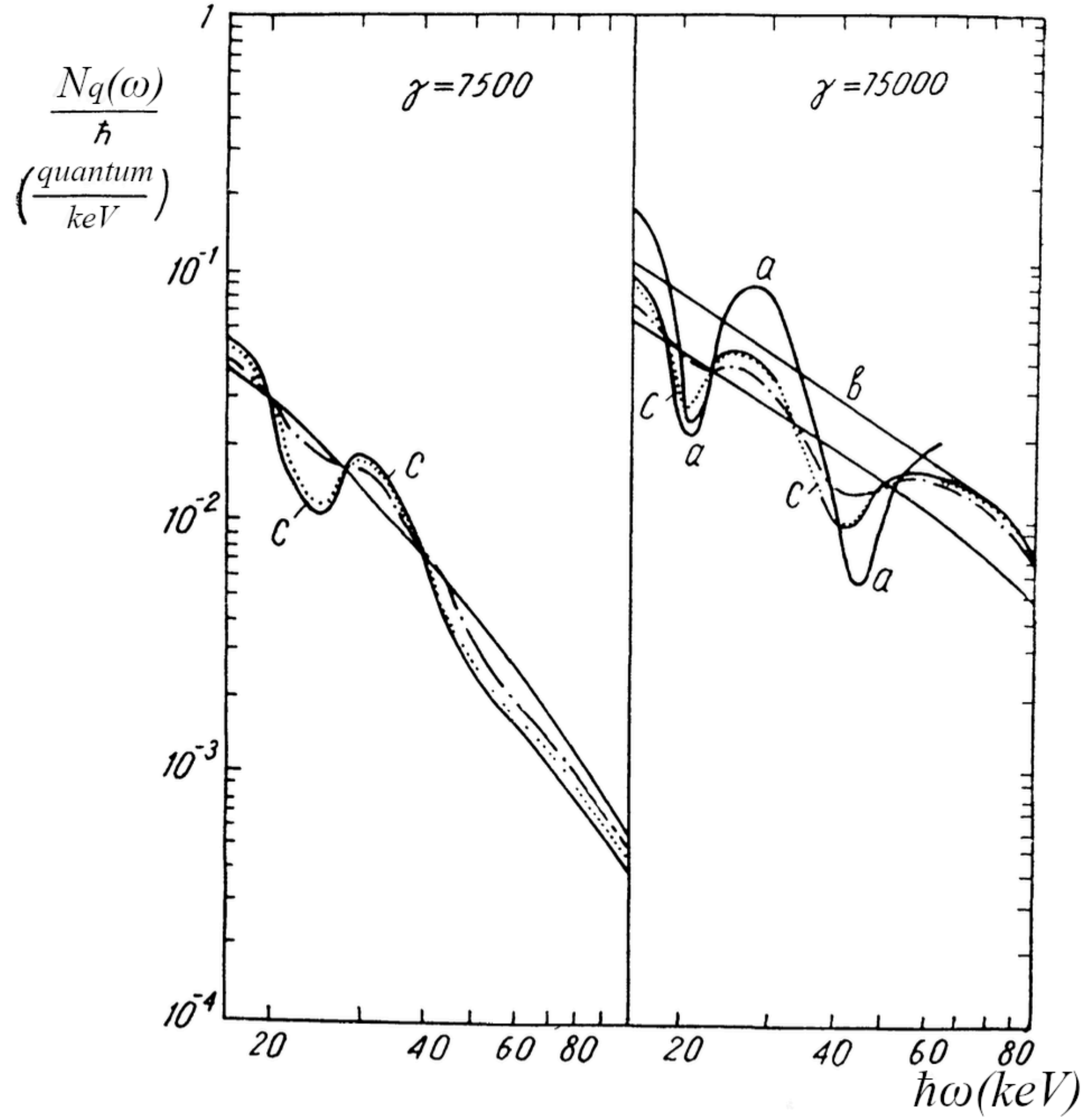

*Fig. 6.5. Same as in Fig. 6.4, for* $\langle a \rangle = 250$ *μm and* $\langle b \rangle = 2000$ *μm. The dash–dotted curves correspond to the case* $\zeta_a = \zeta_b = 14.0\%$*; the solid curves* c *correspond to the case of a regular stack.*

## § 7. BUNCH OF CHARGED PARTICLES

When an interface between media is crossed not by a single particle but by a group (*bunch*) of charged particles, interference occurs between the radiation produced by the individual particles of the bunch. We shall determine under what conditions the particles radiate coherently and how the radiation intensities of the entire bunch and of a single particle are related [**60**.15, **61**.5, **62**.3, **75**.19].

### 7.1. Bunch Form Factor

Let a bunch of $N$ particles with identical charges $e$ and velocities $v$ enter along the $z$-axis from medium 1 into medium 2, perpendicular to the plane interface between the media. The densities of external charges $\rho_{ext}(\boldsymbol{r}, \omega)$ and currents $\boldsymbol{j}_{ext}(\boldsymbol{r}, \omega)$ are written in the following form (cf. (1.12)):

$$\begin{aligned} \rho_{\text{ext}}(\mathbf{r},t) &= e\sum_{i=1}^{N}\delta(\mathbf{r}-\mathbf{r}_i-\mathbf{v}t), \\ \mathbf{j}_{\text{ext}}(\mathbf{r},t) &= \mathbf{v}\,\rho_{\text{ext}}(\mathbf{r},t), \\ \rho_{\text{ext}}(\mathbf{k},\omega) &= \frac{e}{(2\pi)^3 v}\,\delta\Big(k_z-\frac{\omega}{v}\Big)\sum_{m=1}^{N}\exp(i\mathbf{k}\cdot\mathbf{r}_m), \\ \mathbf{j}_{\text{ext}}(\mathbf{k},\omega) &= \mathbf{v}\,\rho_{\text{ext}}(\mathbf{k},\omega). \end{aligned} \tag{7.1}$$

where $\boldsymbol{r}_m$ is the position vector of the $m$th particle, measured, for example, from the center of mass of the bunch; the summation is performed over all particles of the bunch.

The radiation field produced by the bunch is the sum of the radiation fields of the individual particles of the bunch, with the corresponding phase factors taken into account. As a result, for the spectral–angular distribution of the intensity of the resulting radiation we obtain

$$W^{(\pm)}_{\text{bunch}}(\omega,\vartheta) = \frac{1}{2\pi}\,W^{(\pm)}(\omega,\vartheta)\,F(\omega,\vartheta), \tag{7.2}$$

where $W^{(\pm)}(\omega, \vartheta)$ is the spectral–angular distribution of the radiation intensity produced by a single particle upon crossing the interface between media (see Eqs. (1.35) or (1.37) for $W^{(+)}(\omega, \vartheta)$ and analogous equations for $W^{(-)}(\omega, \vartheta)$), and the quantity

$$F(\omega, \vartheta) = \sum_{m,n=1}^{N} \exp\{-i[\varkappa \cdot (\rho_m - \rho_n) + \omega(z_m - z_n)/v\}, \tag{7.3}$$

where

$$\varkappa = \frac{\omega}{c} \sin\vartheta\{\cos\phi, \sin\phi, 0\}, \quad \rho = \{x, y, 0\},$$

describes the character of the interference of the radiation emitted by the particles of the bunch with a given structure and may be called the *bunch form factor*.

If

$$\chi \cdot (\rho_m - \rho_n) \gg 1 \quad \text{or} \quad \omega(z_m - z_n)/v \gg 1, \tag{7.4}$$

holds for all $m \neq n$, then in the quantity $F(\omega, \vartheta)$ only the terms with $m = n$ contribute, i.e., in this case

$$F(\omega, \vartheta) = N$$

and the radiation intensity produced by the bunch is equal to the sum of the radiation intensities produced by the individual particles of the bunch (the particles radiate *incoherently*).

If, on the contrary, the inequalities

$$\chi \cdot (\rho_m - \rho_n) \ll 1 \quad \text{and} \quad \omega(z_m - z_n)/v \ll 1, \tag{7.5}$$

hold for all $m$ and $n$ (from 1 to $N$), then

$$F(\omega, \vartheta) = N^2$$

and the entire bunch radiates as a single particle with charge $Ne$, equal to the sum of the charges of the particles in the bunch [60.15, 61.5] (the particles radiate *coherently*).

In the general case the situation is intermediate, i.e., some of the charges radiate coherently, and overall there is interference of the radiation emitted by individual parts of the bunch.

## 7.2. Averaging over Particle Positions. Coherence Factor

In practice, the exact coordinates of individual particles in the bunch are usually unknown, and only the distribution function of these coordinates is specified. Therefore, it is necessary to perform statistical averaging of the quantity given by Eq. (7.2) over the coordinates of the particles in the bunch. This averaging reduces to averaging the factor $F(\omega, \vartheta)$.

Let us write the quantity $F(\omega, \vartheta)$ in the following form:

$$F(\omega, \vartheta) = N + \sum_{m,n=1}^{N}{}' \exp\{-i[\boldsymbol{\varkappa}\cdot(\boldsymbol{\rho}_m - \boldsymbol{\rho}_n) + \omega(z_m - z_n)/v]\}, \qquad (7.6)$$

where a prime on the summation sign means that $m \neq n$ in the summation.

Averaging expression (7.6) over different positions of the particles in the bunch, we obtain

$$\langle F(\omega, \vartheta)\rangle = N + N(N-1)G_k(\omega, \vartheta), \qquad (7.7)$$

$$G_k(\omega, \vartheta) = \int\int f_2(\boldsymbol{r}, \boldsymbol{r}')\exp\{-i[\boldsymbol{\varkappa}\cdot(\boldsymbol{\rho}-\boldsymbol{\rho}') + \omega(z-z')/v]\}d^3\boldsymbol{r}d^3\boldsymbol{r}', \quad (7.8)$$

where $r = \rho + zn_z$, $r' = \rho' + z'n_z$, and $f_2(r, r')$ is the two-particle distribution function of the particles in the bunch, normalized as follows:

$$\int\int f_2(\boldsymbol{r}, \boldsymbol{r}')d^3\boldsymbol{r}d^3\boldsymbol{r} = 1. \qquad (7.9)$$

The quantity $G_k(\omega, \vartheta)$ can be called the *coherence factor*, since in the case of completely coherent (when conditions (7.5) are satisfied) or completely incoherent (when conditions (7.4) are satisfied) radiation this quantity takes the value 1 or 0, respectively, and in the general case $0 \leq G_k(\omega, \vartheta) \leq 1$. At the same time (see Eq. (7.7)), the structure factor $\langle F(\omega, \vartheta)\rangle$ consists of two terms: the term $N[1 - G_k(\omega, \vartheta)]$, corresponding to the incoherent part of the radiation, and the term $N^2 G_k(\omega, \vartheta)$, corresponding to its coherent part. If the number of particles in the bunch is much greater than some critical value $N_c(\omega, \vartheta)$, characteristic of a given bunch shape, a given frequency, and a fixed emission angle [75.**19**]:

$$N \gg N_{cr}(\omega, \vartheta), \qquad (7.10)$$

where

then the coherent term will dominate over the incoherent term even when the radiation from the entire bunch is not coherent (i.e., conditions given by Eq. (7.5) are not satisfied). Otherwise, the incoherent term will dominate.

We shall assume that the coordinates of the charged particles in the bunch are independent quantities. This means that the two-particle distribution function can be approximately replaced by the product of two single-particle distribution functions:

$$f_2(\boldsymbol{r}, \boldsymbol{r}') \approx f_1(\boldsymbol{r}) f_1(\boldsymbol{r}'). \qquad (7.12)$$

Strictly speaking, this is valid only when the particles are sufficiently far apart. At small distances, due to interactions between the particles, a certain correlation between their positions may arise. However, if the bunch is not too dense, so that the particles are most of the time far from one another, approximation given by Eq. (7.12) is quite adequate.

In addition, we shall also regard the coordinates ρ and z as independent. In the presence of cylindrical symmetry we have

$$f_1(\boldsymbol{r}) = f_\rho(\rho) f_z(z), \qquad (7.13)$$

where

$$2\pi \int_0^\infty f_\rho(\rho)\rho d\rho = 1, \qquad \int_{-\infty}^{\infty} f_z(z) dz = 1.$$

Substituting Eqs. (7.12) and (7.13) into Eq. (7.8), for the coherence factor in the presence of cylindrical symmetry we obtain the following expression:

$$G_k(\omega, \vartheta) = \left| \int_{-\infty}^{\infty} \exp(i\omega z/v) f_z(z) dz \right|^2 \left| 2\pi \int_0^\infty J_0(\varkappa\rho) f_\rho(\rho)\rho d\rho \right|^2, \qquad (7.14)$$

where $J_0(x)$ is the Bessel function of the zeroth order.

For coherence of the radiation ($G_k \approx 1$) it is necessary that both factors in Eq. (7.14) be close to unity. This means that the radiation wavelength must be much larger than both the mean longitudinal distance between the particles in the bunch and the mean transverse distance multiplied by the sine of the emission angle. This is naturally consistent with the general conditions given by Eq. (7.5), where simultaneous fulfillment of both inequalities is assumed.

### 7.3. Particular Distribution Functions

Let us consider two particular distributions of particles in the bunch.

*a) Uniform Distribution.*

$$f_z(z) = \begin{cases} 1/a & \text{for } |z| \le a/2, \\ 0, & \text{for } |z| > a/2, \end{cases}$$
$$f_\rho(\rho) = \begin{cases} 1/\pi b^2, & \text{for } 0 \le \rho \le b, \\ 0, & \text{for } \rho > b, \end{cases} \tag{7.15}$$

where $a$ and $b$ are the longitudinal and transverse dimensions of the bunch.

Substituting Eq. (7.15) into Eq. (7.14), we obtain

$$G_k(\omega, \vartheta) = \left| \frac{4vc\sin(\omega a/2v) J_1((\omega b/c)\sin\vartheta)}{\omega^2 ab\sin\vartheta} \right|^2, \tag{7.16}$$

where $J_1(x)$ is the Bessel function of the first order.

*b) Gaussian Distribution.*

$$f_z(z) = \frac{2}{a\,\pi^{1/2}} \exp\left(-\frac{4z^2}{a^2}\right),$$
$$f_\rho(\rho) = \frac{1}{\pi b^2} \exp\left(-\frac{\rho^2}{b^2}\right). \tag{7.17}$$

In this case

$$G_k(\omega, \vartheta) = \exp\left\{-\frac{\omega^2}{2}\left(\frac{a^2}{4v^2} + \frac{b^2}{c^2}\sin^2\vartheta\right)\right\}. \tag{7.18}$$

If the bunch dimensions are of order 0.1 — 1 cm and the particles' Lorentz factor is $\sim 10^4$, then for the optical frequency range ($c/\omega \sim 10^{-5}$ cm) the coherence factor $G$, in the case of a uniform distribution of particles in the bunch, is of order $10^{-9} - 10^{-13}$, i.e., the critical number $N_{cr} \sim 10^9 - 10^{13}$. In the case of a Gaussian distribution, this number is considerably larger. For the X-ray frequency range the number $N_{cr}$ turns out to be so large that for real beams available in modern accelerators or storage rings the radiation is practically completely incoherent.

## § 8. IONIZATION LOSSES AND THE DENSITY EFFECT

In the previous sections we investigated the radiation arising when a charged particle passes through matter with interfaces. However, a particle loses energy not only to radiation but also to excitation and ionization of the atoms of the medium. Such losses (the so-called ionization losses) of fast particles were first considered classically by N. Bohr and subsequently by Bethe, Bloch, and others within a quantum approach (see Refs. [**48**.1, **80**.14]). Without taking into account the polarization of the medium by the field of the particle, they showed that these losses should grow logarithmically with increasing particle Lorentz factor γ. Then Fermi showed that taking into account the polarization of an infinite homogeneous medium leads to a strong distortion of the particle's field in the medium compared with the field in vacuum and, as a result, after a small logarithmic rise the losses become independent of γ (*the Fermi density effect*).[13]

It is clear that if the layer of matter through which a fast particle passes is sufficiently thick, the influence of the boundaries should be negligible, and the ionization loss per unit path length of the particle in this layer should be the same as in an infinite medium. However, if the layer of matter is thin, i.e., the transit time of the particle through this layer is much shorter than the time required for the polarization of the atoms of the medium to be established, which is of the order of the inverse atomic frequency, then the medium "does not have time" to become polarized by the field of the passing particle. As a result, the ionization is effectively produced by the particle's vacuum field. It will be shown below that, indeed, the density effect is absent in thin layers of matter [**59**.4] (see also Refs. [**62**.7, **65**.11, **65**.12, **66**.10, **69**.8, **81**.1, **82**.12, pp. 3 and 146). In this case, it is important that the layer be thin precisely in the direction of the particle's motion. If, however, the particle moves parallel to the surface of a body having sufficient extent in the direction of its motion, the Fermi density effect in the material always occurs, even if the transverse dimensions of the body are small [**80**.11].

[13] It is assumed that in this case the maximum energy transferable to the electrons of the medium is finite and does not depend on γ, which is usually the case.

### 8.1. Work Done by the Field on a Charged Particle

Following O. Bohr and Landau, we shall calculate the energy loss of a particle passing through a plate of thickness *a* located in vacuum as the work performed by the total field on the particle over the entire path of its motion (along the *z*-axis):

$$W = e\int_{-\infty}^{\infty} E_z(\boldsymbol{r}, t)\, dz \quad (\boldsymbol{r} = \{0, 0, z\},\ t = z/v). \tag{8.1}$$

Let us assume that the losses are small compared with the particle's kinetic energy; therefore, we shall regard the particle velocity $v$ as constant.

The work (Eq. (8.1)) can be split into three terms corresponding to the motion of the particle before the plate $W_{\text{before}}$, inside the plate $W_{\text{inside}}$, and after the plate $W_{\text{after}}$:

$$W = W_{before} = W_{ext} = W_{after} \qquad (8.2)$$

According to Secs. 1 and 2, the total field $E_z$ is the sum of the charge field $E_{\text{ch},\,z}$ and the radiation field $E_{\text{rad},\,z}$. Before and after the plate, the radiation field $E_{\text{rad},\,z}$ is, respectively, $E_z^{(-)}$ and $E_z^{(+)}$. The latter can be calculated from Eqs. (2.2) and (2.1), taking into account Eq. (1.23). Let us also consider that the charge field in vacuum does not perform work (see Ref. [**57**.4], Sec. 84). Therefore:

$$W_{\text{before}} = e\int d^2x\, d\omega\, E_z^{(-)}(\mathbf{x},\omega)\int_{-\infty}^{0}\exp\left[-i(1+u_0)\frac{\omega z}{v}\right]dz, \quad (8.3)$$

$$W_{\text{after}} = e\int d^2x\, d\omega\, E_z^{(+)}(\mathbf{x},\omega)\int_{a}^{\infty}\exp\left[-i(1-u_0)\frac{\omega z}{v}\right]dz. \quad (8.4)$$

Inside the plate, the radiation field has components propagating both forward and backward, i.e., its Fourier component is a superposition of two waves:

$$E_{\text{rad},z}(k,\omega) = \mathcal{E}_z^{(+)}(\chi,\omega)\,\delta(k_z-\lambda) + \mathcal{E}_z^{(-)}(\chi,\omega)\,\delta(k_z+\lambda), \quad (8.5)$$

where $\lambda = \frac{\omega}{c}\left(\varepsilon - \left(\frac{\varkappa c}{\omega}\right)^2\right)^{1/2}$, and the quantities $\mathcal{E}_z^{(+)}$ are determined, for example, through $\mathcal{E}_z^{(-)}$ (see Eqs. (2.2) and (1.23)) and the Fourier components of the charge field (Eq. (1.14)) from the continuity conditions at $z = 0$

$$\mathcal{E}_z^{(+)} = -\left(\frac{u_0}{u} - \frac{1}{\varepsilon}\right)\frac{E_z^{(-)}}{2} + \frac{i\,e\,v^2\varkappa^2}{4\pi^2\omega^3\varepsilon\, u}\left(\frac{\varepsilon+u}{1-u_0^2} - \frac{1}{1-u}\right),$$

$$\mathcal{E}_z^{(-)} = \left(\frac{u_0}{u} + \frac{1}{\varepsilon}\right)\frac{E_z^{(-)}}{2} - \frac{i\,e\,v^2\varkappa^2}{4\pi^2\omega^3\varepsilon\, u}\left(\frac{\varepsilon-u}{1-u_0^2} - \frac{1}{1+u}\right), \quad (8.6)$$

where $u = \lambda v/\omega$, $u_0 = \lambda_0 v/\omega$ (see Eq. (1.20)).

The work done by the field of the charge inside the plate is determined by the well-known equation for ionization losses (see, for example, Ref. [**57**.4]), multiplied by the plate thickness $a$:

$$W_{\text{ch}} = -\frac{\omega_0^2 e^2 a}{v^2}\left(\ln\frac{v\chi_0\gamma}{\omega} - \frac{\beta^2}{2}\right). \quad (8.7)$$

for particle velocities not close to the speed of light, and

$$W_{\text{ch}} = -\frac{\omega_0^2 e^2 a}{c^2}\ln\frac{\chi_0 c}{\omega_0} \equiv W_F. \quad (8.8)$$

for an ultrarelativistic particle (Fermi plateau). Here $x_0^{-1}$ is the minimum distance down to which macroscopic electrodynamics is applicable; $\beta = v/c$, and $\overline{\omega}$ is some mean atomic frequency:

$$\ln \overline{\omega} = \frac{2}{\pi \omega_0^2} \int_0^{\infty} \omega \left| \mathrm{Im} \left( \frac{1}{\varepsilon(\omega)} \right) \right| \ln \omega \, d\omega \ .$$

Taking Eqs. (8.5) and (8.7) (or (8.8)) into account, for the work of the field inside the medium we obtain

$$W_{\mathrm{ext}} = e \int d^2x \, d\omega \left\{ \mathcal{E}_z^{(-)} \int_{-\infty}^{0} \exp\left[ -i(1+u)\frac{\omega z}{v} \right] dz + \right. \\ \left. + \mathcal{E}_z^{(+)} \int_0^a \exp\left[ -i(1-u)\frac{\omega z}{v} \right] dz \right\} + W_{\mathrm{ch}}. \qquad (8.9)$$

Below we shall analyze the quantity $W$ (i.e., the sum of Eqs. (8.3), (8.4), and (8.9)) for small plate thicknesses $a$.

## 8.2. Expansion in Powers of the Plate Thickness

Let us expand the total losses $W$ in a series in powers of $a$:

$$W = W^{(0)} + W^{(1)} + W^{(2)} + \ \dots .$$

Comparing the terms of this series, we obtain the criterion under which it is sufficient to retain only the lowest terms of the expansion.

It is not difficult to verify that, as expected, the zeroth-order term in $a$ vanishes. Indeed, as $a \to 0$, all three terms $W_{before}$, $W_{inside}$, and $W_{after}$ vanish.

The term $W^{(1)}$, linear in $a$, after the corresponding expansions of $E_z^{(\mp)}$, $\mathcal{E}_z^{(\mp)}$, and the integrals over $z$, and after summing $W_{before}$, $W_{inside}$, and $W_{after}$, takes the form

$$W^{(1)} = -\frac{e^2 i v^2 a}{\pi} \int_{-\infty}^{\infty} d\omega \int_0^{x_0} \frac{(\varepsilon - 1)^2 (\beta^4 + u^2 - 1)\, x^3 \, d^2x}{\omega^3 \varepsilon (1 - u^2\beta^2)^2 (1 - u^2)} + W_{\mathrm{ch}}. \qquad (8.10)$$

As a result of integration over ω and χ in Eq. (8.10) by the Landau method [**57**.4], in the case of an ultrarelativistic particle ($\gamma \gg 1$) we obtain

$$W^{(1)} = -\frac{e^2\omega_0^2 a}{v^2}\left(\ln\frac{\omega_0\gamma}{\omega_1} - \frac{1}{2}\right) + W_{ch}. \quad (8.11)$$

where

$$\ln\bar{\omega}_1 = \frac{2}{\omega_0^2\pi}\int_0^\infty x\varepsilon''(x)\ln x dx. \quad (8.12)$$

Substituting Eq. (8.8) into Eq. (8.11), we obtain

$$W^{(1)} = -\frac{e^2\omega_2^0 a}{v^2}\left(\ln\frac{\varkappa_0 v\gamma}{\bar{\omega}_1} - \frac{1}{2}\right). \quad (8.13)$$

The quantity $W^{(1)}$ depends logarithmically on γ (this quantity differs from Eq. (8.7) only in the definition of the mean frequency). Since for sufficiently small thicknesses *a* the quantity $W^{(1)}$ is the leading term, we conclude that the ionization losses of fast particles in sufficiently *thin* layers of matter *increase logarithmically* with increasing γ [**59**.4] (see also Ref. [**62**.7]).[14]

Let us find the term $W^{(2)}$, quadratic in *a*, and compare it with the linear term $W^{(1)}$. After taking into account the next terms in the expansions of Eqs. (8.3), (8.4), and (8.9), we obtain

$$W^{(2)} = -\frac{e^2 v a^2}{2\pi}\int_{-\infty}^{\infty} d\omega \int_0^{\varkappa_0} \frac{(\varepsilon-1)^2(\varepsilon^2 u_0^2 + \gamma^{-4})\varkappa^3 d^2\varkappa}{\omega^2\varepsilon^2 u_0(1-u_0^2)^2}. \quad (8.14)$$

Integrating Eq. (8.14) over ϰ and ω, we obtain

$$W^{(2)} = \frac{\pi e^2\omega_0^4 a^2}{4v^3\bar{\omega}_2}, \quad (8.15)$$

where

[14] In those works, the expressions for the losses linear in *a* differ somewhat from Eq. (8.13). A more accurate equation in the form of Eq. (8.13) was obtained in Refs. [**66**.10, **69**.8].

$$\bar{\omega}_2 = \frac{\pi^2 \omega_0^4}{4} \left[ \int_0^\infty \int_0^\infty \frac{x y}{x+y} \varepsilon''(x) \varepsilon''(y) \, dx \, dy \right]^{-1}. \qquad (8.16)$$

In the expansion of $W$ in powers of $a$, we may retain only the lowest term $W^{(1)}$ if $|W^{(2)}| \ll |W^{(1)}|$, or (for $\gamma \gg 1$)

$$a \ll a_{cr}, \qquad (8.17)$$

where

$$a_{\rm cr} = \frac{4v\bar{\omega}_2}{\pi\omega_0^2}\left(\ln\frac{x_0 v\gamma}{\omega_1} - \frac{1}{2}\right). \qquad (8.18)$$

(the criterion for the applicability of the law of logarithmic increase of losses obtained in Ref. [**62**.7] is incorrect).

As follows from Eq. (8.18), the quantity $a_{cr}$ is a function of $\gamma$. Consequently, one and the same plate can be "thick" or "thin" depending on whether the condition $\gamma \ll \gamma_{cr}$ or $\gamma \gg \gamma_{cr}$ is satisfied, where

$$\gamma_{\rm cr} = \frac{\bar{\omega}_1}{\omega_0}\exp\left(\frac{\pi\omega_0^2 a}{4c\bar{\omega}_2} + \frac{1}{2}\right). \qquad (8.19)$$

From the above it follows that for $\gamma \gg \gamma_{cr}$ inequality (8.17) holds for any plate, and the total losses are determined by Eq. (8.13), i.e., they increase logarithmically with $\gamma$ (Fig. 8.1, segment $A$). In other words, as expected, in a thin plate the Fermi density effect is absent. An analogous situation also occurs for an *oblique* traversal of the particle through the plate: the logarithmic dependence of the losses on the particle energy persists over a wide range of incidence angles of the particle on the plate[15] (except for the case of grazing incidence). The non-perpendicularity of the incidence manifests itself only in that, in a condition of the type (8.17), the role of the plate thickness is played by the *length of the particle's path* in the plate [**80**.10].

[15] An error on this issue was made in Ref. [**68**.9]. For details, see the paper by Avakyan, Ambartsumyan, and Yang Chi [**80**.10].

The absence of the density effect in thin layers of matter was confirmed in experiments by Alikhanyan, Walter, Lorikyan, et al. [**63**.7, **64**.11]. However, in these experiments only relative measurements of the losses were carried out, rather than absolute ones. Absolute measurements of losses were carried out by Ogle et al. [**78**.8]. Unfortunately, relying on not entirely correct theoretical concepts, the authors of this work investigated losses in a plate (silicon) of comparatively large thickness (about 100 μm), in which a logarithmic increase of losses with increasing γ (at electron accelerator energies available) should not be observed (see Ref. [**81**.1]), and naturally, in this experiment they did not observe it.

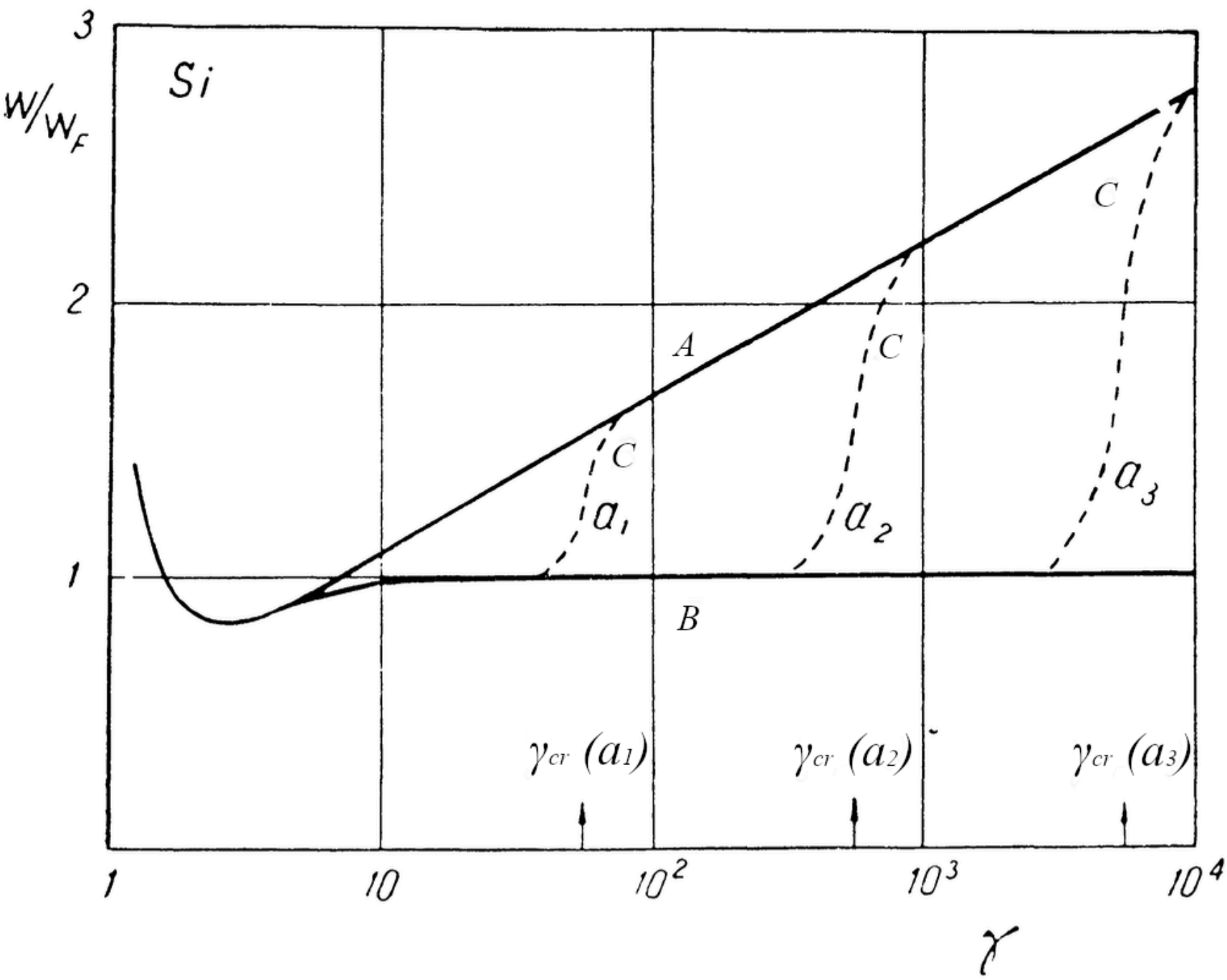

Fig. 8.1. Dependence of the total energy loss of a particle in silicon, normalized to the Fermi plateau $W_F$ (Eq. (8.8)), on the particle Lorentz factor γ. Curve *A* corresponds to the Bohr theory (Eqs. (8.7) or (8.13)), and curve *B* corresponds to the Fermi theory (Eqs. (8.7) and (8.8)). The curves *C* of transition from curve *B* to curve *A* correspond to plates of finite thickness: $a_1 = 0.1$ μm, $\gamma_c = 55$; $a_2 = 0.2$ μm, $\gamma_c = 550$; $a_3 = 0.3$ μm, $\gamma_c = 5500$.

The logarithmic increase in ionization losses was apparently observed in secondary-emission monitors (see Refs. [**65**.12, **67**.10]), where the energy losses are due to effects in the thin surface layer of aluminum oxide.

### 8.3. Plate of Arbitrary Thickness

When the condition $a \ll a_{cr}$ is not satisfied, in the expansion of the total losses in powers of $a$ (Sec. 8.2) one cannot restrict oneself to the lowest-order terms only, and one must use the general equation (8.1).

The expression for the total losses can always be written in the following form (see Eqs. (8.3), (8.4), and (8.9)):

$$W = W_{ch} + W', \qquad (8.20)$$

where $W_{ch}$ is defined by Eqs. (8.7) or (8.8) and corresponds to ionization losses (including losses due to Vavilov–Cherenkov radiation) without taking boundary effects into account, while the quantity $W'$ is due to the presence of boundaries and corresponds, in particular, to losses due to transition radiation, which partly escapes to infinity and is partly absorbed in the plate itself.

Ionization losses are concentrated mainly in the frequency range of the order of atomic frequencies, whereas transition radiation has a broad spectrum extending to frequencies far exceeding atomic ones if the particle Lorentz factor is sufficiently large. If we restrict ourselves to the region of atomic frequencies $\overline{\omega}_1$, then for thicknesses much larger than the formation length of transition radiation at these frequencies (which is of the order of the corresponding wavelength), the quantity $W'$ does not depend on $a$ and, in order of magnitude, is equal to $(e^2\omega_1/c) ln\gamma$.

The ratio of the quantities $W'$ and $W_{ch}$ (for $\gamma \gg 1$) is of the order of

$$\frac{W'}{W_{\text{ch}}} \sim \frac{c\,\overline{\omega}_1 \ln\gamma}{\omega_0^2 a \ln(x_0 c/\omega_0)} \sim \frac{a_{\text{cr}}}{a}.$$

where $a$ is a numerical factor of order unity.

Thus, for $a \gg \alpha a_{cr}$ the indicated ratio is small, and the total losses (in the region of atomic frequencies) are determined mainly by the ionization losses (Eq. (8.8)), i.e., they are practically independent of the particle Lorentz factor $\gamma$ (Fig. 8.1, segment B).

In the region $\gamma \sim \gamma_{cr}$ (or $a \sim a_{cr}$) there is a smooth transition from Eq. (8.8) to Eq. (8.13). It is noteworthy that, in this case, the $\gamma$-dependence may be *steeper* (Fig. 8.1, region *C*) [**81**.1].

In order to estimate, for a given material, $a_{cr}(\gamma)$ (or $\gamma_{cr}(a)$), it is first necessary to calculate $\overline{\omega}_0$, $\overline{\omega}_1$, and $\overline{\omega}_2$ (see Table 8.1). Estimates show that, for example, for $\gamma \sim 10^3 - 10^4$, $a_{cr} \sim 4 - 6 \cdot 10^{-5}$ cm for Li and $2 - 3 \cdot 10^{-5}$ cm for Si.

*Table 8.1.*

Values of the characteristic quantities determining the ionization losses of a particle in matter [81.1] (values are given in eV).

| | $\omega_0$ | $\overline{\omega}_1$ | $\overline{\omega}_2$ |
|---|---|---|---|
| Li | 13.7 | 30.5 | 52.6 |
| Graphite | 30.7 | 66.7 | 116 |
| Si | 31.7 | 105.3 | 171 |

## § 9. TRANSITION RADIATION OF A MAGNETIC CHARGE AND OF ELECTRIC AND MAGNETIC DIPOLES

Transition radiation is produced not only when a charged particle traverses inhomogeneities of the medium. It should also arise when any source of an electromagnetic field moves. Such a source may be an arbitrary electric or magnetic multipole, in particular a hypothetical magnetic monopole.

### 9.1. Dirac Magnetic Monopole

Maxwell's equations possess a remarkable symmetry between the electric and magnetic fields. However, particles carrying magnetic charge $e_m$ have not yet been found. However, no fundamental principle forbidding the existence of particles with magnetic charge is known. To this day, the question of the possible existence of a particle with magnetic charge, often called the Dirac monopole, remains open. Experimental attempts to detect such a particle and theoretical investigations of this problem continue (see, for example, Refs. [**70**.16, **75**.26, **79**.7]). In these experimental searches one can rely on results that follow from Maxwell's equations. It is therefore of interest to consider transition radiation and Vavilov–Cherenkov radiation for a particle with magnetic charge [**63**.3, **71**.9, **79**.7].

Let us denote the magnetic charge density by $\rho_m$ and the magnetic current density by $j_m$. Then the electromagnetic fields of a monopole in a medium are determined by the following equations:

$$\begin{aligned}
&\mathrm{rot}\boldsymbol{H}(\boldsymbol{r},t)=\frac{1}{c}\frac{\partial \boldsymbol{D}(\boldsymbol{r},t)}{\partial t},\\
&\mathrm{rot}\boldsymbol{E}(\boldsymbol{r},t)=-\frac{1}{c}\frac{\partial \boldsymbol{B}(\boldsymbol{r},t)}{\partial t}-\frac{4\pi}{c}\boldsymbol{j}_{\text{м}}(\boldsymbol{r},t),\\
&\mathrm{div}\boldsymbol{B}(\boldsymbol{r},t)=4\pi\rho_{\text{м}}(\boldsymbol{r},t),\\
&\mathrm{div}\,\boldsymbol{D}(\boldsymbol{r},t)=0.
\end{aligned} \tag{9.1}$$

Note that the above equations reduce to the usual Maxwell's equations (Eq. (1.1)) with electric charges and currents under the following replacements: ***B*** by ***D***, ***H*** by ***E***, ρ by $\rho_m$, *t* by $-t$, and vice versa (whence, in particular, it follows that *j* is replaced by $-j_m$ and *v* by $-v$).

When a monopole passes through a medium with constant velocity *v*, we have

$$\begin{aligned}
&\rho_{\text{м}}(\boldsymbol{r},t)=e_{\text{м}}\delta(\boldsymbol{r}-\boldsymbol{v}t),\\
&\boldsymbol{j}_{\text{м}}(\boldsymbol{r},t)=\boldsymbol{v}\rho_{\text{м}}(\boldsymbol{r},t).
\end{aligned} \tag{9.2}$$

Then, by analogy with Eq. (1.14), we obtain

$$H_{\rm M}(\boldsymbol{k},\omega) = -\frac{e_{\rm M} i v}{2\pi^2 \mu \omega^2} \frac{\boldsymbol{\varkappa} + \boldsymbol{n}_z(\omega/v)(1 - \varepsilon\mu\beta^2)}{k^2 v^2/\omega^2 - \varepsilon\mu\beta^2} \delta(k_z - \omega/v), \qquad (9.3)$$

$$\boldsymbol{E}_{\rm M}(\boldsymbol{k},\omega) = -\frac{c}{\omega\varepsilon}[\boldsymbol{k}\times\boldsymbol{H}(\boldsymbol{k},\omega)].$$

The tangential component of the radiation field propagating forward in the second medium is given by an equation analogous to Eq. (1.21):

$$\boldsymbol{H}_t^{(+)}(\boldsymbol{\varkappa},\omega) = \frac{e_{\rm M} i v \boldsymbol{\varkappa}}{4\pi^2\omega^2}\left(\frac{1}{\mu_2(1-u_2)} - \frac{R_{21}^{\rm M}}{\mu_2(1+u_2)} - \frac{P_{12}^{\rm M}}{\mu_1(1-u_1)}\right), \qquad (9.4)$$

where

$$R_{21}^{\rm M} = \frac{\mu_2 u_1 - \mu_1 u_2}{\mu_1 u_2 + \mu_2 u_1}, \qquad P_{12}^{\rm M} = \frac{2\mu_1 u_2}{\mu_1 u_2 + \mu_2 u_1}.$$

From Eq. (9.4) it follows that the transition radiation and Vavilov–Cherenkov radiation produced by a magnetic monopole are polarized such that the radiation magnetic field vector always lies in the plane of observation (cf. Sec. 1.4), i.e., the radiation is linearly polarized in the plane perpendicular to the plane of observation.

The spectral–angular distribution $W_m^{(+)}(\omega,\vartheta)$ of the radiation intensity (transition and Cherenkov) for a magnetic charge is given by an equation analogous to Eq. (1.35):

$$W_m^{(+)}(\omega,\vartheta) = \frac{2e^2\beta^3}{\pi c}\left|\frac{\left[(\mu_1/\mu_2 - 1)(1-u_1) - \beta^2(\varepsilon_1 - \mu_1 - \varepsilon_2\mu_2)\right]u_2^2}{(1-u_2^2)(1-u_1)(\mu_1 u_2 + \mu_2 u_1)}\right|^2 \times$$
$$\times|\varepsilon_2\mu_2|^2 \,\mathrm{Re}\left(\frac{\mu_2}{u_2}\right)\exp\left(-2u_2'' z_{\rm obs}\frac{\omega}{v}\right)\sin^3\vartheta\cos\vartheta, \qquad (9.5)$$
$$u_s = \beta\left(\varepsilon_s\mu_s - \varepsilon_2\mu_2\sin^2\vartheta\right)^{1/2},\ (s = 1,2).$$

By virtue of the symmetry noted above between Eqs. (9.1) and (1.1), Eq. (9.5) can also be obtained from Eq. (1.35) by the replacements $\varepsilon_s \rightleftarrows \mu_s$ and $e \to e_m$.

In the particular case when a magnetic monopole enters from a medium ($\varepsilon_1 = \varepsilon$, $\mu_1 = 1$) into vacuum ($\varepsilon_2 = \mu_2 = 1$), we have [**79**.7]:

$$W_{\text{м}}^{(+)}(\omega, \vartheta) = \frac{2e_{\text{м}}^2\beta^6}{\pi c} \frac{|\varepsilon - 1|^2}{(1-\beta^2\cos^2\vartheta)^2|1-\beta(\varepsilon-\sin^2\vartheta)^{1/2}|^2} \times$$
$$\times \frac{\sin^3\vartheta\cos^2\vartheta}{|\cos\vartheta+(\varepsilon-\sin^2\vartheta)^{1/2}|^2}. \qquad (9.6)$$

As in the case of an electric charge (see Eq. (1.37)), the denominator of this equation contains the factor $|1 - \beta(\varepsilon - sin^2\vartheta)^{1/2}|^2$, whose minimum corresponds to Vavilov–Cherenkov radiation. In addition, in the ultrarelativistic case $(\gamma \gg 1)$ there is also a maximum of transition radiation occurring at the angle $\vartheta \sim \gamma^{-1}$.

The right-hand side of Eq. (9.6) is proportional to $\beta^6$, i.e., there is a stronger dependence on the particle velocity (for $\beta \ll 1$) than in the case of an electric charge. When $\beta \to 1$ $(\gamma \gg 1)$, the frequencies of transition radiation extend into the X-ray region up to frequencies of the order of the cutoff frequency $\omega_{cutoff}$, determined by the same equations as in the case of an electric charge (see Secs. 1.8 and 1.9).

As follows from Eq. (9.6), the spectral–angular and spectral distributions of the XTR intensity produced by an ultrarelativistic magnetic charge coincide with the corresponding distributions for a particle with electric charge $Ze$, where $Z = e_m/e$. On the basis of quantum considerations, Dirac showed [70.16] that $e_m/e = 68.5$. Thus, for an ultrarelativistic magnetic monopole the XTR intensity turns out to be *quite large*. At the same time, the radiation of a magnetic monopole, as already noted above, has a different polarization from that of an electrically charged particle.

### 9.2. Electric and Magnetic Dipoles

An electric dipole is a system of two oppositely charged particles $\pm e$ separated by a distance $d_0 = \{\rho_0, z_0\}$. The transition radiation produced by such a system can be calculated by the method presented in Sec. 7 [**60**.15, **62**.3].

As in the case of a bunch of like charges, if the distance between the charges is sufficiently large, i.e., the two charges radiate (at frequency ω and at angle θ) incoherently; i.e., the dipole radiates twice as much as a single charge *e*.

$$z_0 \gg \frac{v}{\omega} \quad \text{or} \quad \rho_0 \gg \frac{c}{\omega \sin\vartheta}, \qquad (9.7)$$

two charges emit (at frequency ω under angle ϑ) incoherently, i.e. the dipole emits twice as much, as a single charge *e*.

When the conditions in Eq. (9.7) are not satisfied, the two oppositely charged constituents of the dipole radiate coherently and partially cancel each other. When the conditions opposite to those in Eq. (9.7) hold,

$$z_0 \ll \frac{v}{\omega} \quad \text{and} \quad \rho_0 \ll \frac{c}{\omega \sin\vartheta}, \qquad (9.8)$$

the radiation intensity decreases sharply. In the case, for example, when the dipole center crosses the plane interface between two media perpendicularly, the spectral–angular distribution of the forward transition-radiation intensity is given by the following equation:

$$W^{(+)}(\omega,\vartheta) = W^{(+)}_{\text{cutoff}}(\omega,\vartheta)\frac{\omega^2}{4v^2}\left(\rho_0^2\beta^2\sin^2\vartheta + 2z_0^2\right), \qquad (9.9)$$

where $W^{(+)}_{cutoff}(\omega, \vartheta)$ is defined by Eqs. (1.37) or (1.59).

The additional factor $\omega^2(\rho_0^2\beta^2 sin^2\theta + 2z_0^2)/(4v^2)$ not only strongly reduces the radiation intensity but also changes its spectrum. For X-ray frequencies the spectrum now behaves as $\omega^2$ at $\omega \ll \omega_{cutoff}$ and as $\omega^{-2}$ at $\omega \gg \omega_{cutoff}$, i.e., the radiation is mainly concentrated at frequencies of the order of $\omega_{cutoff}$.

A magnetic dipole can be regarded either as a pair of opposite magnetic charges or as being due to an electric current. In these two approaches Maxwell's equations are written differently and, therefore, in general, different transition-radiation fields are obtained. However, at least for X-ray frequencies, the radiation intensities in both approaches turn out to be close and are expressed by an equation analogous to the case of an electric dipole, provided that the conditions in Eq. (9.8) are satisfied.

It is known that the intrinsic magnetic moment of elementary particles is inversely proportional to their mass ($e\hbar/2mc$). Light particles, for example the electron, possess the largest magnetic moments. However, a comparison of the intensity of XTR due to the magnetic moment of the electron with the intensity of XTR due to its charge shows that in the former case the intensity is nevertheless much smaller [**60**.15].

## § 10. PERTURBATION THEORY. POSSIBILITY OF XTR FORMATION IN COSMIC SPACE

Up to now we have considered media whose electrodynamic characteristics ($\varepsilon$ and $\mu$) depended only on a single spatial coordinate z, i.e., media that are homogeneous in the transverse x- and y-directions. When the medium is inhomogeneous in all three spatial directions, the problem of the generation of radiation by a passing charged particle becomes mathematically very difficult and can be solved exactly only in certain particular cases (for example, for a sphere [**69**.6]). However, if the inhomogeneity of the medium is *small*, the problem can be solved approximately by the method of perturbation theory. This is especially justified for frequencies far exceeding atomic ones (in particular, X-ray frequencies), since at such frequencies the dielectric permittivity $\varepsilon$ is very close to unity (see Eq. (1.58)). If, in addition, the dimensions of the region of inhomogeneity are small, the method of perturbation theory proves applicable.

### 10.1. Perturbation Theory

Let us consider the formation of XTR when a fast charged particle passes through a small body located in vacuum, or in the vicinity of such a body (Fig. 10.1) [**75**.32].

The dielectric permittivity of the body (not necessarily homogeneous) at frequencies exceeding atomic ones is determined by

$$\varepsilon(\boldsymbol{r},\omega)=1-\frac{\omega_0^2}{\omega^2}V_0 f(\boldsymbol{r}-\boldsymbol{r}_0), \tag{10.1}$$

where $f(r)$ is the electron density distribution function in the body under consideration, normalized such that the integral over the entire volume of the body gives the total number of electrons inside the body:

$$N=\int f(\boldsymbol{r})d^3\boldsymbol{r}, \tag{10.2}$$

$r_0$ is the radius vector of its "center," $V_0$ is the volume per electron, and $\omega_0 = (4\pi e^2/mV_0)^{1/2}$ is the average plasma frequency of the body. The magnetic permeability of the body is assumed to be equal to unity.

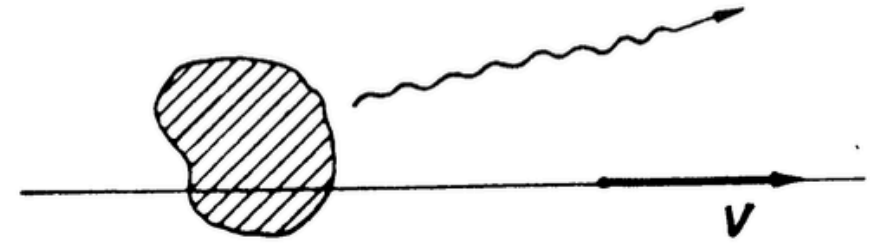

Fig. 10.1. Formation of XTR during the passage of a charged particle through a small body

We start from Eqs. (1.8) for the field potentials, assuming that

$$A(r,\omega)=A_{ch}(r,\omega)+A_{rad}(r,\omega) \tag{10.3}$$

(and correspondingly for $\varphi(r,\omega)$), where $A_{ch}(r,\omega)$ satisfies Eq. (1.8) with $\varepsilon = \mu = 1$.

As a result of substituting Eq. (10.3) into Eq. (1.8), we obtain

$$\left(\Delta+\frac{\omega^2}{c^2}\right)A_{\mathrm{rad}}(\mathbf{r},\omega)=(1-\varepsilon)\frac{\omega^2}{c^2}\left[A_{\mathrm{ch}}(\mathbf{r},\omega)+A_{\mathrm{rad}}(\mathbf{r},\omega)\right] \\ -\varepsilon\, grad\,(1/\varepsilon)\,\mathrm{div}\left[A_{\mathrm{ch}}(\mathbf{r},\omega)+A_{\mathrm{rad}}(\mathbf{r},\omega)\right]. \tag{10.4}$$

The quantity $grad(1/\varepsilon)$ is of the order of $|1-\varepsilon|/\delta$, where $\delta$ is the distance over which $|1-\varepsilon|$ changes substantially. It follows that if

$$\delta \gg \frac{c}{\omega} \tag{10.5}$$

(which can always be regarded as satisfied for the frequencies under consideration), the second term on the right-hand side of Eq. (10.4) can be neglected:

$$\left(\Delta + \frac{\omega^2}{c^2}\right) A_{\rm rad}(\mathbf{r}, \omega) = (1 - \varepsilon(\mathbf{r}, \omega)) \frac{\omega^2}{c^2} [A_{\rm ch}(\mathbf{r}, \omega) + A_{\rm rad}(\mathbf{r}, \omega)]. \tag{10.6}$$

Taking into account Eq. (10.1) and performing a Fourier transform in Eq. (10.6), we obtain the following integral equation:

$$\left(-k^2 - \frac{\omega^2}{c^2}\right) A_{\rm rad}(\mathbf{k}, \omega) = \frac{\omega_0^2 V_0}{(2\pi)^3 c^2} \int d^3k' \\ \left[A_{\rm ch}(\mathbf{k}, \omega) + A_{\rm rad}(\mathbf{k}', \omega)\right] \int f(\mathbf{r}) \exp\left[i(\mathbf{k}' - \mathbf{k}) \cdot \mathbf{r}\right] d^3r. \tag{10.7}$$

where $A_{ch}(k, \omega)$ is determined by Eq. (1.13) with $\varepsilon = \mu = 1$.

From Eqs. (1.6) and (1.7) it is straightforward to express the Fourier components of the field $E(k, \omega)$ in terms of $A(k, \omega)$:

$$E^j(\boldsymbol{k}, \omega) = \frac{i\omega}{c} (\delta^{jl} - k^j k^l c^2/\omega^2) A^l(\boldsymbol{k}, \omega), \tag{10.8}$$

where $j$, $l$ are Cartesian indices ($j, l = x, y, z$), and $\delta^{jl}$ is the Kronecker delta ($\delta^{jl} = 0$ for $j \neq l$, $\delta^{jj} = 1$).

Taking Eqs. (10.7) and (10.8) into account, for $E_{rad}(k, \omega)$ we obtain the following equation:

$$\left(-k^2 + \frac{\omega^2}{c^2}\right) E^j_{\rm rad}(\mathbf{k}, \omega) = \frac{\omega_0^2 V_0}{(2\pi)^3 c^2} \left(\delta_{jl} - \frac{k^j k^l c^2}{\omega^2}\right) \\ \int d^3k' \left[E^l_{\rm ch}(\mathbf{k}', \omega) + E^l_{\rm rad}(\mathbf{k}', \omega)\right] \int f(\mathbf{r}) \exp\left[i(\mathbf{k}' - \mathbf{k}) \cdot \mathbf{r}\right] d^3r. \tag{10.9}$$

where $E_{ch}(k, \omega)$ is defined by Eq. (1.14).

If we assume that the radiation field is much smaller than the field of the charge, then in the first order of perturbation theory one can neglect $E_{rad}(k', \omega)$ on the right-hand side of Eq. (10.9). Then, substituting Eq. (1.14) into Eq. (10.9) and performing the integration over $k'_z$, we obtain

$$E_{\mathrm{rad}}(\mathbf{k}, \omega) = \frac{\omega_0^2 e i V_0}{(2\pi)^4 \pi c^2 v \left(k^2 - \omega^2/c^2\right)} \times$$
$$\times \int G(\boldsymbol{\rho}) \exp\left\{-i\chi \cdot \boldsymbol{\rho} + i\left(\frac{\omega}{v} - k_z\right) z\right\} f(\mathbf{r} - \mathbf{r}_0)\, d^3 r; \tag{10.10}$$

here ρ, □ and $z$, $k_z$ are the parts of vectors $r$ and $k$ transversal and longitudinal to the particle movement direction,

$$G(\rho) = \int \frac{\exp(i\varkappa' \cdot \rho)}{\varkappa'^2 + \omega^2/v^2\gamma^2} \times$$
$$\times \left\{\varkappa' - \frac{kc^2}{\omega^2}(\varkappa \cdot \varkappa') + \frac{\omega}{v\gamma^2}\left(n_z - \frac{kc^2 k_z}{\omega^2}\right)\right\} d^2\varkappa'.$$

After integration we have

$$G(\rho) = \frac{2\pi|\omega|}{v\gamma}\left\{i\left(n - \frac{kc^2}{\omega^2}(\varkappa \cdot n)\right) K_1\left(\left|\frac{\omega\rho}{v\gamma}\right|\right)\right.$$
$$\left. + \gamma^{-1}\left(n_z - \frac{kc^2 k_z}{\omega^2}\right) K_0\left(\left|\frac{\omega\rho}{v\gamma}\right|\right)\right\}, \tag{10.11}$$

where $\mathbf{n} = \rho/\rho$, $K_0(x)$ and $K_1(x)$ are the modified Bessel functions of the second kind (cylindrical functions of imaginary argument; see, for example, Ref. [**71**.13], Sec. 8.4). The second term in expression (10.11) contains the factor $\gamma^{-1}$, which is small for relativistic particles. Therefore, this term is small compared with the first, and henceforth we shall omit it.

If one substitutes Eqs. (1.14) and (10.10) into the right-hand side of Eq. (10.9) and compares the resulting expression (the second approximation of perturbation theory) with Eq. (10.10), one can verify that under the condition

$$\frac{\omega_0^2 a_z}{\omega c} \ll 1 \tag{10.12}$$

($a_z$ is the longitudinal size of the body under consideration), it is quite sufficient to retain the first-order result (10.10).

The spectral distribution of the radiation intensity passing through the plane $z = z_{obs}$ = const over the entire passage of the charged particle is given by the following equation (the observation plane is located far from the body under consideration):

$$W(\omega) = \frac{\omega^2}{(2\pi)^4 c} \int_{(\omega>0)} |E_{\rm rad}(\mathbf{x}, z_{\rm obs}, \omega)|^2 \cos\vartheta \, \sin\vartheta \, d\vartheta \, d\varphi, \qquad (10.13)$$

where $\chi = (\omega/c) \sin \vartheta$,

$$E_{\rm rad}(\chi, z_{\rm obs}, \omega) = \int E_{\rm rad}(\mathbf{k}, \omega) \, \exp(ik_z z_{\rm obs}) \, dk_z. \qquad (10.14)$$

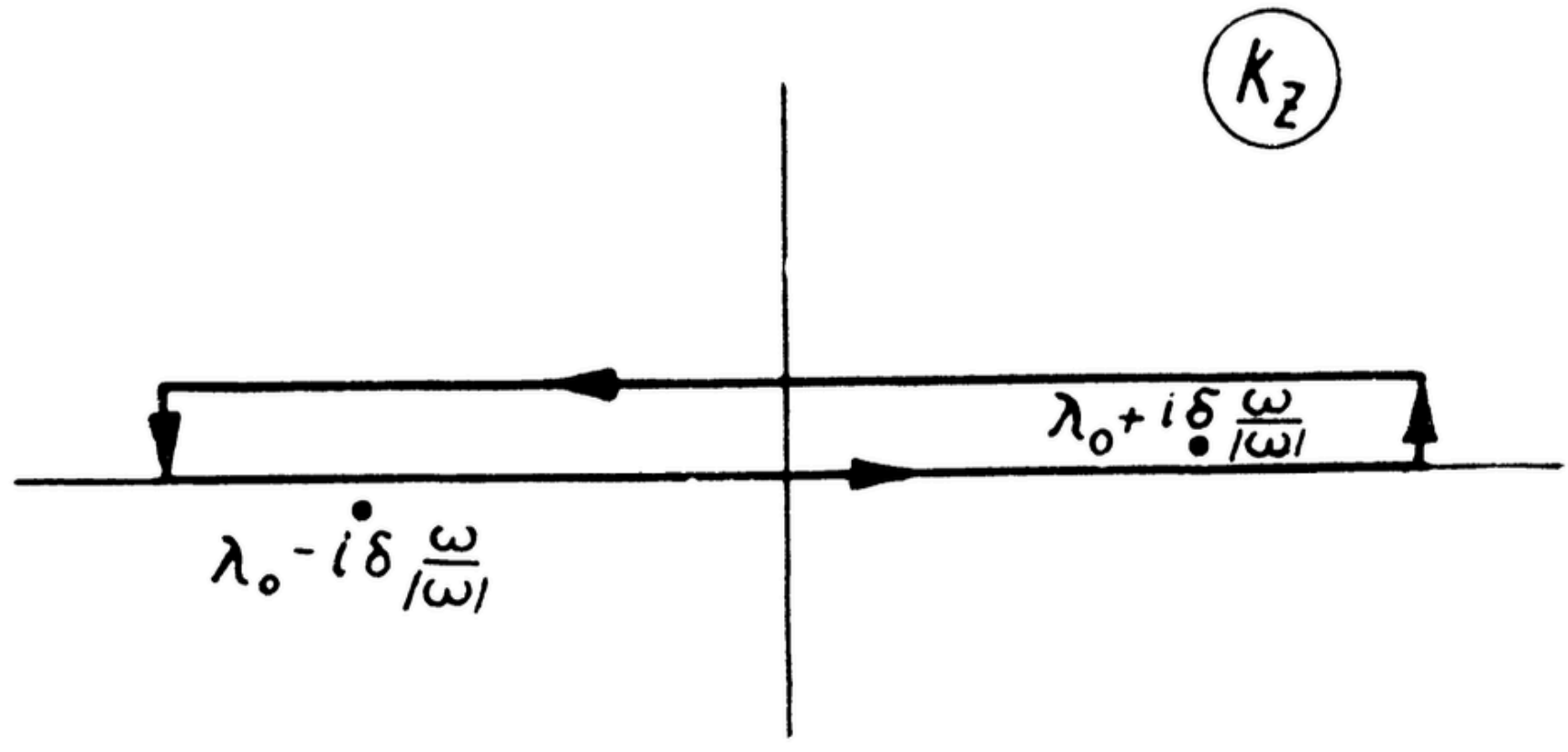

*Fig. 10.2. Integration contour in the complex $k_z$ plane*

The asymptotic expression for this integral as $z_{obs} \to \infty$ (forward radiation) can be evaluated as follows. As seen from Eq. (10.10), the integrand in Eq. (10.14) has two poles at $k_z = \pm\lambda_0$, where $\lambda_0 = (\omega^2/c^2 - \chi^2)^{1/2}$. In evaluating the integral in Eq. (10.14) along the real axis in the complex $k_z$ plane, we displace the indicated poles from the real axis by an infinitesimal amount. At the same time, to obtain the correct asymptotic behavior of the field (Eq. (10.14)) as $z \to \infty$ (so that the field does not grow without bound), it is necessary to shift the pole $k_z = +\lambda_0$ into the upper half-plane and the pole $k_z = -\lambda_0$ into the lower half-plane (for $\omega > 0$). Taking the integral over the contour shown in Fig. 10.2 and noting that the integral over the upper part of the contour vanishes as $z \to \infty$, after letting the lateral parts of the contour go to infinity we obtain

$$E_{\text{rad}}(\mathbf{x}, z_{\text{obs}}, \omega) = -\frac{\omega_0^2 e V_0 \exp(i\lambda_0 z_{\text{obs}})}{(2\pi)^4 c^2 v \lambda_0} \times$$
$$\times \int G(\boldsymbol{\rho}) \exp\left\{-i\mathbf{x}\cdot\boldsymbol{\rho} + i\left(\frac{\omega}{v} - \lambda_0\right)z\right\} f(\mathbf{r} - \mathbf{r}_0)\, d^3r. \qquad (10.15)$$

Equations (10.15), (10.11), and (10.13) give the radiation intensity emitted forward, i.e., into the forward hemisphere with respect to the direction of motion of the charged particle. The equations for the radiation intensity in the backward direction, i.e., into the backward hemisphere, can be obtained from Eqs. (10.15), (10.11), and (10.13) by replacing $v$ with $-v$.

Below we analyze these equations in two limiting cases: when the particle passes through the “center” of the body under consideration (central collision) and when the particle passes outside the body at a distance much larger than its transverse dimensions (noncentral collision).

### 10.2. Central and Noncentral Collisions

Let the distribution function $f(r) = f(\rho, z)$ possess cylindrical symmetry about an axis parallel to the direction of motion of the charge (i.e., the $z$-axis).

Then in the case of a central collision we obtain

$$E^{\text{C}}_{\text{rad}}(\chi, z, \omega) = -\frac{\omega_0^2 e|\omega| \exp\left[i\left(\frac{\omega}{v} - \lambda_0\right)z_0 + i\lambda_0 z\right]}{(2\pi)^2 \omega^3 v \gamma \lambda_0 \chi}$$
$$\left(\chi - \frac{\lambda_0^2 c^2 \mathbf{k}}{\omega^2}\right) P^{\text{C}}, \qquad (k_z = \lambda_0). \qquad (10.16)$$

where

$$P^{\mathrm{C}} = \frac{\omega^2}{c^2} \int p(\rho, \lambda_0)\, J_1(x\rho)\, K_1\left(\frac{\omega\rho}{v\gamma}\right) \rho\, d\rho,$$
$$p(\rho, \lambda_0) = \frac{\omega V_0}{v} \int \exp[i\,(\omega/v - \lambda_0)z]\, f(\rho, z)\, dz. \qquad (10.17)$$

In the case of a non-central collision it follows from Eq. (10.11) that if the minimum distance from the body to the charge trajectory is greater than $v\gamma/\omega$, i.e., if the arguments of the functions $K_0$ and $K_1$ are large, the radiation field is exponentially small. If

$$\frac{\omega a_\perp}{v\gamma} \ll 1$$

($a_\perp$ is the transversal dimension of the body), then from Eqs. (10.11) and (10.15) we obtain

$$E^{\mathrm{NC}}_{\mathrm{rad}}(\chi, z, \omega) = -\frac{\omega_0^2 e|\omega| i \exp[i\,(\omega/v - \lambda_0)z_0 + i\lambda_0 z - i\chi \cdot \boldsymbol{\rho}_0]}{(2\pi)^2 \omega^3 v\gamma\lambda_0} \times$$
$$\times K_1\left(|\frac{\omega\rho_0}{v\gamma}|\right) \left(\mathbf{n}_0 - \frac{kc^2}{\omega^2}(\chi \cdot \mathbf{n}_0)\right) P^{\mathrm{NC}}, \qquad (k_z = \lambda_0). \qquad (10.18)$$

where

$$P^{\mathrm{C}} = \frac{\omega^2}{c^2} \int p(\rho, \lambda_0)\, J_0(x\rho)\, \rho\, d\rho. \qquad (10.19)$$

$\rho_0$ is the transverse component of the radius vector $r_0$ of the center of the body, $n_0 = \rho_0/\rho_0$.

Substituting Eqs. (10.16) and (10.18) into Eq. (10.13), we obtain the frequency spectra of the radiation ($\omega > 0$) [**75**.32]:

$$W^{\mathrm{C}}(\omega) = \frac{\omega_0^4 e^2}{4\pi^8 \omega^4 \gamma^2 c} \int \left|P^{\mathrm{C}}\right|^2 \cos^2\vartheta\, \sin\vartheta\, d\vartheta\, d\varphi, \qquad (10.20)$$

$$W^{\mathrm{NC}}(\omega) = \frac{\omega_0^4 e^2}{4\pi^2 \omega^4 \gamma^2 c}\left[K_1\left(\frac{\omega\rho_0}{v\gamma}\right)\right]^2 \int |P^{\mathrm{NC}}|^2 \times \\ \times \left(1 - \sin^2\vartheta\cos^2\varphi\right)\sin\vartheta\, d\vartheta\, d\varphi, \tag{10.21}$$

where the quantities χ and $\lambda_0$, appearing in $P^C$ and $P^{NC}$, must be replaced, respectively, by $(\omega/c)sin\vartheta$ and $(\omega/c)cos\vartheta$.

In addition, in Eq. (10.21) the azimuthal angle ϕ is measured from the vector $\rho_0$.

The angular distribution of the radiation intensity is determined by the dependence of the integrands in Eqs. (10.20) and (10.21) on ϑ and ϕ. The character of this distribution differs substantially depending on the size of the body.

1. Let the dimensions of the body be much smaller than the wavelength[16]:

$$\frac{\omega a_z}{c} \ll 1 \text{ and } \frac{\omega a_\perp}{c} \ll 1. \tag{10.22}$$

Then the arguments of the functions $J_0$, $J_1$, $K_1$, and of the exponential appearing in Eqs. (10.17) and (10.19) are small, and these functions can be replaced by the lowest terms of their series expansions. As a result we obtain

$$P^{\mathrm{C}} = \frac{\chi\omega^2\gamma V}{4\pi c^2}, \qquad P^{\mathrm{NC}} = \frac{\omega^3 V}{2\pi c^2 v}, \qquad V = V_0 N. \tag{10.23}$$

Upon substituting Eq. (10.23) into Eqs. (10.20) and (10.21), we find

[16] Since we are considering the X-ray region or higher frequencies, conditions in Eq. (10.22) can be satisfied only in the cases of individual atoms or molecules and the frequencies of soft X-rays.

$$W^{\mathrm{C}}(\omega) = \frac{\omega_0^4 e^2 \omega^2 V^2}{64\pi^4 c^7} \int \sin^3\vartheta \cos^2\vartheta \, d\vartheta \, d\varphi,$$

$$W^{\mathrm{NC}}(\omega) = \frac{\omega_0^4 e^2 \omega^2 V^2}{16\pi^4 \gamma^2 c^7} \left[K_1\left(\frac{\omega\rho_0}{v\gamma}\right)\right]^2 \times \qquad (10.24)$$

$$\times \int \left(1 - \sin^2\vartheta \cos^2\varphi\right) \sin\vartheta \, d\vartheta \, d\varphi.$$

It follows that under conditions in Eq. (10.22) the radiation intensity varies very smoothly as the angle $\vartheta$ changes from zero to $\pi/2$. Moreover, the forward and backward radiation do not differ in intensity. The foregoing remains qualitatively valid also when the dimensions of the body are of the order of the wavelength.

2. As the transverse size (radius) of the body increases, i.e.,

$$a_\perp \gg \frac{c}{\omega}, \qquad (10.25)$$

because of the presence of the function $J_1(\rho\omega sin\vartheta/c)$ in $P_\parallel$ and $J_0(\rho\omega sin\vartheta/c)$ in $P_\perp$ small angles play an increasingly important role in the angular distribution. A similar situation occurs when the longitudinal size of the body increases, i.e.,

$$a_z \gg \frac{c}{\omega}, \qquad (10.26)$$

due to the presence of the function $\exp[i(1-\beta\cos\theta)\omega z/v]$ in the quantity $p(\rho, \lambda_0)$. In addition, when Eq. (10.26) is satisfied, because of the function $\exp[-i(1+ +\beta\cos\theta)\omega z/v]$ in $p(\rho, \lambda_0)$, the backward radiation becomes negligibly small compared to the forward radiation.

If we set $f(r) \approx V_0^{-1}$ for $-a_z \le z \le a_z$ and $0 \le \rho \le a_\perp$, then

$$p \approx \frac{2\sin[(1-\beta\cos\vartheta)\omega a_z/v]}{1-\beta\cos\vartheta},$$

$$P^{\mathrm{C}} \approx \frac{p\,|x_1 J_2(x_1) K_1(x_2) - x_2 J_1(x_1) K_2(x_2)|}{\sin^2\vartheta + \gamma^{-2}}, \qquad (10.27)$$

$$P^{\mathrm{NC}} \approx \frac{p\,x_1 J_1(x_1)}{\sin^2\vartheta}.$$

where $x_1 = (\omega a_\perp/c)sin\vartheta$, $x_2 = \omega a_\perp/(v\gamma)$. It follows, in agreement with the above, that when the dimensions of the body are much larger than the wavelength, the functions $P^{\mathrm{II}}$ and $P^{\mathrm{III}}$ have sharp maxima in the region $\theta \ll 1$. The positions and widths of these maxima are determined by the quantities $a_\perp$, $a_z$, $\omega$, and $\gamma$. When $\omega a_\perp/(v\gamma) \gg 1$, the principal maximum of the radiation occurs at an angle of order $\gamma^{-1}$ and has a width of the same order as in the case of a single interface (Sec. 1.8) or a plate (Sec. 2.3) with infinite transverse dimensions. When the transverse size of the body $a_\perp$ is of the order of the "radius" of the particle field $v\gamma/\omega$ or smaller, the direction of the principal radiation maximum and its angular width are determined by interference conditions at the boundaries of the body, namely, by the smaller of the angles $v/(\omega a_\perp)$ and $\sqrt{v/(\omega a_z)}$, similarly to what occurs in the diffraction of light by small bodies.

Let us estimate in more detail the spectral intensities of XTR (Eqs. (10.20) and (10.21)) in the case where the scattering body is a sphere of radius $r$ with a homogeneous density distribution inside.

When the radius of the sphere is much larger than the wavelength ($\omega r/c \gg 1$), from Eqs. (10.20) and (10.21), taking into account estimates of the type (10.27), we obtain

$$W^{\mathrm{C}}(\omega) \approx \frac{e^2\omega_0^4 r^2}{\pi c^3 \omega^2} I^{\mathrm{C}},$$

$$W^{\mathrm{NC}}(\omega) \approx \frac{e^2\omega_0^4 r^4}{\pi c^5 \gamma^2}\left[K_1\left(\frac{\omega\rho_0}{v\gamma}\right)\right]^2 I^{\mathrm{NC}}. \qquad (10.28)$$

where

$$I^{\mathrm{C}} \sim \begin{cases} \ln(\omega r/c), & \text{for } \omega r/c \ll \gamma, \\ \ln(\gamma^2 c/\omega r) - 1, & \text{for } \gamma \ll \omega r/c \ll \gamma^2, \end{cases} \quad (10.29)$$

$$I^{\mathrm{NC}} \sim \begin{cases} 1, & \text{for } \omega r/c \ll \gamma, \\ \ll 1, & \text{for } \gamma \ll \omega r/c. \end{cases} \quad (10.30)$$

When $\omega r/c \gg \gamma^2$, it follows from the condition in Eq. (10.12) for the applicability of perturbation theory that $\omega \gg \omega_0\gamma$, i.e., the XTR intensity is negligibly small.

### 10.3. Possibility of XTR Formation in Cosmic Space

Since fast charged particles exist in cosmic space and cross various inhomogeneities of matter (gaseous and dust clouds, as well as other formations), transition radiation, in particular XTR, should in principle be generated. The question is only how large its intensity is compared with the intensities of other types of radiation (for example, bremsstrahlung). This question attracted attention when X-ray astronomy emerged and systematic extraterrestrial observations of the sky began.

In 1971–1974, a number of papers were published [**71**.12, **72**.25, **72**.26, **73**.24–**73**.26, **74**.20] in which the possibility of realizing in cosmic objects the conditions for the formation of XTR was discussed; in particular, the question of the extent to which the observed diffuse (isotropic) cosmic X-ray emission can be explained by transition radiation produced by high-energy cosmic-ray charged particles on interstellar dust grains. In some works [**71**.12, **72**.26, **73**.26, **74**.20] it was argued that such an explanation is possible, whereas in other works [**72**.25, **73**.24, **73**.25] this possibility was doubted.

It is generally accepted that interstellar dust particles consist predominantly of carbon and have sizes of the order of $10^{-5}$–$10^{-6}$ cm. Then the applicability condition in Eq. (10.12) of the perturbation theory is satisfied in the X-ray and higher-frequency range, and the equations given in Sec. 10.2 are valid.

If, for simplicity, one assumes that the dust particles are spherical with radius $r$, then the spectral intensity (the so-called "birth spectrum" or "generation function") of XTR produced by ultrarelativistic charges on dust particles per unit time and per unit volume has the following form:

$$G_r(\omega) = 2\pi N_p \int f_{\text{ch}}(\gamma)\, d\gamma \int \left(W^{\text{\scriptsize III}}(\omega) + W^{\text{NC}}(\omega)\right)\rho_0\, d\rho_0, \qquad (10.31)$$

where $N_n$ is the number density of dust particles, $f_{ch}(\gamma)$ is the energy distribution of relativistic charges; integration over the impact parameter $\rho_0$ is carried out from zero to $r$ for central collisions and from $r$ to a value of the order of $v\gamma/\omega$ for non-central collisions. In view of estimates in Eqs. (10.28)–(10.30), we obtain

$$G_r(\omega) = \frac{e^2 N_p \omega_0^4 r^4}{\omega^2 c^2} \int \left[\ln \frac{\gamma^2 c}{\omega r} + C_r\right] f_{\text{ch}}(\gamma)\, d\gamma, \qquad (10.32)$$

where $C_r$=0 for $1 \ll \omega r/c \ll \gamma$ and $C_r^2 =- 1$ for $\gamma \ll \omega r/c \ll \gamma^2$. For $\gamma^2 \ll \omega r/c$, the generation function is negligibly small.

It is usually assumed that the energy distribution of cosmic charges follows a power law:

$$f_{\text{ch}}(\gamma) = N_{\text{ch}} \begin{cases} \text{const}, & \text{for } \gamma < \gamma_1, \\ \gamma^{-\alpha_1}, & \text{for } \gamma_1 < \gamma < \gamma_2, \\ \gamma^{-\alpha_2}, & \text{for } \gamma_2 < \gamma, \end{cases} \qquad (10.33)$$

where $\alpha_1$, $\alpha_2$ are certain positive numbers of order 1.5—3, $\gamma_1 \sim 10—100$, $\gamma_2 \sim 10^3$.

The observed spectral energy distribution of XTR passing through a unit area per unit time into a unit solid angle is obtained by integrating the generation function over the volume of space within a cone of unit solid angle whose axis is directed along the line of sight. In doing so, one must take into account the geometric attenuation factor $(4\pi R^2)^{-1}$, as well as the absorption of radiation in outer space. As a result, we have

$$W_{\rm obs}(\omega,\vartheta)=\frac{1}{4\pi}\int G_r(\omega)\exp\left[-\int_0^R \eta(x)\,dx\right]dR. \qquad (10.34)$$

where $\eta(x)$ is an averaged linear absorption coefficient of radiation in that region of outer space located at a distance $x$ from the point of observation (the Earth), and $R$ is the distance from the cosmic dust particle on which the radiation is produced to the observation point.

In Refs. [**73**.26, **74**.20, **81**.11], on the basis of Eq. (10.34), estimates were made of the intensities of the XTR produced on cosmic dust particles both inside and outside our galaxy. The estimates in these works showed that the intensity of intragalactic XTR is many orders of magnitude smaller than the observed intensity of the diffuse cosmic X-ray emission. XTR formed in intergalactic space through the interaction of relativistic electrons with dust particles apparently also makes only a negligible contribution to the diffuse background of the observed X-ray emission [**81**.11].

Relatively recently, the Orion Nebula was studied using the Einstein X-ray satellite. A certain component of soft X-ray emission with a luminosity of $\sim 10^{33}$ erg/s was detected [**79**.12]. It is noteworthy that a finite gamma-ray flux was recorded in the same direction [**80**.12]. Estimates [**81**.11] show that, under reasonable assumptions about the values of the parameters characterizing gas–dust clouds and the flux of relativistic electrons, the luminosity due to XTR in the range 0.5–3.5 keV is of the order of $10^{32}$–$10^{33}$ erg/s, i.e., *very close* to the observed value.

## § 11. PHENOMENOLOGICAL QUANTUM ELECTRODYNAMICS IN TWO MEDIA

In the preceding sections the theory of transition radiation was developed on the basis of the classical Maxwell's equations. It is of interest to formulate a quantum theory of this phenomenon that takes into account the recoil of the fast charged particle. In the case of Vavilov–Cherenkov radiation, the phenomenological quantum theory constructed in the works of Ginzburg [**40**.1], Sokolov [**40**.2], Ryazanov [**57**.6], and others (see the reviews [**57**.7, **59**.8]) gives only a small correction to the usual equation. This is due to the fact that Cherenkov radiation is mainly optical and therefore the recoil imparted to the particle is very small.

As was shown in the preceding sections, transition radiation can contain quanta with frequencies higher than optical ones. Therefore one may expect that the quantum theory of transition radiation will modify the results of the classical theory at high frequencies [60.7]. In addition, the formalism of quantum theory makes it possible to calculate other effects as well that arise when particles or photons pass from one medium to another and have no classical analogs. As an example, in the present section the probability of *conversion of a photon* into an electron–positron pair upon the photon's incidence on a medium boundary or upon its exit from it is calculated. In [77.2, 80.2, 81.2], by means of an analogous formalism, the "*transition*" *production* of pion pairs in collisions of fast nucleons with nuclear matter, as well as *stimulated transition radiation and absorption*, were investigated.

### 11.1. General Scheme of the Theory

To construct a quantum theory of electromagnetic phenomena, it is first of all necessary to consider the free electromagnetic and electron–positron[17] fields. Since the formalism of second quantization of the electron–positron field does not depend on the properties of the medium, we shall make use of the known results of such quantization (see, for example, Ref. [**69**.9]). As for the electromagnetic field, its treatment must be carried out again.

We shall use a gauge for the potentials in which the scalar potential $\varphi = 0$. Then, if in the region $z < 0$ there is a medium with dielectric permittivity $\varepsilon_1(\omega)$, and in the region $z > 0$ a medium with permittivity $\varepsilon_2(\omega)$, the components of the vector potential $A(x)$ satisfy equation [**51**.1] (in this section we set $\hbar = c = 1$):

[17] The results obtained will be valid for particles of any mass with spin 1/2.

$$\left\{\nabla^2 - \left[\theta(-z)\varepsilon_1\left(i\frac{\partial}{\partial t}\right) + \theta(z)\varepsilon_2\left(i\frac{\partial}{\partial t}\right)\right]\frac{\partial^2}{\partial t^2}\right\}A(x) = 0, \qquad (11.1)$$

where θ(z) is the Heaviside step function, equal to zero for negative values of the argument and unity for positive values.

In the case of a homogeneous medium, the general solution of the wave equation is a superposition of plane waves with arbitrary amplitudes. In the presence of two media it is natural to divide all waves into those propagating from the interface into infinity ($z \to \infty$) and those propagating toward the interface from infinity.

In order for these solutions to satisfy the boundary conditions at $z \to 0$, it is necessary to add two accompanying waves to each of the aforementioned primary waves. As a result, we obtain several types of solutions of Eq. (11.1), which are shown in Figs. 11.1 and 11.2. The primary wave is indicated by a bold arrow, and the other two arrows correspond to waves accompanying the primary one. As independent solutions in terms of which the general solution of Eq. (11.1) can be expanded, it suffices to take either waves of type 1 (Fig. 11.1) or waves of type 2 (Fig. 11.2). It is not difficult to verify that they can be expressed in terms of one another through linear combinations.

By expanding the general solution of Eq. (11.1) in terms of waves of type 1, we represent it in the form

$$\boldsymbol{A}(x) = \boldsymbol{A}^{(1a)}(x) + \boldsymbol{A}^{(1б)}(x), \qquad (11.2)$$

where

$$\boldsymbol{A}^{(1a)}(x) = V^{-1/2} \sum_{\varkappa,\omega \geqslant 0,\ l=1,2} (2\omega g^{(l)})^{-1/2} \{\theta(-z)\varepsilon_1^{-1/2} \boldsymbol{e}^{(l)} (a_- e^{ik_{1-}x} + \overset{+}{a}_- e^{-ik_{1-}x}) +$$

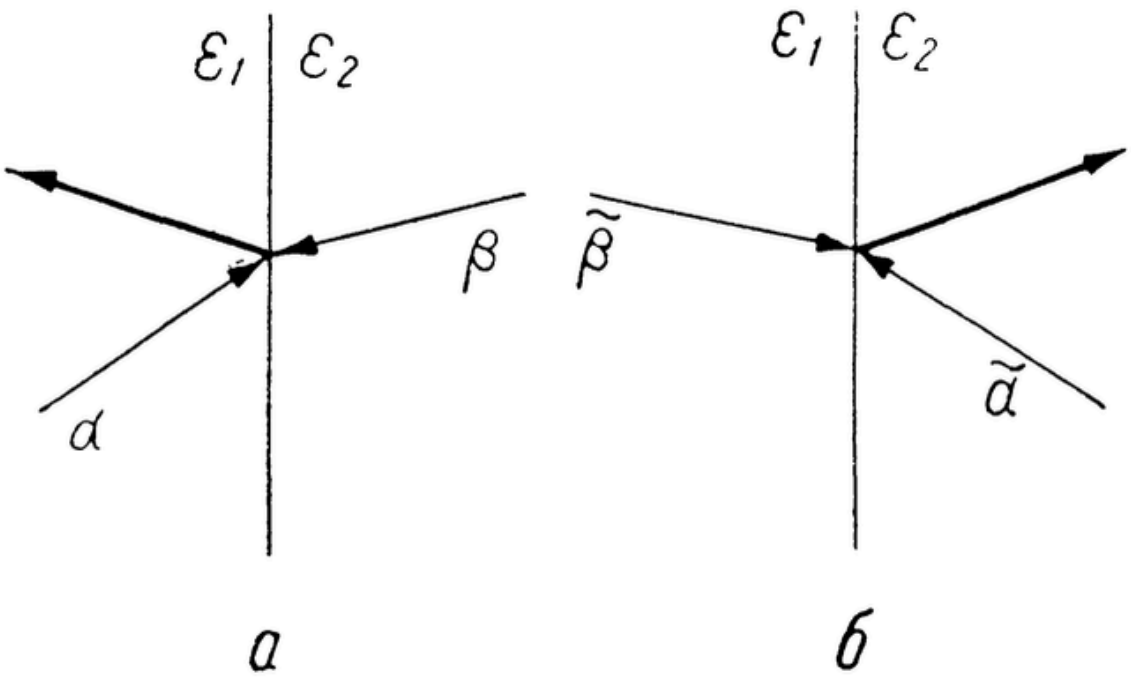

Fig. 11.1. Two principal waves of type 1 (bold arrows) propagating from the interface between the media backward (a) and forward (b), and their accompanying waves (thin arrows).

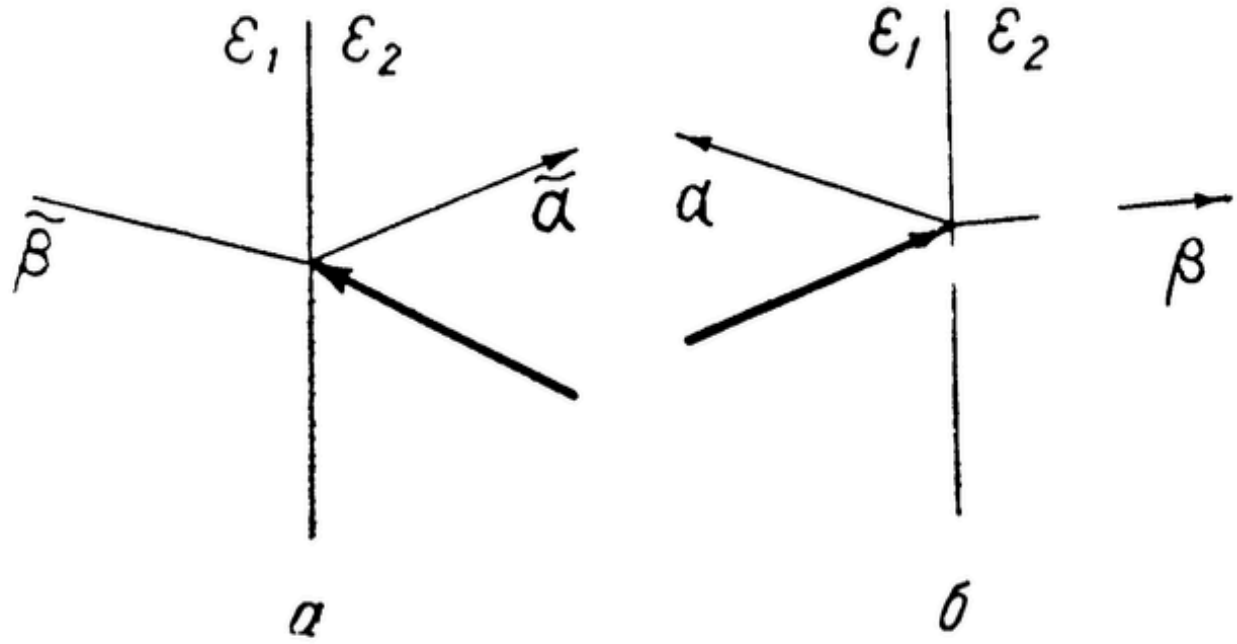

Fig. 11.2. Two principal waves of type 2 (bold arrows) propagating toward the interface between the media backward (a) and forward (b), and their accompanying waves (thin arrows)

$$+\theta(-z)\,\varepsilon_1^{-1/2}\,\boldsymbol{e}_1^{(l)}\alpha^{(l)}\,(a_- e^{ik_{1+}x} + \overset{+}{a}_- e^{-ik_{1+}x}) +$$

$$+\theta(z)\varepsilon_2^{-1/2}\boldsymbol{e}_2^{(l)}\beta^{(l)}(a_- e^{ik_{2-}x} + \overset{+}{a}_- e^{-ik_{2-}x})\}, \qquad (11.3)$$

$$\boldsymbol{A}^{(16)}(x) = V^{-1/2}\sum_{\varkappa,\omega\geqslant 0,\ l=1,2}(2\omega\tilde{g}^{(l)})^{-1/2}\{\theta(z)\varepsilon_2^{-1/2}\boldsymbol{e}'^{(l)}(a_+ e^{ik_{2+}x} + \overset{+}{a}_- e^{-ik_{2+}x}) +$$

$$+\theta(z)\varepsilon_2^{-1/2}\,\boldsymbol{e}_1'^{(l)}\tilde{\alpha}^{(l)}\,(a_+ e^{ik_{2-}x} + \overset{+}{a}_+ e^{-ik_{2-}x}) +$$

$$+\theta(-z)\varepsilon_1^{-1/2}\boldsymbol{e}_2'^{(l)}\tilde{\beta}^{(l)}(a_+ e^{ik_{1+}x} + \overset{+}{a}_+ e^{-ik_{1+}x})\}. \qquad (11.4)$$

To avoid cumbersome notation, the wave amplitudes $a_{\chi,\omega,l,\pm}$ are denoted by $a_\pm$ (regarding the “cross” sign above the amplitudes $a_\pm$, see below). The remaining notation is as follows: $V$ is the normalization volume, $k_s x = \chi \cdot \rho \pm \lambda_s z - \omega t$, $\lambda_s = (\omega^2\varepsilon_s - \varkappa^2)^{1/2}$ (s = 1, 2), $\chi$ and $\rho$ are, respectively, the wave vector of the radiation field and the radius vector of the observation point in the $xy$ plane, $\tilde{g}^{(l)} = 1 + \alpha^{(l)2} + \beta^{(l)2}$, $\tilde{g}^{(l)} = 1 + \tilde{\alpha}^{(l)2} + \tilde{\beta}^{(l)2}$, $e^{(l)}$ (with various indices) are unit polarization vectors of the corresponding waves. Summation over the polarization directions of the waves is carried out over the values $l = 1, 2$, the polarization directions being perpendicular to the direction of wave propagation (the potentials satisfy the Lorenz condition div$A$ = 0). Obviously, the propagation directions of the three waves shown in the figures are coplanar. We choose the polarization vectors with $l = 1$ to lie in the plane of each of the three waves, and the vectors with $l = 2$ to be normal to these planes.

The coefficients $\alpha^{(l)}, ..., \tilde{\beta}^{(l)}$ must be determined from the matching conditions for the radiation fields at the interface between the media. Using the standard method for deriving the Fresnel equations [**57**.4], we obtain the following expressions (the quantities $\alpha^{(l)}$ and $\beta^{(l)}$ represent the corresponding Fresnel coefficients for reflection and transmission for waves of different polarizations; cf. Eq. (1.22)):

$$\alpha^{(1)} = \frac{\lambda_1\varepsilon_2 - \lambda_2\varepsilon_1}{\lambda_1\varepsilon_2 + \lambda_2\varepsilon_1}, \qquad \alpha^{(2)} = \frac{\lambda_1 - \lambda_2}{\lambda_1 + \lambda_2},$$

$$\beta^{(1)} = \frac{2\lambda_1\varepsilon_2}{\lambda_1\varepsilon_2 + \lambda_2\varepsilon_1}, \qquad \beta^{(2)} = \frac{2\lambda_1(\varepsilon_2/\varepsilon_1)^{1/2}}{\lambda_1 + \lambda_2}. \qquad (11.5)$$

The corresponding coefficients with a tilde are obtained from Eqs. (11.5) by interchanging the indices 1⇄2.

As already noted, the general solution of Eq. (11.1) can be expanded in terms of solutions of type 2:

$$\boldsymbol{A}(x) = \boldsymbol{A}^{(2a)}(x) + \boldsymbol{A}^{(2b)}(x). \qquad (11.6)$$

Note that, formally, the solutions $A^{(2a)}(x)$ and $A^{(2b)}(x)$ can be obtained from $A^{(1b)}(x)$ and $A^{(1a)}(x)$, respectively, by the substitutions $k_{s+} \rightleftarrows k_{s-}$, $a_+ \rightleftarrows a_-$, and $\overset{+}{a}_+ \rightleftarrows \overset{+}{a}_-$.

For quantization of the radiation field, we require that the commutators

$$[a_\pm, \overset{+}{a}_\pm] \equiv [a_{\varkappa,\omega,l,\pm}, \overset{+}{a}_{\varkappa',\omega',l',\pm}] = \delta_{\varkappa\varkappa'}\delta_{\omega\omega'}\delta_{ll'}, \qquad (11.7)$$

be nonzero, and all other commutators vanish. From Eq. (11.7) it follows that the amplitude with a superscript plus corresponds to the creation of any one of the three waves shown in Figs. 11.1 and 11.2, and without a plus to annihilation. For the field energy we obtain the following expression:

$$W = \frac{1}{2}\int\left\{\varepsilon\left(\frac{\partial \boldsymbol{A}}{\partial t}\right)^2 + (\operatorname{rot}\boldsymbol{A})^2\right\}d^3\boldsymbol{r} = \sum_{\varkappa,\omega\geqslant 0,\ l=1,2} \omega\left[\left(N_- + \frac{1}{2}\right) + \left(N_+ + \frac{1}{2}\right)\right], \qquad (11.8)$$

where $N_\pm$ are the photon number operators, defined by the equation $N_\pm + 1/2 = (\overset{+}{a}_\pm a_\pm + a_\pm \overset{+}{a}_\pm)/2$, i.e., we have the correct corpuscular picture.

Using Eqs. (11.2)–(11.4) and (11.7), one can compute the commutators and contractions of the field operators, but we shall not do this, since these expressions will not be needed in what follows.

By representing the scattering matrix as a series in powers of the charge, we can compute the probability of any process that occurs in the transition of a charge or a photon from one medium to another. The Feynman diagram corresponding to transition radiation is shown in Fig. 11.3 (the vertical segment of the straight line passing through the vertex of the diagram indicates that an interaction of the aforementioned type takes place here). The main difference in writing the matrix element corresponding to the diagram in Fig. 11.3 from the usual case is that, whereas ordinarily the vertex of the plot is associated with the product of four δ-functions, including $\delta(\sum p_z)$, now this last δ-function is replaced by the sum of the $\delta_+$- and $\delta_-$-functions [**69**.9] with different arguments; that is, in the general case the conservation law for the $z$-components of the momenta of the particles participating in the process does not hold. This occurs because part of the momentum is transferred to a medium that has a density discontinuity along the $z$-axis.

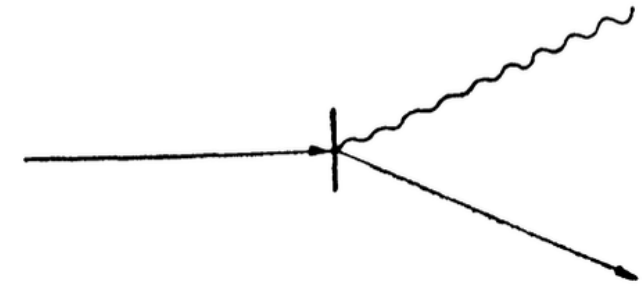

Fig. 11.3. Feynman diagram corresponding to transition radiation

Finally, let us make a remark on the limits of applicability of the theory. Since the theory being developed is macroscopic, its applicability requires that the volume in which the process occurs contains a large number of atoms or, more precisely, dispersion particles. This criterion must be applied to each process separately, thereby establishing the limits of applicability of the equations.

## 11.2. Transition and Cherenkov Radiation

Let a charge, moving along the $z$-axis, impinge perpendicularly on the interface between two media. We shall calculate the probability of emission of a quantum of frequency ω into the second medium, assuming that the particle momentum before emission was $p_1$, and after emission $p_2$. The process under consideration is a first-order effect (see Fig. 11.3). Its matrix element, taking into account that radiation emitted into the second medium corresponds to the creation of photons of type 1b (Fig. 11.1b), is

$$M_{12}=(2\pi)^3 e(2\omega\tilde{g}^{(l)}V)^{-1/2} f_{12}\delta(\mathbf{p}_{2t}+\boldsymbol{\varkappa})\delta(E_1-E_2-\omega), \tag{11.9}$$

where

$$f_{12}=\bar{u}(p_2)\left[\left(\pi\delta(p_1-p_{2z}-\lambda_2)+\mathcal{P}\frac{i}{p_1-p_{2z}-\lambda_2}\right)\frac{\hat{e}'^{(l)}}{\varepsilon_2^{1/2}}+\right.$$
$$+\left(\pi\delta(p_1-p_{2z}-\lambda_1)-\mathcal{P}\frac{i}{p_1-p_{2z}-\lambda_1}\right)\frac{\hat{e}_2'^{(l)}\tilde{\beta}^{(l)}}{\varepsilon_1^{1/2}}+$$
$$\left.+\left(\pi\delta(p_1-p_{2z}+\lambda_2)+\mathcal{P}\frac{i}{p_1-p_{2z}+\lambda_2}\right)\frac{\hat{e}_1'^{(l)}\tilde{\alpha}^{(l)}}{\varepsilon_2^{1/2}}\right]u(p_1). \tag{11.10}$$

Here $u(p_1)$, $u^{-}(p_2)$ are the spinor amplitudes of the electron–positron field, $\hat{e}=\sum_{\nu} e_\nu\gamma_\nu$, where $\gamma_\nu$ are the four-dimensional Dirac matrices [**69**.9]. The symbol $P$ indicates that the integral of the expression following it is to be taken in the sense of the principal value.

We shall confine ourselves to the medium–vacuum case ($\varepsilon_1=\varepsilon$, $\varepsilon_2=1$) and consider radiation in vacuum, i.e., we set $\varkappa=\omega sin\vartheta$, where $\vartheta$ is the angle between the direction of motion of the quantum and the $z$ -axis. Then

$$f_{12}=\bar{u}(p_2)\left[\mathcal{P}\left(\frac{i\hat{e}'^{(l)}}{p_1-p_{2z}-\lambda_2}-\frac{i\hat{e}_2'^{(l)}\tilde{\beta}^{(l)}}{(p_1-p_{2z}-\lambda_1)\varepsilon^{1/2}}+\frac{i\hat{e}_1'^{(l)}\tilde{\alpha}^{(l)}}{p_1-p_{2z}+\lambda_2}\right)+\right.$$
$$\left.+\pi\varepsilon^{-1/2}\hat{e}_2'^{(l)}\tilde{\beta}^{(l)}\delta(p_1-p_{2z}-\lambda_1)\right]u(p_1). \tag{11.11}$$

The first three terms in the last equation, which correspond to the non-conservation of the $z$-component of the momenta of the particles participating in the process, i.e., are due to the presence of the boundary, are naturally associated with transition radiation. As for the term with the δ -function, as we shall see below, it corresponds to Cherenkov radiation generated in the medium and transmitted into vacuum.

Let us first calculate the probability of transition radiation production:

$$dw = \frac{e^2}{2(2\pi)^3 V\omega} |f'_{12}|^2 \delta(p_{2t} + \varkappa)\delta(E_1 - E_2 - \omega) ST d^3p_2 d^3k_{2+}, \qquad (11.12)$$

where $f'_{12}$ is defined by Eq. (11.11), in which only the first three terms are retained, $S$ is the cross-sectional area of the volume in which the process takes place, and $T$ is the duration of the process. As the normalization volume $V$ we take the volume in which the process occurs, i.e., $V = vST$, $d^3k_{2+} = \omega d\omega d^2\Omega$.

Integrating over the electron momenta, we obtain

$$dw = \frac{e^2 \omega d\omega d^2\Omega}{16\pi^3 v} \frac{E_2}{p_{2z}} |f'_{12}|^2, \qquad (11.13)$$

where it should be kept in mind that

$$p_{2z} = (p_1^2 - 2E_1\omega + \omega^2\cos^2\vartheta)^{1/2}, \quad E_2 = E_1 - \omega.$$

Next we must average $|f'_{12}|^2$ over the polarizations of the electron in the initial state and sum over the polarizations of the electron and the photon in the final state. We shall carry out these calculations for two limiting cases. In the first case, without imposing restrictions on the electron velocity, we shall assume that the energy of the emitted quantum is much smaller than the electron energy. In the second case we shall assume the electron to be ultrarelativistic and $\omega \gg \omega_0$.

In the first case, we have

$$\frac{1}{2}\sum |f'_{12}|^2 = -\frac{v^2}{8E^2\omega^2}\sum_{l=1,2} \mathrm{Sp}\{\hat{G}^{\prime(l)}(i\hat{p} - m)\hat{G}^{\prime(l)}(i\hat{p} - m)\},$$

where

$$\hat{G}'^{(l)} = \frac{\hat{e}'^{(l)}}{1-v\cos\vartheta} - \frac{\tilde{\beta}^{(l)}\hat{e}_2'^{(l)}}{\varepsilon^{1/2}(1-v(\varepsilon-\sin^2\vartheta)^{1/2})} - \frac{\tilde{\alpha}^{(l)}\hat{e}_1'^{(l)}}{1+v\cos\vartheta}.$$

Since

$$\mathrm{Sp}\{\hat{e}'^{(l)}(i\hat{p}-m)\hat{e}_1'^{(l)}(i\hat{p}-m)\} = -8(e'^{(l)}\cdot p)(e_1'^{(l)}\cdot p),$$

upon summation over $l$, a nonzero result is obtained only for $l = 1$; hence it follows, in agreement with the classical result, that photons are emitted with the electric field vector lying in the plane passing through the particle's trajectory and the line of sight. As a result we obtain

$$\frac{1}{2}\sum|f_{12}'|^2 = \frac{v^2p^2\sin^2\vartheta}{E^2\omega^2(\varepsilon\cos\vartheta+(\varepsilon-\sin^2\vartheta)^{1/2})^2}\times$$
$$\times\left(\frac{\varepsilon\cos\vartheta+(\varepsilon-\sin^2\vartheta)^{1/2}}{1-v\cos\vartheta} - \frac{2\cos\vartheta}{1-v(\varepsilon-\sin^2\vartheta)^{1/2}} + \frac{\varepsilon\cos\vartheta-(\varepsilon-\sin^2\vartheta)^{1/2}}{1+v\cos\vartheta}\right)^2. \quad (11.14)$$

Substituting Eq. (11.14) into Eq. (11.13), we obtain an equation that coincides with the classical equation for transition radiation emitted forward (cf. (1.37))[18]

$$dw = \frac{e^2v^2(\varepsilon-1)^2(\gamma^{-2}-v(\varepsilon-\sin^2\vartheta)^{1/2})^2\sin^2\vartheta\cos\vartheta\, d\omega\, d^2\Omega}{4\pi^3\omega(\varepsilon\cos\vartheta+(\varepsilon-\sin^2\vartheta)^{1/2})^2(1-v^2\cos^2\vartheta)^2(1-v(\varepsilon-\sin^2\vartheta)^{1/2})^2}, \quad (11.15)$$

where $\gamma = (1-v^2)^{-1/2}$.

Note that, quite analogously, one could obtain the classical equation for transition radiation emitted backward.

In the second limiting case one may set $\tilde{\alpha}^{(l)} = 0$, $\tilde{\beta}^{(l)} = 1$. As a result, we obtain

$$dw = \frac{e^2\omega\, d\omega\, d^2\Omega}{16\pi^3E_1p_{2z}}(E_1E_2-p_1p_2\cos\vartheta-m^2)\left(\frac{1}{p_1-p_{2z}-\lambda_2} - \frac{1}{p_1-p_{2z}-\lambda_1}\right)^2. \quad (11.16)$$

[18] As is evident from Eq. (11.8), in this section we use the Heaviside system of units. To pass to the usual units it is necessary to make the replacement e2/(4)e2.

Restricting ourselves to frequencies satisfying the condition $(1 - \omega/(v_1 p_1)) \gg \gamma_1^{-1}$, where $\gamma_1 = (1 - v_1^2)^{-1/2}$, i.e., not very close to $E_1 - m$, and performing the integration over angles in Eq. (11.16), we obtain

$$w(\omega) = \frac{e^2}{2\pi^2 \omega a_2^2}\left\{\left[\frac{1+a_2^2}{4} + \frac{\omega^2 a_2}{\gamma_1^2 \omega_0^2}\left(\frac{1+a_2^2}{2} - a_1\right)\right]\ln\left(1 + \frac{\gamma_1^2 \omega_0^2}{\omega^2 a_2}\right) - \right.$$
$$\left. - \frac{1+a_2^2}{2} + \frac{a_1}{2} + \frac{a_1}{2(1+\gamma_1^2 \omega_0^2/(\omega^2 a_2))}\right\}, \qquad (11.17)$$

where

$$a_1 = \frac{\omega^2/p_1^2}{2(1-\omega/(v_1 p_1))^2}, \qquad a_2 = 1 + \frac{\omega/p_1}{1-\omega/(v_1 p_1)}.$$

For frequencies $\omega \ll E_1$, when one may take $a_1 \approx 0$, $a_2 \approx 1$, we obtain the usual classical equation for an ultrarelativistic particle (cf. (1.67)). Quantum corrections become significant for $\omega \lesssim E_1$. However, in this case, for ordinary electron densities the frequencies satisfy the relation $\omega \gg \omega_0 \gamma_1$, i.e., the intensity of transition radiation becomes negligibly small, and the quantum corrections are of little interest. The quantum corrections will be substantial if an electron enters from vacuum into a very dense medium, for example, into an atomic nucleus. In this case $\omega_0 \sim 5$ MeV and always $\omega < \omega_0 \gamma_1$. As follows from Eq. (11.17), the quantum calculation makes the probability of emission of frequencies $\omega \approx p_1$ vanish.

Finally, let us consider the limits of applicability of the equations obtained. For frequencies $\omega \lesssim p_1$ the formation length of transition radiation (in ordinary units) $z_{TR} \sim \frac{\hbar \gamma_1}{mc}\left(1 - \frac{\hbar\omega}{v_1 p_1}\right)$. Taking also into account that the radiation is emitted at an angle $\sim \gamma_1^{-1}$, we obtain the condition for the applicability of the theory: $\gamma_1 \gg (\hbar/mc)^{-3} n_e (1 - \hbar\omega/(v_1 p_1))$, where $n_e$ is the number of electrons per unit volume.

Let us now compute the Cherenkov radiation emitted in the medium, i.e., take into account the term with the δ-function in the square brackets of Eq. (11.11). For the probability of emission per unit time we have the expression

$$dw = \frac{e^2\beta^{(l)2}}{2(2\pi)^2\varepsilon\omega}|\bar{u}(p_2)\hat{e}_2^{(l)}u(p_1)|^2\delta(\boldsymbol{p}_1-\boldsymbol{p}_2-\boldsymbol{k}_{1+})\delta(E_1-E_2-\omega)d^3\boldsymbol{p}_2 d^3\boldsymbol{k}_{2\perp}. \tag{11.18}$$

In doing so, it has been taken into account that $\delta^4(p) = 2T \cdot 2V$, where $T$ and $V$ are the time during which the Cherenkov radiation occurs and the volume.

We perform the integration over $p_2$ in the last equation. From the conservation laws (setting $\chi = \omega sin\vartheta$) for the angle $\vartheta'$, at which the Cherenkov radiation propagates with respect to the $z$-axis, we obtain the expression

$$\sin\vartheta' = \frac{1}{v_1}\sqrt{v_1^2\varepsilon - \left(1 + \frac{\omega(\varepsilon-1)}{2E_1}\right)^2}.$$

In the classical case, assuming $p_1 = p_2 = p$, we average the emission probability over the polarizations of the electron in the initial state and sum over the polarizations of the electron and the photon in the final state:

$$\frac{1}{2}\sum|\beta^{(l)}\bar{u}(p)\hat{e}_2^{(l)}u(p)|^2 = \frac{v^2\beta^{(1)2}\sin^2\vartheta}{\varepsilon}.$$

Substituting the last equation into Eq. (11.18) and integrating over the angle $\vartheta$, we obtain

$$w(\omega) = \frac{e^2}{\pi v}\,\frac{(1+v^2(1-\varepsilon))^{1/2}(v^2\varepsilon-1)}{(1+\varepsilon(1+v^2(1-\varepsilon))^{1/2})^2}, \tag{11.19}$$

with the condition $v^2\varepsilon(\omega) \gg 1$ to be satisfied. Equation (11.19) coincides with the corresponding classical equation (1.44), obtained in Sec. 1.6A if in the latter $\varepsilon''$ is taken to zero.

## 11.3. Transition Pair Production by a Photon

As another first-order effect we consider the creation of an electron–positron pair by a γ-quantum moving along the $z$-axis when passing from one medium to another (the interface is the plane z = 0. This process corresponds to the annihilation of a photon of type 2b (Fig. 11.2b), and its diagram is shown in Fig. 11.4. It is evident that this boundary effect may occur because the conservation law of the $z$-components of the momenta of the particles participating in the process is not satisfied. On the other hand, the law of energy conservation requires the quantum to be hard, and therefore we shall assume $\alpha^{(l)} = 0$, $\beta^{(l)} = 1$. Using the expansion (11.6) of the vector potential, we obtain the following expression for the matrix element of the process:

$$M_{12} = \frac{i(2\pi)^3 e}{2(\omega V)^{1/2}} f_{12} \delta(p_{1t} + p_{2t}) \delta(\omega - E_1 - E_2), \qquad (11.20)$$

where

$$f_{12} = (\bar{u}(p_1)\, \hat{e}^{\prime(l)} v(p_2)) \left( \frac{1}{p_{1z} + p_{2z} - \omega\varepsilon_1^{1/2}} - \frac{1}{p_{1z} + p_{2z} - \omega\varepsilon_2^{1/2}} \right). \qquad (11.21)$$

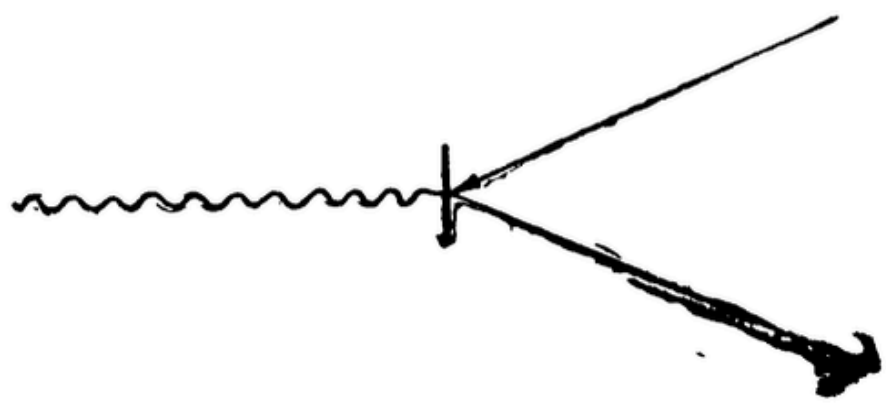

Fig. 11.4. Feynman diagram corresponding to transition production of an electron–positron pair

Here $p_1$ and $p_2$ are the electron and positron momenta, respectively, and $v(p_2)$ is the positron spinor amplitude.

The probability of the process is equal to

$$dw = \frac{e^2}{16\pi^3} |f_{12}|^2 \frac{E_1}{p_{1z}\omega} d^3p_2, \qquad (11.22)$$

with $E_1 = \omega - E_2$, $p_{1z} = (\omega^2 + p_{2z}^2 - 2E_2\omega)^{1/2}$.

After averaging and summing over the polarizations of the photon, the electron, and the positron, we obtain

$$dw = \frac{e^2}{8\pi^2} \frac{(E_1E_2 - p_{1z}p_{2z} + m^2)p_2^2 dp_2 \sin\vartheta d\vartheta}{E_2 p_{1z}\omega} \times$$
$$\times \left( \frac{1}{p_{1z} + p_{2z} - \omega\varepsilon_1^{1/2}} - \frac{1}{p_{1z} + p_{2z} - \omega\varepsilon_2^{1/2}} \right)^2, \qquad (11.23)$$

where $\vartheta$ is the angle between the direction of motion of the positron and the z-axis.

If we restrict ourselves to positron momenta satisfying $m \ll p_2 \ll \omega - m$, then after some transformations and integration over all emission directions of the positrons we obtain the following expression:

$$dw = \frac{e^2 p^2}{16\pi^2 \omega E_2 m^2} (\omega_{01}^2 - \omega_{02}^2)\, dp_2.$$

In the same approximations the pair formation length is $z \sim \hbar p_2 / m^2 c^2$ (in ordinary units). Hence the condition for the applicability of the theory has the form $p_2/mc \gg m^3c^3/n_e\hbar^3$.

Let us compare the probability of pair creation due to the boundary effect with the usual production mechanism in the field of atomic nuclei over the path $z_{min} \sim n_e(mc/\hbar)^2$. Estimates show that when a photon enters from vacuum into a medium with ordinary electron density, the probability of the boundary effect is negligibly small compared with the probability of pair production in the field of atomic nuclei.

In Ref. [**80**.2], stimulated emission by a beam of non-relativistic electrons upon crossing the interface between two different dielectric media in the presence of an external electromagnetic field is considered. Analysis of the total probabilities of emission and absorption of a quantum by a particle implies the possibility of amplification of the external electromagnetic wave by the beam.

In Ref. [**81**.2], the influence of an external (laser) electromagnetic field on the probabilities of Cherenkov and transition radiation is investigated. It is shown that such a field substantially changes the characteristics of both Cherenkov and transition radiation. In the case of Cherenkov radiation, the angular distribution is modified, and an asymmetry appears even in a homogeneous medium. The possibility arises of producing hard (~10 keV) radiation quanta with sufficiently high intensity. The characteristics of transition radiation in a laser field also change substantially. Of particular interest is the appearance of highly collimated hard radiation in the multiphoton absorption regime, with a probability of about 1%.

In Ref. [**77**.2], using the method of the phenomenological quantum theory set forth in the present section, the process of “transition” and “Cherenkov” pion production by relativistic nucleons passing through nuclei is considered. This refers to the emission of pions by the nuclear medium as a result of the interaction of the traversing nucleon with the medium.

# CHAPTER III

## MICROSCOPIC THEORY. QUASI-CHERENKOV RADIATION OF RELATIVISTIC PARTICLES IN CRYSTALS

In the X-ray and higher-frequency range, the radiation wavelength becomes smaller than the sizes of atoms and the distances between atoms in a condensed medium. If the medium is amorphous, the secondary waves of such frequencies emitted by atoms under the action of the field of a passing fast charged particle almost completely cancel one another, except for the waves emitted at the boundaries and propagating at small angles with respect to the particle’s direction of motion (see Sec. 1.8). For such waves the radiation formation length is large (see Eq. (1.30)), and the average dielectric permittivity $\varepsilon(\omega)$ of the medium plays the principal role. Therefore, in this case the macroscopic theory presented in the preceding chapters is fully applicable.

However, a completely different situation occurs in a *crystal*. Owing to the three-dimensional periodic structure of the crystal, the aforementioned secondary waves add up in certain directions determined by the Bragg conditions. The angles between these directions and the particle trajectory are, as a rule, not small; therefore, the corresponding radiation formation lengths are of the order of the wavelength, i.e., they are smaller than atomic sizes and interatomic distances. For such waves, the macroscopic theory is no longer applicable, and a treatment based on the microscopic structure of the crystal is necessary.

## § 12. FUNDAMENTALS OF THE THEORY

To describe the electromagnetic fields arising when a charged particle passes through matter, we shall use the microscopic Maxwell's equations. These equations contain the total current density, which consists of the free current (external sources) and the current $\boldsymbol{j}_{bound}$ of bound (or induced) charges. The latter is the quantum-statistical average of the bound-charge current density operator, which in turn depends on the electromagnetic fields in the given problem. An explicit expression for $\boldsymbol{j}_{bound}$ can be found using standard perturbation theory, bearing in mind that the electromagnetic interaction of matter with the field is proportional to the small parameter—the fine-structure constant $e^2/\hbar c$. It can be shown [71.4] that in the linear (in the field) approximation

$$j^l_{\rm bound}(\mathbf{k},\omega) = \frac{\omega\Omega_0}{2i(2\pi)^4}\int\sum_n \exp\left[i(\mathbf{k}'-\mathbf{k})\cdot\mathbf{R}_n\right]\times\chi^{lm}(\mathbf{k},\mathbf{k}')E^m(\mathbf{k}',\omega)\,d^3k' \quad (l,m=x,y,z). \tag{12.1}$$

Here and henceforth, summation over repeated Cartesian indices is implied. In Eq. (12.1) $\boldsymbol{R}_n$ is the radius vector of the $n$-th unit cell, $\Omega_0$ is the volume of the unit cell,

$$\chi^{lm}(\mathbf{k},\mathbf{k}') = -\frac{\omega_0^2}{z_{\rm cell}\,\omega^2}\sum_q \exp\left[i(\mathbf{k}'-\mathbf{k})\cdot\mathbf{r}_q - M_q(\mathbf{k}'-\mathbf{k})\right]f_q^{lm}(\mathbf{k},\mathbf{k}') \tag{12.2}$$

(summation is carried out over the atoms of the crystal unit cell), $\omega\delta=4\pi Ne^2/m_0$ is the plasma frequency of the medium, $N$ is the number of electrons per unit volume of the substance, $e$ and $m_0$ are the charge and the rest mass of the electron, $z_{cell}$ is the number of electrons in the unit cell, $M_q(\boldsymbol{k'}-\boldsymbol{k})$ determines the Debye–Waller factor.

The tensor $f_q^{lm}(k, k')$ has a rather complicated form [**68**.8,**71**.4]. If one considers frequencies far exceeding atomic ones, but such that the process of electron–positron pair creation does not yet play a significant role, then the main contribution to the indicated tensor comes from Thomson scattering of photons by atoms. Then

$$f_q^{lm}(\boldsymbol{k}, \boldsymbol{k}') = \delta^{lm} f_q(\boldsymbol{k}' - \boldsymbol{k}), \tag{12.3}$$

where $f_q(\boldsymbol{k'}-\boldsymbol{k})$ is the form factor of the $q$-th atom, $\delta_{ln}$ is the Kronecker symbol.

Accounting for dissipative processes (for example, due to the photoelectric effect) leads to the appearance of an imaginary part in the quantity given by Eq. (12.3).

The field of a particle in a medium can be represented as a sum

$$\boldsymbol{E}(\boldsymbol{k}, \omega) = \boldsymbol{E}_{ch}(\boldsymbol{k}, \omega) + \boldsymbol{E}_{scat}(\boldsymbol{k}, \omega),$$

where $\boldsymbol{E}_{ch}(\boldsymbol{k}, \omega)$ is the Fourier component of the field of the charge in vacuum (see Sec. 1.3), and $\boldsymbol{E}_{scat}(\boldsymbol{k}, \omega)$ is the additional part due to interaction with the medium and determined from the equation

$$\left(k^2 - \frac{\omega^2}{c^2}\right) E^l_{scat}(\mathbf{k}, \omega) = \frac{\Omega_0}{(2\pi)^3} \int \sum_n \exp\left[i(\mathbf{k}' - \mathbf{k}) \cdot \mathbf{R}_n\right] \left[\frac{\omega^2}{c^2}\chi^{lm}(\mathbf{k}, \mathbf{k}') - k^l k^j \gamma^{jm}(\mathbf{k}, \mathbf{k}')\right] E^m(\mathbf{k}', \omega)\, d^3k'. \tag{12.4}$$

It is not difficult to verify that in the region of optical and lower frequencies, when the wavelength is much larger than the interatomic distances, the quantity $\lambda^{lm}$ (see Eq. (12.2)) coincides with the polarizability of the medium used in macroscopic electrodynamics. In the general case, the tensor (Eq. (12.2)) also describes the polarizability of the medium for wavelengths comparable to atomic dimensions and interatomic distances, or smaller than them. In Eqs. (12.2) and (12.4), the microscopic structure of the medium and of the atoms is explicitly taken into account.

## § 13. THIN CRYSTALLINE PLATE. KINEMATICAL THEORY

For simplicity, we shall assume that Eq. (12.3) holds, i.e.

$$f^{lm}(\boldsymbol{k},\ \boldsymbol{k}') = \delta^{lm}\chi(\boldsymbol{k},\ \boldsymbol{k}'). \tag{13.1}$$

In the X-ray and higher-frequency range

$$|\chi(\boldsymbol{k},\ \boldsymbol{k}')| \ll 1 \tag{13.2}$$

and Eq. (12.4) contains a small parameter. We shall solve this equation by the method of perturbation theory. As the first approximation, we take the solution $\boldsymbol{E}_{\text{scat}}$ obtained from Eq. (12.4) if, on its right-hand side, the quantity $\boldsymbol{E}$ is set equal to $\boldsymbol{E}_{\text{ch}}$.

Let the particle move along the $z$-axis, and the crystal have infinite dimensions in the $x$ and $y$ directions and be located between the planes $z = -z_0N_z/2$ and $z = z_0N_z/2$, where $z_0$ is the spacing between atomic planes. After integrating over $\boldsymbol{k}'$ on the right-hand side of Eq. (12.4) and subsequently performing the inverse Fourier transform, we obtain the field in the first approximation:

$$E^{(1)}_{\text{scat}}(\mathbf{r},t) = -\frac{ez_0}{4\pi^2 v}\int_{\lambda_0}\frac{\omega^2}{c^2}F\left(\frac{\omega}{v}-\lambda_0\right)\exp[i\,(\mathbf{x}\cdot\boldsymbol{\rho}+\lambda_0 z-\omega t)]\times$$
$$\times\sum_{\mathbf{k}_h}\chi(\mathbf{k},\mathbf{k}_h)\left\{\frac{\omega v}{c^2}-x_h-\frac{\omega n_z}{v}+\frac{c^2}{\omega^2}(x+\lambda_0 n_z)\times\mathbf{x}\cdot\mathbf{x}_h+\frac{i_0\omega}{v}\gamma^{-2}\right\} \tag{13.3}$$
$$\left(x_h^2+\frac{\omega^2}{v^2}-\frac{\omega^2}{c^2}\right)^{-1}d^2x\,d\omega.$$

where $\lambda_0 = (\omega^2/c^2 - \varkappa^2)^{1/2}$, $\boldsymbol{\varkappa}_h = \boldsymbol{\varkappa} + \boldsymbol{K}_{h\perp}$, $\boldsymbol{K}_{h\perp}$ is the projection of the reciprocal lattice vector $\boldsymbol{K}_h$ (multiplied by $2\pi$) onto the $(x, y)$ plane, $\boldsymbol{n}_z$ is the unit vector in the $z$ direction, $\boldsymbol{\rho} = \{x, y, 0\}$, and

$$F(x) = \frac{\sin[x z_0 (N_z + 1)/2]}{\sin(x z_0/2)}.$$

The function $|F(x)|$ for $N_z \gg 1$ has sharp maxima at values of the argument

$$x = \frac{2\pi h}{z_0} \quad (h\text{—}integer)$$

(13.4)

The maximum corresponding to the value $h = 0$ occupies a special position. The argument $\omega/v - \lambda_0$ of the function under consideration cannot be equal to zero for any real values of $\varkappa$ and $\omega$. However, in the case of ultrarelativistic particles the value $\omega/v - \lambda_0$ can become very close to zero. For a crystal of finite thickness (i.e., for a finite value of $N_z$) the function $F(\omega/v - \lambda_0)$ can then attain a sufficiently large value, comparable with its maximum value $F(0) = N_z + 1$. Physically, this reflects the fact that a crystal of finite thickness has boundaries at which transition radiation is generated. It propagates at small angles relative to the trajectory of an ultrarelativistic particle and forms a *central radiation spot*.

In addition to this central spot, *side radiation spots* are also formed in the crystal, emitted at the *Bragg angles*. Indeed, it follows from Eq. (13.3) that when the following conditions are satisfied:

$$\omega/v - \lambda_0 = 2\pi h/z_0 \equiv K_{hz} \qquad (13.5)$$

($h$ is an integer) and

$$|\boldsymbol{\varkappa} + \boldsymbol{K}_{h\perp}|^2 \approx \frac{\omega^2}{v^2} - \frac{\omega^2}{c^2} \qquad (13.6)$$

the radiation is large.

Let us introduce the radiation angle $\vartheta$, measured from the particle trajectory and defined by the relations:

$$\lambda_0 = \frac{\omega}{c}\cos\vartheta, \qquad \varkappa = \frac{\omega}{c}\sin\vartheta. \qquad (13.7)$$

From conditions (13.5) and (13.6) for ultrarelativistic particles, one readily obtains that

$$\vartheta \approx 2\vartheta_B \qquad (13.8)$$

where $\vartheta_B$ is the Bragg angle defined by the well-known expression

$$sin\vartheta_B = |K_{hz}|/K_h \qquad (13.9)$$

Equation (13.3) gives only the result of the first approximation. We can restrict ourselves to this approximation if the higher-order corrections are small. To obtain the next approximation, we substitute the first-order result into the right-hand side of Eq. (12.4) and then perform the inverse Fourier transform. Comparing the expression for the field obtained in this way with Eq. (13.3), we find that the correction is small for a sufficiently thin crystalline plate, i.e., when the following condition is satisfied:

$$|\chi(\boldsymbol{k}, \boldsymbol{k}')|\frac{\omega a}{c} \ll 1, \qquad (13.10)$$

where $a = N_z z_0$ is the thickness of the crystal.

Let us now calculate the spectral–angular and spectral distributions of the radiation intensity in the central and side spots of the Laue pattern formed during the passage of a fast particle through a thin crystalline plate. The total energy flux passing through the plane z = const, parallel to the crystal boundaries and sufficiently far from the crystal, is given by

$$W = \frac{c}{4\pi}\int |E_{scat}(r,t)|^2 |cos\vartheta| d^2\rho dt. \qquad (13.11)$$

Here the absolute value of $cos\vartheta$ is taken, since spots may also be formed in the backward hemisphere with respect to the direction of motion of the charged particle.

First, let us find the radiation intensity in the central spot. To this end, in Eq. (13.3) we take the term corresponding to the values $K_{h\perp} = K_{hz} = 0$ (i.e., $h = 0$ in Eq. (13.4)) and substitute it into Eq. (13.11). Bearing in mind that $\varkappa = (\omega/c)sin\vartheta$ and $\omega > 0$, we obtain

$$W = \frac{8e^2}{\pi v} \int |\chi_{00}|^2 \frac{\sin^2[\omega a(\vartheta^2 + \gamma^{-2})/4v]}{(\vartheta^2 + \gamma^{-2})^4} \vartheta^3 d\vartheta\, d\omega, \qquad (13.12)$$

where

$$\chi_{00} = \chi\left(\frac{\omega \boldsymbol{n}}{c}, \frac{\omega \boldsymbol{n}}{c}\right), \qquad (13.13)$$

$n$ is the unit vector in the direction of radiation.

Let us now find the radiation intensity in a side spot corresponding to a nonzero reciprocal lattice vector $K_h$ (multiplied by 2π) with $K_{hz} \neq 0$. Substituting Eq. (13.3) into Eq. (13.11), we find

$$W_h = \frac{e^2}{\pi^2 c^3} \int |\chi_{0h}|^2 \frac{(\sin^2\varphi' + \cos^2\varphi' \cos^2 2\vartheta_Б)}{(\vartheta'^2 + \gamma^{-2})^2} \times$$
$$\times \frac{\sin^2(a\xi/2)}{\xi^2} \vartheta'^3 d\vartheta'\, d\varphi'\, \omega^2 d\omega, \qquad (13.14)$$

where

$$\chi_{0h} = \chi\left(\frac{\omega \mathbf{n}}{c}, \frac{\omega \mathbf{n}}{c} + \mathbf{K}_h\right),$$
$$\xi = \frac{\omega_B}{c}\left[\frac{2(\omega - \omega_B)\sin^2\vartheta_B}{\omega_B} + \frac{\gamma^{-2}}{2} - \frac{2\vartheta' \sin 2\vartheta_B \cos\varphi' - \vartheta'^2}{2\cos 2\vartheta_B}\right], \qquad (3.15)$$
$$\omega_B = cK_h^2/2|K_{hz}|.$$

$\vartheta'$ is the radiation angle in the side spot measured from the Bragg direction

$$\mathbf{n}_B = \sin 2\vartheta_B \frac{\mathbf{K}_{h\perp}}{|K_{h\perp}|} + \cos 2\vartheta_B \mathbf{n}_z, \qquad (13.16)$$

and $\varphi'$ is the angle between the vectors $\boldsymbol{\varkappa} = \boldsymbol{\varkappa}_h - \mathbf{K}_{h\perp}$ and $\mathbf{K}_{h\perp}$ (Fig. 13.1).

The integration in Eq. (13.14) is straightforward provided that the following inequality holds:

$$\frac{2a\omega_B}{c}\sin^2\vartheta_B \gg 1. \qquad (13.17)$$

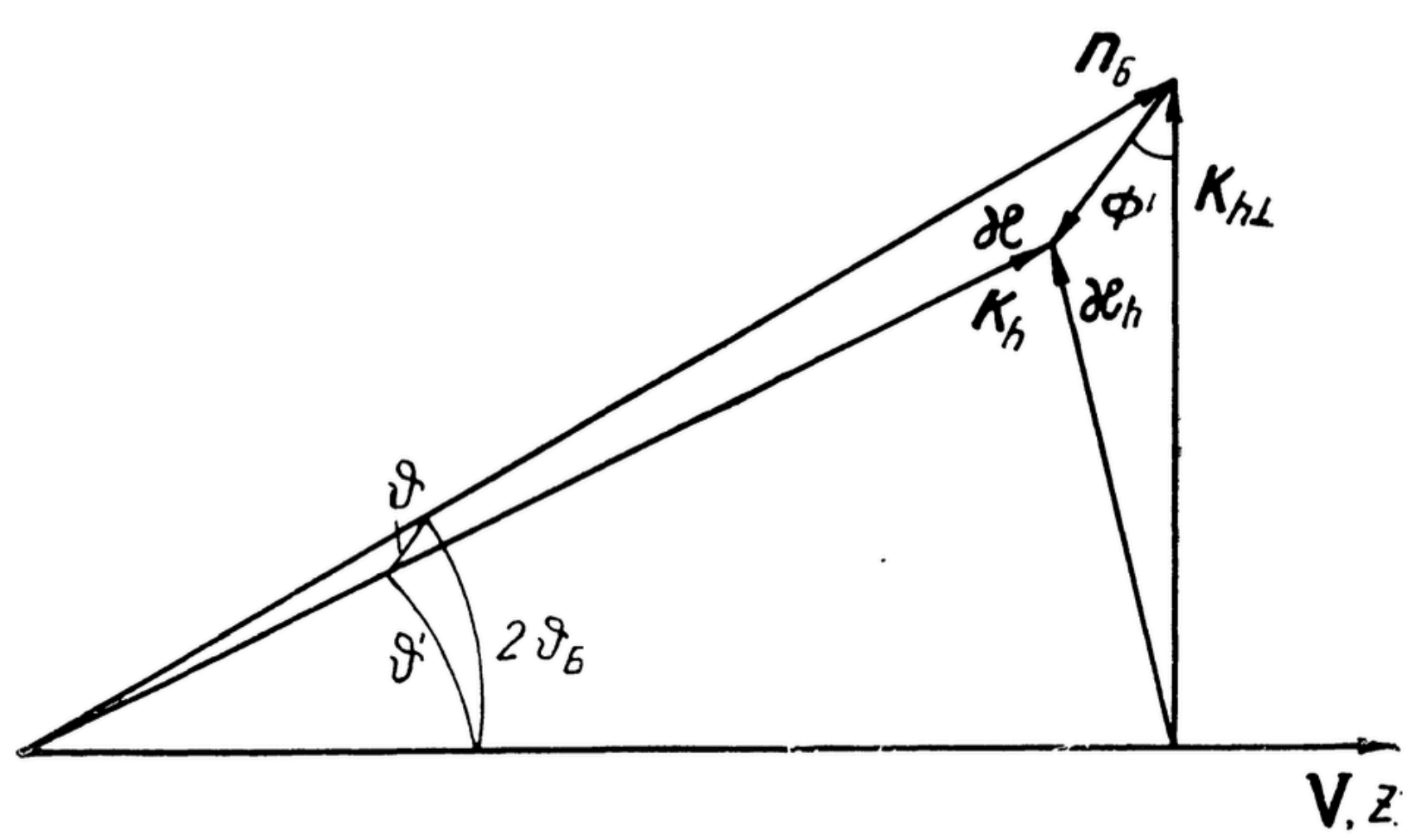

Fig. 13.1. Bragg direction $n_B$ and the radiation angles $\vartheta'$ and $\varphi'$ in the side spot

The Bragg frequencies $\omega_B$ are usually of the order of several or tens of keV, and the Bragg angles $\vartheta_B$ are of the order of several or tens of degrees, i.e., inequality (13.17) is well satisfied for $a > 10^{-6} - 10^{-5}$ cm. the range of ω for which the expression $\sin^2(a\xi/2)/\xi^2$ is appreciable is very narrow: $|\omega - \omega_B| \ll \omega_B$. After the substitution

$$\frac{\sin^2(a\xi/2)}{\xi^2} \approx \frac{\pi a c}{4\sin^2\vartheta_B}\,\delta(\omega - \omega_B).$$

and after subsequent integration over ω, φ' and ϑ', from Eq. (13.14) we obtain

$$W_h = \frac{e^2 a\omega_B^2}{8c^2}|\chi_{0h}|^2 \frac{1+\cos^2 2\vartheta_B}{\sin^2\vartheta_B}(2\ln\gamma - 1). \qquad (13.18)$$

where the quantity $|\chi_{0h}|^2$ is taken at $\omega = \omega_B$.

In the above Eqs. (13.12), (13.14), and (13.18), due to condition (13.10), the absorption of radiation in the medium can be neglected.

It is not difficult to verify that Eq. (13.12) for the central spot, within the limits of its applicability, coincides with Eq. (2.21) for XTR, obtained in the macroscopic theory without taking into account the crystalline structure of matter.

Let us now consider Eq. (13.14), which determines the spectral–angular distribution of the intensity of the side spot corresponding to the vector $K_h$. The integrand of Eq. (13.14) contains the product of two functions: $sin^2(a\xi/2)/\xi^2$ and $\vartheta'^3/(\vartheta'^2 + \gamma^{-2})^2$, each of which has a sharply pronounced maxima. It is clear that the greatest value of the product is obtained when the maxima of these two functions coincide. This occurs for such a value of the azimuthal angle $\phi'$ that the principal maximum of the function $sin^2(a\xi/2)/\xi^2$ (for a given value of $\omega$ close to $\omega_B$) falls at the angle $\vartheta' \sim \gamma^{-1}$. In this case, the angular width with respect to $\vartheta'$ is determined by the smaller of the widths of the maxima of these functions:

$$\Delta\vartheta' \sim \min\{\gamma^{-1}, \frac{c\gamma\cos 2\vartheta_B}{a\omega_B}\} \qquad (13.19)$$

(the angle $\vartheta_B$ is not close to $\pi/4$). According to condition (13.10), the second of the quantities (13.19) is less than the first only when

$$\gamma^{-2} \gg |\chi(\boldsymbol{k},\ \boldsymbol{k}_h)|.$$

However, in this case the radiation intensity is small. If the value of $\gamma$ is sufficiently large ($\gamma^2|\chi(k,\ k_h)| \gg 1$), then $\Delta\vartheta' \sim \gamma^{-1}$.

The angular width in $\phi'$ and the frequency width of the radiation spectrum in the side spot are determined by the width of the function $sin^2(a\xi/2)/\xi^2$:

$$\Delta\varphi' \sim \frac{\gamma c}{a\omega_B \tan 2\vartheta_B}, \qquad \Delta\omega \sim \frac{c}{2a\sin^2\vartheta_B}. \qquad (13.20)$$

The total radiation intensity in the side spot is given by Eq. (13.18). From this equation it follows that, in the case of a thin crystal for which the present equation is applicable, the total intensity of the side spot is directly proportional to the thickness of the crystal. At the same time, the positions of the side spot and their intensities depend significantly on the structural parameters of the crystal.

The direct proportionality of the total intensity of the side spot to the crystal thickness indicates that the radiation in these directions is emitted over the entire length of the particle's path inside the crystal. This radiation closely resembles Vavilov–Cherenkov radiation (Sec. 1.6.A). In a homogeneous amorphous medium, Vavilov–Cherenkov radiation does not arise in the X-ray frequency region where $Re[\varepsilon(\omega)] < 1$; however, the periodic structure of a crystal leads to the appearance of X-ray radiation in certain Bragg directions, which is also called *"quasi-Cherenkov"* radiation (similar radiation in a periodic layered medium is possible in other frequency ranges as well, and it was originally called "parametric Cherenkov radiation" [**57**.1]).

Coherent X-ray radiation of a charged particle moving uniformly in a straight line in a crystal was first considered by Ter-Mikaelian [**61**.13, **69**.1] using perturbation theory; however, the criterion of its applicability (Eq. (13.10)) was not taken into account. In doing so, the author considered an infinite crystal, for which the first approximation of perturbation theory is insufficient. As a result, the spectral–angular distribution of the radiation intensity was not entirely correct: instead of (13.14), an expression containing δ-functions was obtained, i.e., the distribution had an infinitely narrow angular (or frequency) width. In reality, the correct expressions are given by Eqs. (13.14) and (13.12) for thin crystalline plates that satisfy condition in Eq. (13.10) (see also Ref. [**72**.29]). In the case of plates of arbitrary thickness, however, perturbation theory must be abandoned, as discussed in detail in the next section.

# § 14. CRYSTALLINE PLATE OF ARBITRARY THICKNESS. DYNAMICAL THEORY

For a sufficiently large plate thickness, when condition in Eq. (13.10) is not satisfied, the field of the scattered waves becomes comparable to the charge field or even larger. In this case, the problem of radiation generation cannot be solved using only the first approximation. For an amorphous plate, the problem can readily be solved, neglecting the diffuse background, by averaging Eq. (12.4) over the positions of the atoms in the material, as a result of which one obtains the usual X-ray transition radiation. For a crystalline plate, however, the problem is substantially more complicated due to the presence of side spots of the Laue pattern, caused by coherent Bragg scattering (see Sec. 13). In the vicinity of the Bragg frequencies, in addition to the primary transmitted wave, *strong waves* arise, related to the primary wave by the Bragg condition (or, as is often said, lying together with it near the same Ewald sphere). The simplest situation is when two waves are located near the Ewald sphere. Then, in solving the problem, these two waves are the dominant ones, while the others can be regarded as negligible. This corresponds to the so-called *two-wave approximation* in the dynamical theory of X-ray scattering in a crystal (see, e.g., Refs. [**67**.1, **72**.6, **72**.7]).

## 14.1. Charge Field in Two-Wave Approximation

Let us find the field of an ultrarelativistic charged particle moving inside a crystalline plate of arbitrary thickness *a*. For this purpose, we calculate the component of the charge field, additional to the vacuum field $E_{\mathrm{ch}}$, which is due to the interaction with the crystal (the "scattered" charge field $E_{\mathrm{scat}}$). Equation (12.4) for the Fourier components of the scattered field in the microscopic theory, after replacing the sums over the crystal lattice sites by the corresponding δ- functions, takes the following form:

$$\left(k^2 - \frac{\omega^2}{c^2} - \frac{\omega^2}{c^2}\chi_{00}\right)\mathbf{E}^{\mathrm{scat}}(\mathbf{k}) - \sum_{h\neq 0}\frac{\omega^2}{c^2}\gamma_{0h}\mathbf{E}^{\mathrm{scat}}(\mathbf{k}_h) +$$
$$+\mathbf{k}\left[\chi_{00}\,\mathbf{k}\cdot\mathbf{E}^{\mathrm{scat}}(\mathbf{k}) + \sum_{h\neq 0}\chi_{0h}\,\mathbf{k}\cdot\mathbf{E}^{\mathrm{scat}}(\mathbf{k}_h)\right] = \chi_{00}\left[\frac{\omega^2}{c^2}\mathbf{E}_{\mathrm{ch}} - \mathbf{k}(\mathbf{k}\cdot\mathbf{E}_{\mathrm{ch}})\right], \tag{14.1}$$

$$\mathbf{E}^{\mathrm{scat}}(\mathbf{k}) = \mathbf{E}_{\mathrm{scat}}(\mathbf{k},\omega),$$
$$\mathbf{E}_{\mathrm{ch}} = \mathbf{E}_{\mathrm{ch}}(\mathbf{k},\omega) = \frac{ei}{2\pi^2}\frac{\omega\mathbf{v} - \mathbf{k}c^2}{k^2c^2 - \omega^2}\delta(\omega - \mathbf{k}\cdot\mathbf{v}). \tag{14.2}$$

Here

$$\gamma^{ql}(\boldsymbol{k}_i,\ \boldsymbol{k}_j) = \chi_{ij}\delta^{ql} \quad (q,\, l = x,\ y,\ z),$$

indices $i, j = 0$ or $h$, respectively, for $k$ or $k_h$.

In Eq. (14.1) the unknowns are $E^{\mathrm{scat}}(\boldsymbol{k})$ and $E^{\mathrm{scat}}(\boldsymbol{k}_h)$, where $h$ runs over the values corresponding to the nodes of the reciprocal lattice, except for the point $\boldsymbol{k}_h = 0$. To determine these quantities, it is necessary to write an analogous equation for $E^{\mathrm{scat}}(\boldsymbol{k}_h)$ and to solve the resulting system of equations simultaneously.

Let there now be, in the vicinity of the Ewald sphere, only two waves with wave vectors $\boldsymbol{k}$ and $\boldsymbol{k}_h = \boldsymbol{k} + \boldsymbol{K}_h$, where $\boldsymbol{K}_h$ is a certain fixed reciprocal-lattice vector (multiplied by $2\pi$). Then, restricting ourselves to these two waves, we obtain a system of two equations whose solution has the following form:

$$\mathbf{E}^{\mathrm{scat}}(\mathbf{k}) = [\chi_{00}(\eta_h - \chi_{hh}) + \chi_{0h}\chi_{h0}]\mathbf{E}_{\mathrm{ch}} - (\eta_h - \chi_{hh})\chi_{00}\,\mathbf{E}_{\mathrm{ch}}\cdot\mathbf{k} - \chi_{0h}a_{0h}\frac{kc^2}{\omega^2} -$$
$$-\chi_{0h}\chi_{h0}(\mathbf{E}_{\mathrm{ch}}\cdot\mathbf{k}_h + a_{h0})\frac{k_hc^2}{\omega^2}/D_n, \tag{14.3}$$

$$\mathbf{E}^{\mathrm{scat}}(\mathbf{k}_h) = \chi_{h0}\left\{\eta_{00}\mathbf{E}_{\mathrm{ch}} - (\chi_{00}\mathbf{E}_{\mathrm{ch}}\cdot\mathbf{k} + \chi_{0h}a_{ch})\frac{kc^2}{\omega^2} -\right.$$
$$\left.-(\eta_{h0} - \chi_{00})(\mathbf{E}_{\mathrm{ch}}\cdot\mathbf{k}_h + a_{h0})\frac{k_hc^2}{\omega^2}\right\}/D_n,$$

where

$$
\begin{gathered}
a_{0h} = \left[(\zeta^2\chi_{00} + \eta_0 - \chi_{00})\mathbf{E}_{\rm ch}\cdot\mathbf{k} - \eta_0\mathbf{E}_{\rm ch}\cdot\mathbf{k}_h\right]\chi_{h0}/D_p, \\
a_{h0} = \left[-\zeta(\chi_{0h}\chi_{h0} + \chi_{00}(\eta_h - \chi_{hh}))\mathbf{E}_{\rm ch}\cdot\mathbf{k} + (\zeta^2\chi_{0h}\chi_{h0} + \chi_{00}(\eta_h - \chi_{hh}))\mathbf{E}_{\rm ch}\cdot\mathbf{k}_h\right]/D_p, \\
D_p = (\eta_0 - \chi_{00})(\eta_h - \chi_{hh}) - \zeta^2\chi_{0h}\chi_{h0}, \\
D_n = (\eta_0 - \chi_{00})(\eta_h - \chi_{hh}) - \chi_{0h}\chi_{h0}, \\
\chi = k_i^2c^2/\omega^2 - 1, \qquad \zeta = k\cdot k_hc^2/\omega^2.
\end{gathered} \tag{14.4}
$$

Expressions (14.3), together with the Fourier component of the charge field in vacuum (Eq. (14.2)), determine the Fourier components of the charge field inside the crystal. Far from the Bragg frequencies, when $|\gamma_h| \gg |\gamma_0|$ and $|\gamma_h| \gg |\chi_i|$, we have

$$
|E^{scat}(k_n)| \ll |E^{scat}(k)|,
$$

and

$$
E^{scat}(k) \approx \frac{\chi_{00}[E_{ch} - (E_{ch}\cdot k)kc^2/\omega^2]}{\gamma_0 - \chi_{00}}.
$$

After straightforward calculations we obtain the usual expression for the field of the charge in an amorphous medium (see Sec. 1.3).

## 14.2. Free Radiation Field

Let us now find the free radiation field inside the crystal. To this end, we proceed from the homogeneous equations obtained from Eq. (14.1) by setting the right-hand side equal to zero and replacing $E^{scat}(k)$ and $E_{scat}(k_h)$ by $E^{free}(k)$ and $E^{free}(k_h)$, respectively. In the two-wave approximation it suffices to retain two equations for the two strong waves lying near the Ewald sphere.

A nontrivial solution of the system of two homogeneous equations is obtained when the following conditions are satisfied:

1) the vectors $E^{free}(k)$ and $E^{free}(k_h)$ are collinear and both perpendicular to the vectors $k$ and $k_h$ (the case of “normal” polarization), and

$$
\begin{gathered}
D_p = (\eta_0 - \chi_{00})(\eta_h - \chi_{hh}) - \chi_{0h}\chi_{h0} = 0, \\
\mathbf{E}^{\rm free}(\mathbf{k}_h) = \frac{\eta_0 - \chi_{00}}{\chi_{0h}}\mathbf{E}^{\rm free}(\mathbf{k});
\end{gathered} \tag{14.5}
$$

2) the vectors $E^{free}(k)$, $E^{free}(k_h)$, $k$, and $k_h$ are coplanar (the case of “parallel” polarization), and

$$\begin{gathered} D_p = (\eta_0 - \chi_{00})(\eta_h - \chi_{hh}) - \zeta^2 \chi_{0h}\chi_{h0} = 0, \\ \mathbf{E}^{\text{free}}(\mathbf{k}) \parallel (\zeta\mathbf{k} - \mathbf{k}_h), \qquad \mathbf{E}^{\text{free}}(\mathbf{k}_h) \parallel (\mathbf{k} - \zeta\mathbf{k}_h), \\ |\mathbf{E}^{\text{free}}(\mathbf{k}_h)| = \left| \frac{\eta_0 - \chi_{00}}{\chi_{0h}\zeta} \mathbf{E}^{\text{free}}(\mathbf{k}) \right|. \end{gathered} \tag{14.6}$$

The dispersion equations (14.5) and (14.6) establish a relation between $k$ and ω. Each of these equations has two roots for $k_z$ in the vicinity of the Ewald sphere.

As a result, the four-dimensional Fourier components of the free radiation field inside the crystal can be written as follows:

$$\begin{aligned} \mathbf{E}^{\text{free}}(\mathbf{k}) &= \mathbf{e}_n\left[E_{n1}^{\text{free}}(k_z - \lambda_{n1}) + E_{n2}^{\text{free}}(k_z - \lambda_{n2})\right] - \mathbf{e}_p\left[E_{p1}^{\text{free}}(k_z - \lambda_{p1}) + E_{p2}^{\text{free}}(k_z - \lambda_{p2})\right], \\ \mathbf{E}^{\text{free}}(\mathbf{k}_h) &= \mathbf{e}_n\left[E_{n1}^{\text{free}}(k_{hz} - K_{hz} - \lambda_{n1}) + E_{n2}^{\text{free}}(k_{hz} - K_{hz} - \lambda_{n2})\right]\frac{\eta_0 - \chi_{00}}{\chi_{0h}} \\ &-\mathbf{e}_{hp}\left[E_{p1}^{\text{free}}(k_{hz} - K_{hz} - \lambda_{p1}) + E_{p2}^{\text{free}}(k_{hz} - K_{hz} - \lambda_{p2})\right]\frac{\eta_0 - \chi_{00}}{\zeta\chi_{0h}}. \end{aligned} \tag{14.7}$$

Here $e_n$ (parallel to the vector $k \times k_h$) is the unit vector of “normal” polarization; $e_p$ (parallel to the vector $\zeta k - k_h$) and $e_{hp}$ (parallel to the vector $k - \zeta k_h$) are unit vectors of “parallel” polarization; $E^{\text{CB}}_{\alpha i}(\alpha = n, p; i = 1, 2)$ are amplitudes to be determined from the boundary conditions. The quantities $\lambda_{\alpha i}$ are the roots of the dispersion equations (14.5) (for $\alpha = n$) or (14.6) (for $\alpha = p$) with respect to $k_z$, which we seek in the form

$$k_z = \lambda_{\alpha i} = \frac{\omega}{c}(1 - \Delta_{\alpha i})\cos\vartheta,$$

assuming that the new unknown roots $\Delta_{\alpha i}$ are small: $|\Delta_{\alpha i}| \ll 1$. Here $\vartheta = \varkappa c/|\omega|$ is the radiation angle in the central spot.

In the previous section we introduced the Bragg angle $\vartheta_B$ (Eq. (13.9)) and the Bragg frequency $\omega_B$ (Eq. (13.15)), which satisfy the kinematic conditions

$$K_{hz} = \frac{\omega_B}{c}(\cos 2\vartheta_B - 1), \qquad K_{h\perp} = \frac{\omega_B}{c}\sin 2\vartheta_B. \qquad (14.8)$$

From the first relation it follows that $K_{hz} \leq 0$.

Let

$$|\omega| = \omega_B(1 + \nu),$$

where $\nu$ is the small relative deviation of the frequency from the kinematic Bragg frequency $\omega_B$, $|\nu| \ll 1$. Solving Eq. (14.5), we obtain

$$\begin{gathered}\Delta_{ni} = \left\{\nu - \chi_{00}\cos 2\vartheta_B - \chi_{hh} \pm \left[(\chi_{00}\cos 2\vartheta_B - \chi_{hh} + \nu)^2 - 4\chi_{0h}\chi_{h0}\cos 2\vartheta_B\right]^{1/2}\right\}/(4\cos\vartheta_B), \\ \nu = -2(2\nu - \vartheta^2)\sin^2\vartheta_B + 2\vartheta(1-\nu)\sin 2\vartheta_B\cos\varphi.\end{gathered} \qquad (14.9)$$

The solutions $\Delta_{pi}$ of Eq. (14.6) are obtained from $\Delta_{ni}$ by replacing $\chi_{0h}\chi_{h0}$ with $\chi_{0h}\chi_{h0}cos^2 2\vartheta_B$. In the expression for $\nu$, the angle $\phi$ is the azimuthal angle between the vectors $\varkappa$ and $K_{h\perp}$ in the $(x, y)$ plane, so that

$$2\vartheta(1-\nu)\sin 2\vartheta_B\cos\varphi = 2\mathbf{x}\cdot\mathbf{K}_{h\perp}\frac{c^2}{\omega\omega_B}.$$

Equations (14.7) and (14.9) are valid for all values of the angle $\vartheta_B$, except for angles close to $\pi/4$.

### 14.3. Continuity Conditions at the Boundaries

In the X-ray frequency range under consideration, the continuity conditions at the boundaries can be approximately written as the continuity conditions for the electric field. For simplicity, we consider the case of normal incidence of a charged particle on the crystal surface. Then the continuity of the field must hold at $z = 0$ and $z = a$, and for arbitrary $t$ and ρ. This means that the continuity conditions must be satisfied for each of the waves corresponding to the vectors $k$ and $k_h$.

Depending on whether the quantity $k_{hz}$ is greater or less than zero, two cases must be distinguished: (1) when the diffracted wave corresponding to the wave vector $k_h$ leaves the crystal in the direction of motion of the charged particle, or, in other words, when $2\vartheta_B < \pi/2$ (Laue case), and (2) when the diffracted wave is directed in the opposite direction, i.e., $2\vartheta_B > \pi/2$ (Bragg case). Let us consider these cases separately.

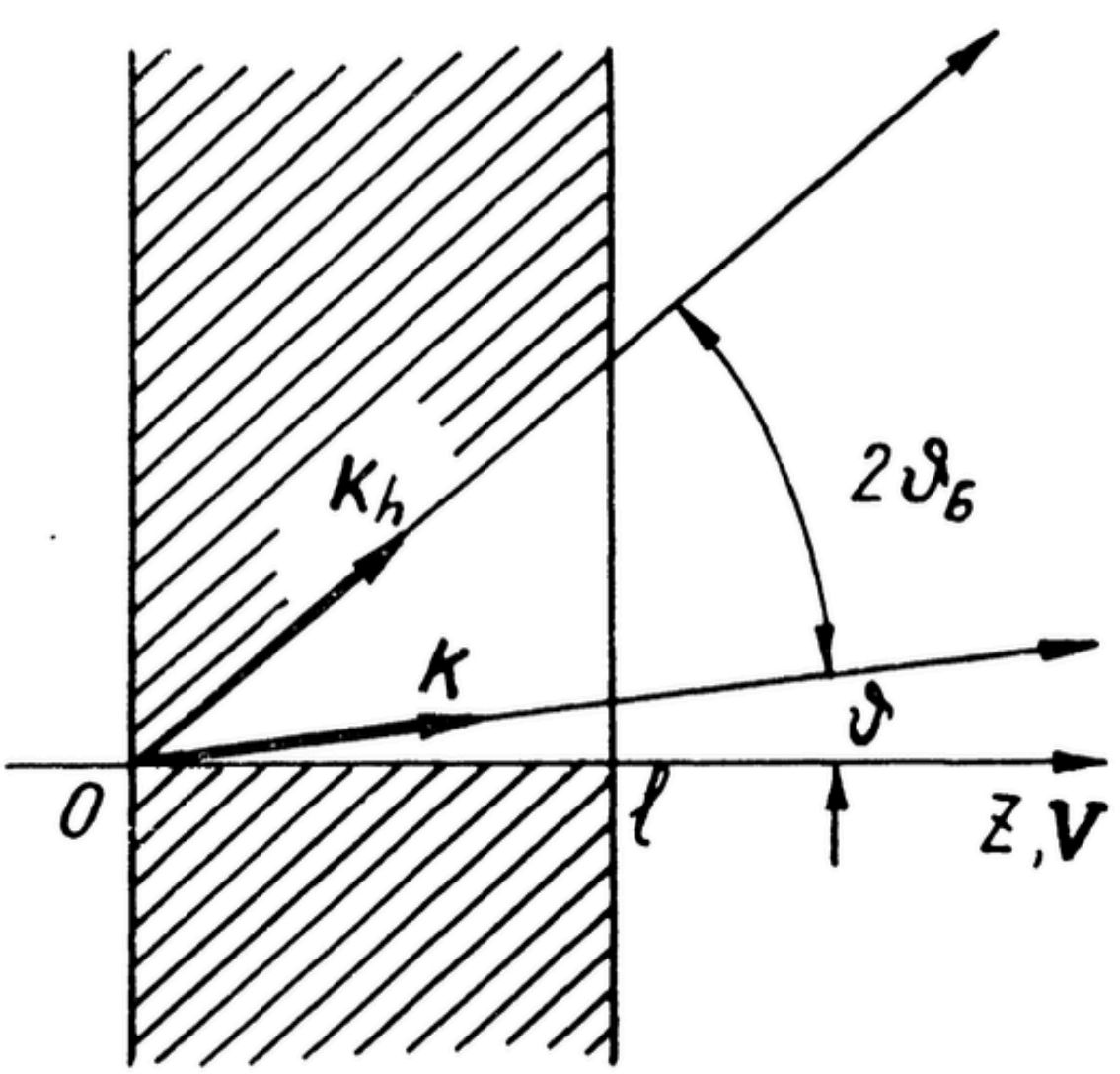

Fig. 14.1. Laue Case

(1) *Laue Case* (Fig. 14.1). For the Fourier components of the free field in vacuum in the region of space before the crystal ($z < 0$) we have

$$E^{vac}(k) = E^{vac}(k_h) = 0,$$

and in the region beyond the crystal ($z > a$) we have

$$\begin{aligned} \mathbf{E}^{\text{vac}}(\mathbf{k}) &= \left(E_n^{\text{vac}}\mathbf{e}_n + E_p^{\text{vac}}\mathbf{e}_p\right)\delta(k_z - \lambda_0), \\ \mathbf{E}^{\text{vac}}(\mathbf{k}_h) &= \left(E_{hn}^{\text{vac}}\mathbf{e}_n + E_{hp}^{\text{vac}}\mathbf{e}_{hp}\right)\delta(k_{hz} - \lambda_h), \end{aligned} \tag{14.10}$$

Here

$$\lambda_0 = \frac{\omega}{c}\left(1 - \frac{\chi^2 c^2}{\omega^2}\right)^{1/2}, \qquad \lambda_h = \frac{\omega}{c}\left(1 - \frac{\chi_h^2 c^2}{\omega^2}\right)^{1/2},$$

$E_a^{vac}$ and $E_{ha}^{vac}$ (α= n, p) are amplitudes to be determined.

By substituting expressions (14.3), (14.7), and (14.10) into the continuity conditions, one can determine the amplitudes $E_{ai}^{free}$, $E_a^{vac}$, *and* $E$. (a=1, 2) For normal polarization (a = n) we obtain

$$E_{n1}^{\mathrm{free}} = \frac{-E_n^{\mathrm{scat}}(2\Delta_{n2} + \gamma_{00}) - \gamma_{0h}E_{hn}^{\mathrm{scat}}}{2(\Delta_{n2} - \Delta_{n1})},$$
$$E_{n2}^{\mathrm{free}} = \frac{E_n^{\mathrm{scat}}(2\Delta_{n1} + \gamma_{00}) + \gamma_{0h}E_{hn}^{\mathrm{scat}}}{2(\Delta_{n2} - \Delta_{n1})}, \tag{14.11}$$
$$E_n^{\mathrm{vac}} = E_n^{\mathrm{scat}}\exp[i(\omega/v - \lambda_0)a] + E_{n1}^{\mathrm{free}}\exp[-i\omega a\Delta_{n1}/c] + E_{n2}^{\mathrm{free}}\exp[-i\omega a\Delta_{n2}/c],$$

$$E_{hn}^{\mathrm{vac}} = E_{hn}^{\mathrm{scat}}\exp[i(\omega/v - \lambda_0)a] - E_{n1}^{\mathrm{free}}\exp[-i\omega a\Delta_{n1}/c]\,(2\Delta_{n1} + \gamma_{00})/\gamma_{0h}$$
$$-E_{n2}^{\mathrm{free}}\exp[-i\omega a\Delta_{n2}/c]\,(2\Delta_{n2} + \gamma_{00})/\gamma_{0h}, \tag{14.12}$$

where $E_a^{scat}$ and $E_{ha}^{scat}$ are amplitudes of scattered charged field in direction $e_n$:

$$E_n^{\mathrm{scat}} = \frac{ei(\mathbf{v}\cdot[\mathbf{k}\times\mathbf{k}_h])}{2\pi^2 v\omega|\mathbf{k}\times\mathbf{k}_h|}\,\frac{\chi_{00}(\eta_h - \chi_{hh}) + \chi_{0h}\chi_{h0}}{\eta_0 D_n},$$
$$E_{hn}^{\mathrm{scat}} = \frac{ei(\mathbf{v}\cdot[\mathbf{k}\times\mathbf{k}_h])}{2\pi^2 v\omega|\mathbf{k}\times\mathbf{k}_h|}\,\frac{\chi_{h0}}{D_n}, \tag{14.13}$$

The quantities $\eta_0$, $\eta_h$ *and* $D_n$ arise from $\eta_0$, $\eta_h$ *and* $D_n$ by substituting $k_z$ to ω/υ, so that

$$\tilde{\eta}_0 = \vartheta^2 + \gamma^{-2}, \quad \tilde{\eta}_h = \vartheta^2 + \gamma^{-2}\cos 2\vartheta_Б - 4\vartheta\sin^2\vartheta_Б + 2\vartheta\sin 2\vartheta_Б\cos\phi.$$

For parallel polarization, one can obtain amplitudes of free radiation fields inside and outside the crystal ($E_{pi}^{free}$, $E_{p}^{free}$, $E_{hp}^{vac}$) from Eqs. (14.11) and (14.12) by substituting index n in all quantities with index p, as well as substituting quantities $Z_{0h}$ with $Z_{0h}\cos 2\vartheta_B$.

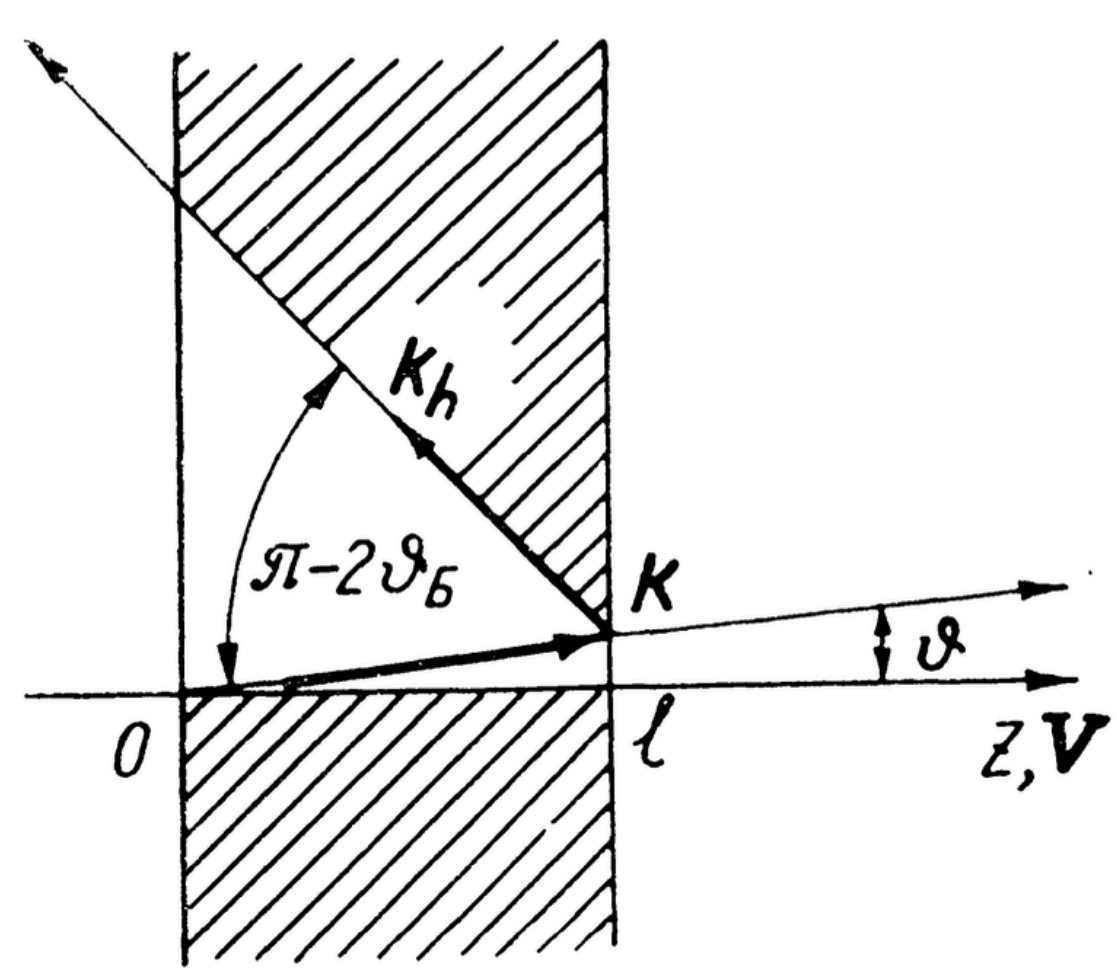

*Fig. 14.2. Bragg Case*

(2) *Bragg Case* (Fig 14.2). In this case, the Fourier components of the radiation field in vacuum in the region before the crystal have the following form:

$$\mathbf{E}^{\mathrm{vac}}(\mathbf{k}) = 0, \qquad \mathbf{E}^{\mathrm{vac}}(\mathbf{k}_h) = \left(E_{hn}^{\mathrm{vac}}\mathbf{e}_n + E_{hp}^{\mathrm{vac}}\mathbf{e}_{hp}\right)\delta(k_{hz} - \lambda_h),$$

and in the region beyond the crystal:

$$\mathbf{E}^{\mathrm{vac}}(\mathbf{k}) = \left(E_{n}^{\mathrm{vac}}\mathbf{e}_n + E_{p}^{\mathrm{vac}}\mathbf{e}_{p}\right)\delta(k_{z} - \lambda_0), \qquad \mathbf{E}^{\mathrm{vac}}(\mathbf{k}_h) = 0,$$

From the boundary conditions of field continuity at the interfaces one can obtain the field amplitudes. For normal polarization we have

$$\begin{aligned} E_{n1}^{\mathrm{free}} = &\left\{-E_n^{\mathrm{scat}}(2\Delta_{n2} + \chi_{00})\exp[-i\omega a\Delta_{n2}/c] - E_{hn}^{\mathrm{scat}}\chi 0h\exp[i(\omega/v - \lambda_0)a]\right\} \\ &\times\left\{(2\Delta_{n2} + \chi_{00})\exp[-i\omega a\Delta_{n2}/c] - (2\Delta_{n1} + \chi_{00})\exp[-i\omega a\Delta_{n1}/c]\right\}^{-1}, \end{aligned} \tag{14.14}$$

$$E_{n2}^{\rm free} = \left\{E_n^{\rm scat}(2\Delta_{n1} + \chi_{00})\exp[-i\omega a\Delta_{n1}/c] + E_{hn}^{\rm scat}\chi_{0h}\exp[i(\omega/v - \lambda_0)a]\right\}$$
$$\times\left\{(2\Delta_{n2} + \chi_{00})\exp[-i\omega a\Delta_{n2}/c] - (2\Delta_{n1} + \chi_{00})\exp[-i\omega a\Delta_{n1}/c]\right\}^{-1}, \quad (14.15)$$
$$E_n^{\rm vac} = E_n^{\rm scat}\exp[i(\omega/v - \lambda_0)a] + E_{n1}^{\rm free}\exp[-i\omega a\Delta_{n1}/c] + E_{n2}^{\rm free}\exp[-i\omega a\Delta_{n2}/c],$$

where $E_n^{scat}$ and $E_{hn}^{scat}$ are defined by Eqs. (14.13).

The amplitudes $E_{pl}^{free}$, $E_p^{vac}$, and $E_{hp}^{vac}$ for parallel polarization, as in the Laue case, are obtained from Eqs. (14.14) and (14.15) by replacing the index n with p and $\chi_{0h}$ with $\chi_{0h}\cos 2\vartheta_B$.

## 14.4. Dynamical Maxima

As was already noted at the end of Sec. 14.1, far from the Bragg frequencies the quantity $|E^{scat}(\mathbf{k}_h)|$ is much smaller than $|E^{scat}(\mathbf{k})|$, and the field of a charge in the crystal reduces to the field of a charge in a medium determined by the typical equation of macroscopic electrodynamics. This means, in particular, that far from the Bragg frequencies the side radiation spot $E^{vac}(\mathbf{k}_h)$ will be weak, while the central spot $E^{vac}(\mathbf{k})$ will be determined by the usual equation for transition radiation that follows from the macroscopic theory, which does not take into account the crystalline structure of the medium. The above also follows from Eqs. (14.12) and (14.15). Indeed, far from the Bragg frequencies, when $|\nu| \gg |\chi_{ij}|$, from Eq. (14.9) we obtain $|\Delta_{\alpha2}| \approx -\chi_{00}/2$, $\Delta_{\alpha1} \approx \nu/2 \cos 2\vartheta_B$, i.e., $|\Delta_{\alpha1}| \gg |\Delta_{\alpha2}|$. Then from Eqs. (14.11) and (14.14) it follows that, both in the Laue case and in the Bragg case, $|E_{\alpha1}^{free}| \ll |E_{\alpha2}^{free}|$ and $E_{\alpha2}^{scat} = -E_\alpha^{scat}$, and according to Eqs. (14.12) and (14.15) – $|E_{h\alpha}^{vac}| \ll |E_\alpha^{vac}|$ and

$$E_a^{vac} = E_a^{scat}\{exp[i(\omega/v\ -\ \lambda_0)a]\ -\ exp[i\omega a\chi_{00}/2c]\}.$$

Summing both polarizations, we obtain an expression for $E^{vac}(\mathbf{k})$ which, as one can verify, exactly coincides with the equation for transition radiation that follows from the macroscopic theory.

Thus, deviations from the macroscopic equation can occur only at frequencies ω close to the Bragg frequency.

Taking various reciprocal lattice vectors $\mathbf{K}_h$, we obtain a set of Bragg frequencies corresponding to different Bragg angles:

$$\vartheta_B = arctg\frac{|K_{hz}|}{K_{h\perp}}.$$

Depending on whether $2\vartheta_B < \pi/2$ or $2\vartheta_B > \pi/2$, the radiation will be determined, respectively, by the equations for the Laue case or the Bragg case.

The total energy flux passing through the plane z = const, parallel to the crystal boundaries and located sufficiently far from it, for the central and side spots is determined, respectively, by the equations:

$$W = c(2\pi)^2 \int \left\{|E_n^{\mathrm{vac}}|^2 + |E_p^{\mathrm{vac}}|^2\right\} d^2x\, d\omega, \qquad (14.16)$$

$$W_h = c(2\pi)^2 \int \left\{|E_{hn}^{\mathrm{vac}}|^2 + |E_{hp}^{\mathrm{vac}}|^2\right\} |\cos 2\vartheta_B|\, d^2x\, d\omega. \qquad (14.17)$$

Here and below, in the equations for W and $W_h$, integration is performed over frequencies $\omega > 0$.

Let us first consider the limiting cases.

I. Let the crystal be sufficiently thin, i.e.,

$$\left|\frac{\omega\, a\, \Delta_{ai}}{c}\right| \ll 1 \qquad (14.18)$$

for all values of $\varkappa$ (or $\varkappa_h$) and ν for the given spot, which means (see Eq. (14.9)), generally speaking, the simultaneous fulfillment of the conditions

$$\left|\frac{\omega\, a\, \chi_{ij}}{c}\right| \ll 1, \quad \left|\frac{\omega\, a\, \nu}{c}\right| \ll 1$$

and

$$\left|\frac{\omega a}{c\gamma}\right| \ll 1 \; for \; K_{h\perp} \neq 0,$$
$$\left|\frac{\omega a}{c\gamma^2}\right| \ll 1 \; for \; K_{h\perp} = 0.$$
(14.20)

Then from Eqs. (14.12) and (14.15), both in the Laue case and in the Bragg case, we obtain

$$E_{\alpha}^{vac} = \frac{e\, a\, \chi_{00}\, k_{\alpha}}{4\pi^2 \omega \tilde{\gamma}_0} \quad (14.21)$$

and

$$E_{hn}^{vac} = \frac{e\, a\, \chi_{h0}\, k_{hx}}{4\pi^2 |cos2\vartheta_B|\, \omega \tilde{\gamma}_0},$$
$$E_{hp}^{vac} = \frac{e\, a\, \chi_{h0}\, (k_{hy} + \varkappa_h) cos2\vartheta_B}{4\pi^2 |cos2\vartheta_B|\, \omega \tilde{\gamma}_0}. \quad (14.22)$$

In Eq. (14.21) $k_{\alpha} = k^{x}$ for normal polarization ($\alpha = n$), and $k_{\alpha} = k_{y}$ for parallel polarization ($\alpha = p$).

Substituting Eq. (14.21) into Eq. (14.16), we find that, in the thin-crystal case under consideration, the equation for the energy flux in the central spot coincides with Eq. (13.12) obtained using perturbation theory.

For the side spot, from Eqs. (14.22) and (14.17) we have

$$W_h = \frac{e^2 a^2}{4\pi^2 c^3} \int \omega^2 |\gamma_{h0}|^2 d\omega \int \frac{(sin^2\phi + cos^2\phi\, cos^2 2\vartheta_B)}{(\vartheta^2 + \gamma^{-2})^2} \vartheta^3 d\vartheta\, d\phi,$$
(14.23)

where $\phi$ is the azimuthal angle between the vectors $\varkappa_h - K_{h\perp}$ and $K_{h\perp}$, and $\vartheta$ is the emission angle in the side spot, equal to the angle formed by the vector $K_h$ and the “kinematic” Bragg direction $\{0, -\, sin2\vartheta_B, cos2\vartheta_B\}$.

It is easy to see that Eq. (14.23) coincides with Eq. (13.14) if one takes into account conditions in Eqs. (14.19) and (14.20).

II. Let us now consider the case of a thick crystal, when

$$\left|\frac{\omega\, a\, \mathrm{Im}\, \Delta_{\alpha i}}{c}\right| \gg 1. \qquad (14.24)$$

As in the previous section, let us consider the Laue and Bragg cases separately.

(1) *Laue Case.* Since $cos2\vartheta_B > 0$ and the quantities $\gamma_i$ have small positive imaginary parts, it follows from Eq. (14.9) that

$$\mathrm{Im}\, \Delta_{\alpha 1} < \mathrm{Im}\, \Delta_{\alpha 2} < 0. \qquad (14.25)$$

Therefore, under the conditions in Eq. (14.24), from Eqs. (14.12) we obtain

$$\begin{aligned} E_\alpha^{\mathrm{vac}} &= E_\alpha^{\mathrm{scat}} \exp\{i(\omega/v - \lambda_0)a\}, \\ E_{h\alpha}^{\mathrm{vac}} &= E_{h\alpha}^{\mathrm{scat}} \exp\{i(\omega/v - \lambda_0)a\}, \qquad (\alpha = n, p). \end{aligned} \qquad (14.26)$$

Substituting these expressions into Eqs. (14.16) and (14.17), we find

$$\begin{aligned} W &= c(2\pi)^2 \int \{|E_n^{\mathrm{scat}}|^2 + |E_p^{\mathrm{scat}}|^2\}\, d^2x\, d\omega, \\ W_h &= c(2\pi)^2 \int \{|E_{hn}^{\mathrm{scat}}|^2 + |E_{hp}^{\mathrm{scat}}|^2\} |\cos 2\vartheta_B|\, d^2x\, d\omega. \end{aligned} \qquad (14.27)$$

The quantities $E_\alpha^{scat}$ and $E_{h\alpha}^{scat}$ contain characteristic denominators $\widetilde{D}_\alpha$, which become small at certain frequencies $\omega_a$ and the corresponding angles. The frequencies $\omega_a$ obtained in the dynamic theory under consideration differ somewhat from the kinematic Bragg frequencies $\omega_B$ and may be called “dynamic” Bragg frequencies. At such frequencies and angles the quantities $|E_\alpha^{scat}|^2$ and $|E_{h\alpha}^{scat}|^2$, and consequently $W$ and $W_h$, reach maxima which in Ref. [72.5] were called “dynamic maxima.”

Substituting the expressions for $E_\alpha^{scat}$ and $E_{h\alpha}^{scat}$ into (14.27) and bearing in mind the case $K_{h\perp} c/\omega \gg \gamma^{-1}$, we obtain

$$W = \frac{e^2}{\pi^2 c} \int \left\{ \left| \frac{\chi_{00}(\widetilde{\eta}_h - \chi_{hh}) + \chi_{h0}\chi_{h0}}{\widetilde{D}_n} \right|^2 \sin^2\phi + \right.$$

$$+\left|\frac{\chi_{00}\left(\tilde{\eta}_h-\chi_{hh}\right)+\chi_{0h}\chi_{h0}\cos^2 2\vartheta_B}{\tilde{D}_p}\right|^2 \cos^2\phi\Big\}\frac{\vartheta^3\, d\vartheta\, d\phi\, d\omega}{(\vartheta^2+\gamma^{-2})^2}, \quad (14.28)$$

$$W_h=\frac{e^2}{\pi^2 c}\int|\chi_{0h}|^2\left\{\frac{\sin^2\phi}{|\tilde{D}_h|^2}+\frac{\cos^2 2\vartheta_B\cos^2\phi}{|\tilde{D}_p|^2}\right\}\cos 2\vartheta_B\,\vartheta^3\, d\vartheta\, d\phi\, d\omega. \quad (14.29)$$

As can be seen from the above equations, the radiation intensities $W$ and $W_h$ at given values of $\vartheta$ and $\phi$ generally contain two dynamic maxima corresponding to the minima of the denominators $|\tilde{D}_n|^2$ and $|\tilde{D}_p|^2$. The positions, widths, and heights of these maxima depend on $\vartheta$ and $\phi$. Upon integration over $\vartheta$ and $\phi$, the above maxima add together, which leads to a broadening of the maximum in the frequency dependence of the resulting radiation intensity.

(2) *Bragg Case.* From Eq. (14.9) we obtain that

$$Im\,\Delta_{a1}+Im\,\Delta_{a2}=-\left(\chi''_{00}\cos 2\vartheta_B+\chi''_{hh}\right)/4\cos 2\vartheta_B\geq 0$$

(since $\cos 2\vartheta_B<0$ and $\chi''_{00}\sim\chi''_{hh}>0$) and

$$Im\,\Delta_{a2}<0<Im\,\Delta_{a1}.$$

Then, from Eq. (14.15), taking into account Eq. (14.14), we obtain

$$E_n^{\text{vac}}=\left[E_n^{\text{scat}}+\frac{E_{hn}^{\text{scat}}\chi_{0h}}{2\Delta_{n1}-\chi_{00}}\right]\exp\left[i\left(\frac{\omega}{v}-\lambda_0\right)a\right],$$
$$E_{hn}^{\text{vac}}=E_{hn}^{\text{scat}}+\frac{(2\Delta_{n2}+\chi_{00})E_n^{\text{scat}}}{\chi_{0h}}. \quad (14.30)$$

Analogous expressions for $E_p^{vac}$ and $E_{hp}^{vac}$ can be obtained from Eq. (14.30) by replacing $n$ with $p$ and $\chi_{0h}$ with $\chi_{0h}\cos 2\vartheta_B$.

III. Finally, let us consider a crystal of intermediate thickness, when for $E_n^{vac}$ and $E_{h\alpha}^{vac}$ the general equations (14.12) or (14.15) must be used. In this case, as a rule, complex interference patterns arise that depend on the specific characteristics $\chi_{ij}$ of the crystal. Without carrying out a detailed analysis of the spectral–angular and spectral distributions of the radiation in this case, we make the following remarks.

(1) *Laue Case.* As already noted, in the Laue case the inequalities (14.25) hold. At the same time, the quantity $|Im\,\Delta_{\alpha 2}|$, as follows from Eq. (14.9), may be very small for $|\gamma| \approx |\chi_{ti}|$ if the value $|\chi_{0h}| \approx |\chi_{h0}|$ is close to $|\chi_{00}| \approx |\chi_{hh}|$. This means that the term containing $exp[-\,i\omega a\Delta_{\alpha 2}/c]$ in the amplitude of the radiation field (Eq. (14.12)) will attenuate only weakly as $a$ increases.

The existence of a weakly attenuating electromagnetic wave in the crystal near the Bragg frequencies, predicted by the dynamical theory, is known as *anomalous transmission* or the Borrmann effect [**67**.1, **72**.6, **72**.7]. When a charged particle passes through a sufficiently thick crystal near the Bragg frequencies, dynamical effects arise, and the appearance of an “anomalously transmitted component” in the resulting radiation is natural.

From Eqs. (14.12) it follows that the radiation both in the central spot and in the side spots is the result of the superposition of three wave fields: the scattered field of the charge and two fields of free radiation emerging from the crystal. Each of these fields has its own phase factor that depends oscillatory on the crystal thickness $a$. For such thicknesses $a$, when inequality (14.18) is no longer satisfied but $|\omega a\,Im\,\Delta_{\alpha i}/c|$ is still much smaller than unity, i.e., the absorption of radiation in the crystal is negligible, the radiation intensity is the result of interference of all three fields. For larger thicknesses such that $|\omega a\,Im\,\Delta_{\alpha 2}/c| \ll 1 \leq |\omega a\,Im\,\Delta_{\alpha 1}/c|$, i.e., the “normally transmitted part” of the radiation is completely absorbed in the crystal while the “anomalously transmitted part” is still absorbed only slightly, the radiation intensity is determined by the interference of only two fields: the scattered field of the charge and the “anomalously transmitted” free field. Finally, for still larger thicknesses $a$ (inequality (14.24)), when both parts of the free field are absorbed, the radiation intensity is given by equations (14.26), analyzed above.

The nature of the above-mentioned oscillatory dependences of the radiation intensities on the crystal thickness $a$ is the same as in the well-known Pendellösung effect (see, for example, Refs. [**72**.6, **72**.7]). However, in the case under consideration this phenomenon is complicated by the presence of the scattered charge field.

(2) *Bragg Case.* The remarks made above for the Laue case are in principle also valid for the Bragg case. However, now the anomalously transmitted wave in the side spot is absent. Near the boundaries of the crystal the radiation field also interferes with the charge's own field $E_{ch}$. Owing to the difference in the phase velocities of the two fields, the region in which these fields interfere is limited by the formation length. In vacuum the formation length for the central spot is of order $c/[\omega(\gamma^{-2} + \chi_{00})]$ and is a macroscopic quantity, whereas for the side spots it is $c/[\omega(1 - \beta|cos2\vartheta_B|)] \sim c/\omega$, i.e., much smaller (for $2\vartheta_B \neq \pi$).

The theory presented can also be applied in the case when scattering occurs on nucleons. For this it is sufficient to use the corresponding expressions for the quantities $\chi_{ij}$ [**70**.15, **71**.14, **72**.28, **75**.25, **76**.14].

### 14.5. Calculation of the Number of Quanta

Let us now examine the spectral distribution of the numbers of radiation quanta in the central and side spots, and also compute the total numbers of quanta [**75**.8, **75**.33].
From Eqs. (14.16) or (14.17) we obtain that the spectral distribution of the numbers of quanta is given by the following equations:

$$\frac{d\Lambda}{d\omega} = \frac{(2\pi)^2\omega}{\hbar c} \int \left\{|E_n^{\text{vac}}|^2 + |E_p^{\text{vac}}|^2\right\} d\vartheta d\varphi,$$
$$\frac{dN_h}{d\omega} = \frac{(2\pi)^2\omega}{\hbar c} \int \left\{|E_{hn}^{\text{vac}}|^2 + |E_{hp}^{\text{vac}}|^2\right\} |\cos 2\vartheta_B| \, d\vartheta d\varphi. \quad (14.31)$$

The general equations (14.12) (Laue case) or (14.15) (Bragg case) for $E_\alpha^{vac}$ and $E_{h\alpha}^{vac}$ $(\alpha = n, p)$ are rather cumbersome. We shall restrict ourselves to the limiting case of a thick crystal, when conditions in Eq. (14.24) are satisfied and one can use Eqs. (14.26) or (14.30).

Let us consider the quantities (14.31) in the Laue case for large and small $|\nu| = |\omega - \omega_B|/\omega_B$.

Let

$$|\nu| \gg \max\{\gamma^{-2},\ |\chi'|,\ |\chi_{0h}|,\ |\chi_{h0}|\}. \qquad (14.32)$$

Then it is easy to see that dynamical maxima exist only at angles

$$\vartheta \gg |\nu \,\mathrm{tg}\,\vartheta_{Б}|. \qquad (14.33)$$

After straightforward integration over φ and θ in this case (i.e., when Eq. (14.32) holds) we obtain

$$\frac{dN}{d\nu} = N(\nu) = N_{\mathrm{TR}}(\nu) + N'(\nu),$$
$$N'(\nu) = \frac{|\chi_{0h}|^4(1 + 5\cos^4 2\vartheta_B)}{32 \cdot 137\,\chi'' \sin 2\vartheta_B\, |\tan \vartheta_B|^5}. \qquad (14.34)$$

$$\frac{dN_h}{d\nu} = N^h(\nu) = \frac{|\chi_{0h}|^2 \cot 2\vartheta_B (1 + \cos^2 2\vartheta_B)}{4 \cdot 137\,\chi''\, |\tan \vartheta_B|}. \qquad (14.35)$$

where $N_{TR}(\nu)$ is the spectral distribution of the number of XTR quanta, which is easily obtained from the right-hand side of Eq. (1.59) by dividing by $\hbar$.

Expressions (14.34) and (14.35) have different dependences on $|\nu|$. In the central spot this dependence is much sharper. However, it should be borne in mind that the indicated expressions are valid only for $|\nu| \ll 1$.

Let now

$$|\nu| \ll \gamma^{-2} + |\chi'|.$$

Then after integrating over φ and θ we obtain

$$\frac{dN'(\nu)}{d\nu} \approx \frac{|\chi_{0h}|^2}{4 \cdot 137\,\chi'' \sin 2\vartheta_B} \frac{(A - A_h)^2}{A\, A_h\, (A + A_h)}. \qquad (14.36)$$

$$\frac{dN_h(\nu)}{d\nu} \approx \frac{|\chi_{0h}|^2 \cot 2\vartheta_B}{4 \cdot 137\,\chi''\, A}, \qquad (14.37)$$
$$A = (\gamma^{-2} + |\chi_{00}|)^{1/2}, \qquad A_h = (\gamma^{-2} + |\chi_{00} - \chi_{0h}|)^{1/2}.$$

From Eq. (14.34)–(14.37) it follows that the “edges” of the spectral distributions (Eqs. (14.34) and (14.35)) do not depend on the Lorentz factor γ of the particle, while the central parts (14.36) and (14.37) depend on γ through the quantities A and $A_h$; that is, initially they grow linearly with γ (for $\gamma^{-2} \gg |\square_{00}|$), and then gradually approach saturation.

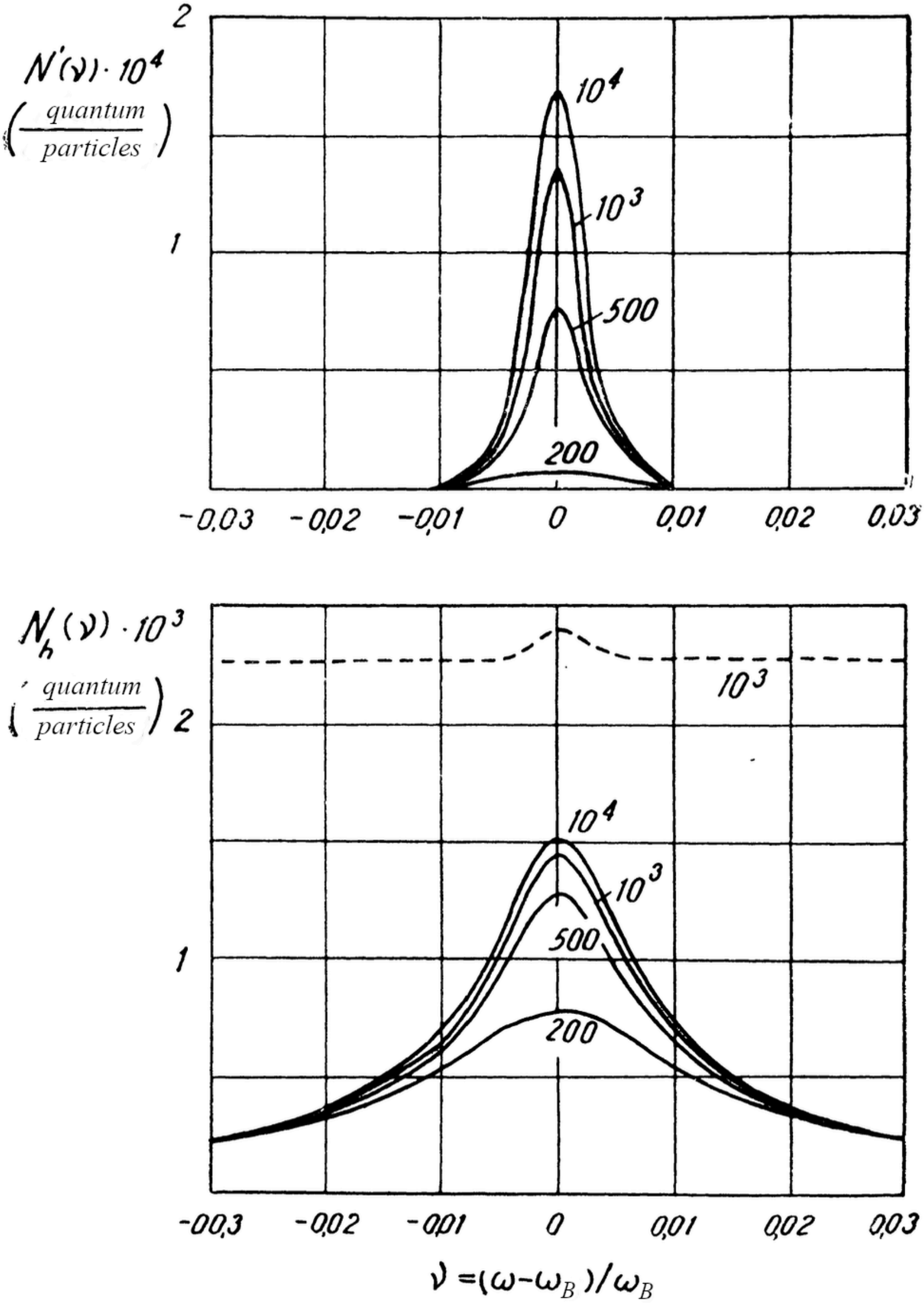

*Fig. 14.3. Spectral distributions $N'(\nu)$ and $N_h(\nu)$ of the numbers of quasi-Cherenkov radiation quanta in the Laue case. The numbers on the curves indicate the values of the γ-factor. The dashed curve in the lower figure corresponds to $N(\nu)$ for $\gamma$=1000; $\hbar\omega_B$=5.268 keV*

The properties of the frequency distributions of the number of quanta discussed above are illustrated in Fig. 14.3, which shows curves obtained by numerical integration of formulas (14.31).

It follows from the figure that in the central region the quantity $N_h(\nu)$ is approximately an order of magnitude larger than $N'(\nu)$ and amounts to about 62% of the quantity $N_{TR}(\nu)$.

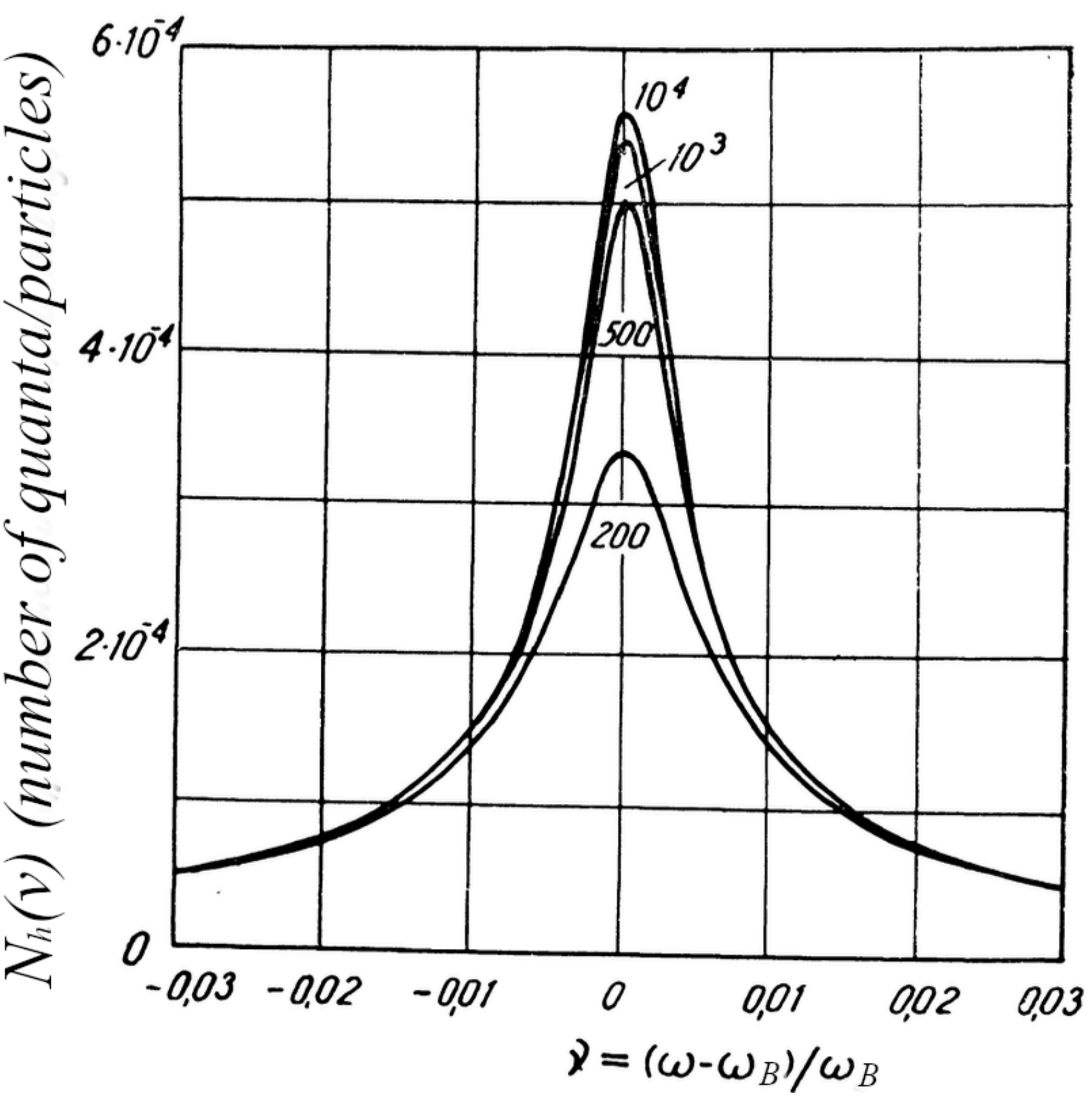

*Fig. 14.4. The same as in Fig. 14.3, for the Bragg case*

In the Bragg case, there are no dynamical maxima in the central spot. The spectral distributions of the numbers of quanta for the side spot in this case (Fig. 14.4) are quite similar to the corresponding curves in the Laue case (Fig. 14.3).

The total numbers of quanta in the central and side spots can be obtained by numerical integration of the corresponding spectral distributions (Fig. 14.5). It follows from the figure that for $\gamma^{-2} \ll |\chi_{00}|$ the total numbers of quanta in the dynamical maxima practically do not depend on $\gamma$. In the central spot (transition radiation) the total number of quanta in the dynamical maximum constitutes only a small fraction of the number of transition-radiation quanta in the same frequency range. As for the total number of quanta in the side spot, it can be of the order of $10^{-4}$ quanta per charged particle.

## § 15. STACK OF CRYSTALLINE PLATES

Let us now consider X-ray radiation (transition and quasi-Cherenkov) produced when a relativistic charged particle passes through a stack of regularly spaced $N$ identical crystalline plates.

We shall assume that the number $N$ is such that the following inequality holds:

$$|Nr_1|^2 \ll 1, \tag{15.1}$$

where $r_1 \sim \omega_0^2/(4\omega^2)$ is the reflection coefficient of X-ray radiation at a plate boundary.

Because of the smallness of $r_1$, the above inequality is satisfied up to $N$ of order $10^4$. When this inequality is satisfied, the radiation not Bragg-reflected at the plate boundaries is weak, and we will neglect it.

The problem of the formation of X-ray radiation (transition and quasi-Cherenkov) in a stack of crystalline plates (taking into account Eq. (15.1)) can be solved in general form in the two-wave approximation. This yields quite cumbersome equations, and we don't show them here (see Refs. [**75**.5, **81**.3]). Let us only report the physical findings following from the analysis of general equations.

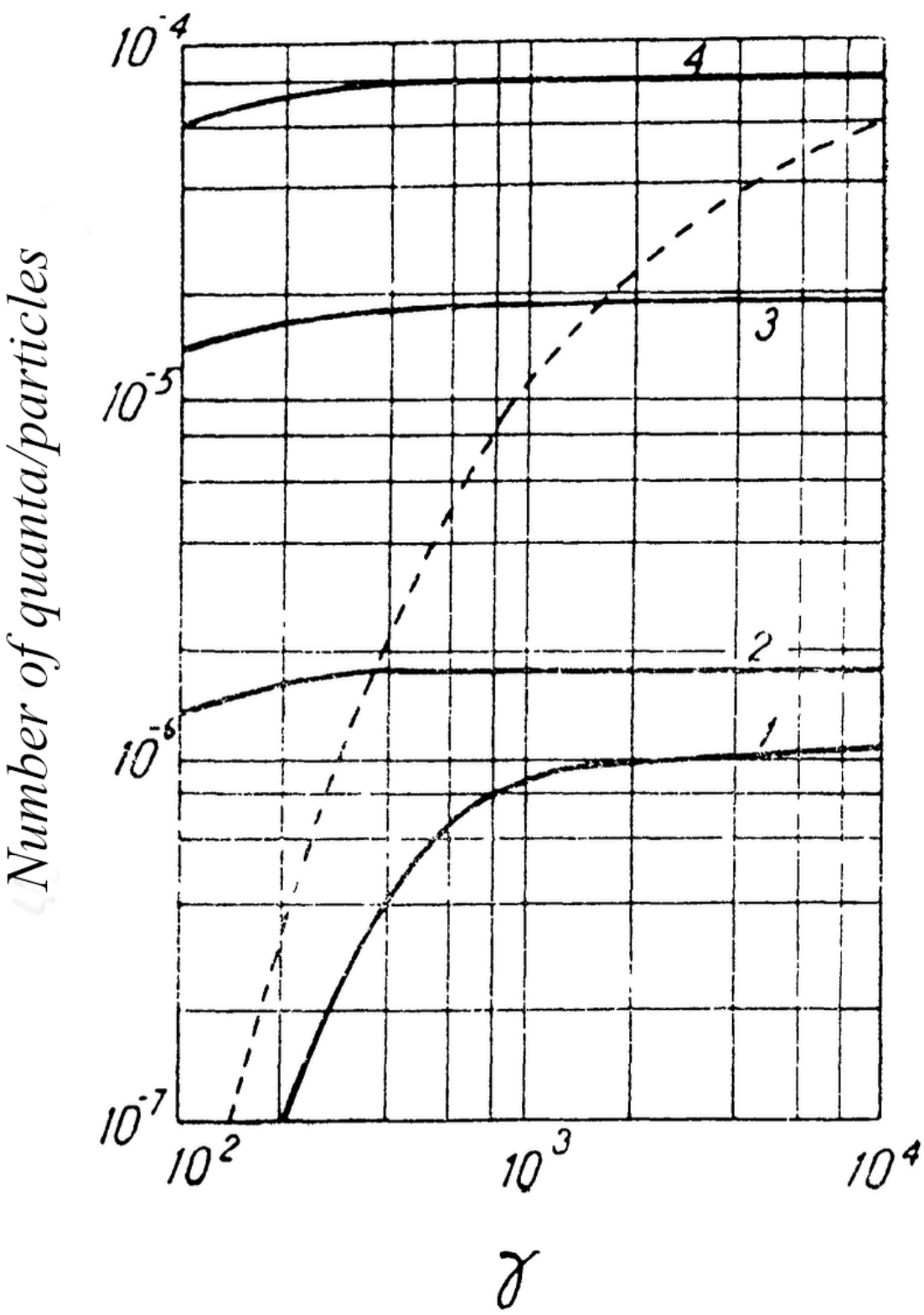

Fig. 14.5. Dependence of the total number of quasi-Cherenkov radiation quanta in the central and side spots on the value of the particle's gamma factor. Curve 1 corresponds to $N'$ (Laue case, $\hbar\omega_B = 5.268$ keV), 2 — $N_h$ (Bragg case, $\hbar\omega_D = 3.226$ keV), 3 — $N_h$ (Bragg case, $\hbar\omega_D = 5.268$ keV), 4 — $N_h$ (Laue case, $\hbar\omega_B = 5.268$ keV). The dashed curve corresponds to the number of transition-radiation quanta in the range $\Delta\nu = 5 \cdot 10^{-3}$ at $\hbar\omega = = 5.268$ keV (i.e., $\hbar\Delta\omega = 26.3$ eV in absolute units)

It is natural that if the thickness of each crystalline plate exceeds the absorption lengths of the normally and anomalously transmitted waves, then the pattern in the stack should not differ in any way from the analogous pattern in a single such thick plate considered above (see Sec. 14).

When the plates of the stack are sufficiently “transparent,” the radiation (both transition and quasi-Cherenkov) generated in different plates will interfere with each other. As a result, the Fourier components of the radiation fields, both in the Laue case and in the Bragg case, contain in the denominator the quantity

$$X_{cr} = \cos\varphi - \cos x,$$

$$\varphi = \left(\frac{\omega}{v} - \lambda_0\right)(a+b) + \frac{(\Delta_{\alpha 1} - \Delta_{\alpha 2})\lambda_0 a}{2},$$

$$x = \frac{(\Delta_{\alpha 2} - \Delta_{\alpha 1})\lambda_0 a}{2} \quad (\alpha = n,\ p),$$

$$\lambda_0 = \frac{\omega}{c}\left(1 - \frac{\varkappa^2 c^2}{\omega^2}\right)^{1/2}, \qquad (15.2)$$

where $a$ and $b$ are the thickness of the plate and the distance between the plates, and $\Delta_{\alpha i}$ is determined by Eq. (14.9).

When $|X_{st}|^2 \ll 1$, a maximum appears in the spectral-angular distribution of the radiation intensity. By varying $a$ and $b$, one can arrange for this condition to be satisfied at the dynamical maximum, i.e., so that the dynamical maxima of the radiation arising in different crystalline plates would interfere constructively. In this case, analysis of the equations shows that if the entire stack is sufficiently “transparent,” the intensity of the maximum will be proportional to the square of the number of plates in the stack, $N^2$, and the width of the maximum will be smaller by approximately a factor of $N$. As a result, the total intensity (or the number of quanta), integrated over the entire maximum, will be greater by approximately a factor of $N$. A similar result is obtained also when the stack is not sufficiently transparent. Then the total intensity (or the total number of quanta) of the maximum is greater by the corresponding factor $N_{eff}$.

# CHAPTER IV

## MULTIPLE SCATTERING OF A PARTICLE: BREMSSTRAHLUNG AND THE EDGE EFFECT

In the preceding chapters we assumed that a charged particle moves uniformly and rectilinearly in a medium. However, in reality the particle always undergoes *scattering* by the atoms of the medium. This scattering primarily leads to the emission of bremsstrahlung. In addition, as was first pointed out by Garibian and Pomeranchuk [59.6] and by Pafomov [60.9], if in the vicinity of the boundaries of a medium, over the formation length of transition radiation, the particle undergoes sufficiently strong multiple scattering, then the formation of "transition radiation" will proceed somewhat differently. Obviously, what is directly observable is always the *total radiation* emitted from the layer of matter under consideration. The intensity of this observable total radiation differs from the intensity of bremsstrahlung arising in an unbounded medium over a length equal to the thickness of the layer of matter under consideration (bremsstrahlung neglecting boundaries). The difference is due to the influence of the boundaries (for simplicity, in the present chapter we consider only frequencies for which $\varepsilon' \ll 1$, so that Vavilov–Cherenkov radiation is absent); therefore the difference between the two intensities mentioned above is a natural quantitative characteristic of this influence. We shall call this quantity the *"edge effect"* [**76**.3, **76**.4, **77**.7, **80**.3, **82**.14].[19] Thus, the concept of the edge effect is, in a certain sense, a generalization of the transition radiation to a case where the multiple scattering of the particle is taken into account.

[19] The aforementioned quantity was used already in Refs. [**60**.10–**60**.12] to characterize the influence of boundaries. However, the authors of those works unsuccessfully identified this quantity with transition radiation taking into account multiple scattering of the particle. Pafomov [**64**.2] noted that this quantity can become negative under certain conditions and therefore cannot be the intensity of any radiation. In fact, a negative value of the edge effect simply means that under these conditions the bremsstrahlung without accounting for boundaries is greater than the total radiation, whose intensity, of course, is always positive.

## § 16. PLATE

Consider the formation of radiation by a fast charged particle passing through a plate of arbitrary thickness $a$ located in vacuum. In the limiting case $a \to \infty$ we obtain an infinite or semi-infinite medium, previously considered by Migdal [54.2,57.5], Garibian and Pomeranchuk [59.6], and others.

### 16.1. Spectral–Angular Distribution of the Radiation Intensity

Let the particle move with velocity $v_0$ along the $z$-axis from $-\infty$ and at the point $r_0 = 0$ enter perpendicularly into a plate with dielectric permittivity $\varepsilon = \varepsilon(\omega) = \varepsilon' + i\varepsilon''$.

The spectral–angular distribution of the intensity of the radiation emitted by the particle forward relative to its motion, at large distances from the plate, is determined by the equation (see Ref. [**73**.32])

$$W(\omega, \vartheta_0) = \frac{e^2\omega^2}{4\pi^2 c^3} |\boldsymbol{B}|^2, \tag{16.1}$$

where the quantity $B$ is proportional to the Fourier component of the magnetic-field strength vector and consists of two parts ($B_1$ and $B_0$), corresponding to the motion of the particle inside the plate and outside it:

$$\boldsymbol{B} = \boldsymbol{B}_1 + \boldsymbol{B}_0,$$

$$\boldsymbol{B}_1 = \exp[i(\lambda - \lambda_0)a] \int_0^T \boldsymbol{n} \times \boldsymbol{v} \exp\{i(\omega t - \boldsymbol{k}\cdot\boldsymbol{r})\} dt,$$

$$\boldsymbol{B}_0 = \frac{-2i\boldsymbol{n}\times\boldsymbol{v}_0\exp[i(\lambda-\lambda_0)a]}{\omega(\gamma^{-2}+\vartheta_0^2)} + \tag{16.2}$$

$$+\int_T^\infty \boldsymbol{n}\times\boldsymbol{v}_1\exp\{i[\omega t-\boldsymbol{k}_0\cdot(\boldsymbol{r}_1+\boldsymbol{v}_1(t-T))]\}dt.$$

In the equations presented, it is taken into account that the frequencies under consideration are far greater than atomic frequencies, so that, in particular, the reflection coefficient is small and the reflected waves can be neglected. In Eq. (16.2) $r$ and $v$ are the coordinate and velocity of the particle in the plate at an arbitrary instant $t$, and $r_1$ and $v_1$ are the corresponding quantities at the instant $T$ when the particle leaves the plate. The two-dimensional vector $\vartheta_0$ lies in the plane perpendicular to the photon emission direction $n$ and is defined by the relation

$$\boldsymbol{\vartheta}_0 = \frac{\boldsymbol{n}(\boldsymbol{n}\cdot\boldsymbol{v}_0)-\boldsymbol{v}_0}{v_0}. \tag{16.3}$$

The quantities $\lambda$ and $\lambda_0$ have already been defined earlier (see Eq. (1.16)):

$$\lambda = \frac{\omega}{c}(\varepsilon-\sin^2\vartheta)^{1/2} = \lambda' + i\lambda'', \quad \lambda_0 = \frac{\omega}{c}\cos\vartheta.$$

The vectors $k$ and $k_0$ have identical components in the $x, y$ plane equal to $\varkappa$ ($\varkappa = (\omega/c)sin\vartheta$), and longitudinal components equal to $\lambda$ and $\lambda_0$, respectively. The quantity $exp[i(\lambda - \lambda_0)a]$ represents the transmission coefficient through the plate [**57**.4].

A charged particle moves in the plate along a trajectory determined by multiple scattering on the atoms of the medium. In doing so, we assume that the magnitude $v$ of the velocity remains unchanged, while the changes in the direction of the velocity are small. After emerging from the plate, the charge again moves in vacuum with constant velocity $v_1$.

In order to obtain the observable radiation intensity, it is necessary to average the expression (16.1) over all possible particle trajectories:

$$W(\omega, \vartheta_0)=\frac{e^2\omega^2}{4\pi^2c^3}\{<|\boldsymbol{B}_1|^2>+<|\boldsymbol{B}_0|^2>+2\,\mathrm{Re}<\boldsymbol{B}_0^*\cdot\boldsymbol{B}_1>\}. \tag{16.4}$$

Let us introduce the conditional probability $w(r, v; r', v'; t' - t)$ that the coordinate and velocity of the particle at time $t'$ take the values $r'$ and $v'$, provided that at time $t$ they had the values $r$ and $v$. Then

$$<|\boldsymbol{B}_1|^2>=v^2\exp(-2\lambda''a)\int d^3\boldsymbol{r}\,d^2\boldsymbol{\vartheta}\,d^3\boldsymbol{r}'\,d^2\boldsymbol{\vartheta}'\int_0^T dt'\times$$
$$\times\int_0^T dt\,w(\boldsymbol{r}_0, \boldsymbol{v}_0; \boldsymbol{r}, \boldsymbol{v}; t)\,w(\boldsymbol{r}, \boldsymbol{v}; \boldsymbol{r}', \boldsymbol{v}'; t'-t)\times$$
$$\times\boldsymbol{\vartheta}\cdot\boldsymbol{\vartheta}'\exp\{i[\omega(t'-t)-\boldsymbol{k}\cdot\boldsymbol{r}'+\boldsymbol{k}^*\cdot\boldsymbol{r}]\}, \tag{16.5}$$

where the two-dimensional vectors $\vartheta$ and $\vartheta'$ are related to $v$ and $v'$ by relations analogous to Eq. (16.3).

The second term of Eq. (16.4) has the following form:

$$<|\boldsymbol{B}_0|^2>=\frac{4v^2}{\omega^2}\exp(-2\lambda''a)\frac{\vartheta_0^2}{(\gamma^{-2}+\vartheta_0^2)^2}+$$
$$+v^2\int d^3\boldsymbol{r}_1\,d^2\boldsymbol{\vartheta}_1\int_0 dt'\int_0^\infty dt\,\vartheta_1^2\,w(\boldsymbol{r}_0, \boldsymbol{v}_0; \boldsymbol{r}_1, \boldsymbol{v}_1; T)\times$$
$$\times\exp\{i(\omega-\boldsymbol{k}_0\cdot\boldsymbol{v}_1)(t-t')\}+$$
$$+\frac{4v^2}{\omega}\exp(-2\lambda''a)\,\mathrm{Re}\,i\int d^3\boldsymbol{r}_1\,d^2\boldsymbol{\vartheta}_1\times$$
$$\times\int_0^\infty dt\frac{\boldsymbol{\vartheta}_0\cdot\boldsymbol{\vartheta}_1}{\gamma^{-2}+\vartheta_0^2}w(\boldsymbol{r}_0, \boldsymbol{v}_0, \boldsymbol{r}_1, \boldsymbol{v}_1; T)\times$$
$$\times\exp\{i[\omega T-\boldsymbol{k}\cdot\boldsymbol{r}_1+(\omega-\boldsymbol{k}_0\cdot\boldsymbol{v}_1)t\}. \tag{16.6}$$

Here the two-dimensional vector $\vartheta_1$ is associated with the vector $v_1$. In addition, in integrals (Eq. 16.6)), for simplicity, we changed the time record starting point.

Finally, the third term of Eq. (16.4) can be written in the following form:

$$2\mathrm{Re}\langle \boldsymbol{B}_0^*\cdot\boldsymbol{B}_1\rangle=\frac{4v^2}{\omega}\exp(-2\lambda''a)\mathrm{Re}\{i\int d^3\boldsymbol{r}d^2\boldsymbol{\vartheta}\times$$

$$\times\int_0^{\infty}dt\frac{\boldsymbol{\vartheta}_0\cdot\boldsymbol{\vartheta}}{\gamma^{-2}+\vartheta_0^2}w(\boldsymbol{r}_0,\boldsymbol{v}_0;\ \boldsymbol{r},\boldsymbol{v};t)\exp[i(\omega t-\boldsymbol{k}\cdot\boldsymbol{r})]\}+$$

$$+2v^2\mathrm{Re}\{\exp[i(\lambda-\lambda_0)a]\int d^3\boldsymbol{r}d^2\boldsymbol{\vartheta}d^3\boldsymbol{r}_1d^2\boldsymbol{\vartheta}_1\int_0^{\infty}dt\int_0^{\infty}dt'\boldsymbol{\vartheta}\cdot\boldsymbol{\vartheta}_1\times$$

$$\times\exp\{i[\omega(t-t'-T)-\boldsymbol{k}\cdot\boldsymbol{r}+\boldsymbol{k}_0\cdot(\boldsymbol{r}_1+\boldsymbol{v}_1t')]\}\times$$

$$\times w(\boldsymbol{r}_0,\boldsymbol{v}_0;\ \boldsymbol{r},\boldsymbol{v};\ t)w(\boldsymbol{r},\boldsymbol{v};\ \boldsymbol{r}_1,\boldsymbol{v}_1;\ T-t)\}. \qquad (16.7)$$

The individual terms of Eqs. (16.5)–(16.7) are conveniently represented as the diagrams shown in Fig. 16.1.

Let us note that in all the terms written above the averaging is performed in such a way that the coordinates and velocities $r$, $v$ or $r'$, $v'$ at times $t$ or $t'$ (later than the time $t_0$ corresponding to the particle entering the plate) are successively connected by conditional probabilities with the values of the particle coordinate $r_0$ and velocity $v_0$ at the initial time $t_0$ (see also the diagrams in Fig. 16.1).

Let us introduce the following notation:

$$u(\boldsymbol{\vartheta}_0,\boldsymbol{\vartheta};t)=\int w(\boldsymbol{r}_0,\boldsymbol{v}_0;\boldsymbol{r},\boldsymbol{v};t)\exp\{i[\omega t-\boldsymbol{k}\cdot(\boldsymbol{r}-\boldsymbol{r}_0)]\}d^3\boldsymbol{r},$$

$$u_1(\boldsymbol{\vartheta}_0,\boldsymbol{\vartheta};\ t)=\int w(\boldsymbol{r}_0,\boldsymbol{v}_0;\boldsymbol{r},\boldsymbol{v};t)\exp(2\lambda''z)\,d^3\boldsymbol{r}, \qquad (16.8)$$

$$u_0(\boldsymbol{\vartheta}_0,\boldsymbol{\vartheta};t)=\int w(\boldsymbol{r}_0,\boldsymbol{v};\boldsymbol{r},\boldsymbol{v};t)\ d^3\boldsymbol{r}.$$

Then the Eqs. (16.5)–(16.7) can be written in the following form:

$$
\langle |\boldsymbol{B}_1|^2 \rangle = 2\,v^2 \exp(-\eta\, a)\,\mathrm{Re} \int\limits_0^T d\,t \int\limits_0^{T-t} d\,\tau \int \boldsymbol{\vartheta}\cdot\boldsymbol{\vartheta}'\, u_1\,(\boldsymbol{\vartheta}_0, \boldsymbol{\vartheta}; t) \times
$$

$$
\times\, u\,(\boldsymbol{\vartheta}, \boldsymbol{\vartheta}'; \tau)\, d^2\boldsymbol{\vartheta}\, d^2\boldsymbol{\vartheta}',
$$

$$
\langle |\boldsymbol{B}_0|^2 \rangle = \frac{4\,v^2 \exp(-\eta\, a)}{\omega^2}\, \frac{\vartheta_0^2}{(\gamma^{-2}+\vartheta_0^2)^2} +
$$

$$
+\, v^2 \int\limits_0^\infty dt \int\limits_0^\infty dt' \int \vartheta_1^2 \exp\left[\frac{i\,\omega}{2}(\gamma^{-2}+\vartheta_1^2)\,(t-t')\right] \times
$$

$$
\times\, u_0\,(\boldsymbol{\vartheta}_0, \boldsymbol{\vartheta}_1; T)\, d^2\boldsymbol{\vartheta}_1 + \frac{4\,v^2}{\omega} \exp(-\eta\, a) \times
$$

Fig. 16.1. Diagrams corresponding to the individual terms of Eqs. (16.5)–(16.7) for radiation arising during the passage of a particle through a plate. Straight lines correspond to the motion of the particle in vacuum, zigzag lines correspond to the motion in the plate in the presence of multiple scattering, dashed lines correspond to relative probabilities integrated over the arguments indicated at their right ends. Diagram a corresponds to $\langle |B_1|^2 \rangle$, diagrams b, c, and d correspond to the three terms of the quantity $\langle |B_0|^2 \rangle$, and diagrams e and f correspond to the terms of the quantity $2Re\langle B_0 * B_1 \rangle$ (see Eq. 16.7))

$$\times \operatorname{Re} i \int_0^\infty dt \int \frac{\vartheta_1 \cdot \vartheta_0}{\gamma^{-2}+\vartheta_0^2} \exp\left[\frac{i\omega t}{2}(\gamma^{-2}+\vartheta_1^2)\right] u(\vartheta_0, \vartheta_1; T)\, d^2\vartheta_1, \tag{16.9}$$

$$2 \operatorname{Re} \langle B_0^* \cdot B_1 \rangle = \frac{4v^2}{\omega} \exp(-\eta a) \operatorname{Re} i \int_0^T dt \int \frac{\vartheta \cdot \vartheta_0}{\gamma^{-2}+\vartheta_0^2} u(\vartheta_0, \vartheta; t)\, d^2\vartheta +$$

$$+ 2v^2 \exp(-\eta a) \operatorname{Re} \int_0^T dt \int_0^\infty dt' \int \vartheta \cdot \vartheta_1 \exp\left[-\frac{i\omega t'}{2}(\gamma^{-2}+\right.$$

$$\left.+\vartheta_1^2)\right] u_1(\vartheta_0, \vartheta_1; t)\, u^*(\vartheta, \vartheta_1; T-t)\, d^2\vartheta\, d^2\vartheta_1.$$

Here $\eta = 2\lambda''$ is the linear absorption coefficient of radiation in the medium.

Let us compute the quantities in Eq. (16.8). To this end, following Migdal, we start from the kinetic equation for the conditional probability $w(r_0, v_0; r, v; t)$:

$$\frac{\partial w}{\partial t} + v \cdot \frac{\partial w}{\partial r} = v \int S(\vartheta', \vartheta)\, [w(r_0, v_0; r', v'; t) -$$

$$- w(r_0, v_0; r, v, t)]\, d^2\vartheta', \tag{16.10}$$

where $S(\vartheta', \vartheta)$ is the probability of scattering of a particle through the angle $\vartheta' - \vartheta = (v - v')/v$ per unit path length. The function $w$ appearing in Eq. (16.10) must satisfy the initial condition

$$w(r_0, v_0; r, v; 0) = \delta(r - r_0)\, \delta(\vartheta - \vartheta_0). \tag{16.11}$$

Multiplying Eq. (16.10) by $exp\{i[\omega t - k \cdot (r - r_0)]\}$ and integrating over $r$, we obtain an equation for $u(\vartheta_0, \vartheta; t)$:

$$\frac{\partial u}{\partial t} - \frac{i\omega}{2}(g + \vartheta^2)\, u = v \int S(\vartheta', \vartheta)\, [u(\vartheta_0, \vartheta'; t) -$$

$$- u(\vartheta_0, \vartheta; t)]\, d^2\vartheta', \quad g = \gamma^{-2} + 1 - \varepsilon, \tag{16.12}$$

with the initial condition

$$u(\vartheta_0, \vartheta; 0) = \delta(\vartheta - \vartheta_0). \tag{16.13}$$

If in Eq. (16.12) one expands $[u(\vartheta_0, \vartheta'; t) - u(\vartheta_0, \vartheta; t)]$ in a power series in $\vartheta' - \vartheta$ and discards terms higher than quadratic (Fokker–Planck approximation), then we obtain

$$\frac{\partial u}{\partial t} - \frac{i\omega}{2}(g + \vartheta^2)\, u = q\, \Delta_\vartheta\, u, \tag{16.14}$$

where $\Delta_\vartheta$ is the two-dimensional Laplace operator with respect to the variable $\vartheta$,

$$\begin{gathered} q = \frac{v}{4}\int S(\vartheta)\,\vartheta^2\, d^2\vartheta = q_0\gamma^{-2}, \\ q_0 = \left(\frac{E_s}{m_0c^2}\right)^2 \frac{c}{4L_{\mathrm{rad}}}. \end{gathered} \tag{16.15}$$

$m_0$ is the rest mass of the traversing particle, $L_{rad}$ is the radiation length, $E_s = 21$ MeV.

The solution of Eq. (16.14) satisfying the initial condition in Eq. (16.13) has the following form:

$$\begin{aligned} u(\vartheta_0, \vartheta; t) = \sigma(\pi \operatorname{sh} x)^{-1} \exp\{-\sigma[(\vartheta^2 - 2\vartheta\cdot\vartheta_0/\operatorname{ch} x + \\ + \vartheta_0^2)\operatorname{cth} x + gx]\}, \end{aligned} \tag{16.16}$$

where

$$\sigma = \frac{(1-i)}{4}\sqrt{\frac{\omega}{q}}, \qquad x = (1-i)\sqrt{\omega q}\, t. \tag{16.17}$$

The quantity $u_0(\vartheta_0, \vartheta; t)$ can be obtained from Eq. (16.16) by letting $\omega \to 0$:

$$u_0(\vartheta_0, \vartheta; t) = (4\pi q t)^{-1} \exp\left\{-\frac{(\vartheta - \vartheta_0)^2}{4qt}\right\}. \tag{16.18}$$

As for the quantity $u_1(\vartheta_0, \vartheta; t)$, for not very thick plates the exponential in Eq. (16.8) can be taken outside the integral sign, and as a result we obtain

$$u_1(\vartheta_0, \vartheta; t) = u_0(\vartheta_0, \vartheta; t)\exp(\eta v t). \qquad (16.19)$$

Substituting Eqs. (16.16), (16.18), and (16.19) into Eq. (16.9) and performing the integration over the intermediate angles $\vartheta$, $\vartheta'$, and $\vartheta_1$ (bearing in mind that, for example, $d^2\vartheta = \vartheta d\vartheta d\phi$, $0 \leq \vartheta < \infty$, $0 \leq \phi \leq 2\pi$). For this it is convenient to complete the square in the exponents so that it contains the integration angle. As a result, taking into account Eq. (16.4), we obtain the spectral–angular distribution of radiation in the following form [**76.3, 76.4**]:[20]

$$W_{tot}(\omega, \vartheta_0) = W_1 + W_2, \qquad (16.20)$$

$$W_1 = \frac{2e^2}{\pi^2 c} Q \,\mathrm{Re}\,\sigma \int_0^\infty \exp(hx)\,dx \int_0^\infty \exp(-s_2 y) \times$$

$$\times \frac{\partial}{\partial y} \frac{\exp[-s_0 \,\mathrm{th}\, y/(1 + x \,\mathrm{th}\, y)]}{1 + x \,\mathrm{th}\, y}\,dy\,, \qquad (16.21)$$

$$W_2 = \frac{2e^2}{\pi^2 c} \mathrm{Re}\,\sigma \left\{ \frac{Q s_0}{2(s_0 + s_1)^2} - \frac{1}{2} \int_0^\infty dx \int_x^\infty \exp(-s_1 y) \times \right.$$

$$\times \frac{\partial}{\partial y} \frac{\exp[-s_0 y/(1 + x_a y)]}{1 + x_a y}\,dy + \frac{Q s_0 \exp(-s_2 x_a)}{s_0 + s_1} \times$$

$$\times \int_0^x \exp(-s_1 x) \frac{\partial}{\partial x} \exp\left[ -\frac{s_0 (x + \mathrm{th}\, x_a)}{1 + x \,\mathrm{th}\, x_a} \right] dx +$$

$$+ \frac{Q s_0}{s_0 + s_1} \int_0^{x_a} \exp(-s_2 x) \frac{\partial}{\partial x} \exp(-s_0 \,\mathrm{th}\, x)\,dx +$$

$$+ \int_0^{x_a} \exp[-(s_2 + h) x]\,dx \int_0^\infty \exp(-s_1 y) \times$$

---

[20] In the analogous equation in the work of Pafomov [**65**.5], the terms corresponding to diagrams *c* and *f* (the second and the last terms in the quantity W2 (16.22)) are obtained incorrectly.

$$\times \frac{\partial}{\partial y} \frac{\exp[-s_0 \xi/(1+(x_a-x)\xi)]}{1+(x_a-x)\xi} dy \Big\}. \qquad (16.22)$$

The following notation is introduced:

$$s_0 = \sigma \vartheta_0^2, \; s_1 = \sigma \gamma^{-2}, \; s_2 = \sigma g, \; h = 2i\sigma\eta c/\omega,$$

$$g = \gamma^{-2} + \frac{\omega_0^2}{\omega^2} - i\frac{\eta c}{\omega}, \quad x_a = (1-i)\sqrt{\omega q}\, a/c, \qquad (16.23)$$

$$Q = \exp(-hx_a), \; \xi = (y + \mathrm{th} x)/(1 + y\,\mathrm{th} x).$$

The above Eqs. (16.20)—(16.22) are valid when the mean square angle of multiple scattering of a particle in the plate is small, i.e.,

$$\langle \vartheta^2 \rangle_a = \frac{4qa}{v} = \frac{x_a}{\sigma} \ll 1. \qquad (16.24)$$

In addition, the product $\eta a \langle \vartheta^2 \rangle_a$ is also small.

When the plate thickness is much smaller than the absorption length, i.e.,

$$hx_a = \eta a \ll 1, \qquad (16.25)$$

and, moreover, the imaginary part of the polarizability of the medium is much smaller (in absolute value) than its real part,

$$|\varepsilon''| \ll |1 - \varepsilon'|, \qquad (16.26)$$

in equations (16.21) and (16.22) one may set $h = 0$. Then these equations reduce to the corresponding equations obtained neglecting absorption in the medium (see Ref. [**76**.3]).

From the spectral–angular distribution of the intensity of the total radiation (Eq. (16.20)) it follows that, in contrast to ordinary transition radiation (without taking multiple scattering into account), in the present case (for $q \neq 0$) the radiation intensity per unit solid angle at small angles ($\vartheta_0 \to 0$) does not tend to zero because of bremsstrahlung (the terms corresponding to diagrams *a*, *d*, and *e* in Fig. 16.1).

The distribution in Eq. (16.20) depends on five dimensionless parameters: $x_a$, $s_0$, $s_1$, $s_2$, and $h$. These parameters may be regarded as ratios of certain lengths. For example, the parameter

$$x_a = (1 - i)\frac{a}{z_{\text{brem}}}, \qquad (16.27)$$

represents (up to the factor $(1 - i)$) the ratio of the plate thickness $a$ to the length

$$z_{\text{brem}} = \frac{c\gamma}{(\omega q_0)^{1/2}}, \qquad (16.28)$$

over which *bremsstrahlung is produced* (with multiple scattering taken into account). For the parameters $s_1$, $s_2$, and $h$ we have

$$s_1 = \frac{\pi(1 - i)}{4}\frac{z_{\text{brem}}}{z_{\text{vac}}}, \qquad (16.29)$$

$$s_2 = \frac{\pi(1 - i)}{4}\frac{z_{\text{brem}}}{z_{\text{med}}}, \quad h = \frac{1 + i}{2}\frac{z_{\text{brem}}}{\gamma^{-1}}, \qquad (16.30)$$

where $\gamma_i^{-1}$, $\bar{z}_{vac}$, and $\bar{z}_{med}$ are, respectively, the absorption length and the formation lengths of transition radiation in vacuum and in the medium:

$$\bar{z}_{\text{vac}} = \frac{\pi c\gamma^2}{\omega}, \qquad \bar{z}_{\text{med}} = \frac{\pi c}{\omega g}. \qquad (16.31)$$

The parameter $s_0$ is determined by the radiation angle $\vartheta_0$ (see Eq. (16.23)).

Let us consider the case when the following conditions are satisfied: (1) the thickness of the plate is much smaller than the bremsstrahlung formation length $z_{\text{brem}}$,

$$|x_a| \ll 1; \qquad (16.32)$$

(2) the mean square of the multiple-scattering angle of the particle in the plate ($\langle\vartheta^2\rangle_a = 4qa/\tau$) is much smaller than the square of the characteristic scattering angle, i.e.

$$\left|\frac{x_a}{s_1}\right| \ll 1. \qquad (16.33)$$

Under these conditions, the integrands in Eqs. (16.21) and (16.22) can be expanded in a series in powers of the small quantities in Eqs. (16.32) and (16.33), and one can retain only the lowest-order terms. Then, after integration, we obtain

$$W_{\text{tot}}(\omega,\vartheta_0)=\frac{2e^2\vartheta_0^2}{\pi c}\left\{\left|\frac{1}{\gamma^{-2}+\vartheta_0^2}-\frac{1}{g+\vartheta_0^2}\right|^2 F_{\text{tot}}+4qT\frac{\gamma^{-4}+\vartheta_0^4}{(\gamma^{-2}+\vartheta_0^2)^4}\right\}. \qquad (16.34)$$

The first term in Eq. (16.34) corresponds to the ordinary transition radiation produced in the plate by a uniformly and rectilinearly moving particle (i.e., neglecting multiple scattering, see Sec. 2.3). The second term corresponds to bremsstrahlung, also without taking into account the influence of multiple scattering. It is clear that such a result is quite natural.

Let us now consider a plate of large thickness. Note that for a sufficiently thick non-absorbing plate, the radiation intensity must contain a component directly proportional to the thickness of the plate. This part corresponds to bremsstrahlung produced over the entire length of the particle's path inside the plate. Such a situation occurs provided that the plate thickness is much greater than the bremsstrahlung formation length, taking multiple scattering into account, i.e.,

$$|x_a| \gg 1. \qquad (16.35)$$

The part proportional to the thickness $a$ must be separated from $W_1$ (see diagram $a$ in Fig. 16.1). If in the expression for $W_1$ the upper limit of integration $x_a - x$ is replaced by $\infty$, one obtains the quantity corresponding to bremsstrahlung from a path length $a$ in an infinite non-absorbing medium [**54**.2, **57**.5],

$$W_{\text{brem}}(\omega,\vartheta_0)=\frac{4e^2\vartheta_0}{\pi c}\operatorname{Re}\left\{\int_0^{x_a}dx\int_0^{\infty}\exp(-s_2 y)\frac{\partial}{\partial y}\frac{\exp[-s_1\tanh y(1-x\tanh y)]}{1-x\tanh y}\,dy\right\}. \qquad (16.36)$$

The difference between the total radiation (Eq. (16.20)) and the bremsstrahlung (Eq. (16.36)) is due to the presence of the plate boundaries. This difference

$$W_{\rm edge}(\omega, \vartheta_0) = W_{\rm tot}(\omega, \vartheta_0) - W_{\rm brem}(\omega, \vartheta_0). \qquad (16.37)$$

is natural to be referred to as the *edge effect* for a non-absorbing plate.

If at the same time the mean square of the multiple-scattering angle over the length $a$ is still smaller than the square of the radiation angle $\gamma^{-1}$ (i.e., condition in Eq. (16.32) holds), then it can be shown that Eq. (16.37) reduces (for $\vartheta_0^2 \gtrsim \gamma^{-2}$) to the equation for the spectral–angular distribution of ordinary transition radiation produced in a plate.

## 16.2. Frequency Spectrum of Radiation

To obtain the frequency spectrum of the radiation, the spectral–angular distribution (Eq. (16.20)) must be integrated over the angles $\phi_0$ and $\vartheta_0$ (or $s_0$). Integration over $\phi_0$ is trivial because of the axial symmetry of the problem. Integration over $\vartheta_0$ will be performed from zero to some maximum value $\vartheta_m$ of order unity. Since the characteristic angles in the problem under consideration ($\langle\vartheta^2\rangle_a^{1/2}$ and $|g|^{1/2}$) are small, one should expect that the result of the above integration over $\vartheta_0$ does not depend on the upper limit of integration. In other words, formally one may let $\vartheta_m$ to infinity. In this case, the logarithmic divergences that arise in the individual terms cancel each other. As a result, we obtain the following equation for the frequency spectrum of the total radiation [**76**.4]:

$$W_{\rm tot}(\omega) = \frac{2e^2}{\pi c}{\rm Re}\left\{\frac{1+Q}{2}\left(\ln\frac{s_1}{s_2} - \frac{s_2}{h}\ln\frac{s_2+h}{s_2} - 1\right) + \right.$$

$$+Q\exp(-s_2x_a)\left[\frac{s_2+h}{h}G((s_2+h)x_a)-\frac{s_2}{h}G(s_2x_a)-G(s_1\tanh x_a)\right]-$$
$$-s_1Q\exp(-s_2x_a)\int_0^\infty e^{-s_1x}G\left(\frac{s_1(x+\tanh x_a)}{1+x\tanh x_a}\right)dx- \tag{16.38}$$
$$-\int_0^{x_a}\left[(s_2+h)e^{-hx}+s_2Q\right]e^{-s_2x}G(s_1\tanh x)\,dx+$$
$$+\frac{1}{h}\int_0^{x_a}\left[(s_2+h)e^{-hx}-s_2Q\right]e^{-s_2x}\left(\coth x-\frac{1}{x}\right)dx\Bigg\}.$$

where $G(z)=exp(z)Ei(-z)$, and $Ei(z)$ is the exponential integral [**71**.13].

From a comparison of the spectral intensity (Eq. (16.38)) of the total radiation $W_{tot}(\omega)$ with the spectral intensity (Eq. (2.36)) of ordinary transition radiation without taking multiple scattering of the particle into account (see Figs. 16.2 and 16.3), it is seen that under certain conditions these quantities differ significantly from each other. This is evidently due to the appearance of bremsstrahlung caused by multiple scattering of the particle in the plate.

At the end of Sec. 16.1 it was pointed out that in the case of a weakly absorbing medium, from the spectral–angular distribution of the total radiation intensity one can naturally single out the part (16.36) corresponding to bremsstrahlung. After integrating the expression (16.36) over the radiation angle, we obtain

$$W_{\mathrm{brem}}(\omega)=\frac{a}{z_{\mathrm{brem}}}W_M(\omega), \tag{16.39}$$

where $z_{brem}$ is the bremsstrahlung formation length (see Eq. (16.28)), and $W_{\mathrm{M}}(\omega)$ is given by the following equation:

$$W_{\mathrm{M}}(\omega)=\frac{2e^2}{\pi c}\,\mathrm{Re}\Big\{(1-i)(s_2+h)\times$$

$$\times \int_0^{\infty} \exp[-(s_2+h)x]\left(\operatorname{cth}x-\frac{1}{x}\right)dx\Big\}. \qquad (16.40)$$

The quantity $W_M(\omega)/z_{brem}$ describes the spectrum of bremsstrahlung emitted per unit path length in an infinite medium. For $h = 0$, this quantity coincides with Migdal's equation [**54**.2, **57**.5].

In the case when absorption in the medium cannot be regarded as weak, one can introduce the concept of an effective plate thickness

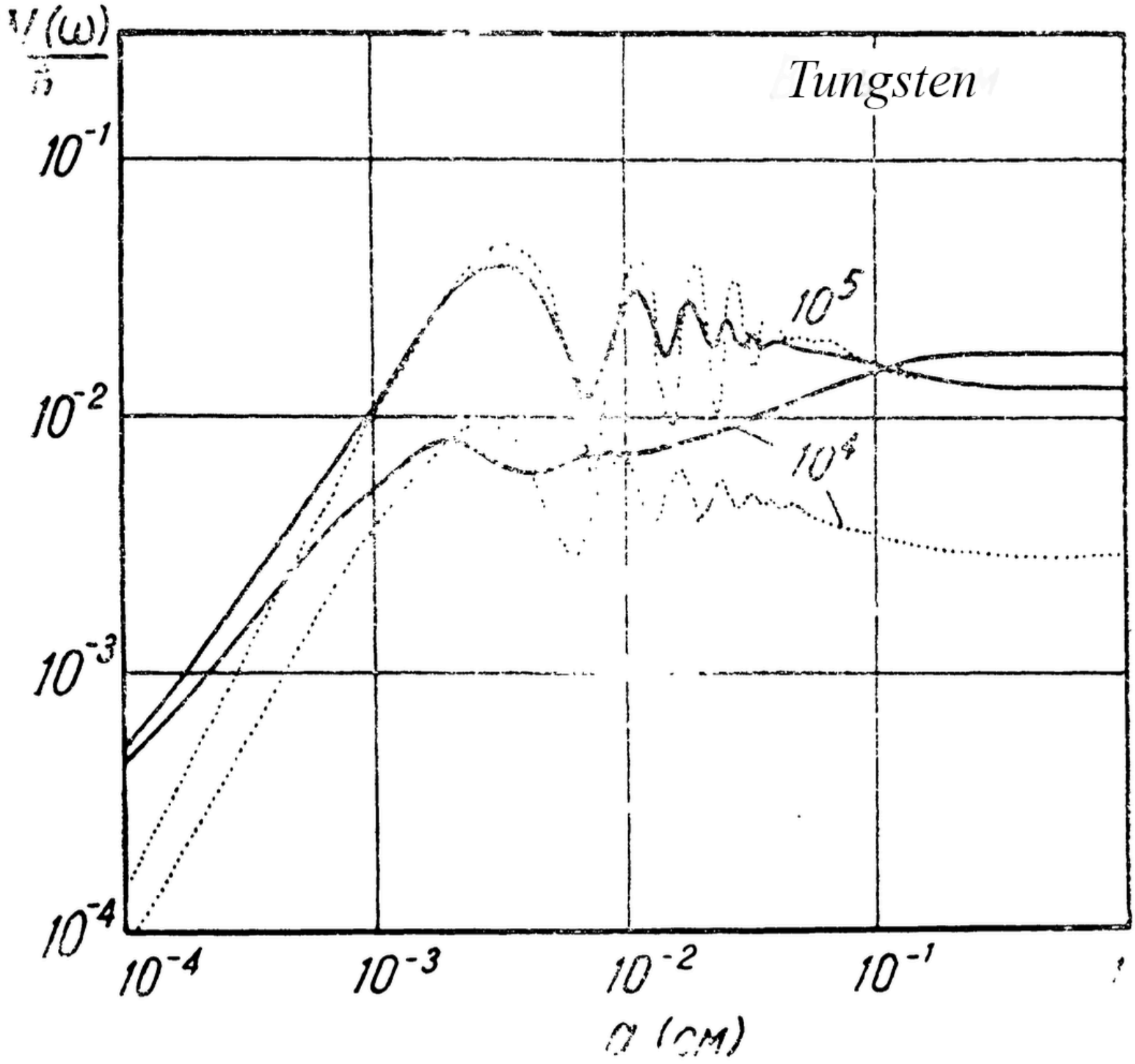

Fig. 16.2. Dependence of the intensities of the total (solid curves) and ordinary transition radiation (dotted curves) at $\hbar\omega = 200$ keV on the thickness of a Mylar plate ($\hbar\omega0 = 20$ eV, $\gamma^{-1} = 10$ cm, $L_{rad} = 27.5$ cm). The numbers by the curves indicate the values of the Lorentz factor

$$a_{\mathrm{eff}} = \frac{1 - \exp(-\eta a)}{\eta}. \qquad (16.41)$$

For $\eta a \ll 1$ the quantity $a_{eff}$ coincides with the plate thickness $a$, while for $\eta a \gg 1$ it coincides with the absorption length $\gamma^{-1}$.

Then, in the general case, the spectral intensity of bremsstrahlung emerging from the plate is given by the following equation:

$$W_{\mathrm{brem}}(\omega) = \frac{a_{\mathrm{eff}}}{z_{\mathrm{brem}}} W_M(\omega). \qquad (16.42)$$

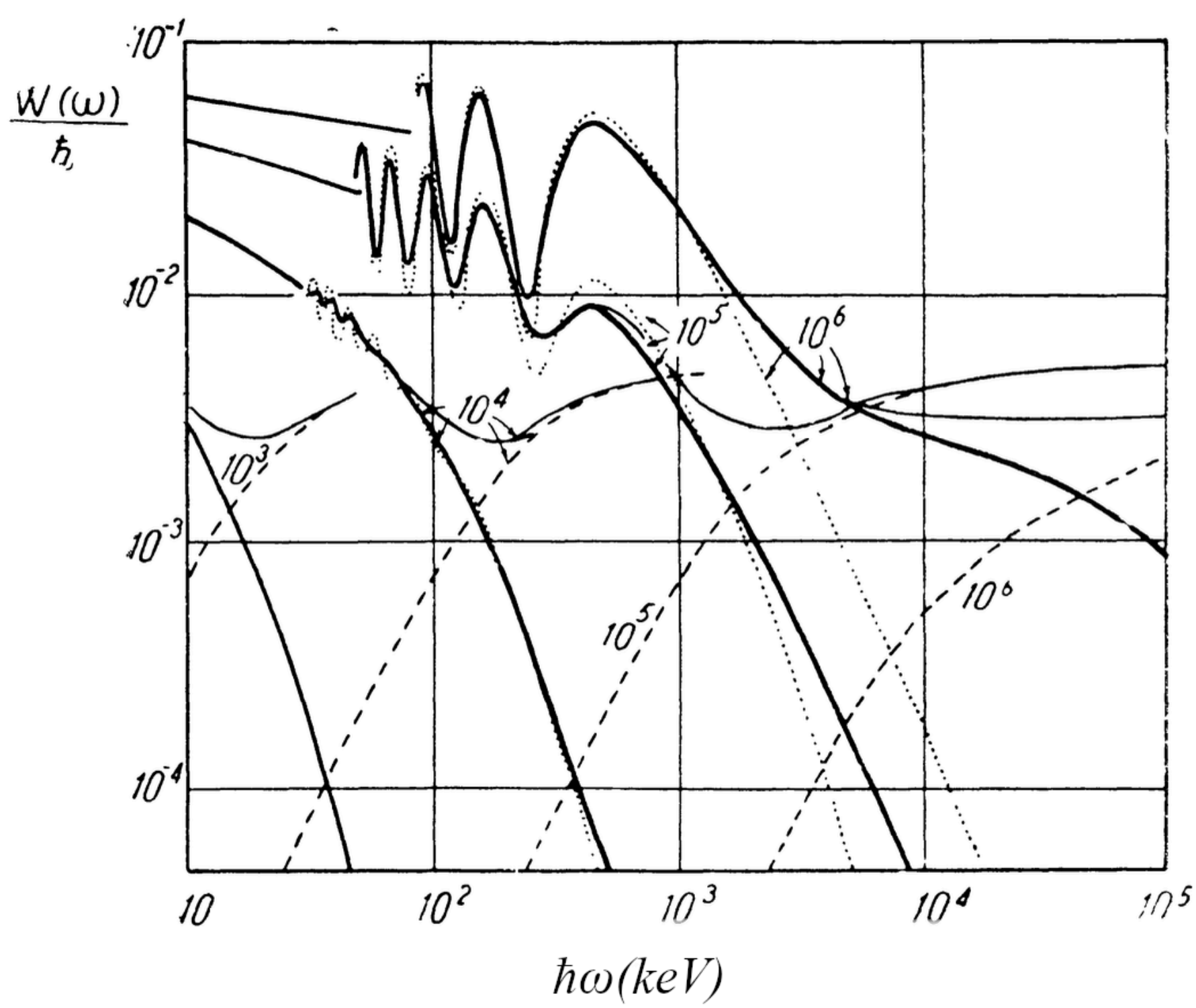

Fig. 16.3. The same as in Fig. 16.2, for tungsten ($\hbar\omega_0 = 80.3$ eV, $\gamma^{-1} = 0.08$ cm, $L_{rad} = 0.35$ cm).

By analogy with Eq. (16.37), the difference between the intensities of the total radiation and the bremsstrahlung, due to the influence of the boundaries, will be called the edge effect:

$$W_{\text{edge}}(\omega) = W_{\text{tot}}(\omega) - W_{\text{brem}}(\omega). \qquad (16.43)$$

In the limiting case where particle scattering is absent ($q \to 0$), the bremsstrahlung $W_{brem}(\omega)$ vanishes and the edge effect $W_{edge}(\omega)$ coincides with the total radiation $W_{total}(\omega)$, which, in turn, as follows from (16.38), coincides with ordinary transition radiation (cf. Eq. (2.36)). Thus, in the general case ($q \neq 0$) the edge effect could be called "transition radiation" with allowance for multiple scattering. However, such a designation can only be used conditionally, since under certain conditions the magnitude of the edge effect may become negative (see Sec. 16.3), whereas the intensity of the total radiation, naturally, always remains positive.

When absorption is absent and the plate thickness is much larger than the formation length of transition radiation $\overline{|z_{med}|}$, the quantity (Eq. (16.43)), after averaging over a small frequency range, coincides with twice the edge effect obtained in Ref. [**60**.11] for a semi-infinite non-absorbing medium, which is quite natural.

The frequency spectra of the total radiation (Eq. (16.38)) and of the edge effect (Eq. (16.43)) generally depend in a complicated way on the radiation frequency ω, the Lorentz factor γ of the charged particle, and the plate thickness $a$. To clarify this dependence, note that the behavior of the edge effect is largely determined by the ratio $z_{brem}/\overline{|z_{med}|}$, i.e., by the magnitude $|s_2|$. For insufficiently large values of γ ($\gamma < \gamma_c$), the curve $z_{brem}(\omega)$ lies entirely above the curve $|\bar{z}_{med}|$. As γ increases, the curves approach each other. The value $\gamma_c$ is determined by the condition that the indicated curves touch. If a light particle (an electron) is considered, then

$$\gamma_{\text{cr}} \sim \frac{\omega_0}{q_0}. \qquad (16.44)$$

For $\gamma > \gamma_c$, the curves $z_{brem}(\omega)$ and $|z^{-}_{med}(\omega)|$ intersect at the points $\omega_1$ and $\omega_2$, so that in the region $\omega_1 < \omega < \omega_2$ we have $|z_{brem}| < |z^{-}_{med}|$, whereas outside this region $|z_{brem}| > |z^{-}_{med}|$.

In the case when $\omega_1 \ll \omega_0\gamma \ll \omega_2$ and the absorption is small, simple expressions are obtained for $\omega_1$ and $\omega_2$:

$$\omega_1 \sim \left(\frac{\omega_0^4\,\gamma^2}{q_0}\right)^{1/3}, \quad \omega_2 \sim q_0\,\gamma^2. \tag{16.45}$$

When absorption plays a significant role, then

$$\omega_1 \sim (\eta\,c\,\gamma)^2/4q_0, \tag{16.46}$$

while $\omega_2$ is given by the same equation as before.

Below we analyze Eq. (16.43) for different values of $|s_2|$.

### 16.3. Analysis of the Total Radiation Frequency Spectrum

A. *Case* $|s_2| \gg 1$. In this case, in the integrals of equations (16.38) and (16.40), which contain the factors $exp(-\ s_2x)$ or $exp(-\ (h\ +\ s_2)x)$, small values of $x$ are essential. After expanding the pre-exponential factors of the integrands in powers of $x$, these integrals can be evaluated explicitly. As a result, from Eq. (16.39) we obtain the following expression:

$$W_{\text{brem}}(\omega) = \frac{2e^2}{3\pi c}\text{Re}\,\frac{x_a}{s_2}. \tag{16.47}$$

which, in the absence of absorption, coincides with the Ter-Mikaelian equation [54.3, 55.1] for the bremsstrahlung intensity taking into account the polarization of the medium.

As for the edge effect (Eq. (16.43)), after the above-mentioned expansion we obtain

$$W_{\text{edge}}(\omega) = W_{\text{TR}}(\omega) + W_{\text{int}}(\omega), \tag{16.48}$$

where the first term does not depend on $q_0$ and represents the intensity of ordinary transition radiation (without taking multiple scattering into account) emitted from a plate of given thickness, whereas the second term, which vanishes for $q_0 \to 0$, describes the influence of multiple scattering on the transition radiation. Taking Eqs. (16.43) and (16.48) into account, one may regard this term as, in a certain sense, describing the "interference" between transition radiation and bremsstrahlung. Let us present explicit expressions for the terms of Eq. (16.48) in two limiting cases.

(1) Let the thickness of the plate be much greater than the formation length of transition radiation in the medium $|z^{-}_{med}|$, i.e.

$$|x_a s_2| \ll 1. \qquad (16.49)$$

In this case

$$W_{\mathrm{TR}}(\omega) = \frac{e^2}{\pi c}(1+Q)\,\mathrm{Re}\left\{\frac{s_2+s_1}{s_2-s_1}\ln\frac{s_2}{s_1} - \frac{s_2}{h}\ln\frac{s_2+h}{s_2} - 1\right\}, \qquad (16.50)$$

$$W_{\mathrm{int}}(\omega) = \frac{2e^2}{\pi c}\mathrm{Re}\left\{\frac{1-Q}{s_2^2}\left[-\frac{2}{3}+\frac{p}{2}F(1;4;5;1-p)\right]+\right.$$
$$\left.+\frac{2}{s_2^4}\left[-\frac{14}{15}-\frac{11}{15}Q-2p(1+Q)\left(1-\frac{2}{7}(2+3p)F(1;7;8;1-p)\right)\right]\right\}. \qquad (16.51)$$

where $p = s_1/s_2$, and $F(\alpha;\ \beta;\ \gamma;\ z)$ is the hypergeometric series (see, for example, Ref. [**71**.13]).

Let us compare the quantities in Eqs. (16.50) and (16.51). In the "pre-cutoff" frequency region, when

$$\left|\frac{\bar{z}_{\mathrm{vac}} - \bar{z}_{\mathrm{med}}}{\bar{z}_{\mathrm{vac}}}\right| \gtrsim 1, \qquad (16.52)$$

the intensity of transition radiation $W_{TR}(\omega)$ is of order $e^2/c$ and is much larger than $W_{int}(\omega)$ due to the condition $|s_2| \gg 1$.

However, in the frequency range the cutoff, when

$$\left|\frac{\bar{z}_{\mathrm{vac}} - \bar{z}_{\mathrm{med}}}{\bar{z}_{\mathrm{vac}}}\right| \ll 1, \qquad (16.53)$$

the quantity (16.50) decreases sharply and can become comparable to (16.51). Expanding these quantities in powers the small parameter (16.53), we obtain

$$W_{\text{TR}}(\omega) = \frac{e^2\gamma^4}{6\pi c}(1-Q)\left[\left(\frac{\omega_0}{\omega}\right)^4 - \left(\frac{\gamma c}{\omega}\right)^2\right], \qquad (16.54)$$

$$W_{\text{int}}(\omega) = \frac{8e^2\gamma^4 q_0}{15\omega\pi c}\left[7(1-Q)\frac{\gamma c}{\omega} + \frac{32q_0}{7\omega}(8-13Q)\right]. \qquad (16.55)$$

As is seen from these expressions, at high frequencies the magnitude of the edge effect (Eq. (16.48)) can become negative; however, as already noted, the intensity of the total radiation (Eq. (16.38)) remains positive.

(2) Let now

$$|x_a s_2| \ll 1. \qquad (16.56)$$

If, in addition, condition in Eq. (16.32) holds, then we obtain

$$W_{\text{TR}}(\omega) = \frac{e^2}{2\pi c}|(s_2 - s_1)x_a|^2(1 - 2C - 2\ln|s_1 x_a|), \qquad (16.57)$$

$$W_{\text{int}}(\omega) = \frac{2e^2}{3\pi c}\text{Re}\,\frac{(s_2 - s_1)x_a}{s_1 s_2}, \qquad (16.58)$$

where $C = 0.5772$ is the Euler constant.

Comparing Eqs. (16.57) and (16.58), we note that the intensity of transition radiation in the case under consideration is proportional to $a^2$, whereas the interference term is proportional to $a$. Therefore, for sufficiently small thickness $a$ the latter term may become dominant. Then the intensity of the total radiation (the sum of Eqs. (16.47) and (16.58)) is given by the Bethe–Heitler formula:

$$W_{\text{tot}}(\omega) = \frac{2e^2}{3\pi c}\text{Re}\,\frac{x_a}{s_1}, \qquad (16.59)$$

Thus, when the condition $|s_2| \gg 1$ is satisfied, i.e.,

$$|z_{\text{med}}| \ll z_{\text{brem}}, \qquad (16.60)$$

the transition mechanism plays a decisive role when considering the pre-cutoff frequency region (Eq. (16.52)) and sufficiently thick plates. In this case, multiple scattering leads to a dephasing of the phase relations of the fields as a result of the bending of the trajectory of the charged particle inside the medium. This influence manifests itself as a *smoothing* of the interference maxima and minima in the frequency spectrum of X-ray transition radiation. The smoothing becomes more pronounced the lower the energy of the charged particle (see Fig. 16.2 for $a \sim 10^{-1}$ cm and Fig. 16.3 for $a \sim 10^{-2}$ cm).

When the frequency exceeds the cutoff and condition (16.53) holds, or when the plate is sufficiently thin, as already noted, the terms in Eqs. (16.55) or (16.58), proportional to $q_0$ and caused by scattering, become significant.

B. *Case* $|s_2| \ll 1$. In this case, the bremsstrahlung intensity (Eq (16.40)) is readily computed explicitly:

$$W_{\mathrm{brem}}(\omega) = \frac{2e^2}{\pi c}\frac{a_{\mathrm{eff}}}{z_{\mathrm{brem}}}. \qquad (16.61)$$

As for the edge effect, the situation depends substantially on the ratio of the mean square of the multiple-scattering angle over the plate thickness ($\langle\vartheta^2\rangle_a = 4q_0\gamma^{-2}a/v$) to the square of the characteristic radiation angle $\gamma^{-2}$.

1) Let

$$\langle\vartheta^2\rangle_a \gg \gamma^{-2}, \qquad \text{i. e. } |x_a| \gg |s_1|. \qquad (16.62)$$

The first integral (from zero to infinity) in Eq. (16.38) can be estimated as follows. We divide the integral into two parts so that in the first part (from zero to some $x_1$) in the argument of the function $G$ the quantity $x\tanh x_a$ can be neglected compared with unity, whereas in the second part (from $x_1$ to infinity) unity can be neglected compared with $x\tanh x_a$, and $\tanh x_a$ can also be neglected compared with $x$. Then both parts of the integral are readily computed. The remaining integrals are estimated in a similar manner.

As a result, we obtain that if $|x_a| \lesssim 1$, then

$$W_{\mathrm{edge}}(\omega) = \frac{2e^2}{\pi c}\mathrm{Re}\left\{\ln\frac{\sinh x_a}{s_1} - 1 - C - x_a\right\}. \qquad (16.63)$$

If, however, $|x_a| \gg 1$, then

$$W_{\mathrm{edge}}(\omega) = \frac{2e^2}{\pi c}\left\{\frac{1-Q}{2}\ln\frac{2}{|s_2|} + \frac{1+Q}{2}\ln\frac{1}{2|s_1|} - \frac{1+Q}{2} - \frac{1-Q}{2}\mathrm{Re}\left(\frac{s_2}{h}\ln\frac{s_2+h}{s_2}\right) - C\right\}. \qquad (16.64)$$

In particular, in the absence of absorption it follows that

$$W_{\text{edge}}(\omega) = \frac{2e^2}{\pi c}\left\{\ln\frac{1}{2|s_1|} - 1 - C\right\}. \qquad (16.65)$$

From Eqs. (16.63) and (16.64) it is seen that if in the case under consideration one again uses the representation (16.48), then the second ("interference") term due to scattering turns out to be of the same order as, or larger than, the first.

2) Let now

$$\langle\vartheta^2\rangle_a \ll \gamma^{-2}, \qquad \text{i. e. } |x_a| \ll |s_1|. \qquad (16.66)$$

Then, by virtue of the inequality $|s_1| \ll |s_2| \ll 1$, the conditions $|x_a| \ll 1$ and $|s_1 x_a| \ll 1$ are also satisfied. Therefore, in the first integral (from zero to infinity) of Eq. (16.38) the quantity $x \tanh x_a$ can be regarded as small compared with unity, whereas in the remaining two integrals (from zero to $x_a$) the variable $x$ is small. Expanding the integrands in powers of $x$ and retaining only the leading terms, we obtain

$$W_{\text{edge}}(\omega) = \frac{2e^2}{3\pi c}\mathrm{Re}\,\frac{x_a}{s_1}. \qquad (16.67)$$

From the standpoint of expression (16.48), the edge effect in this case consists mainly of the interference term. However, the term written in Eq. (16.67) is also the leading one in the intensity of the total radiation, since in the case under consideration the bremsstrahlung intensity (Eq. (16.61)) has the form $(2e^2/\pi c) Re x_a$ and is small compared with Eq. (16.67). This means that when the conditions $|x_a| \ll |s_1| \ll |s_2| \ll 1$ are simultaneously satisfied, the intensity of the total radiation is again determined by the Bethe–Heitler formula (Eq. (16.59)).

Summarizing, we see that when the following condition is satisfied:

$$|s_2| \ll 1, \qquad \text{i. e. } z_{\text{brem}} \ll |z_{\text{med}}|. \qquad (16.68)$$

the bremsstrahlung mechanism is dominant in the edge effect. This means that if one represents the edge effect in the form of Eq. (16.48), then the interference term plays a significant role and is sometimes the leading one. In this connection, we also note that in the above Eqs. (16.63), (16.65), and (16.67) the plasma frequency of the substance $\omega_0$, which characterizes the polarization properties of the medium, is altogether absent.

## 16.4. Spectral Hardening of the Edge Effect

Thus, for $|s_2| \ll 1$ and $|x_a| \ll |s_1|$, or $a \ll z^2_{brem}/z^-_{vac}$, the spectral intensity of the total radiation is described by the Bethe–Heitler formula (Eq. (16.59)). One may suppose (see Ref. [77.1], p. 30) that when the plate is thick, radiation of the same kind is produced in boundary layers of thickness of order $z^2_{brem}/z^-_{vac}$, whereas the radiation produced in the interior of the plate is described by Migdal's equation (16.61). Since in the case $|s_2| \ll 1$ the Bethe–Heitler formula gives a larger intensity than Migdal's equation for the same path length, an "excess" of radiation is effectively produced in the boundary layers, which gives rise to the edge effect in the case under consideration. Indeed, the quantity (Eq. (16.65)), obtained from the analysis of the general equation (16.38), is of order $e^2/c$. A quantity of the same order is obtained from the Bethe–Heitler formula (Eq. (16.67)) if one sets $a \approx z^2_{brem}/z^-_{vac}$ (i.e., $x_a \approx s_1$).

As stated at the end of Sec. 16.2, the region $|s_2| \ll 1$ lies between the frequencies $\omega_1$ and $\omega_2$ for $\gamma \gg \gamma_c$. The cutoff frequency $\omega_{cutoff}$ of ordinary transition radiation (without accounting for multiple scattering) is also located between $\omega_1$ and $\omega_2$. For $\omega \gtrsim \omega_{cutoff}$ the spectral intensity of ordinary transition radiation, even in the case of a sufficiently thick plate, becomes very small compared with its intensity at $\omega < \omega_{cutoff}$, where it is of order of $e^2/c$ until frequencies of order of $\omega_2 \approx q_0\gamma^2$ (see 16.45)). In other words, in this frequency range there is a kind of hardening of the edge-effect spectrum at high frequencies. This hardening, as noted above, is due to the fact that a charged particle in the boundary layers of the plate scatters somewhat differently than in the bulk of the plate.

Figures 16.4–16.6 show the curves of the spectral intensities of the total radiation $W_{tot}(\omega)$ and of the edge effect $W_{edge}(\omega)$ for different plate thicknesses $a$ and different values of the electron Lorentz factor $\gamma$. For comparison, the frequency spectra $W_{TR}(\omega)$ of ordinary transition radiation are also shown.

The figures clearly show the smoothing of interference maxima and minima, and the enrichment in high frequencies in the spectrum of the edge effect compared with the spectrum of ordinary transition radiation (without accounting for multiple scattering). The degree of smoothing in the case $\gamma = 10^5$ is noticeably greater than in the case $\gamma = 10^6$, and leads to a change in the intensity by almost a factor of two. For the same value of the $\gamma$-factor, the degree of smoothing increases with increasing plate thickness (as long as the thickness remains smaller than the absorption length).

Hardening sets in when $\gamma \gg \gamma_c$. For tin $\gamma_c \sim 1.6 \cdot 10^4$; therefore for $\gamma = 10^4$ (Fig. 16.5) we practically do not observe hardening, whereas for $\gamma = 10^5$ (Fig. 16.6) the hardening effect becomes quite significant (see also Ref. [**77**.18]).

Let us now consider the dependence of the magnitude of the edge effect on the plate thickness $a$. When

$$a \gg |z_{\text{med}}| \quad \text{for} \quad |\bar{z}_{\text{med}}| \ll z_{\text{brem}}, \qquad (16.69)$$

the edge effect (the sum of Eqs. (16.50) and (16.51)) depends on the thickness only through the combination Q=exp(--η*a*), i.e., it does not depend on *a* both for a weakly absorbing plate ($\eta a \ll 1$), and for strongly absorbing plate ($\eta a \gg 1$). The similar situation occurs when

$$a \gg z_{\text{brem}} \quad \text{for} \quad |\bar{z}_{\text{med}}| \gg z_{\text{brem}}, \qquad (16.70)$$

and the edge effect is defined by the Eq. (16.64). Combining the cases (16.69) and (16.70), we conclude that the quantity of our introduced edge effect  (16.43) does not depend on the thickness of the plate (both for $\eta a \ll 1$ and $\eta a \gg 1$), if

$$a \gg \min\{z_{\text{brem}}, |z_{\text{med}}|\}. \qquad (16.71)$$

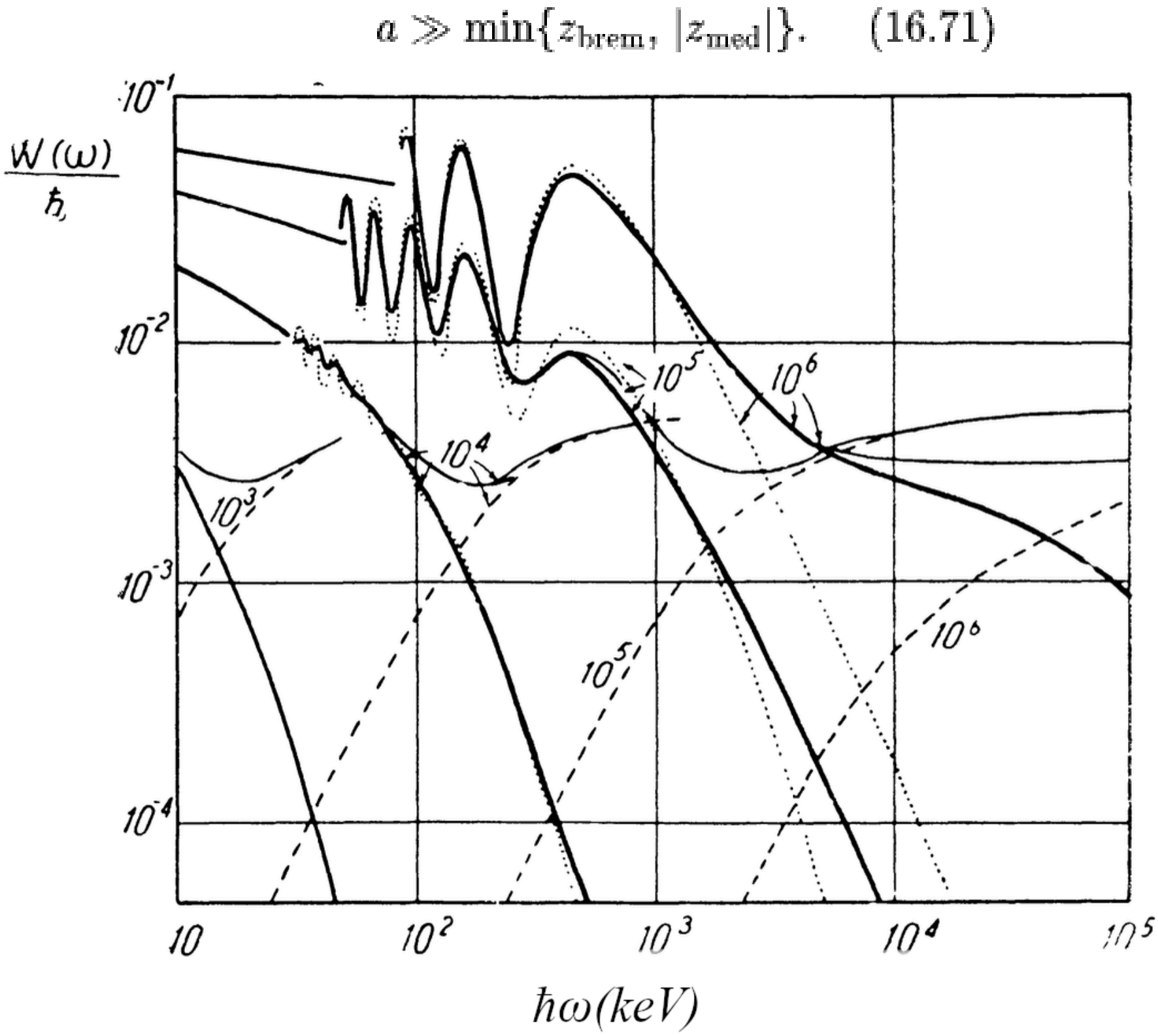

Fig. 16.4. Frequency spectra of the radiation produced in a 1-mm-thick Mylar plate. Thin solid curves: total radiation (according to Eq. (16.38)); dashed curves: bremsstrahlung from a 1 mm path length in an infinite medium (according to Eqs. (16.42) and (16.40)); bold solid curves: the difference between these quantities, i.e., the edge effect in the plate (Eq. (16.43)); dotted curves: ordinary transition radiation (without taking multiple scattering of the particle into account) in the plate (according to Eq. (2.36)). At low frequencies, owing to strong oscillations of the curves, only averaged frequency spectra are shown in the figure. The numbers by the curves indicate the electron Lorentz factor $\gamma$

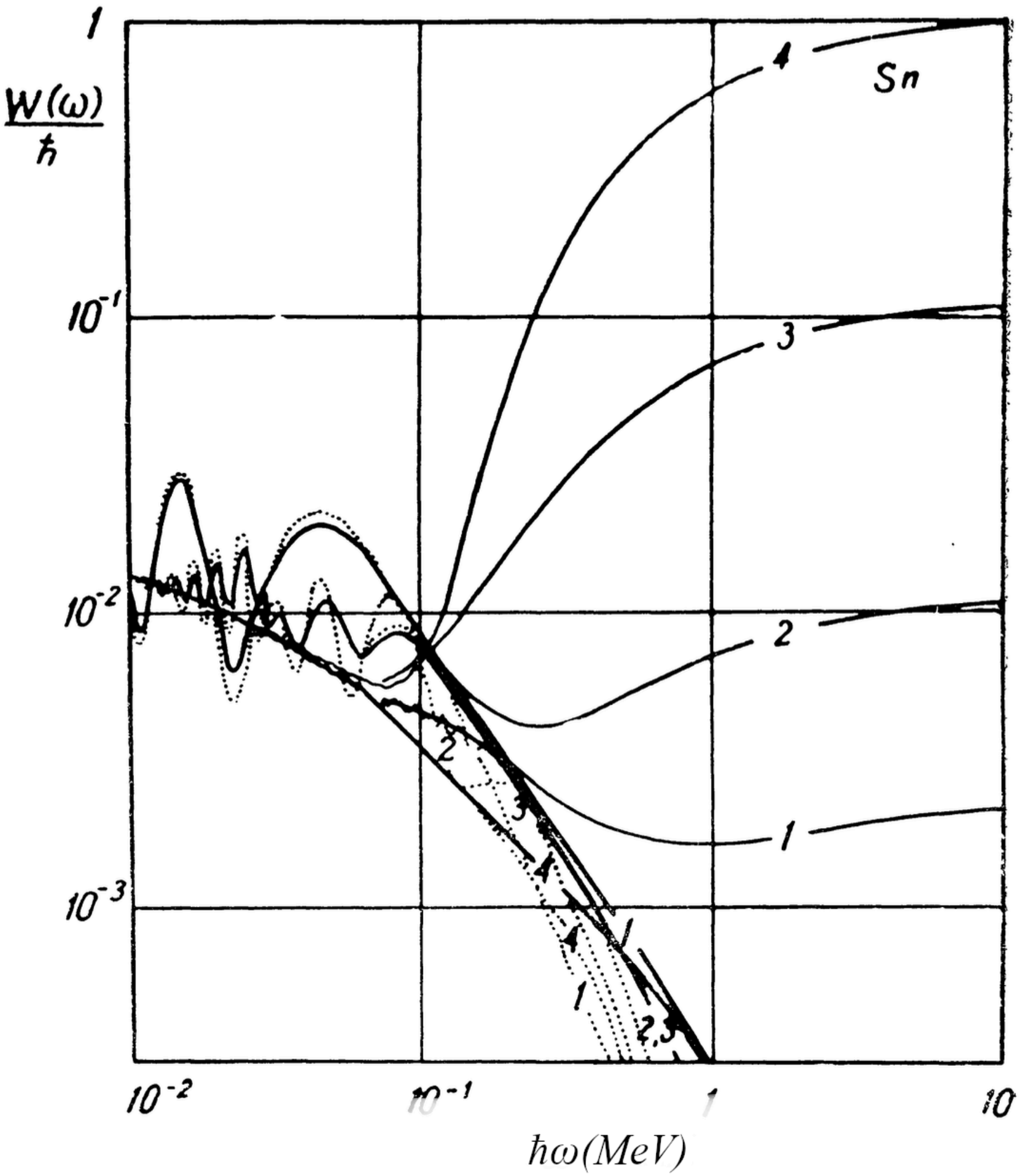

Fig. 16.5. Frequency spectra of the total radiation $W_{tot}(\omega)$ (thin solid curves), produced in a tin plate ($\hbar\omega_0 = 51.2$ eV, $L_{rad} = 1.21$ cm) by an electron with Lorentz factor $\gamma = 10^4$. Bold solid curves: edge effect; dotted curves: ordinary transition radiation. In the low-frequency range, the curves of total radiation and the edge effect coincide. The numbers on the curves correspond to different values of the plate thickness $a$: 1—20; 2—100; 3—$10^3$; 4—$10^4$ μm.

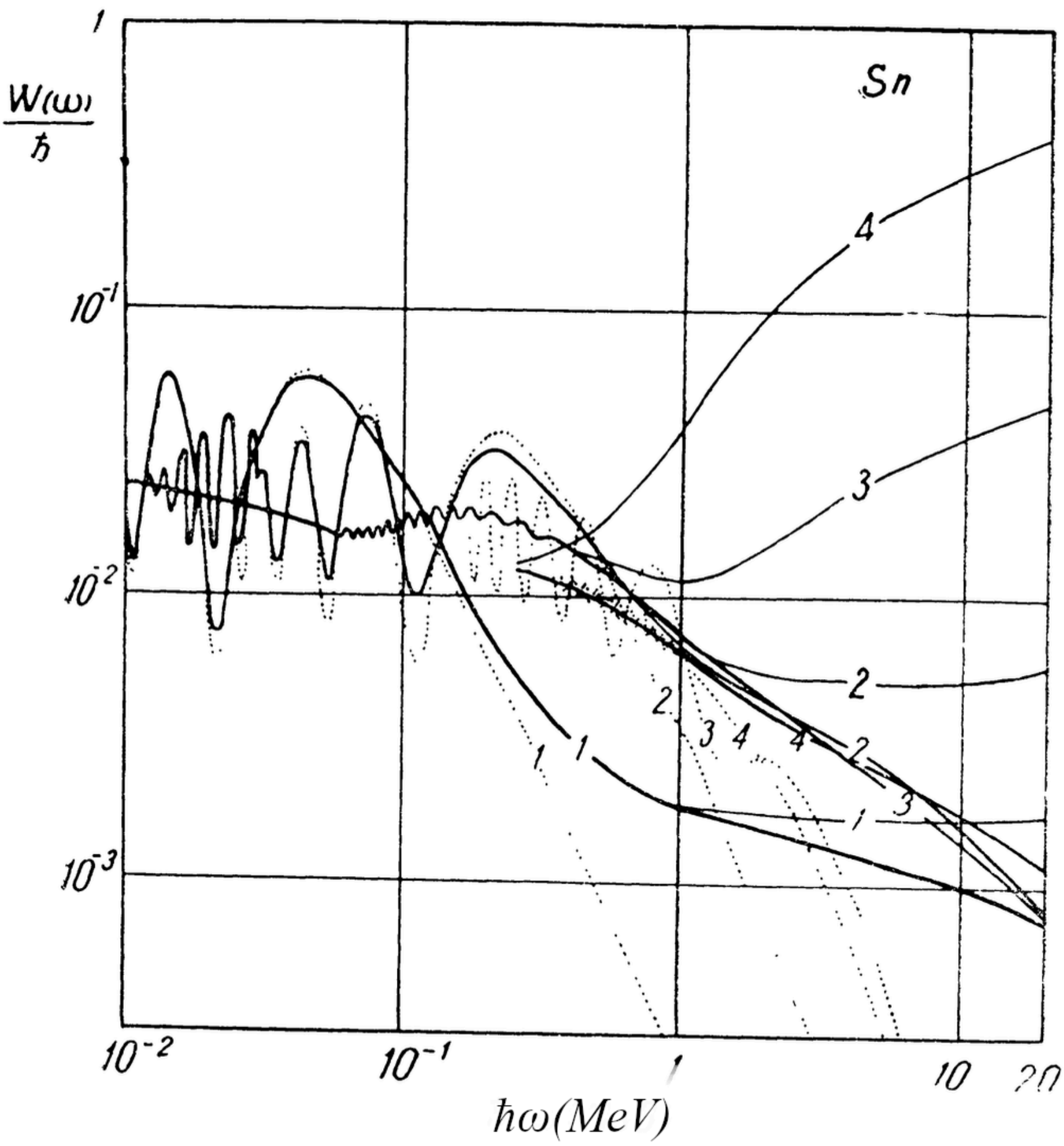

Fig. 16.6. Same as in Fig. 16.5, for $\gamma = 10^5$.

The foregoing is clearly seen in Figs. 16.2–16.6.

On the other hand, it is known that the intensity (averaged over a small frequency range) of ordinary transition radiation in a plate ceases to depend on the thickness $a$ when $a \gg |z^{-}_{med}|$. It is easy to see that condition in Eq. (16.71) is a generalization of the indicated inequality to the case where multiple scattering is taken into account.

## § 17. TWO PLATES

To clarify the role of interference effects between the radiation generated in different plates, let us consider the case where a relativistic particle with Lorentz factor γ passes through two identical plates of thickness $a$ and dielectric permittivity ε(ω), arranged in vacuum parallel to each other and separated by a distance $b$.

The spectral–angular distribution of the intensity of the total radiation is determined by Eq. (16.1), where the quantity **C** consists of contributions from both plates and three vacuum segments. In addition to the terms pertaining to each individual plate (see the diagrams in Fig. 16.1), there are also terms that pertain simultaneously to two plates, i.e., those due to the interference between the radiations formed in different plates (Fig. 17.1). After averaging over the particle trajectories (see Sec. 16.1), we obtain an equation for $W_{\text{tot}}(\omega, \vartheta_0)$, which is not given here because of its cumbersomeness (see Ref. [**80**.3]).

### 17.1. Frequency Spectrum of the Total Radiation

After integrating the spectral–angular distribution $W_{\text{tot}}(\omega, \vartheta_0)$ over the radiation angle $\vartheta_0$, we obtain an equation for the frequency spectrum of the total radiation emitted from two plates:

$$W_{2a}^{\text{tot}}(\omega) = N_{\text{eff}} W_{1a}^{\text{tot}}(\omega) + Q\, W^{\text{int}}(\omega), \qquad (17.1)$$

where $N_{\text{eff}} = 1 + Q$ is the effective number of plates, and $W_{1,a}^{tot}(\omega)$ is the frequency spectrum of the total radiation (Eq. (16.38)) for a plate of thickness $a$. The second term in Eq. (17.1) describes the interference of the radiations emitted from different plates:

$$W^{\text{int}}(\omega) = \frac{2e^2}{\pi c}\text{Re}\left\{\exp(-s_1 x_b)\left[W_0^{\text{int}} + W_a^{\text{int}} + W_b^{\text{int}}\right]\right\}, \qquad (17.2)$$

where the remaining notations are given in Eq. (16.23).

$$W_0^{\text{int}} = 1 + (s_1 x_b - 1)G(s_1 x_b) + \ln x_b + Q\exp(-2s_2 x_a)\times$$

$$\times \left[ \ln(x_b + 2\tanh x_a) - G\left( s_1 \frac{x_b + 2\tanh x_a}{1 - (x_b + \tanh x_a)\tanh x_a} \right) \right] - \qquad (17.3)$$

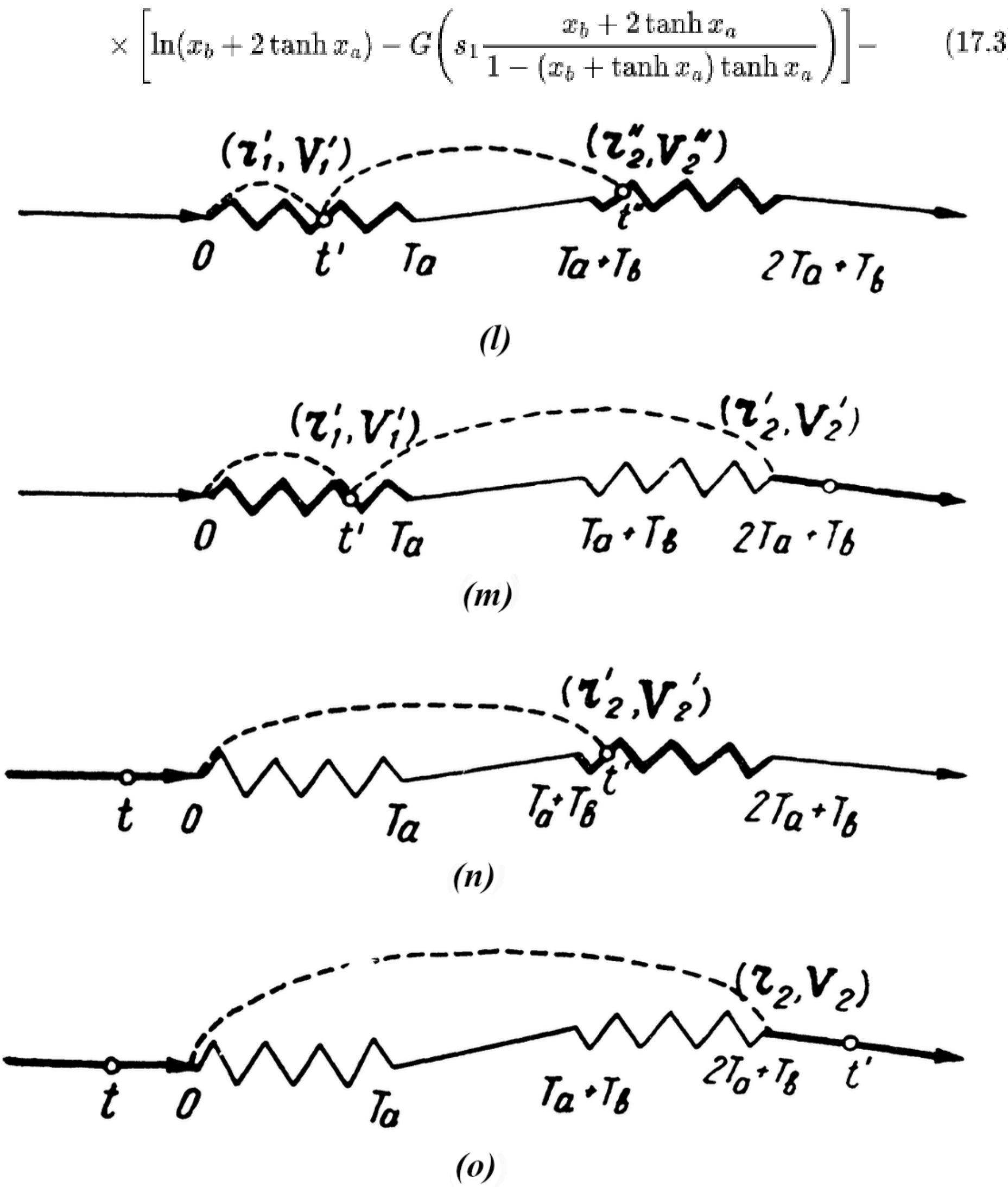

Fig. 17.1. Diagrams due to interference between the radiations formed in different plates. Straight lines correspond to particle motion in vacuum, zigzag lines correspond to motion in the plates. The bold portions of the diagrams indicate the regions of integration. Dashed lines correspond to relative probabilities integrated over the arguments indicated at their right ends

$$-(1+Q)\exp(-s_2x_a)\left[\ln(x_b+\tanh x_a)-G(s_1(x_b+\tanh x_a))\right],$$

$$W_a^{\rm int}=\int_0^\infty \exp(-s_2x)\left\{s_2(h+s_2)\int_0^\infty \exp[-(h+s_2)y]\ln(\tanh x+\tanh y+x_b)\,dy\right.$$
$$\left.+[s_2+(h-s_2)\exp(-hx)]\left(G\left(s_1\frac{\tanh x+\tanh x_a+x_b}{1+(x_b+\tanh x)\tanh x_a}\right)-\ln(\tanh x+\tanh x_a+x_b)\right)\right\}dx, \quad (17.4)$$

$$W_b^{\rm int}=s_1\exp(-s_2x_a)\left\{(1+Q)\exp(-s_1x_b)\int_{x_b}^\infty \exp(-s_1x)\,G\left(s_1\frac{x+\tanh x_a}{1+x\tanh x_a}\right)dx\right.$$
$$\left.-Q\exp(-s_2x_a)\int_0^\infty \exp(-s_1x)\,G\left(s_1\frac{\tanh x_a+a_0+x_b}{1+(x_b+a_0)\tanh x_a}\right)dx\right\}, \quad (17.5)$$

$$x_b=(1-i)\,\gamma^{-1}\sqrt{\omega q_0}\;b/c,$$
$$\alpha_0=(x+\operatorname{th}x_a)/(1+x\operatorname{th}x_a), \quad (17.6)$$

In the limiting case of the absence of multiple scattering ($q_0 \to 0$), Eq. (17.1) yields the spectral distribution of the intensity of ordinary transition radiation for two absorbing plates:

$$W_{2,a}^{\rm TR}(\omega)=N_{\rm eff}W_{1,a}^{\rm TR}(\omega)+\frac{2e^2}{\pi c}Q\,{\rm Re}\left\{\exp(-s_1x_b)\left[G_b(x_b)-\exp(-s_2x_a)(1+Q)G_b(x_a+x_b)+\right.\right.$$
$$\left.\left.+\exp(-2s_2x_a)Q\,G_b(2x_a+x_b)\right]\right\}. \quad (17.7)$$

Here $W_{1,a}^{TR}(\omega)$ is the frequency spectrum of transition radiation from a plate of thickness $a$ with absorption taken into account (see Eq. (2.36)). The auxiliary function $G_b(z)$ has the following form:

$$G_b(z) = 1 + [s_1 z - (d + d^*)] G(s_1 z) + d G(s_2 z) + d^* G((s_2 + h) z), \quad (17.8)$$

where $d = - s_2[1/h + 1/(s_2 - s_1)]$.

If in Eq. (17.1) the width $b$ of the vacuum gap is taken to zero, then, naturally, one obtains the spectral distribution for a single plate with doubled thickness, $W^{tot}_{1,2a}(\omega)$.

Finally, under the condition $|s_1 x_b| \gg 1$, i.e., $b > c\gamma^2/\omega$ (the plates are separated by a distance much larger than the formation length $z^-_{vac}$ of transition radiation in vacuum), the term $W_{int}(\omega)$ oscillates rapidly and, after averaging over a small frequency range, makes no contribution to expression (17.1). As a result, for the frequency spectrum of the total radiation we obtain the corresponding expression for a single plate of thickness $a$, multiplied by the effective number of plates, $N_{eff} W^{tot}_{1,a}(\omega)$.

## 17.2. Edge Effect for Two Plates

As it passes through a stack of plates, a charged particle moves freely in the vacuum gaps and undergoes multiple scattering on the atoms of the material inside the plates. Therefore, the total radiation produced in a stack of absorbing plates consists of bremsstrahlung over the path length $N_{eff} a_{eff}$ ($N_{eff} = (1 - Q^N)/(1 - Q)$, $a_{eff} = (1 - Q)/\eta$) and the edge effect, or "transition radiation," with multiple scattering taken into account.

The concept of the edge effect introduced in Section 16 for a single plate can be generalized to the case of an arbitrary number $N$ of absorbing plates:

$$W^{\rm edge}_{N,a}(\omega) = W^{\rm tot}_{N,a}(\omega) - \frac{N_{\rm eff} a_{\rm eff}}{z_{\rm brem}} W_M(\omega), \quad (17.9)$$

where $W^{tot}_{N,a}(\omega)$ is the frequency spectrum of the total radiation emitted from a stack consisting of $N$ plates of thickness $a$, and the quantity $W_M(\omega)$ is defined by Eq. (16.40). In particular, in the case under consideration $N = 2$, and taking into account Eqs. (17.1) and (16.43), we obtain

$$W_{2,a}^{\text{edge}}(\omega) = (1+Q)W_{1,a}^{\text{edge}}(\omega) + Q\,W^{\text{int}}(\omega). \qquad (17.10)$$

It follows that the analysis of edge effect for a single plate carried out in Section 16 is fully transferable to the case of two plates. It is necessary to conduct a study of the second interference in Eq. (17.10).

### 17.3. Analysis of the Interference Term

As in Sec. 16.3, we consider the interference term for two limiting values of the quantity $|s_2|$.

A. Let

$$|s_2| \gg 1 \qquad (|\bar{z}_{\text{med}}| \ll z_{\text{brem}}), \qquad (17.11)$$

Then, by virtue of the inequality $|s_2| \gg \max\{|s_1|, |s_2+h|\}$, the main contribution to the integrands of Eqs. (17.4) and (17.5) comes from small x. Expanding in the small parameter $x$ and substituting the leading term of the expansion into Eq. (17.1), we obtain expression (17.7) for ordinary transition radiation in the case of two absorbing plates, $W_{2,a}^{TR}(\omega)$. This result is quite natural, since condition (17.11) defines the region where the transition mechanism (having a shorter formation length) dominates over the bremsstrahlung mechanism of radiation.

B. Let the opposite condition now be satisfied:

$$|s_2| \ll 1 \qquad (|\bar{z}_{\text{med}}| \gg z_{\text{brem}}). \qquad (17.12)$$

We distinguish two cases.

1. When $|x_a| \leq 1$, one can show that

$$\begin{aligned} W^{\text{int}}(\omega) = \frac{2e^2}{\pi c}\text{Re}\Big\{ &\exp(-s_1 x_b)\Big[1 + 2G(s_1/x_a) - \\ &-G\Big(s_1\frac{x_b+1/x_a}{2+x_a x_b}\Big) + \big[1+(h+2s_2)x_b\big]\ln\frac{x_b(2x_a+x_b)}{(x_a+x_b)^2} - \\ &-\big[(h+2s_2-s_1)x_b+1\big]G(s_1 x_b) - \big[(2s_2+h)(2x_a+x_b)+1\big]\times \\ &\times G\big(s_1(2x_a+x_b)\big) + 2\big[(h+2s_2)(x_a+x_b)+1\big]G\big(s_1(x_a+x_b)\big)\Big]\Big\}. \end{aligned} \qquad (17.13)$$

for $|x_a| \gg |s_1|$;

$$W^{\text{int}}(\omega) = \frac{2e^2}{\pi c}\text{Re}\left\{ \exp(-s_1 x_b)\Big[\big[1+(h+2s_2)x_b\big]\ln\frac{x_b(2x_a+x_b)}{(x_a+x_b)^2} - \right.$$
$$-\big[(h+2s_2-s_1)x_b+1\big]G(s_1x_b) - \big[(2s_2+h-s_1)(2x_a+x_b)+1\big]\times \qquad (17.14)$$
$$\left.\times G\big(s_1(2x_a+x_b)\big) + 2\big[(2s_2+h-s_1)(x_a+x_b)+1\big]G\big(s_1(x_a+x_b)\big)\Big]\right\}$$

for $|x_a| \gg |s_1|$;

If in Eq. (17.13) we set the vacuum gap width $b = 0$, we immediately obtain the following expression:

$$W^{\text{int}}(\omega) = \frac{2e^2}{\pi c}\text{Re}\left\{1 + C + \ln\left|\frac{2s_1}{x_a}\right|\right\}, \qquad (|x_a| \gg |s_1|). \qquad (17.15)$$

which, as expected, represents the difference between the magnitude of the edge effect in a plate of doubled thickness $2a$ and twice the magnitude of the edge effect in a plate of thickness $a$. If, however, one lets $b$ tend to zero in Eq. (17.14), then $W^{int}(\omega)$ becomes zero, since in this case the bremsstrahlung mechanism of radiation is dominant, and the aforementioned difference for bremsstrahlung tends to zero.

2. When $|x_a| \gg 1$, then

$$W^{\text{int}}(\omega) = \frac{2e^2}{\pi c}\text{Re}\left\{ \exp(-s_1 x_b)\Big[1 + (s_1 x_b - 1)G(s_1 x_b) + \right.$$
$$\left. + 2G\big(s_1(x_b+1)\big) - \ln\frac{x_b(2x_a+x_b)}{(x_a+x_b)^2}\Big]\right\}, \qquad \text{for } |x_a| \gg |s_1|. \qquad (17.16)$$

Here the form of $W^{int}(\omega)$ is given only for the case $|x_a| \gg |s_1|$, since the conditions $|x_a| \ll |s_1|$ and $|x_a| \gg 1$ are incompatible.

Let us analyze Eq. (17.16) in greater detail. Here one can obtain the explicit behavior of the interference term as a function of the width $b$ of the vacuum gap.

(1) Let $|x_b| \ll 1$ ($b \ll z_{brem}$). Then

$$W^{\text{int}}(\omega) \approx \frac{2e^2}{\pi c}\text{Re}\left\{1 + C + \ln|2s_1|\right\}. \qquad (17.17)$$

Note that this expression, up to an overall sign, coincides with the quantity $W^{edge}_{1,a}(\omega)$ defined by Eq. (16.65) under the conditions $|s_2| \ll 1$ and $|x_a| \gg 1$.

(2) When $|x_b| \sim 1$ ($b \sim z_{brem}$), we obtain

$$W^{\text{int}}(\omega) \approx \frac{2e^2}{\pi c}\text{Re}\left\{1 + C + \ln\left|s_1\frac{x_b(2+x_b)}{(1+x_b)^2}\right|\right\}. \qquad (17.18)$$

(3) If $|x_b| \gg 1$ ($b \gg z_{brem}$), the interference term has the following form:

$$W_{\text{int}}(\omega) = \begin{cases} \frac{2e^2}{\pi c}\text{Re}\{1 + C + \ln|s_1 x_b|\}, & \text{for } |s_1 x_b| \ll 1, \\ \frac{2e^2}{\pi c}\text{Re}\{\exp(-s_1 x_b)[1 + (s_1 x_b - 1)G(s_1 x_b)]\}, & \text{for } |s_1 x_b| \sim 1, \\ 0, & \text{for } |s_1 x_b| \gg 1, \end{cases} \qquad (17.1)$$

From the above analysis it follows that the interference pattern of radiation from individual plates in the stack depends substantially on the formation length $z^{-}_{vac}$ of transition radiation in vacuum. Moreover, in the region where the bremsstrahlung mechanism is dominant (the condition $|s_2| \ll 1$ holds), the situation is complicated by the fact that the bremsstrahlung formation length $z_{brem}$ also plays a significant role.

Figures 17.2 and 17.3 show the spectra of the total radiation $W^{tot}_{2,a}(\omega)$ and of the edge effect $W^{edge}_{2,a}(\omega)$ for two tin plates of thickness $a = 100$ μm, placed in vacuum at different separations $b$ and for different values of the electron Lorentz factor $\gamma$. For comparison, the spectra $W^{TR}_{2,a}(\omega)$ of ordinary transition radiation are also shown. Note that in Fig. 17.2 the spectra are shown in the frequency range 0.03–1 MeV because for $\gamma = 10^4$ the cutoff frequency of the spectrum is of the order of 0.3 MeV, and therefore the interference phenomena in the spectrum disappear.

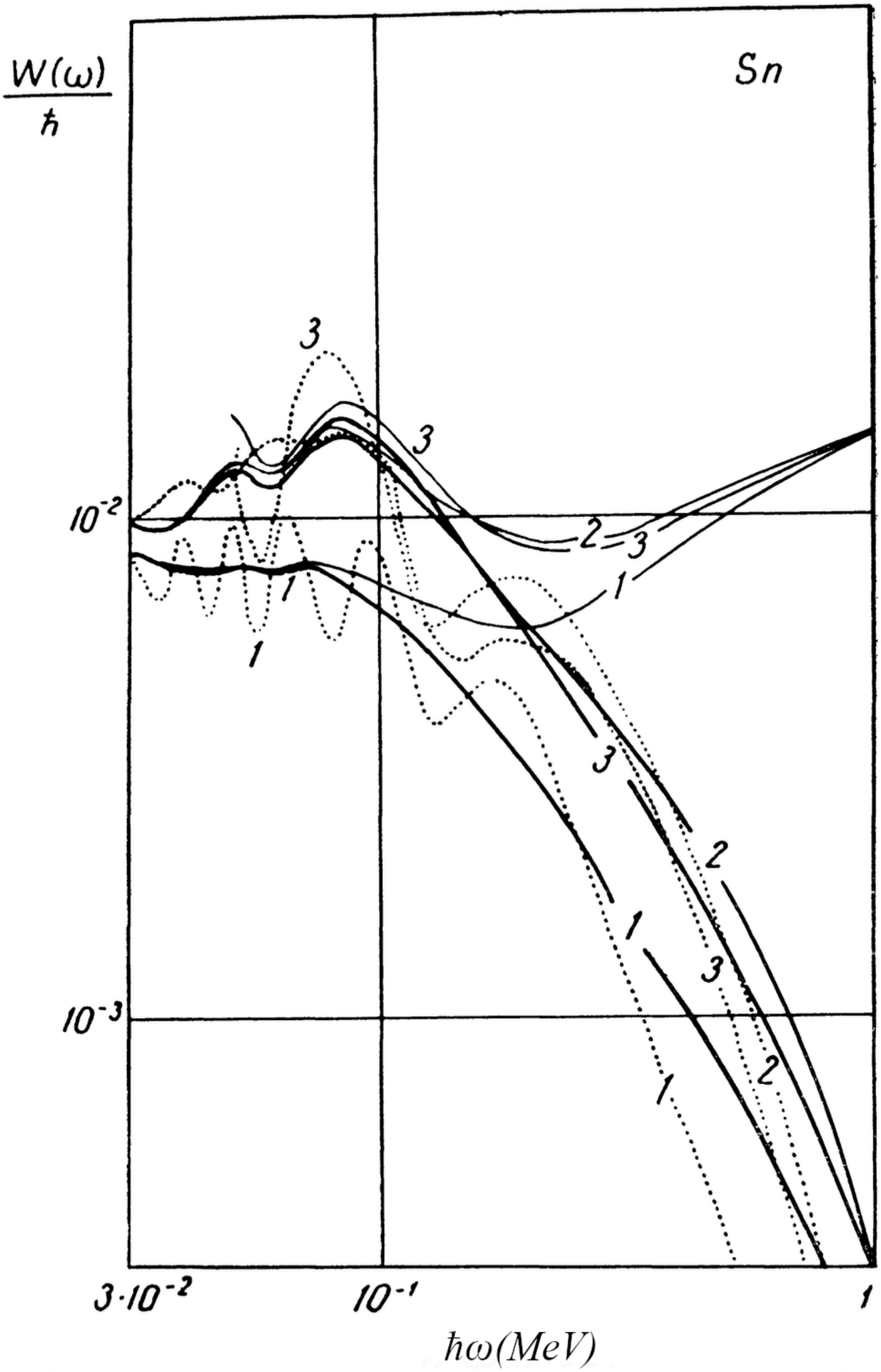

Fig. 17.2. Frequency spectra of the total radiation $W_{2,a}^{tot}(\omega)$ (thin solid curves) produced in two tin plates of thickness $a = 100$ μm by an electron with Lorentz factor $\gamma = 10^4$. Bold solid curves: edge effect; dotted curves: ordinary transition radiation. The numbers by the curves correspond to different values of the vacuum gap length $b$: —1, 2—100, 3—$10^4$ μm.

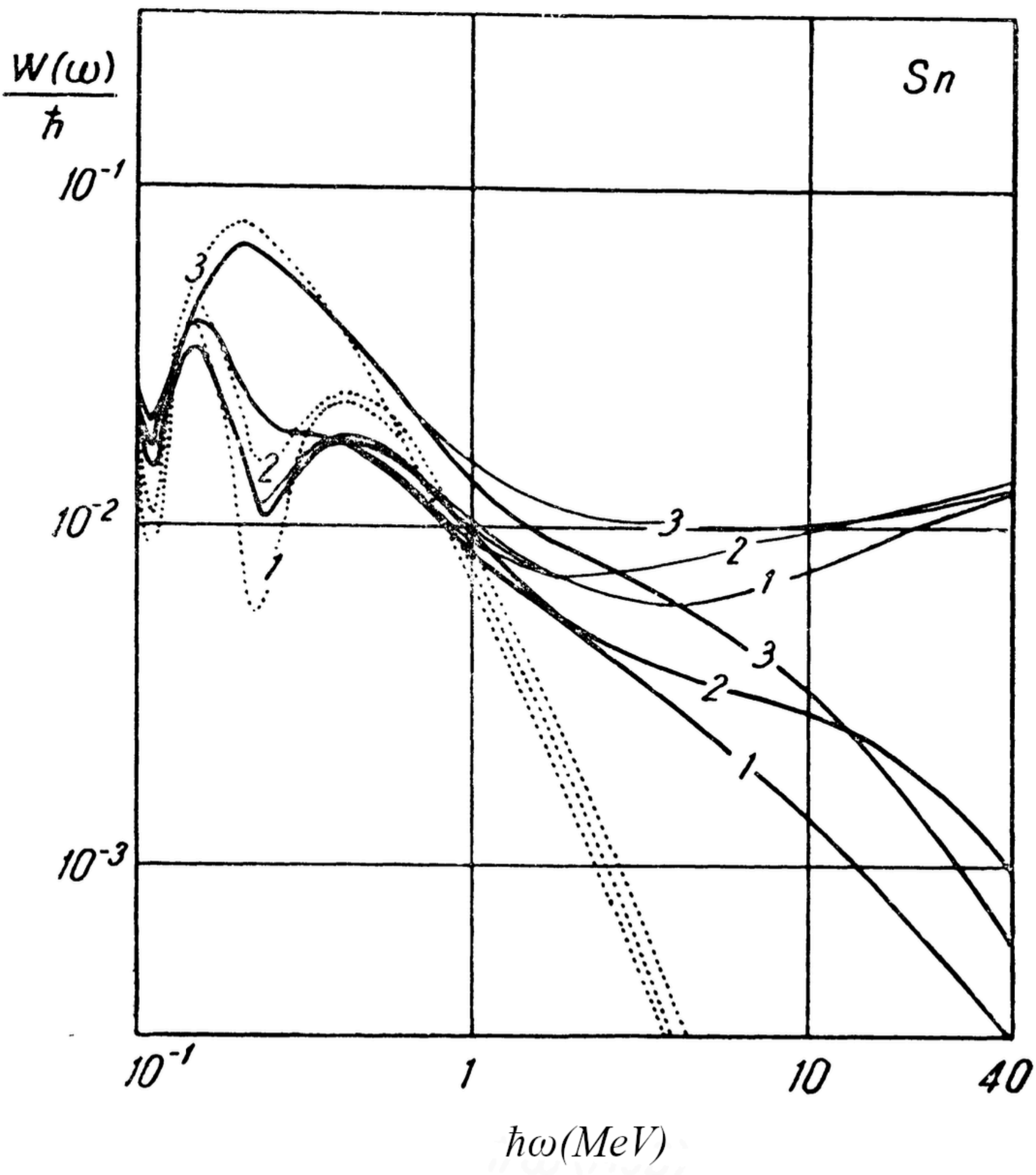

*Fig. 17.3. Same as in Fig. 17.2, for* $\gamma = 10^5$.

The figures clearly show the smoothing of interference extrema and spectral hardening of the edge effect compared to the spectrum of ordinary transition radiation. It can also be seen that when $b \ll z^{-}_{vac}$, these spectra coincide with the corresponding spectra for a single plate with doubled thickness $2a$. This is manifested in the fact that the oscillations become twice as frequent as in the case of a single plate of thickness a (see

Figs. 16.5, 16.6). With increasing b, the “extra” minima in the spectra gradually rise and eventually disappear when $b \sim z^{-}_{vac}$. When $b > z^{-}_{vac}$, these spectra practically coincide with the corresponding spectra for a plate of thickness $a$, multiplied by $N_{eff}$.

For $\gamma = 10^{6}$, the quantity $z^{-}_{vac}$ becomes very large (for example, at $\omega = 0.1$ MeV we have $z^{-}_{vac} \approx 6$ m); however, numerical calculation shows that at $b = 100$ μm the intensity is $N_{eff} W^{tot}_{1,a}(\omega)$, rather than $W^{tot}_{1,2a}(\omega)$. This is explained by the fact that, due to the contribution of large effective radiation angles $\sim (\omega_0^2/\omega^2 \cdot \gamma^{-2})^{1/2}$ (see Sec. 1.8), only when the condition $b \ll |z^{-}_{ext}|$ is satisfied does the radiation spectrum for two plates merge with the corresponding spectrum for a single plate with doubled thickness.

It is interesting to note that, in addition to the smoothing of extrema, there is also a certain shift of the maxima and minima of the spectrum of the total radiation (or the edge effect) relative to the spectrum of ordinary transition radiation. Such a shift was absent in the case of a single plate (Figs. 16.4–16.6). Apparently, the presence of this shift is due to interference between the radiation emitted from different plates.

## § 18. STACK OF AN ARBITRARY NUMBER OF PLATES

On the basis of the general Eq. (16.1), one can find the spectral–angular distribution of the intensity of the total radiation emitted by a particle traversing a stack consisting of an arbitrary number N of regularly spaced plates. In this case, all N + 1 vacuum regions contribute to the quantity $B$. Averaging over all possible trajectories is carried out essentially as in Sec. 16.1. However, performing such averaging in the general case is difficult. Explicit expressions can be obtained in two special, but sufficiently important for practical purposes, cases.

## 18.1. Large Spacings Between Plates

The spectral–angular and spectral distributions of the intensity of the total radiation for a stack, as for two plates (see Sec. 17.1), contain interference terms corresponding to diagrams similar to those shown in Fig. 17.1. These terms are proportional to the quantity $exp(-s_1 x_b)$ or to its higher powers, and therefore oscillate rapidly for separations between the plates much larger than the formation length $z^{-}_{vac}$ of transition radiation in vacuum:

$$|s_1 x_b| \gg 1 \qquad (b \gg z_{\mathrm{vac}}). \qquad (18.1)$$

After averaging over a small frequency range, the interference terms vanish.

As a result, for the spectral distribution of the total emission of the stack we have

$$W^{\mathrm{tot}}_{N,a}(\omega) = N_{\mathrm{eff}} W^{\mathrm{tot}}_{1,a}(\omega), \qquad (18.2)$$

where $N_{eff} = (1 - Q^N)/(1 - Q)$, $Q = exp(-\gamma a)$, $\gamma = \varepsilon'' \omega/c$ is the linear absorption coefficient, and $W^{tot}_{1,a}(\omega)$ is determined by Eq. (16.38).

## 18.2. Nearest-Neighbor Approximation and the Possibility of its Refinement

When condition (18.1) is not satisfied, it is necessary to take into account the interference between different plates and gaps. The terms corresponding to interference between regions separated from each other by one plate or more (for example, diagrams *m*, *n*, *o* in Fig. 17.1) contain the factor $sh^{-1} x_a$ (or its higher powers). This complex factor is small in modulus for $|x_a| \gg 1$. An analogous situation also occurs when the plate thickness is much larger than the absorption length, i.e., $\gamma a \gg 1$. In the indicated cases one can restrict oneself to accounting only for interference between "nearest neighbors," for example, diagram *l* in Fig. 17.1.

In the nearest-neighbor approximation we obtain

$$W_{N,a}^{\text{tot}}(\omega) = N^{\text{eff}} W_{1,a}^{\text{tot}} + \frac{2e^2}{\pi c}(N_{\text{eff}} - 1)\,[\text{ci}(x_\gamma) - \cos x_\gamma + x_\gamma \sin x_\gamma$$

$$+ x_\gamma \sin x_\gamma - Q\,\text{Re}\,\{\exp(ix_\gamma)\int_0^{x_a} dx_1 \int_0^{x_a} dx_2 \exp[-s_2(x_1 + x_2) - hx_1] \times$$

$$\times [\text{sh}\,(x_1 + x_2) + x_b\,\text{ch}\,x_1\,\text{ch}\,x_2]^{-2}\} + \text{Re}\,\{\exp(ix_\gamma)\int_0^{x_a} [\exp[-(s_2 + h)x] +$$

$$+ \exp(-s_2 x)]\,[(\text{th}\,x + x_b)^{-1} + s_1 \exp[s_1(\text{th}\,x + x_b)] \times$$

$$\times \text{Ei}(-s_1(\text{th}\,x + x_b))]\,(\text{ch}\,x)^{-2} dx\}, \qquad b \gg c/\omega. \qquad (18.3)$$

When $|x_a|$ (or $\eta a$) is of order unity or, even more so, smaller than unity, the nearest-neighbor approximation becomes inadequate. To refine it, one must take into account terms due to interference between increasingly distant neighbors. First of all, these are the terms corresponding to regions separated from each other by one plate, then by two plates, and so on. As a result, we obtain a certain series that converges quite rapidly, starting from terms for which $m|x_a| \sim 1$ (or $m\eta a \sim 1$), where $m$ is an integer (the order of the approximation). Physically, this means that in the presence of multiple scattering the wave phases become dephased (and in the presence of absorption the wave amplitudes decrease) and therefore the interference between waves arising in segments of the particle trajectory that are sufficiently far apart becomes negligible.

### 18.3. Edge Effect in a Stack of Plates and Experimental Separation of the Edge Effect from the Total Radiation

As stated in Sec. 17, the edge effect in a stack of plates can be defined as follows:

$$W^{\mathrm{edge},N,a}(\omega) = W_{\mathrm{tot},N,a}(\omega) - \frac{N_{\mathrm{eff}}a_{\mathrm{eff}}}{z_{\mathrm{brem}}}W_M(\omega). \quad (18.4)$$

If the distance $b$ between the plates is much greater than the formation length of transition radiation in vacuum, then, as noted in Sec. 18.1, the intensity of the total radiation from the entire stack is equal to the intensity of the total radiation from a single plate multiplied by $N_{\mathrm{eff}}$. An analogous relation also holds for the edge effect:

$$W_{N,a}^{\mathrm{edge}}(\omega) = N_{\mathrm{eff}}W_{\mathrm{edge},1,a}(\omega), \qquad \text{for } b \gg z_{\mathrm{vac}}. \quad (18.5)$$

In other words, for a stack with large spacings between the plates, all the results obtained in Sec. 16 concerning total radiation and the edge effect in a single plate hold.

The influence of multiple scattering of a particle on the formation of radiation in a non-absorbing periodic medium was considered by Ter-Mikaelian in a model where the dielectric permittivity of the medium has a periodic inhomogeneity, and the probability of scattering of the particle is assumed to be the same throughout the medium [**69**.1]. The equation for the spectral distribution of the intensity of the total radiation in this model also consists of two terms. The first term corresponds to bremsstrahlung without taking into account the inhomogeneity of the dielectric permittivity, while the second describes the effect due to the presence of inhomogeneity in the dielectric permittivity, in particular, the presence of boundaries. In this sense, the second term corresponds to the boundary effect in the equal-probability scattering model.

Figure 18.1 schematically shows the trajectory of a particle in the isotropic scattering model (lower curve). The same figure shows the trajectory in a more realistic theory, in which the particle undergoes scattering only inside the plates (upper curve).

Let us compare the results of these two theories. For the comparison it is necessary to assume that the mean square of the multiple-scattering angle over one period of the medium (equal to the sum of the plate thickness *a*a and the vacuum gap b) has the same value in both theories. Let us first consider the total radiation. One can verify that, up to small correction terms, the frequency spectra of the total radiation emitted by a single period of the medium agree in both theories. Indeed, in the range of values of ω and γ where bremsstrahlung is the dominant contribution to the total radiation, the intensities of the latter in both theories practically coincide, because in both the bremsstrahlung intensity is the same when boundary effects are neglected. In the range of values of ω and γ where bremsstrahlung is small, the total radiation practically coincides with conventional transition radiation, which depends only on the distribution of the dielectric permittivity, the same in both theories.

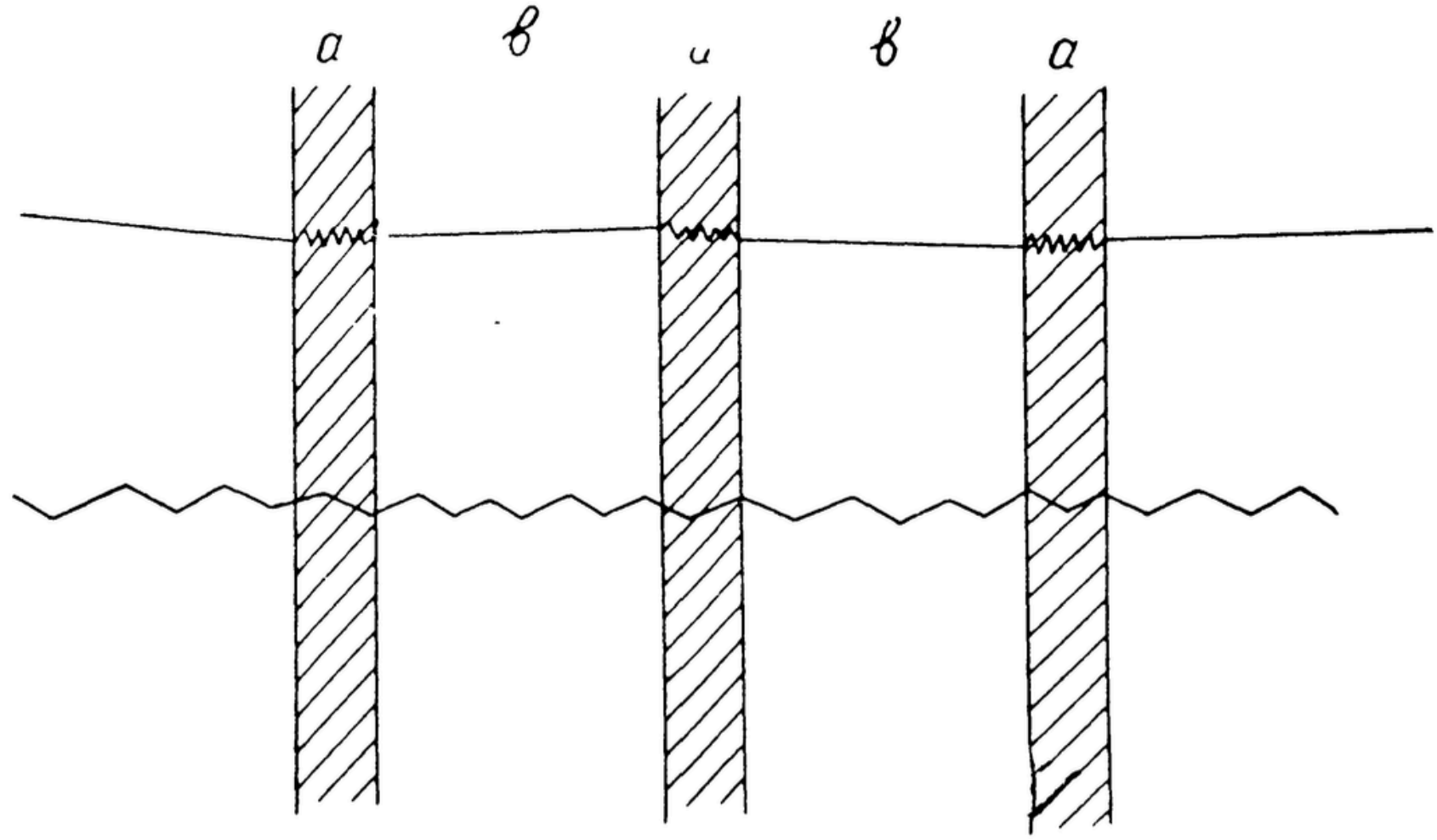

Fig. 18.1. Trajectory of a charged particle passing through a stack of plates (upper curve), and the trajectory in the model of uniformly distributed nuclei (lower curve)

Let us now consider the edge effect. As noted in Section 16, for $\gamma \ll \gamma_c$ the formation length of bremsstrahlung for all frequencies exceeds the corresponding formation length of transition radiation in matter, i.e., the inequality $|s_2| \gg 1$ holds. In this case, the magnitude of the edge effect is essentially close to the spectrum of ordinary transition radiation. Therefore, the results of both theories should practically coincide in the indicated range of γ. Some difference may arise in the degree of smoothing of the interference maxima and minima, as well as in the region of high frequencies ω ≫ ω0γ. The results of the theories should also coincide for $\omega \ll \omega 1 = (\omega 0^4/q)^{1/3} and \gamma > \gamma_c$ (in this region we still have |S2| ≫ 1). However, for $\gamma \gg \gamma c$ and

$\omega > \omega 1$ the results of the theories will be different, since in the equiprobable scattering model the particle is scattered uniformly everywhere (Fig. 18.1) and therefore the peculiar enrichment of the spectrum (Sec. 16.4) that occurs when scattering takes place only inside the plates is absent.

This is also clearly seen from the results of a numerical calculation of the frequency dependence of the edge effect (Fig. 18.2). In the case of a light material (Mylar) the plotted curves in both theories practically coincide[21]. In the case of tungsten, however, for $\gamma \gg 10^4$ and $\omega > 100$ keV the difference between the two approaches is quite significant.

Experimentally, what is directly observable is always the total radiation; therefore, the question arises of experimentally isolating the edge effect from the total radiation.

As was noted in Sec. 16.4, when the thickness $a$ of the plate is much greater than the formation length of the bremsstrahlung or the transition radiation, i.e.,

$$a \gg \min\{z_{\text{brem}}, |z_{\text{med}}|\}. \qquad (18.6)$$

the edge effect becomes independent of $a$. The quantity $W_{M(\omega)}$ is also independent of $a$. It follows that, in order to experimentally isolate the edge effect from the total radiation for a plate of thickness $a$ under study, it is generally necessary to perform three measurements. Two measurements must be performed on a plates with thicknesses $a_1$ and $a_2$, satisfying the condition in Eq. (18.6), for determining the quantity $W_M(\omega)/z_{brem}$ from Eq. (16.39), while the third measurement is performed on the plate under study. Then, using the same Eq. (16.39), the quantity $W^{edge}_{1,a}(\omega)$ is determined from the measured value $W^{tot}_{1,a}(\omega)$ and the obtained value $W_M(\omega)/z_{brem}$.

[21] An attempt was made by Arutyunyan and collaborators to compare their experimental results [**66**.11, **67**.11, **72**.31], obtained at the electron accelerator of the Physics Institute of the Academy of Sciences, with theoretical assumptions (with and without accounting for multiple scattering). However, the aforementioned work [**72**.31] presents erroneous numerical results of the calculation [**69**.12] using the Ter-Mikaelian equation [**69**.1, p. 368], which accounts for multiple scattering (see Ref. [**77**.7]). The excesses of the experimental data (in some cases by tens of times) over the theory are in fact explained by causes of instrumental origin (see Refs. [**73**.35, **74**.12, **77**.1, p. 41]).

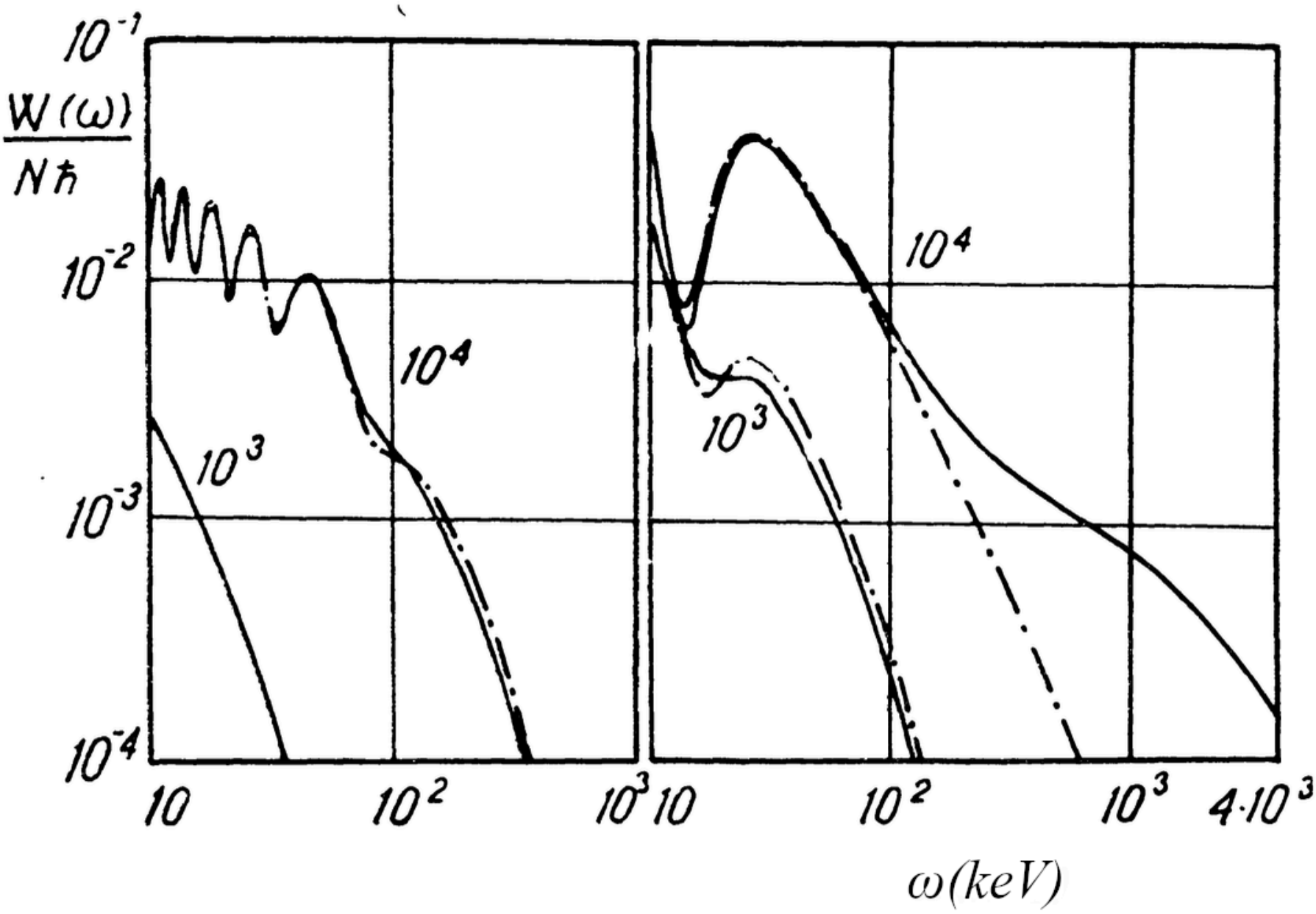

Fig. 18.2. Frequency spectra of the edge effect in a stack of Mylar ($\hbar\omega_0$ = 24. 4 eV, *L*$_{rad}$ = 28.7 cm, left-hand curves) and tungsten ($\hbar\omega_0$ = 80. 3 eV, *L*$_{rad}$ = 0.35 cm, right-hand curves) plates. The solid curves are calculated according to Eqs. (18.5) and (16.38), the dash-dotted curves are calculated according to the model of equiprobable particle scattering [69.1, p. 368]. The numbers by the curves indicate the value of the γ-factor. For $\gamma = 10^3$ in the case of Mylar both curves coincide. In all cases the spectrum of ordinary transition radiation (without taking multiple scattering of the particle into account) coincides with the corresponding dash-dotted curves to within a few percent

To isolate the edge effect in a stack of plates it is first necessary, in an analogous way, to determine $W_M(\omega)/z_{brem}$ from two measurements on individual plates, and then to perform a third measurement on the stack of plates under study.

Sometimes, however, the edge effect ("transition radiation" with allowance for multiple scattering) is taken to be the difference between the total radiation produced in the stack under study and that produced in a plate of thickness $Na$, i.e., the quantity

$$W_{N,a}^{\text{tot}}(\omega) = W_{1,Na}^{\text{tot}}(\omega). \qquad (18.7)$$

This difference obviously cannot describe the edge effect in the limiting cases: $N \to 1$, $b \to 0$, as well as $\gamma_1 a \gg 1$. It does not always correspond to the edge effect defined by Eq. (18.4), even when $N_{eff} \gg 1$ (i.e., $N \gg 1$ and $\gamma_1 a \ll 1$) and $b \gg z_{vac}$. Indeed, substituting Eq. (18.4) into Eq. (18.7), and also bearing in mind Eq. (18.5), we obtain that the difference in Eq. (18.7) is equal to

$$N^{\text{eff}} W_{1,a}^{\text{edge}}(\omega) - W_{1,Na}^{\text{edge}}(\omega). \quad (18.8)$$

This quantity approximately corresponds to the edge effect in a stack, i.e., to the first term, if the second term is much smaller than the first. However, this is not the case, for example, when $a \ll |z_{med}|$.

# CHAPTER V

## EXPERIMENTAL STUDIES

Modern accelerator technology and experimental high-energy particle physics deal with energies on the order of hundreds of GeV, and in the coming years energies on the order of thousands and, possibly, tens of thousands of GeV will be mastered. Somewhat earlier, experiments with cosmic rays were initiated in this energy domain. At such energies, traditional methods of particle registration and identification (recognition) (for example, using Cherenkov detectors or by measuring ionization losses in the region of relativistic velocities) become practically inapplicable.

Fundamentally new possibilities are provided by the use of X-ray transition radiation detectors (TRDs)[22]. It is quite obvious that, in order to develop such detectors with optimal performance characteristics, it is necessary, alongside theoretical calculations, to carry out a comprehensive experimental study of the properties of XTR.

## § 19. EXPERIMENTAL STUDIES OF THE XTR PROPERTIES

As stated in the Introduction, X-ray transition radiation in a layered medium produced by muons in cosmic rays was first observed by Arutyunyan, Ispiryan, and Oganesyan of the Yerevan Physics Institute in 1964 [**64**.3, **65**.6].

The first measurements of the XTR intensity (from 6 to 200 keV) emitted by positrons with energy 2 GeV ($\gamma$ ~ 4,000) in a regular stack consisting of $N$=231 aluminum plates (plate thickness $a$=25 μm, spacing between plates $b$=0.3 mm) were carried out by Yuan and collaborators [**69**.4] at the Cambridge Electron Accelerator (USA) in 1969. In this facility the positrons were deflected by a magnet, and the XTR was recorded with a *germanium detector*. A little later [**70**.2–**70**.4] these authors, using the same method, for the first time investigated the dependence of the XTR spectrum on the positron energy. They found that in the range $\gamma$=1,000–8,000 the total XTR energy (from 3 to 270 keV) depends almost linearly on $\gamma$.

By the end of 1969, Alikhanyan, Lorikyan, and others proposed and implemented [**70**.6, **70**.7] at the electron accelerator of the Yerevan Physics Institute a new method of XTR detection using a *streamer chamber* with an admixture of the heavy gas xenon (Xe). The use of a streamer chamber enabled separate observation of both the radiation and the particle, while the presence of Xe led to high efficiency in recording XTR quanta.

[22] A review of the current state of particle detectors, in particular TRDs, is given in Refs. [**80**.15, **82**.6] (see also [**75**.33, **77**.1, **82**.12]).

This facility is shown schematically in Fig. 19.1. The electrons passed through the radiator and entered the streamer chamber located 11 m from the radiator, as a result of which the photoelectrons produced by XTR quanta in the streamer chamber were spatially separated from the track of the primary ionizing particle (Fig. 19.2). The value of the proposed method lies primarily in its clarity and simplicity. At the same time, for the quantitative determination of the XTR intensity, the method required further development.

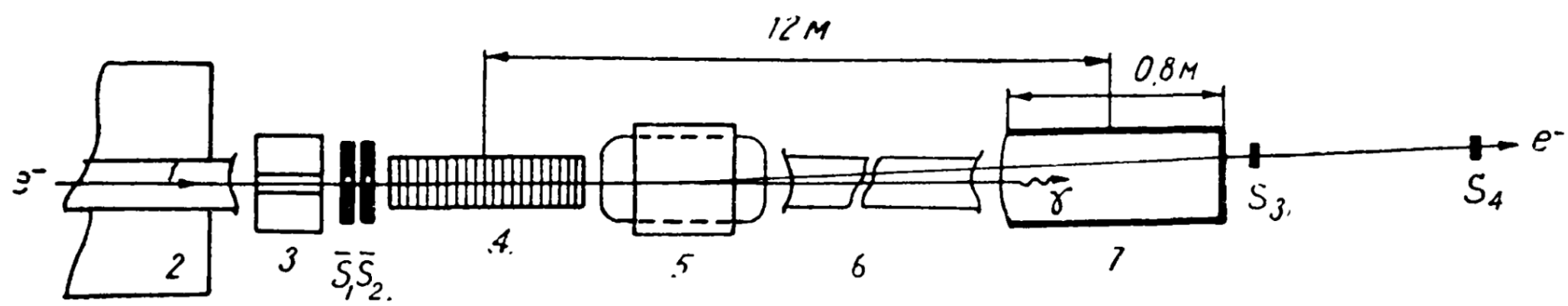

Fig. 19.1. Layout of the facility for investigating XTR using a streamer chamber, developed at the Yerevan Physics Institute [**70**.6, **70**.7]: 1—electron beam, 2—shielding, 3—collimator, 4—XTR radiator, 5—deflecting magnet, 6—vacuum pipe, 7—streamer chamber, $\bar{S}_1$, $\bar{S}_2$, $S_3$, $S_4$—scintillation counters.

An important result of the cited work was that, at Lorikyan's suggestion, an irregular porous medium—*plastic foam*—was used for the first time as an XTR radiator, and it was experimentally shown that XTR is produced also in an irregular medium provided it contains a sufficient number of interfaces. In this case, the total number of XTR photons emitted from the irregular medium is comparable in order of magnitude to the number of XTR photons emitted from a similar regular stack. In addition, that work also measured the dependence of the number of XTR photons on the electron energy (1.2 and 2 GeV) and found that this dependence is almost linear.

At the same time Alikhanyan, Ispiryan, Oganesyan, et al. [**70**.5] investigated XTR produced by electrons with energies from 0.4 to 4 GeV in a regular stack consisting of polyethylene layers. XTR was recorded by a CsI(Tl) scintillator crystal with a hole in the middle. According to the authors' design, the electrons were to pass through the hole, while the XTR photons emitted within a certain angle range. However, the actual situation was considerably more complex due to scattering of electrons in the plates of the stack leading to bending of their trajectories.

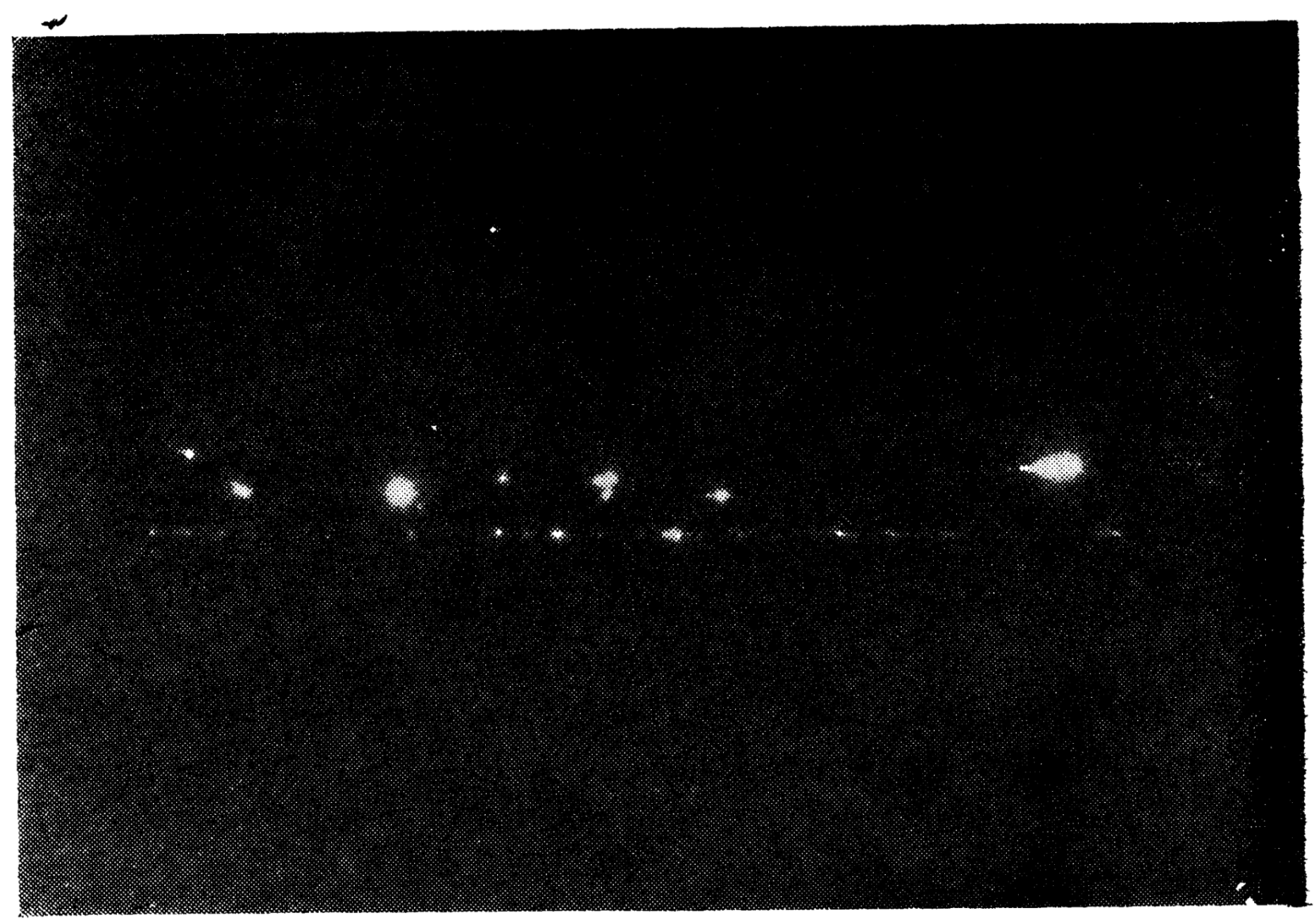

Fig. 19.2. Photograph of an electron track (straight segment) and the traces of XTR photons absorbed in the gas of a streamer chamber [**73**.9]. Owing to the presence of a magnet after the XTR radiator (see Fig. 19.1) the particle track is slightly deflected, and therefore the photons are located only on one side of the track. Electron energy is of order 5 GeV

In 1971, Yuan's group (USA) first used *multiwire proportional chambers* filled with a high-Z gas for the detection of XTR on a 10 GeV Cornell synchrotron [**71**.6]. Energy deposition occurred in the chambers (see Ref. [**61**.4]), due both to ionization of the gas caused by the traversing charged particle and to the absorption of quanta of the XTR accompanying the particle. Further development of this method was carried out in works [**72**.9, **73**.13, **74**.2, **74**.13, **75**.11, **75**.12, **75**,15]. In these works, in addition to regular stacks, irregular layered media were also used as radiators of transition radiation. At present, the method based on the use of multiwire proportional chambers is widely employed in TRDs (see Sec. 20).

### 19.1. Regular Stacks

A detailed experimental study of the properties of the XTR formed in regular stacks was carried out in the works [**70**.3, **70**.4, **71**.2, **73**.12, **74**.13, **75**.27–**75**.29, **77**.8, **77**.11, **78**.2, **80**.13]. The frequency spectra of X-ray transition radiation were measured in the aforementioned work [**70**.3], as well as in [**73**.12] for a stack of organic films and an electron energy of 3 GeV (Fig. 19.3). Measurements were carried out using a multisection proportional chamber (at $\hbar\omega$=8–20 keV) and a NaI(Tl) scintillator (at $\hbar\omega$=20–100 keV). Comparison with theory showed good agreement.

In Refs. [**70**.4, **74**.2] the dependence of the XTR intensity on the plate thickness $a$ and the spacing $b$ between the plates in the stack (Fig. 19.4) was investigated. From the figure it follows, in agreement with the theory (Secs. 2 and 3), that when the plate thickness $a$ is smaller than the formation length of XTR in matter, the production of radiation is suppressed, and when the spacing $b$ between the plates exceeds the formation length of XTR in vacuum, the radiation intensity ceases to depend on $b$ (Fig. 19.4b). Some decrease in the radiation intensity (Fig. 19.4a) at large $a$ is explained by the increased absorption of radiation in the stack, i.e., by a decrease in $N_{\mathrm{eff}}$ (Sec. 3).

In [**75**.29] the dependence of the XTR intensity (together with ionization losses) on the particle Lorentz factor was measured for different plate thicknesses $a$ of the radiator and different lengths of the proportional chamber (Fig. 19.5). It can be seen that changing these parameters leads to a change not only in the magnitude of the XTR intensity but also in its $\gamma$-dependence.

In the study of the frequency spectrum of XTR, great importance is attached to the identification of interference maxima and minima of the spectrum is the spectral resolving power of the facility. From this point of view, the works of Cherry, Mueller, et al. [**77**.1, p. 137, **77**.8, **78**.2], in which a *single-crystal Bragg spectrometer* was used (Fig. 19.6), deserve attention.

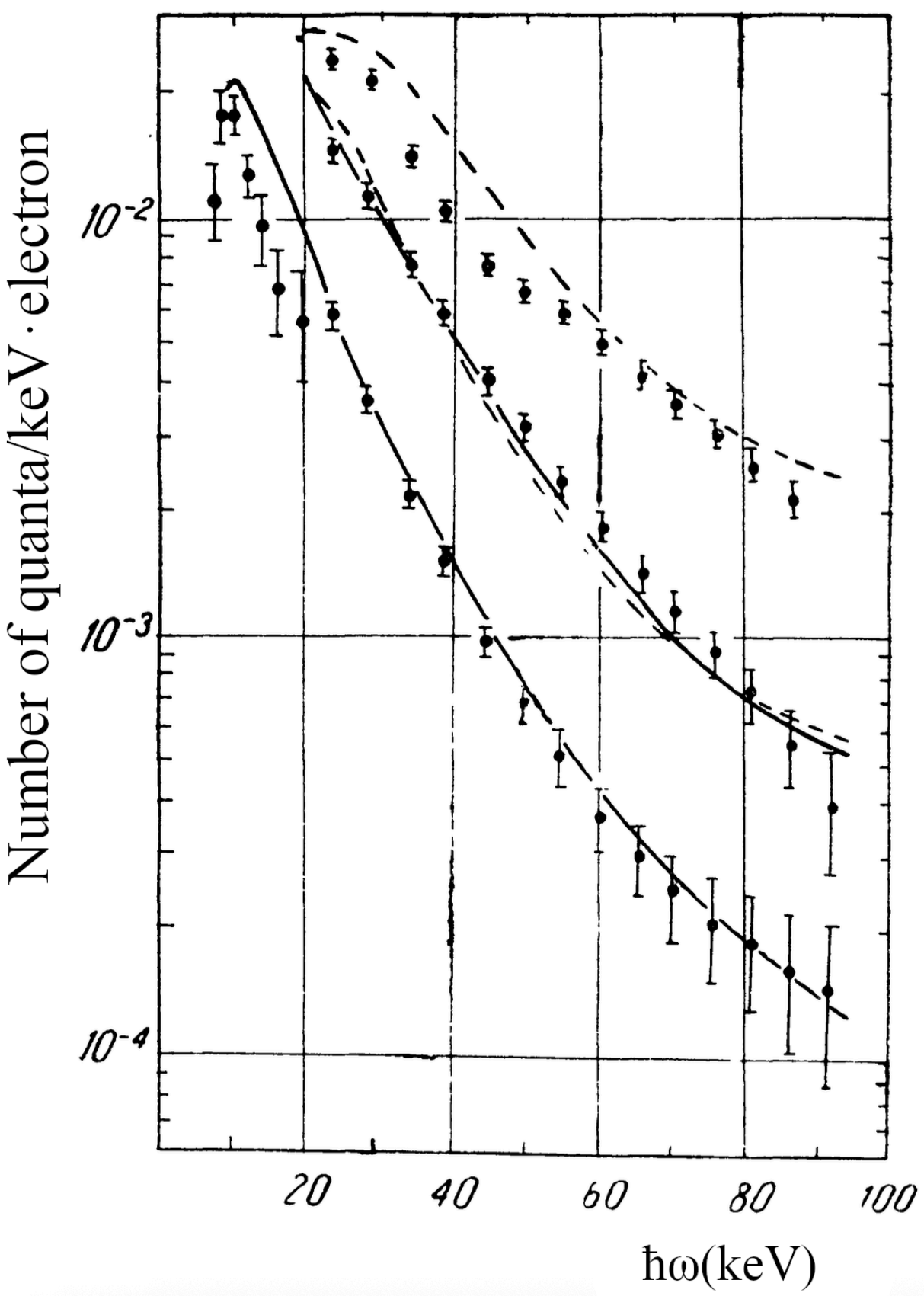

Fig. 19.3. Frequency spectrum of XTR produced by electrons with energy 3 GeV in a regular stack of organic foils ($a$=20 μm, $b$=500 μm, $N$=*32*, 125, and 595 for the lower, middle, and upper groups of experimental points, respectively) [**73**.12]. Solid curves correspond to theoretical calculations. Dashed curves were obtained by rescaling the experimental data to the case $N$=32

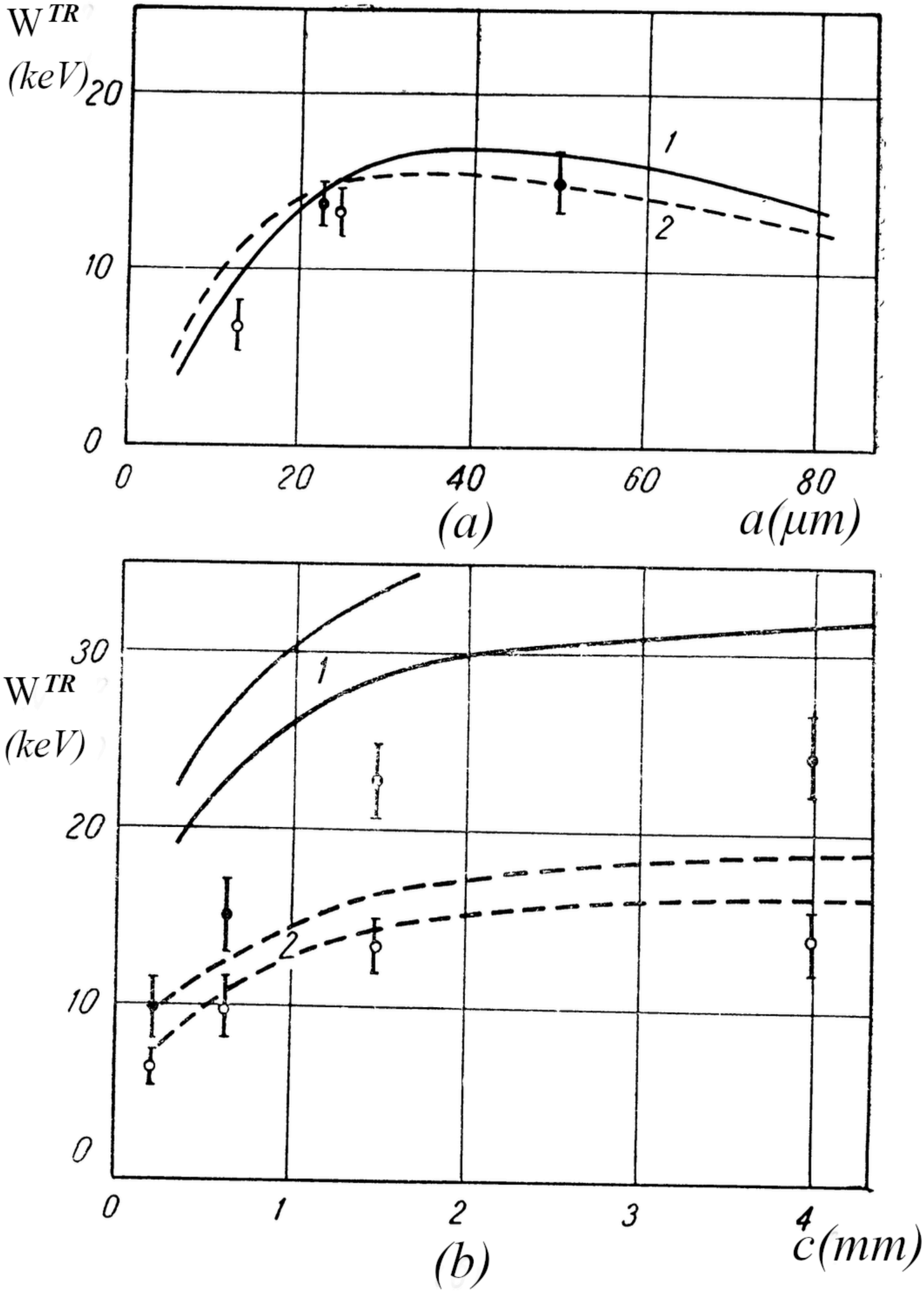

Fig. 19.4. Dependence of the energy deposition by XTR quanta absorbed in a 4 cm layer of krypton on the stack parameters $a$ and $b$ [**74**.2] (curves 1 and black dots—for polypropylene; curves 2 and open circles—for Mylar): (a) dependence on $a$, $b$ = 1.5 mm, number of layers $N$ = 188, electron energy is 10 GeV; (b) dependence on $b$, $a$ = 25 μm,

$N = 188$, electron energy 15 GeV; the region between curves with the same number corresponds to uncertainties in the theoretical calculations

The XTR radiator consisted of layers of polypropylene ($a$ = 16–82 μm, $b$ = 1.4 mm). The particles (electrons with energies of 5 and 9 GeV) were deflected by a magnet and subsequently recorded by a system of scintillation counters, while the XTR quanta passed through a collimator and were then scattered in the crystal and were recorded by a proportional detector, which excluded the simultaneous registration of two or more quanta. The spectral distributions of the number of XTR quanta, experimentally measured in the range from 4 to 30 keV for three different radiators on the facility described, are given in Fig. 19.7. The number of layers in these radiators was chosen so that the total amount of material and, consequently, the absorbing power of the radiator, as well as the mean square of the electrons' multiple scattering angle, would remain approximately the same. This figure also shows theoretical spectra that take into account absorption in the radiators, calculated using the Monte Carlo method. According to the theory (Sec. 2.4), with increasing *a* (as long as $a<c\gamma/\omega_0$), progressively harder quanta should appear in the frequency spectrum of XTR. This is clearly seen in Fig. 19.7. Moreover, with increase in *a,* the number of interference maxima and minima increases. The shapes of the measured spectra in this frequency interval, in particular the positions of the maxima and minima, agree well with theory; however, the intensity is 15–20% lower than the theoretical value. The authors of Refs. [**77**.8, **78**.2] believe that this discrepancy is due to a number of technical reasons not taken into account in the theoretical calculations.

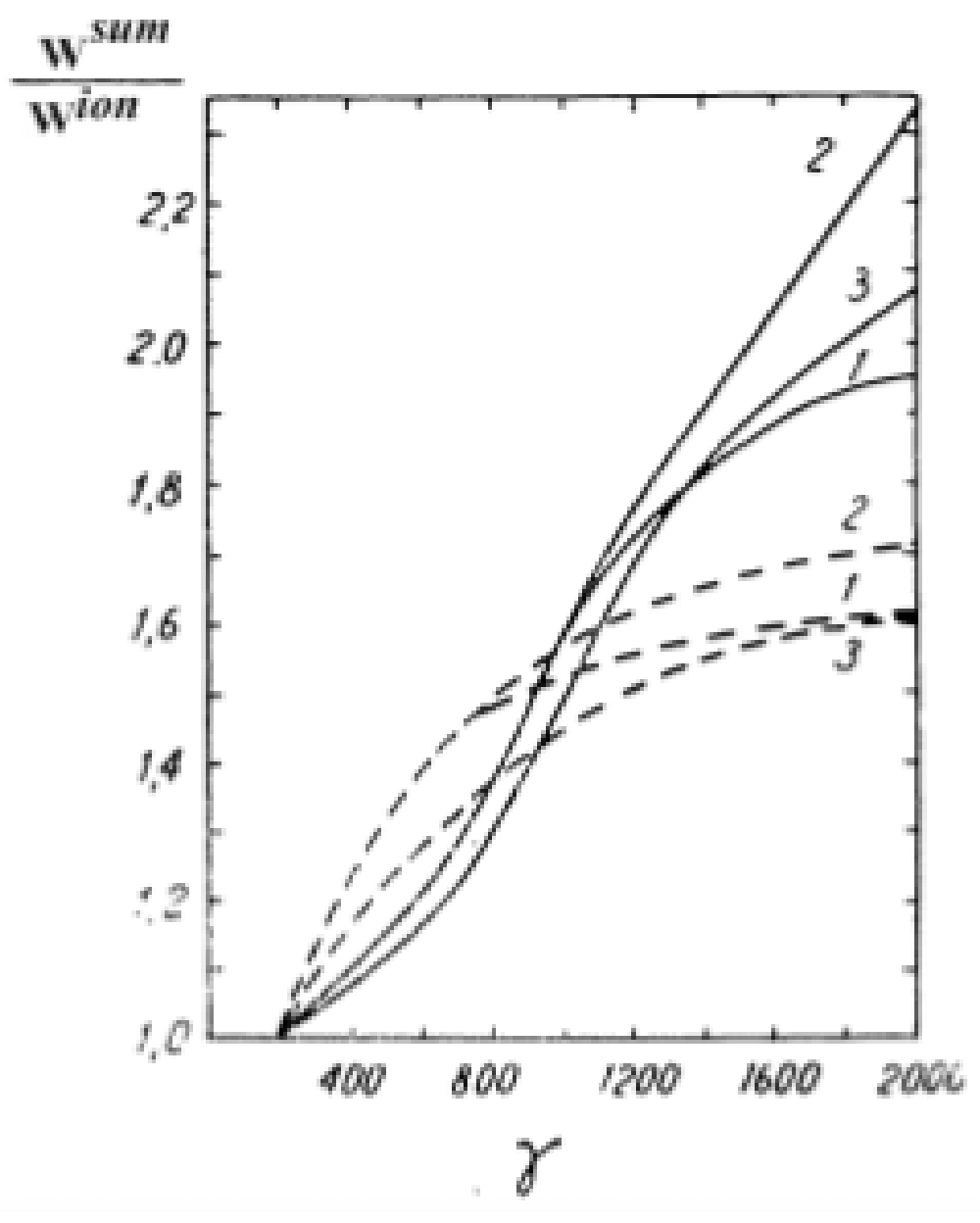

Fig. 19.5. Dependence of the ratio of the total energy deposition $W^{sum}$ to the ionization energy deposition $W^{ion}$ on the Lorentz factor γ for different thicknesses of a proportional chamber filled with xenon (1–3.2 mm, 2–4.8 mm, 3–6.4 mm) and different thicknesses of Mylar layers (dashed curves—3 μm, solid curves—10 μm) [80.13].

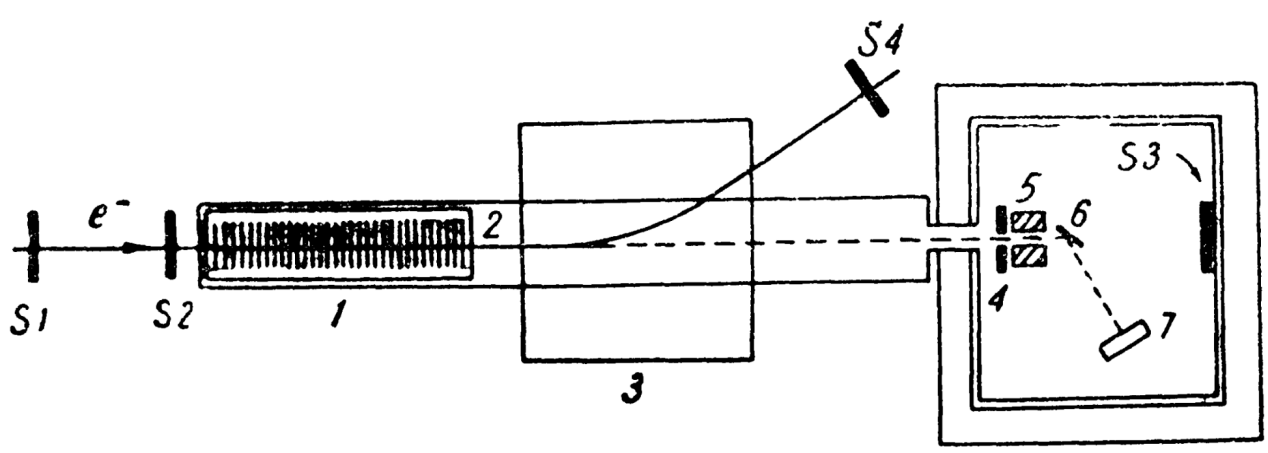

Fig. 19.6. Schematic of the experimental facility for the study of XTR at the University of Chicago [**74**.13]: *S1, S2, S3, S4*—scintillation counters, 1—radiator, 2—helium beam pipe, 3—deflecting magnet, 4—anticoincidence counter, 5—collimator, 6—Bragg single crystal, 7—proportional detector, 8—lead shield

To extend the range of the investigated XTR quantum energies to 100 keV, Cherry [**7**.2] also carried out a series of measurements in which, instead of a Bragg spectrometer, a 0.37 mm thick NaI scintillator was used as the radiation detector. The measurements were performed at an electron energy of 6.4 GeV on two different polypropylene radiators ($N$=200, $a$=244 μm, $b$=0.75 cm and $N$=100, $a$=244 μm, $b$=1.5 cm). In Fig. 19.8, the

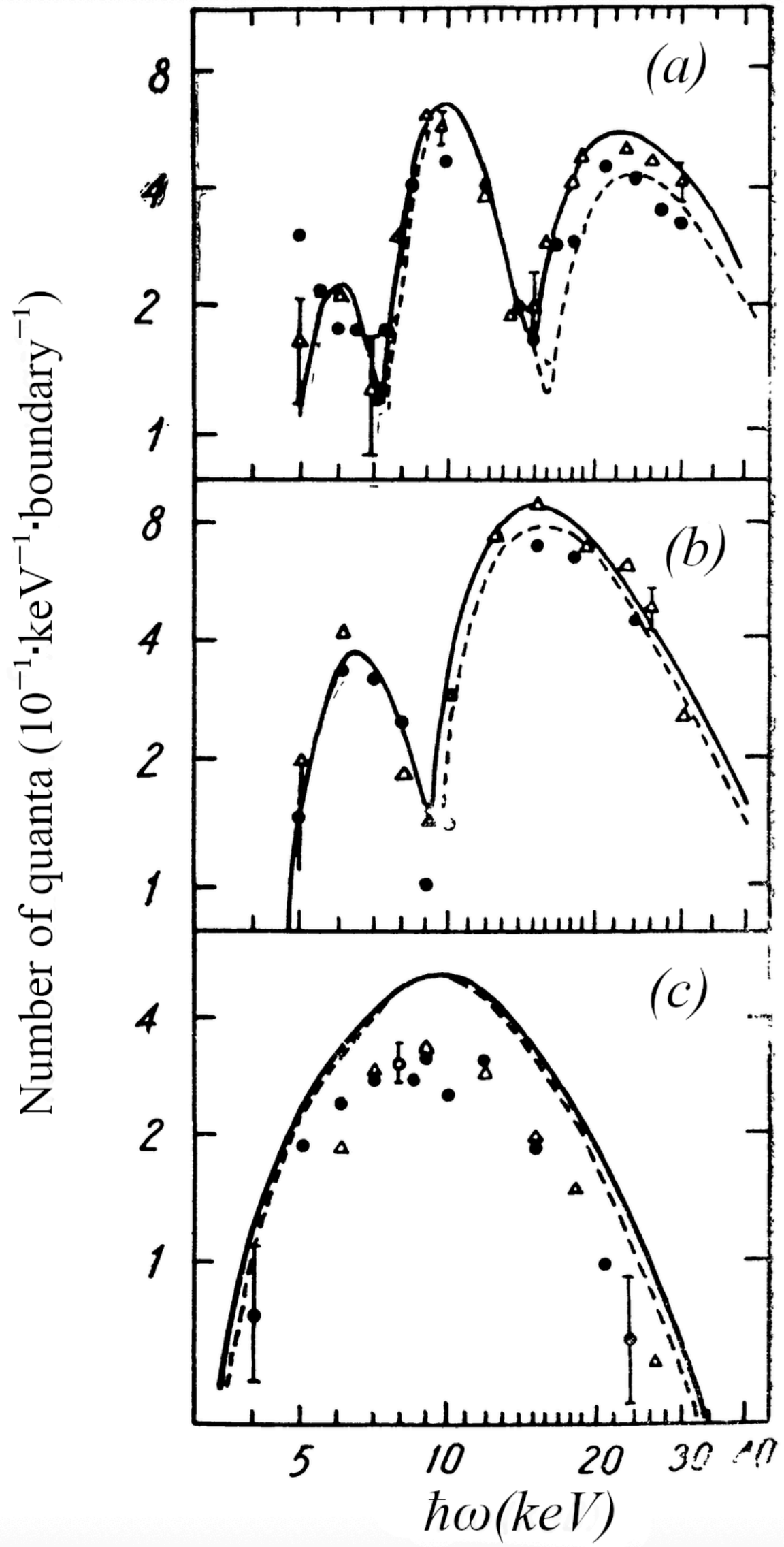

Fig. 19.7. Frequency spectra of XTR from electrons in various layered radiators [**77**.8]: (a) $a = 82$ μm, $N = 200$; (b) $a = 50$ μm, $N = 250$; (c) $a$ = 16 μm, N = 1000; the solid curves and triangular data points correspond to 9 GeV, the dashed curves and black points correspond to 5 GeV

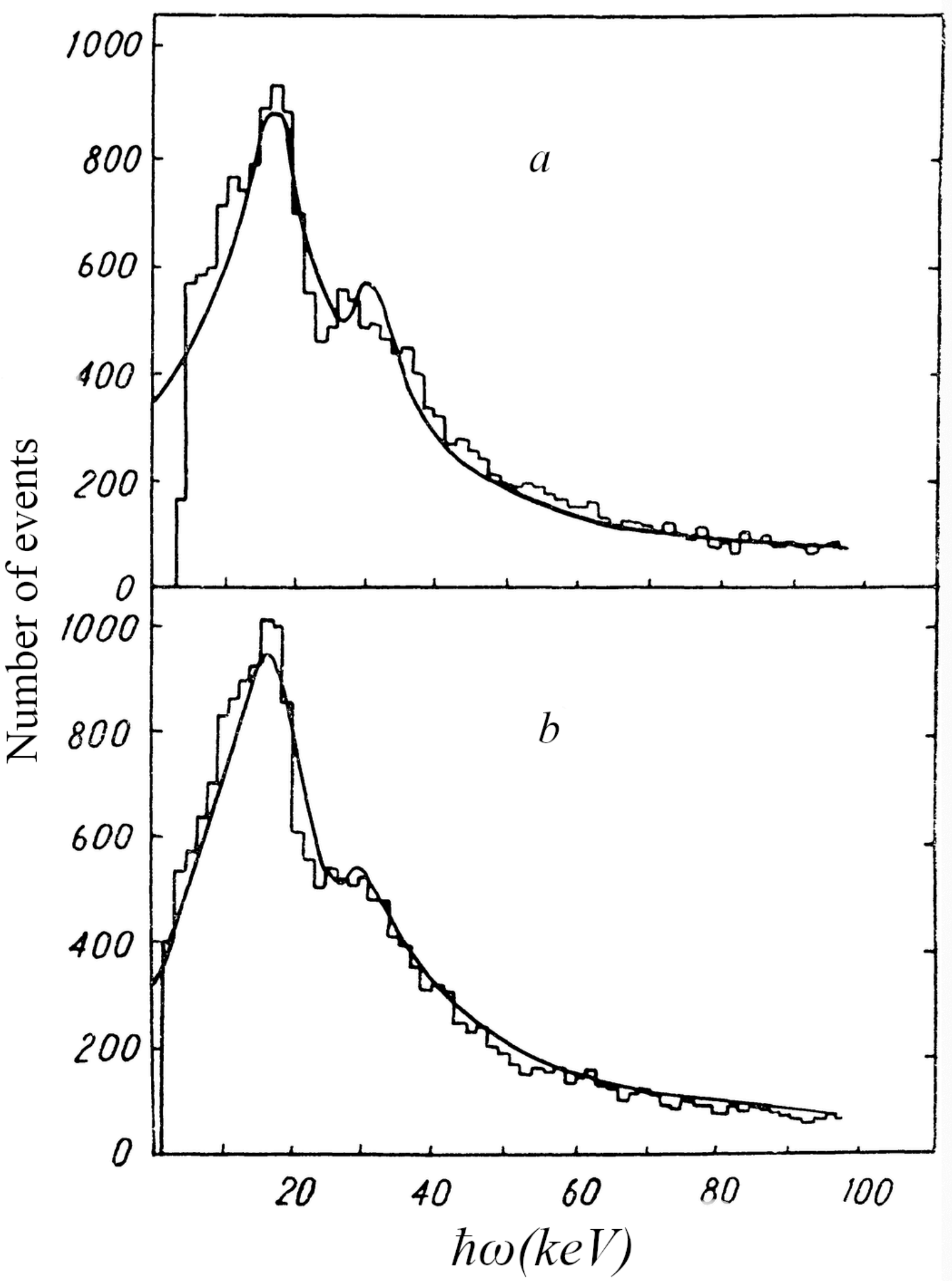

Fig. 19.8. Frequency spectra of electron XTR measured with a scintillation counter [**78**.2]: (a) $a = 244$ μm, $b$ = 1.5 cm, $N = 100$; (b) $a$ = 244 μm, $b$ = 1.5 cm, $N = 200$; solid curves—theoretical calculation, histograms—experimental data

XTR spectra measured experimentally (histograms) and calculated by the Monte Carlo method (solid curves) are shown. Although the resolution of the NaI scintillator is worse than that of the Bragg spectrometer, in this case the agreement between theory and experiment is quite good. The interference maxima at 28 and 17 keV are seen rather clearly

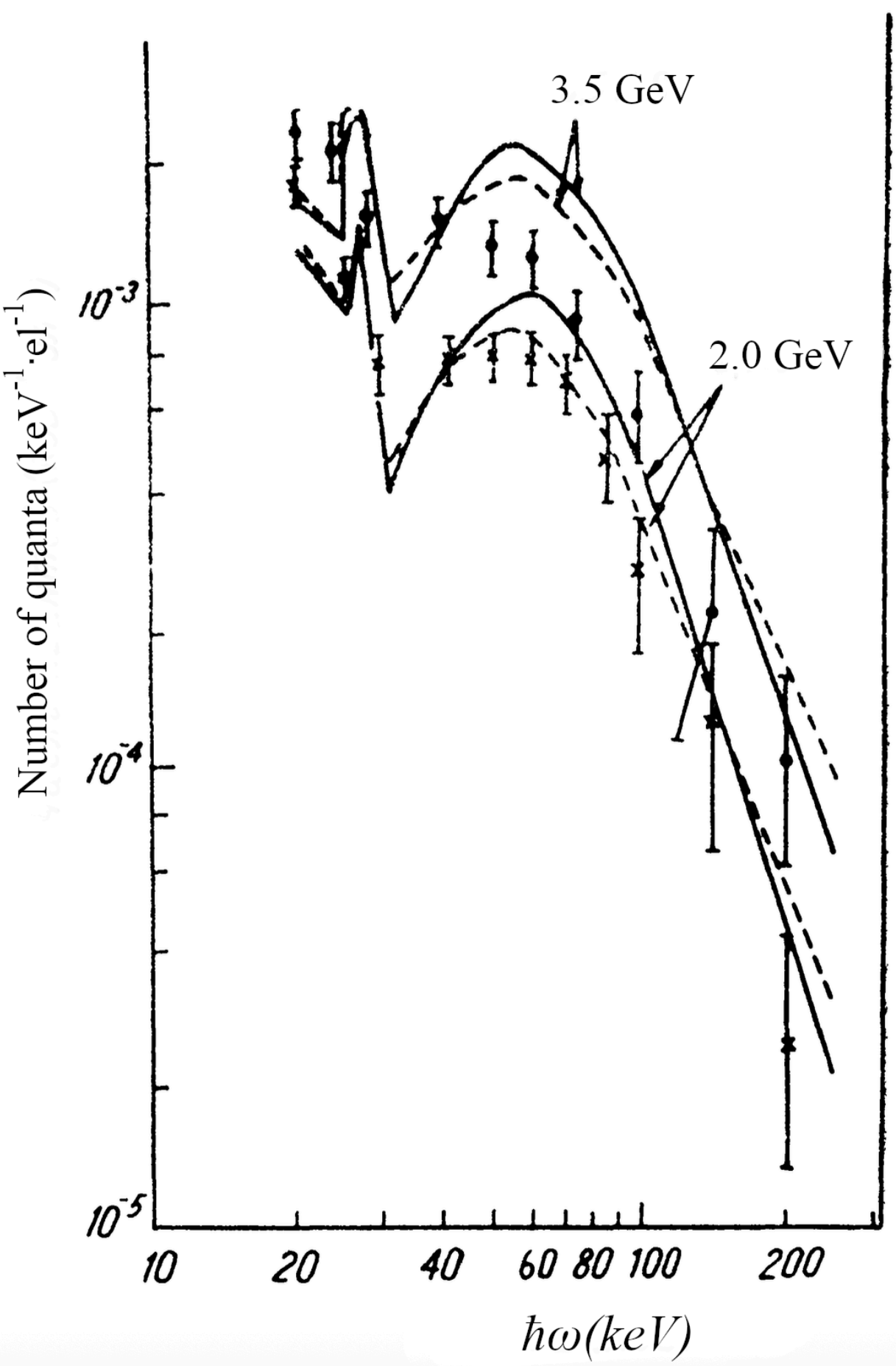

Fig. 19.9. XTR spectra from electrons in tin foils [**80**.13]. Black points correspond to an energy 3.5 GeV, crosses correspond to 2 GeV. Solid curves are calculated without taking into account multiple scattering of electrons, while dashed curves include multiple scattering

A study of XTR in copper and tin foils conducted by Lorikyan et al. [**80**.13], aimed at investigating the effect of multiple electron scattering on the formation of XTR, is of considerable interest. In particular, the XTR spectrum was examined in the photon energy range 20–200 keV for electrons with energies 1–3.5 GeV passing through a stack of 20 tin foils ($a = 20$ μm, $b = 1$ mm). The radiation was recorded with a 2-cm-thick NaI(Tl) crystal. Figure 19.9 shows the XTR spectra measured experimentally (dots and crosses) and calculated taking into account multiple scattering (dashed curves) and without it (solid curves). In the range $40 \leq \hbar\omega < 100$ keV, the experimental results are closer to the calculated curves that include multiple scattering. These measurements, in agreement with theory (see Chapter IV), show that for the electron energies available in this experiment, in the region $\hbar\omega < 10^2$ keV, multiple scattering has little effect on the integral intensity of XTR.

## 19.2. Irregular Media

Irregular (in particular, porous) media are of great interest for the theory and practical applications of XTR. The properties of XTR emitted by porous media (such as plastic foam) have been studied by many authors [**71**.1, p. 542, **71**.6, **71**.7, **73**.10, **73**.11, **73**.13, **75**.15, **77**.8, **77**.10, **78**.2]. In Ref. [**71**.1, p. 542] the dependence of the number of XTR quanta emitted from plastic foam on the electron energy was measured for the first time using a streamer chamber. It was found that at electron energies of 0.6, 1.3, and 2.5 GeV, the number of recorded quanta per electron was 0.35, 0.70, and 1.05, respectively (plastic foam thickness was 280 cm with density equal to 0.042 g/cm³)[23]. It can be seen that in the indicated range of electron energies the dependence turned out to be almost linear. Fluctuations in the number of transition radiation quanta were also obtained in this work. In studies [**73**.10, **73**.11], the spectral distributions of transition radiation emitted from foams of various densities were measured for electron energies of 3 GeV (Fig. 19.10). Measurements were taken with a multisection proportional chamber (at $\hbar\omega$ = 8—20 keV) with a gas filling of 90% Ar + 10% $CH_4$ and a NaI(Tl) scintillator (at $\hbar\omega$ = 20–100 keV). The experimental results are in satisfactory agreement with the theory (Sec. 6).

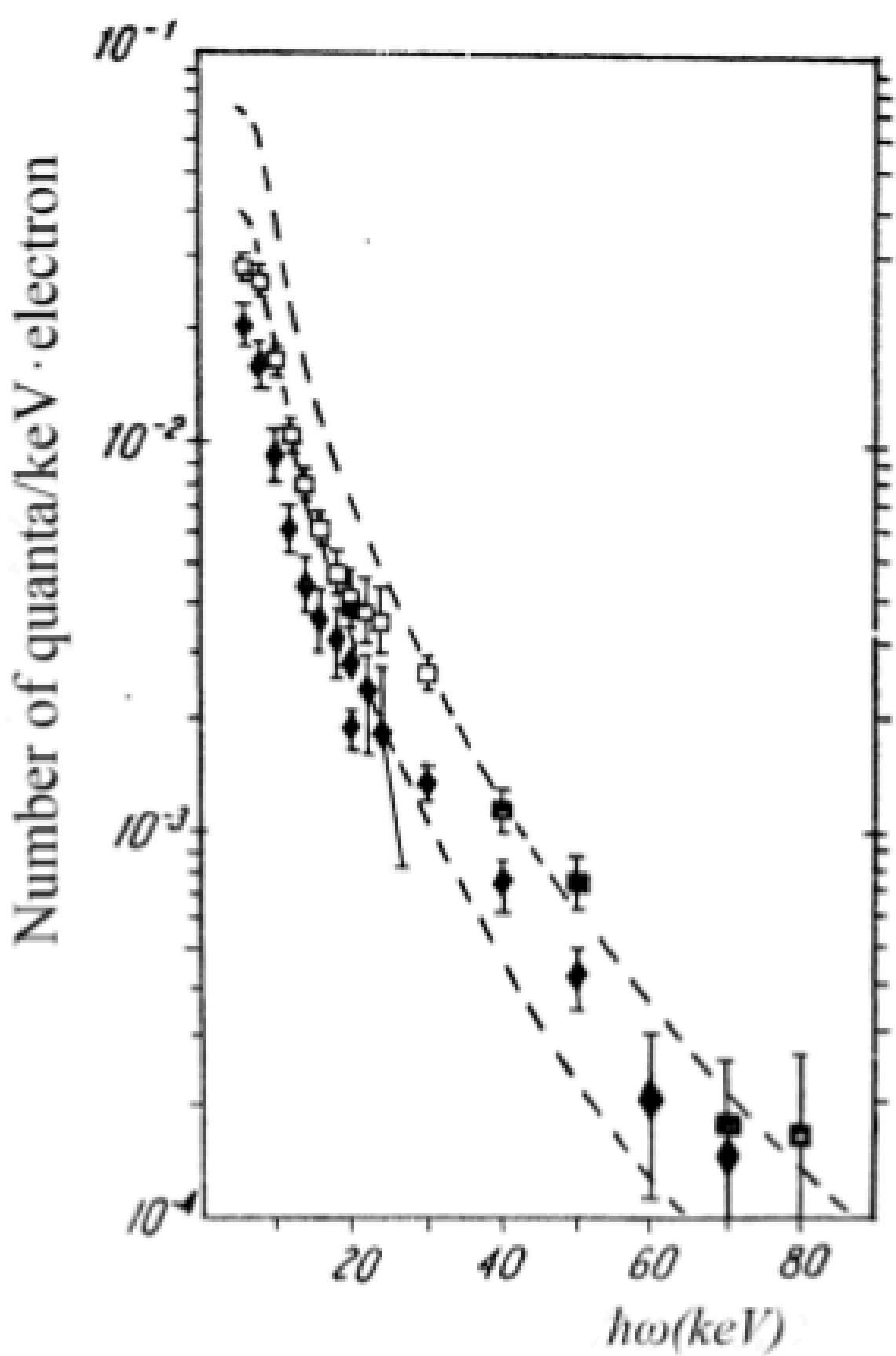

Fig. 19.10. Frequency spectrum of XTR produced by 3 GeV electrons in plastic foam 1 cm thick with a density of 0.09 g/cm³ (□) and 0.04 g/cm³ [73.11]. The dashed curves [75.27] are calculated from the theory (Sec. 6) under the assumption that the wall thicknesses and pore sizes are distributed according to a

[23] The values of the numbers of quanta given in the work for crystals are greatly overestimated due to, possibly, incorrect subtraction of the bremsstrahlung background.

In Ref. [**75**.14], the angular distributions of the number of XTR quanta from plastic foam were measured, also in good agreement with the theory. In this case, it is necessary to take into account that the experimental angular distributions of XTR are broadened due to multiple scattering of particles in the radiator material [**75**.14] (see also Sec. 16). From this point of view, the angular distributions given in Ref. [**70**.3] in fact do not correspond to the true angular distributions of XTR in the absence of multiple scattering of particles.

The total intensity and frequency spectrum of XTR emitted from various porous materials have also been investigated in Refs. [**71**.2, **71**.6, **75**.15, **77**.8, **77**.10]. In particular, Ref. [**75**.15] showed that the total intensity of XTR depends significantly on the electron Lorentz factor $\gamma$ in the range $\gamma$ from $10^3$ to $10^4$, whereas for $10^4 \lesssim \gamma \lesssim 3{\cdot}10^4$ this dependence becomes much weaker. The frequency spectrum of XTR produced by electrons with an energy of 1.38 GeV ($\gamma = 2.7{\cdot}10^3$) in porous materials was measured in Ref. [**77**.10]. The agreement between experiment and theory (Sec. 6) turned out to be good (Fig. 19.11).

Ispiryan and Oganesyan with collaborators [**72**.11] used a radiator made of lithium hydride (LiH) powder, which has a low absorption coefficient.

Müller, Cherry, et al. [**75**.15, **77**.8, **78**.2] compared the energy deposition (in 4-cm layers of xenon and krypton gases) of XTR emitted from a regular stack of Mylar plates ($a$ = 25 μm, $b$ = 1.5 mm, $N$ = 188) with that from certain porous materials 4 cm thick at an electron energy of 9 GeV. From this comparison it follows that the radiation yields of various porous materials differ greatly from one another. This is due both to differences in densities, pore sizes and wall thicknesses (and their distributions), and to chemical composition, i.e., to the absorbing power of the radiator material.

Furthermore, another irregular medium—a material made of polymer *fibers*—was investigated experimentally [**81**.7,**82**.1]. For example, Müller et al. [**82**.1] studied a material produced by U.S. industry consisting of polyolefin fibers of diameter 2–6 or 17 μm, with an average spacing between fibers of about 70 or 400 μm. The authors

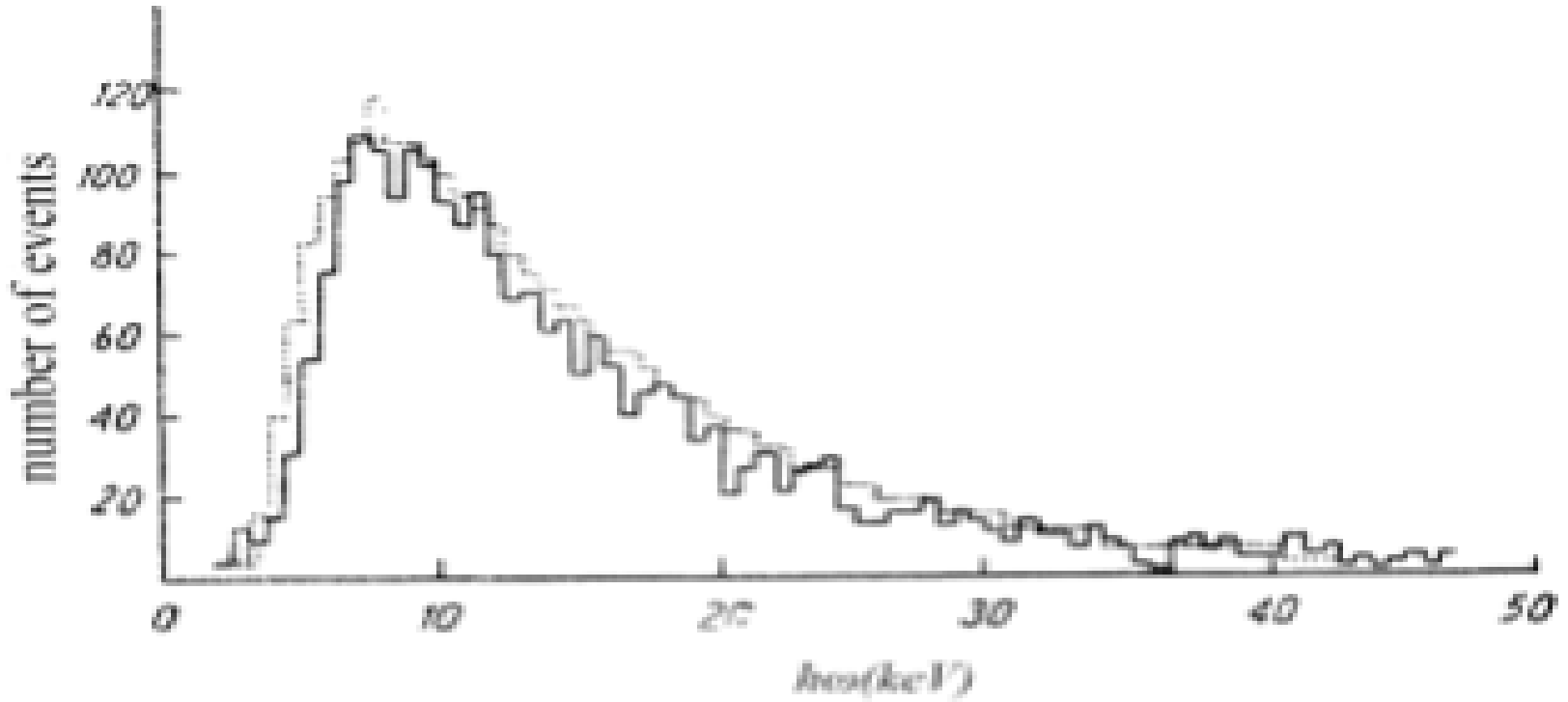

Fig. 19.11. Spectrum of XTR of electrons in a porous material [77.10] (Ethafoam-220 plastic foam). The solid and dotted histograms correspond to the experimental data and the theoretical calculation

note that in the region of small values of the particle Lorentz ($\gamma \ll 10^3$), when for optimal production of X-ray transition radiation (XTR) a thickness of the material layer of order $10^{-4}$ cm is required, it is technically very difficult to fabricate a stack of such thin plates with spacings between the plates of order $10^{-2}$ cm, especially with an area of the order of several square meters and larger. Fabricating plastic foam with the corresponding values of the mean wall thickness and the mean pore size is also difficult. From this point of view, a material made of thin fibers is quite suitable, and large-area radiators can be easily made from it. Experimental results show that the intensities of XTR from fibrous material and from plastic foam are comparable, and in the region $\gamma \sim 10^3$–$10^4$ in the former case the intensity is somewhat larger.

## § 20. TRANSITION RADIATION DETECTORS

A device intended for the identification of charged particles of different masses having the same energy, and possibly for determining the Lorentz factor of a particle by measuring the intensity (or the number of quanta) of XTR produced by the particle in a radiator, is commonly called a *transition radiation detector (TRD)*. It should be borne in mind that XTR is emitted at very small angles relative to the particle's direction of motion (see Sec. 1.6). Therefore, in detection it is necessary to separate the XTR quanta from the particle itself.

At present, there are various ways to spatially separate the particle and the radiation. The previous section considered several of these methods: by means of a magnet deflecting the particle; by angular discrimination of the charge (only radiation quanta at angles $\vartheta \gg \vartheta_1 \neq 0$ are recorded); and by using a streamer chamber followed by visual analysis of the photograph.

In addition to these methods, the particle and the radiation can be spatially separated by recording secondary radiation produced by an XTR quantum and propagating at a large angle with respect to the particle's trajectory, for example, characteristic radiation [**61**.4, **64**.3, **65**.6, **74**.11] or radiation due to Compton scattering [**71**.7, **74**.14]. Such radiation is already not difficult to record separately from the particle.

Meanwhile, in TRDs developed to date, another method of separating XTR quanta from the particle is widely used—the *energy-deposition method* [**61**.4], employing multiwire proportional chambers as the recording device.

### 20.1. Energy-Deposition Method. Rejection Factor and Detection Efficiency

The essence of the energy-deposition method is as follows. The energy $W^{\mathrm{sum}}$ deposited in the detector, which consists of ionization losses $W^{\mathrm{ion}}$ and the absorbed energy of X-ray transition radiation $W^{TR}$, depends on the Lorentz factor γ mainly because of the γ-dependence of $W^{TR}$ (at large γ the quantity $W^{ion}$ reaches the Fermi plateau; see Sec. 8). Obviously, the larger the ratio $W^{TR}/W^{ion}$ and the stronger the γ-dependence of $W^{TR}$, the more sensitive the detector is to the value of γ (in other words, the higher its resolving power).

However, it must be borne in mind that both $W^{ion}$ and $W^{TR}$ are subject to fluctuations. The fluctuations of $W^{ion}$ are described by the Landau distribution [**44**.1, **80**.14] and depend only on the thickness $\boldsymbol{l}_r$ and the density $\rho_r$ of the gas in the proportional chamber. The fluctuations of $W^{TR}$, however, depend not only on the parameters of the proportional chamber, but also on the parameters of the radiator and on the particle Lorentz factor.

Characteristic distributions of $W^{sum}$ at different electron energies, measured in Ref. [**75**.16], are shown in Fig. 20.1, from which it follows that with increasing electron energy the mean value of the energy deposition $W^{sum}$ increases monotonically due to the increase in the intensity of XTR. At the same time, the width of the distribution also increases. Fluctuations of the energy deposition lead to an overlap of the distribution curves of $W^{sum}$ corresponding to particles with different γ (Fig. 20.2), which limits the resolving power of the detector.

Let there be, for simplicity, particles of two types ($\gamma_1<\gamma_2$) with the corresponding energy-deposition distribution functions $f_I(W^{sum})$ ($i$=1,2), normalized to unity:

$$\int_0^{\infty} f_i(W^{sum})dW^{sum} = 1.$$

Particles are usually identified as follows. A particle passing through the detector is identified as a particle of type 2 if its energy deposition $W^{sum}$ is greater than or equal to a specified threshold value $W_{thr2}$, and as a particle of type 1 if

$$W^{sum} \ll W_{thr1},$$

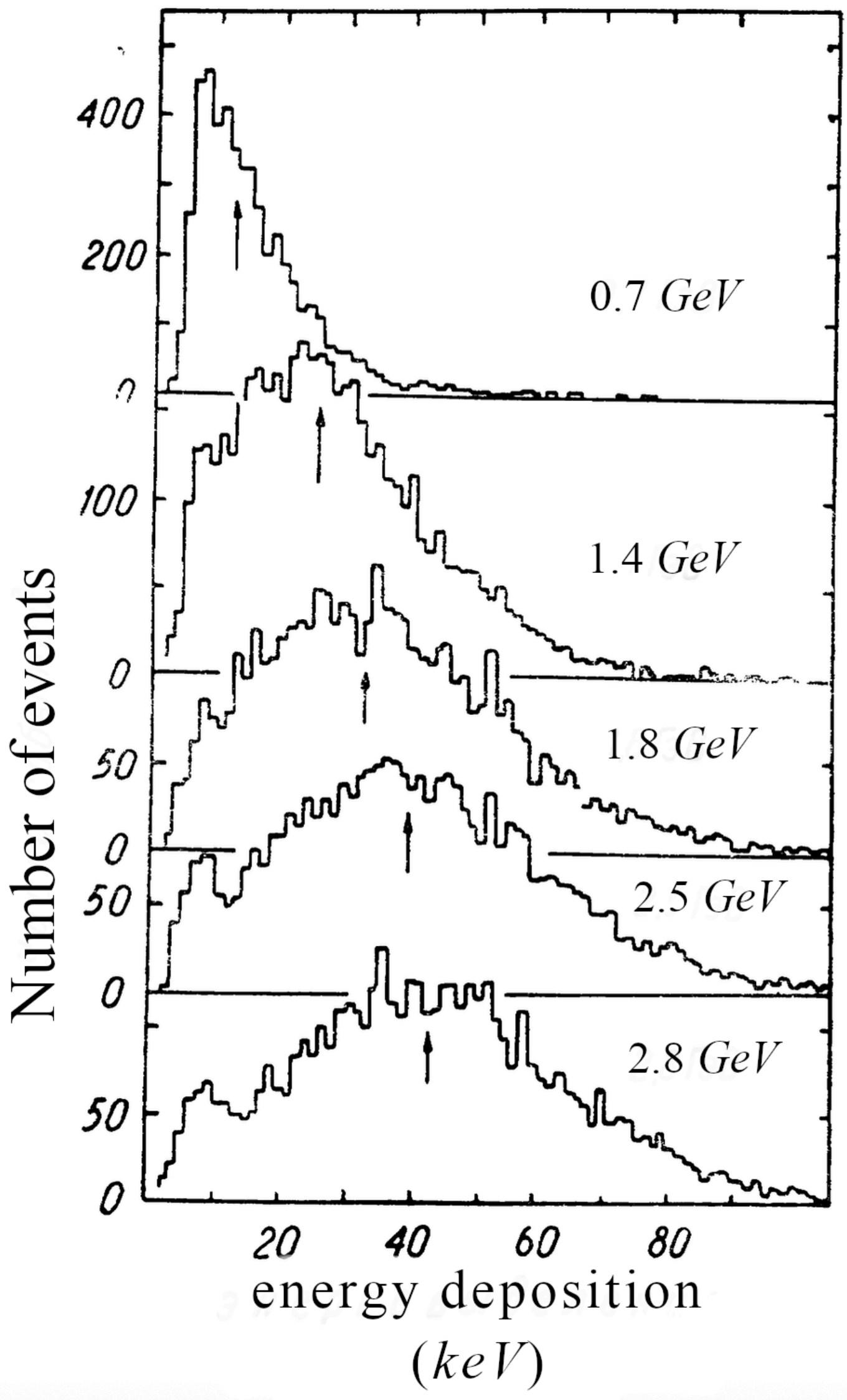

Fig. 20.1. Distributions of the energy losses of electrons of different energies passing through a lithium radiator. The arrows indicate the mean energy deposition [**75**.16]

where $W_{\text{thr1}} < W_{\text{thr2}}$. If $W_{\text{thr1}} < W^{sum} < W_{\text{thr2}}$, then the event is considered indefinite and is disregarded.

Due to the overlap of distribution curves 1 and 2, a particle of type 1 may also produce an energy deposition $W^{sum} > W_{\text{thr1}}$ and thereby be identified as a particle of type 2. Similarly, a particle of type 2 may produce $W^{sum} \leq W_{\text{thr2}}$ and be identified as a particle of type 1. The quantities

$$\delta_1 = \int_{W_{thr2}}^{\infty} f_1(W^{sum})dW^{sum} \; and \; \delta_2 = \int_{0}^{W_{thr1}} f_2(W^{sum})dW^{sum}$$

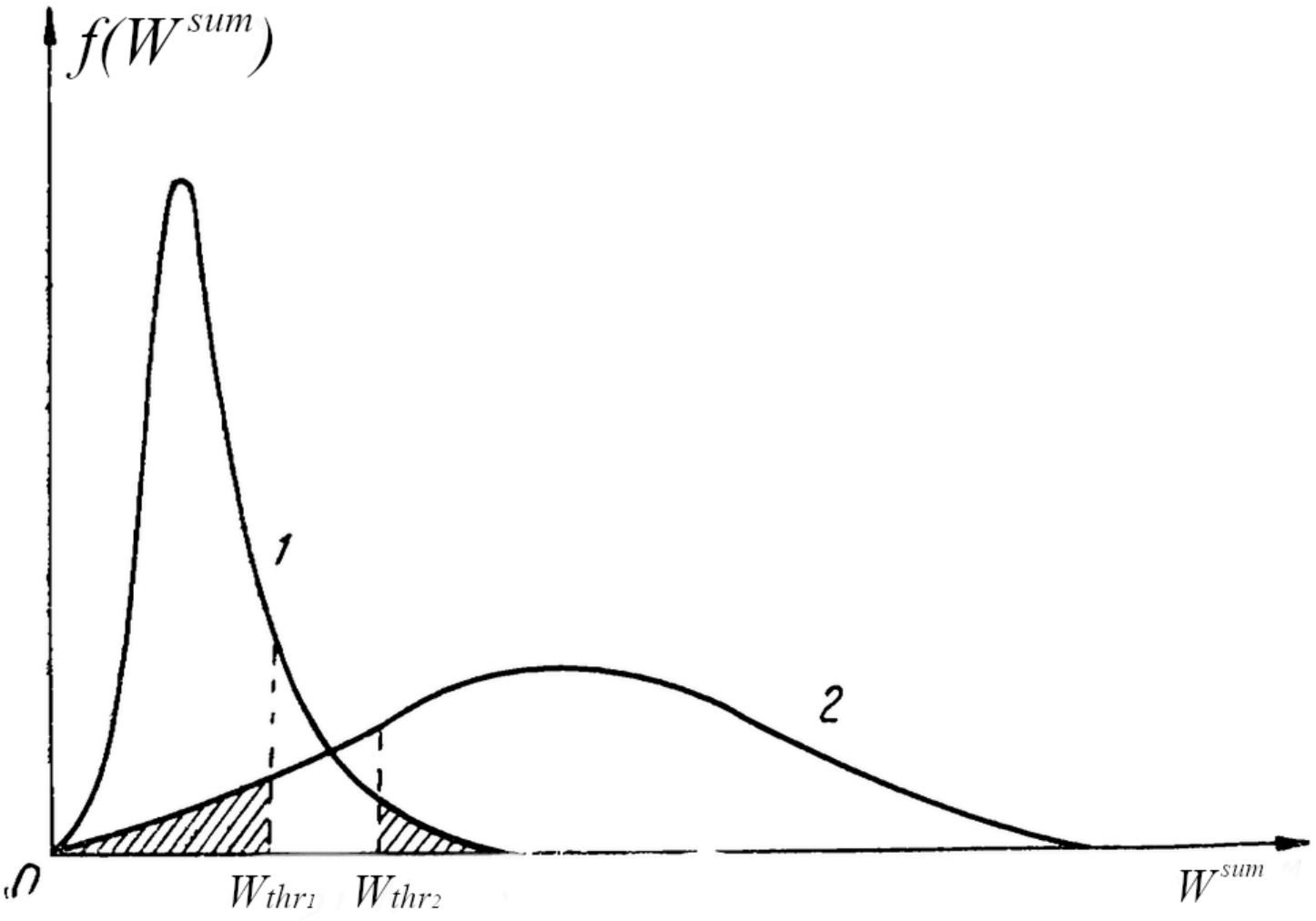

Fig. 20.2 Energy-deposition distribution functions for two types of particles (curves 1 and 2) with Lorentz factors $\gamma_1 < \gamma$.

(the shaded regions in Fig. 20.2) characterize the relative fractions of such misidentified particles and are called *rejection factors*. The quantities

$$\eta_{p1} = \int_{0}^{W_{thr1}} f_1(W^{sum})dW^{sum} \; and \; \eta_{p2} = \int_{W_{thr1}}^{\infty} f_2(W^{sum})dW^{sum}$$

characterize the relative numbers of correctly recorded particles (of types 1 and 2, respectively) and are referred to as the detection efficiencies. It is clear that the larger $W_{thr1}$ and the smaller $W_{thr2}$, the lower the rejection coefficients; however, the detection efficiencies decrease accordingly. A typical dependence of $\delta_1$ on $\eta_{p2}$ (or $\delta_2$ on $\eta_{p1}$) for a modern TRD detector is shown in Fig. 20.3.

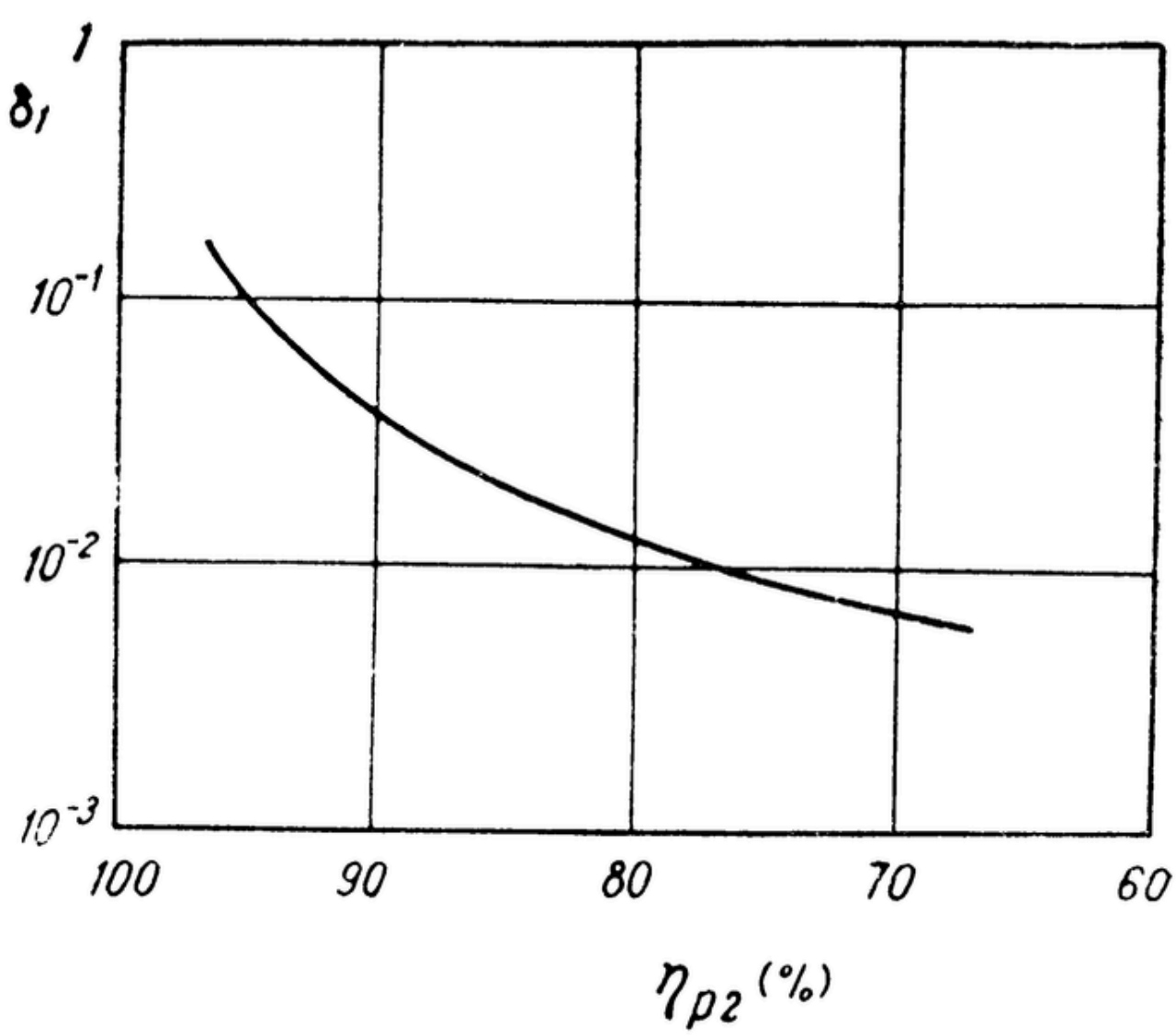

Fig. 20.3 Dependence of the rejection factor $\delta_1$ for type-1 particles on the detection efficiency $\eta_{p2}$ for type-2 particles

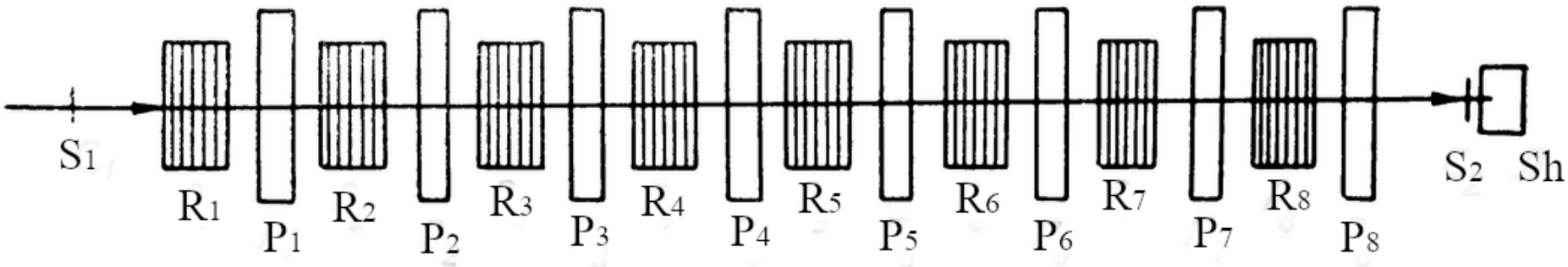

Fig. 20.4. Schematic representation of a TRD consisting of eight consecutively arranged sections (modules): R1—R8—XTR radiators, P1—P8—multiwire proportional chambers, S1, S2—scintillation counters, Sh—shower detector. The arrows indicate the direction of motion of the charged particle

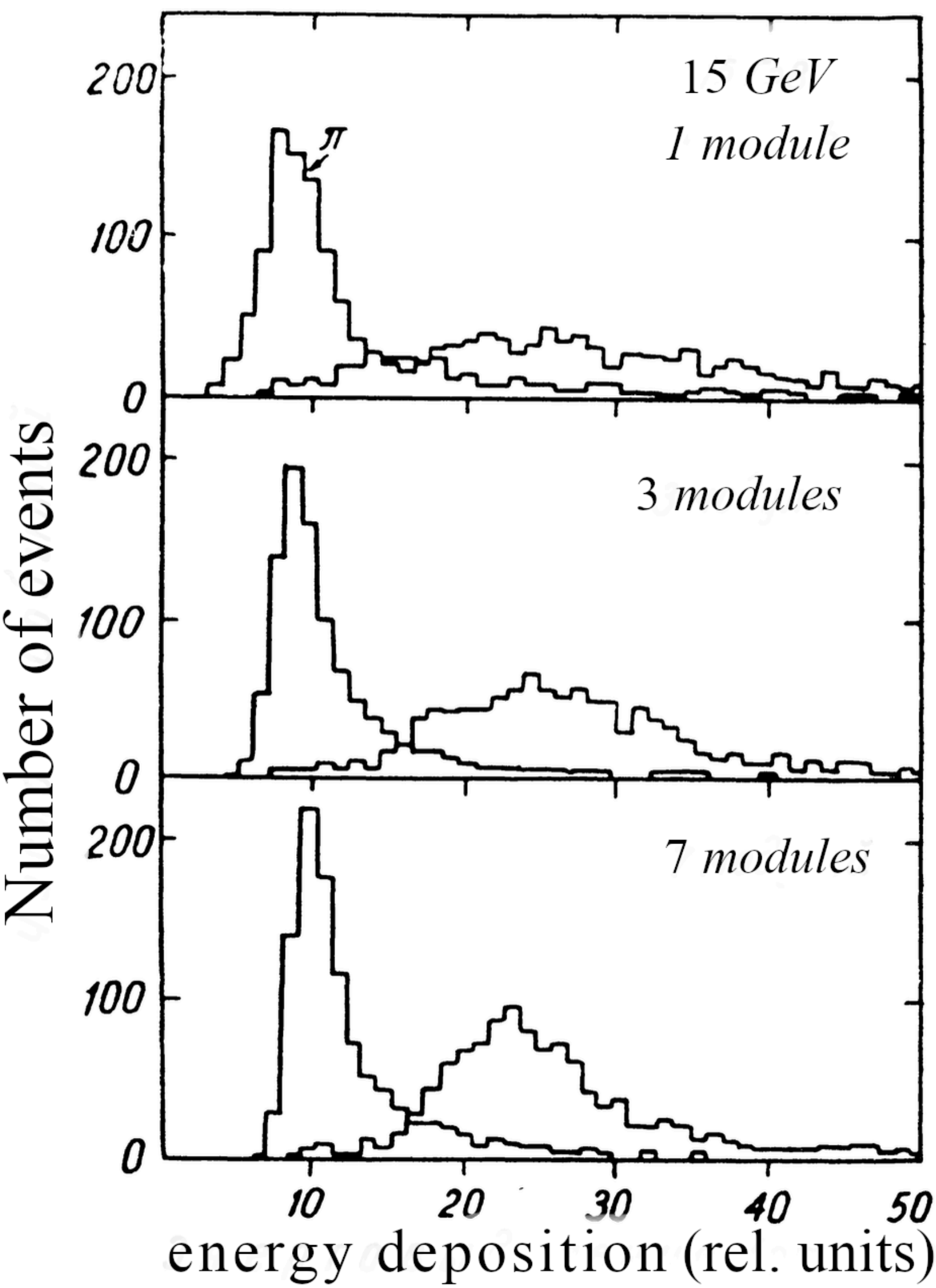

Fig. 20.5. Dependences of the energy-deposition distribution function for electrons and pions with the same energy of 15 GeV on the number of TRD modules [**74**.13]

Thus, to increase the resolving power of the TRD, it is necessary to reduce the overlap of the distribution curves, i.e., to reduce fluctuations in the energy deposition. To achieve this, it is necessary, first of all, to increase the yield of transition radiation and to optimize its γ-dependence by selecting appropriate values of $N$, $a$, and $b$, as well as the radiator material [**75**.30, **76**.15, **81**.12]. In addition, a significant reduction of fluctuations can be achieved by using a series of sequentially arranged sections (modules) consisting of transition radiation radiators and multiwire proportional chambers (Fig. 20.4). As shown clearly in Fig. 20.5, an increase in the number of modules leads to a narrowing of the energy-loss distribution, i.e., to an improvement in the detector resolution.

## 20.2. Experiments Employing Transition Radiation Detectors

After the first attempt to use an XTR detector in an experiment to measure the energy spectrum of muons of the horizontal flux of cosmic rays with energies above 700 GeV [**65**.6], a number of experiments employing TRDs has by now already been carried out.

Almost simultaneously, in 1973–1976, at the Yerevan Physics Institute (USSR) [**73**.14, **74**.16, **74**.17, **76**.7] and at the University of Maryland (USA) [**75**.1, p. 2538], facilities for cosmic-ray studies using TRDs were constructed.

### *A. Pion Facility*

The facility of the Yerevan Physics Institute (the Pion facility) is located at the high-altitude cosmic-ray station (Mt. Aragats, 3250 m above sea level) and is intended for the study of the interaction cross section of cosmic pions and protons with iron nuclei at energies above 300 GeV. The schematic diagram of this facility is shown in Fig. 20.6. The facility consists of a horizontally arranged hodoscope, nine identical independent four-module[24] TRDs installed vertically beneath the hodoscope, and an ionization calorimeter located below the TRDs.

[24] To improve the resolving power (without a significant reduction in light-gathering power), the number of modules has now been increased to five.

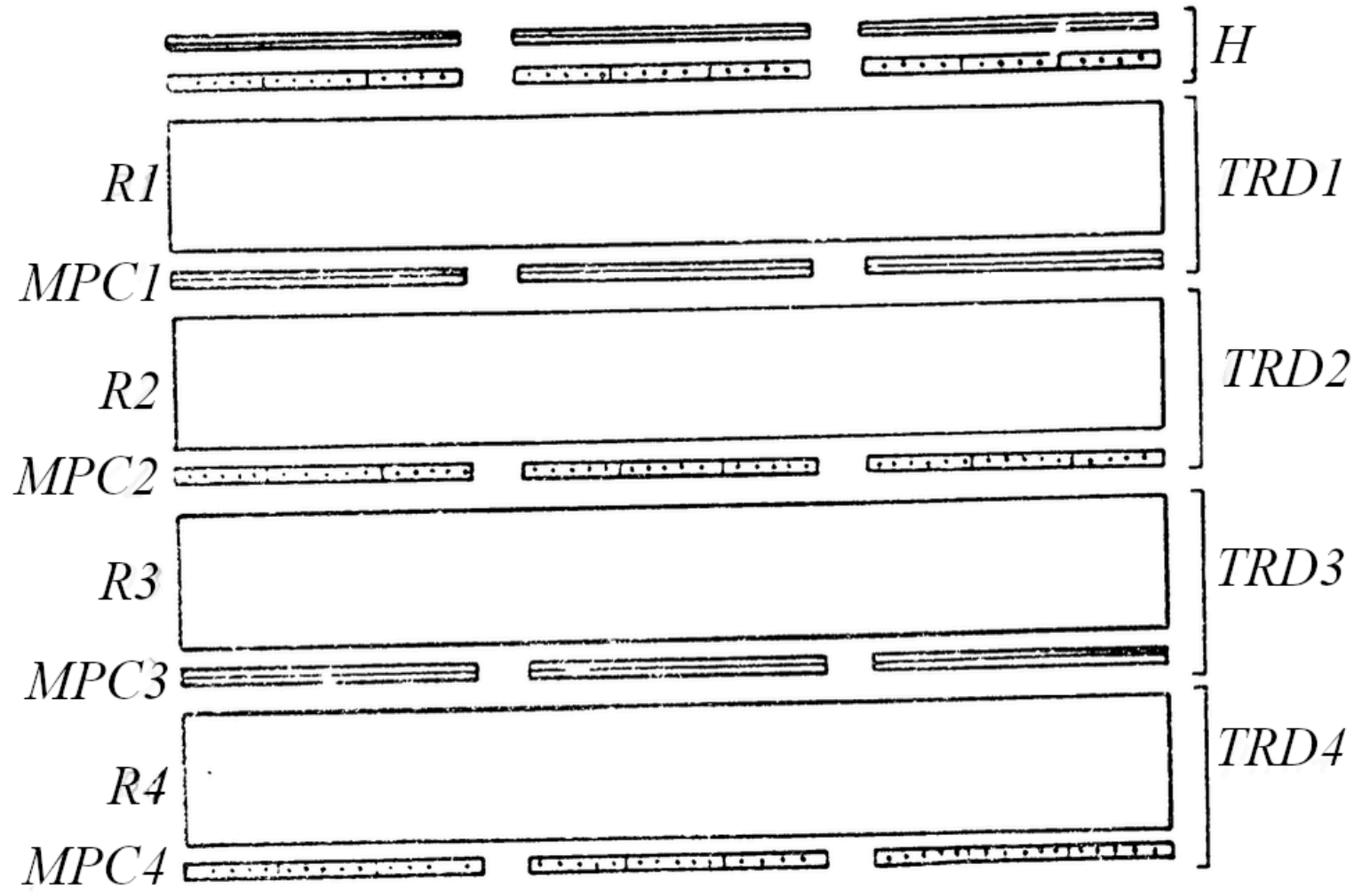

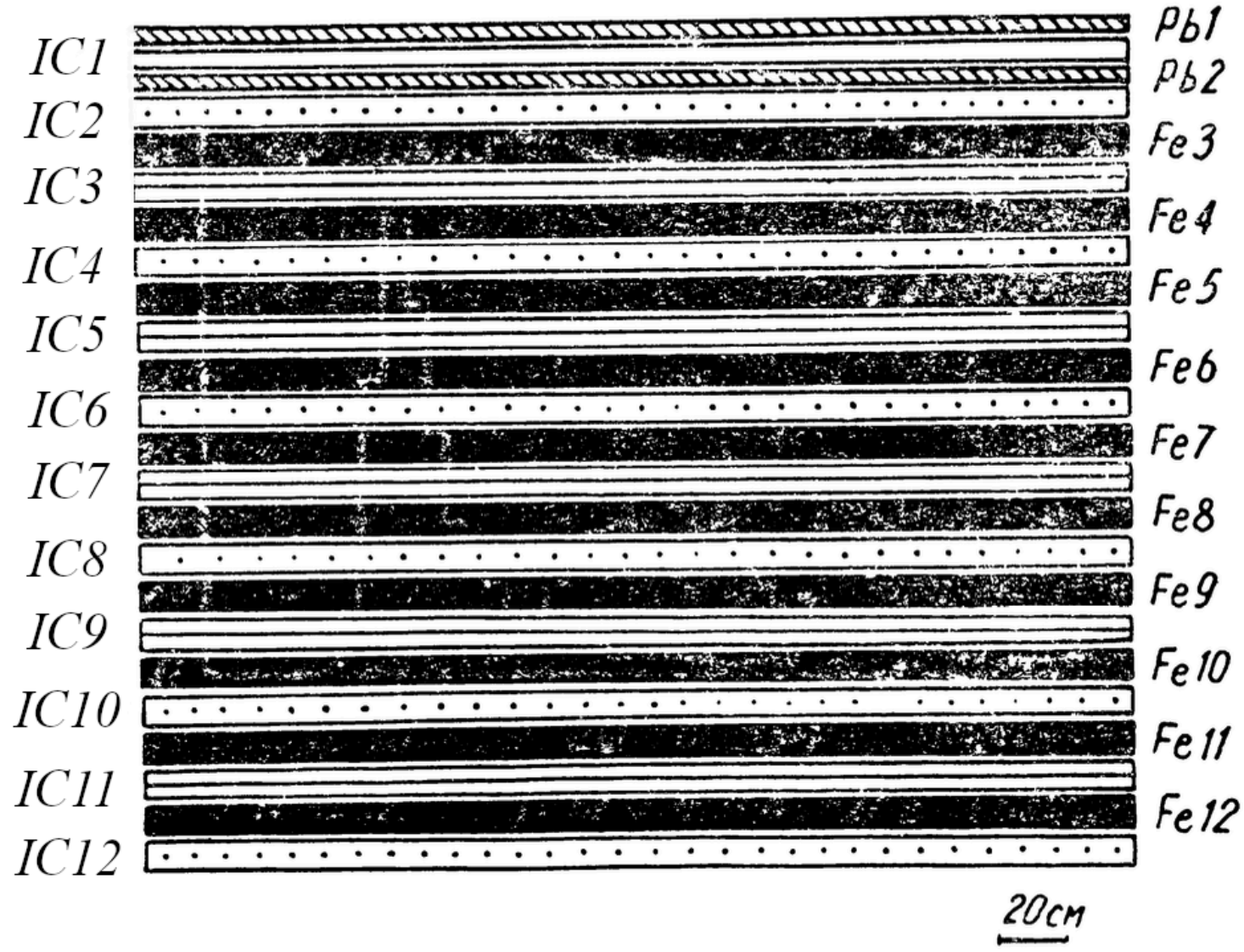

Fig. 20.6. Layout of the Pion facility [**74**.17]: H—hodoscope, TRD1–TRD4—TRDs, R1–R4—radiators, MPC1–MPC4—multiwire proportional chambers, Pb1, Pb2, Fe3–Fe12—absorbers of the ionization calorimeter made of lead and iron, respectively, IC1–IC12—rows of ionization chambers

The hodoscope consists of two layers of multiwire proportional chambers, which determine the initial coordinates of the particle trajectory with an accuracy of up to 2.8 cm. With the aid of the hodoscope one can also obtain a pattern of the distribution of accompanying particles in the horizontal plane.

The TRDs of the facility distinguishes pions from protons and consist of four modules, each of which in turn includes a radiation radiator (a stack of 120 layers of Mylar ($C_5H_4O_2$) films of thickness $a = 22$ μm with an air gap between the films $b = 3$ mm) and a multiwire proportional chamber (active area about 1 m², thickness 4 cm, gas filling 90% Ar + 10% $C_3H_8$). The wire directions of the chambers in neighboring modules are mutually perpendicular to reduce the background from accompanying air showers, as well as from particles traveling from bottom to top (the so-called albedo particles).

The calorimeter measures the energy of a traversing particle, has an area of 10 m², and consists of two lead plates 3 and 2 cm thick, which exclude events induced by electrons and photons, and 10 iron plates 10 cm thick each, in which nuclear interactions occur. Between all plates (lead or iron) there are ionization chambers with a diameter of 10 cm each. The total amount of material in the calorimeter is 900 g/cm². The design of the calorimeter makes it possible to follow the development of the hadronic–electromagnetic cascade and to determine the direction of the traversing particle with good accuracy. The accuracy of the energy measurement is ~15%.

The facility records events when the total energy released in the calorimeter is equal to 300 GeV or greater, and the hadronic–electromagnetic cascade spans at least four layers of iron. When these conditions are met, a trigger pulse is formed and information from all ionization chambers, the sections of the multiwire proportional chambers of the TRDs, and the hodoscope is transmitted to the computer for preliminary selection and accumulation of events. Final processing of the statistical material is carried out on a BESM-6 computer.

During the first two years of operation of the Pion installation, 2700 hadronic events were selected and processed, and the ratios $N_\pi / N_p$ of the number of pions to the number of protons were evaluated in different energy ranges, as well as the corresponding cross sections for interactions of hadrons with iron nuclei at hadron energies up to 2000 GeV.

### *B. University of Maryland Facility*

The University of Maryland facility is also designed for the study of cosmic hadrons. A schematic diagram of the facility is shown in Fig. 20.7. The facility consists of a TRD and a calorimeter. The TRD, with an area of 1 m$^2$, contains 24 identical modules, each consisting of a radiator (plastic foam, 13.6 cm thick) and a multiwire proportional chamber (thickness 5 cm, gas filling: 90% Ar + 10% $CH_4$). The calorimeter has an area of 4 m$^2$ and contains eight iron plates, each 15 cm thick. Between the plates are scintillation counters and wide-gap streamer chambers, which make it possible to record events caused by hadrons with energies ≥ 100 GeV that are not accompanied by dense extensive air showers. Hadrons without accompaniment (“single” hadrons) were selected using a special streamer chamber located between the TRDs and the calorimeter.

The ratios $N_\pi / N_p$ obtained from the analysis of events recorded with this installation are consistent with the data obtained with the Pion installation. However, the statistical errors in the University of Maryland facility are larger. At present, work at this facility has already been terminated.

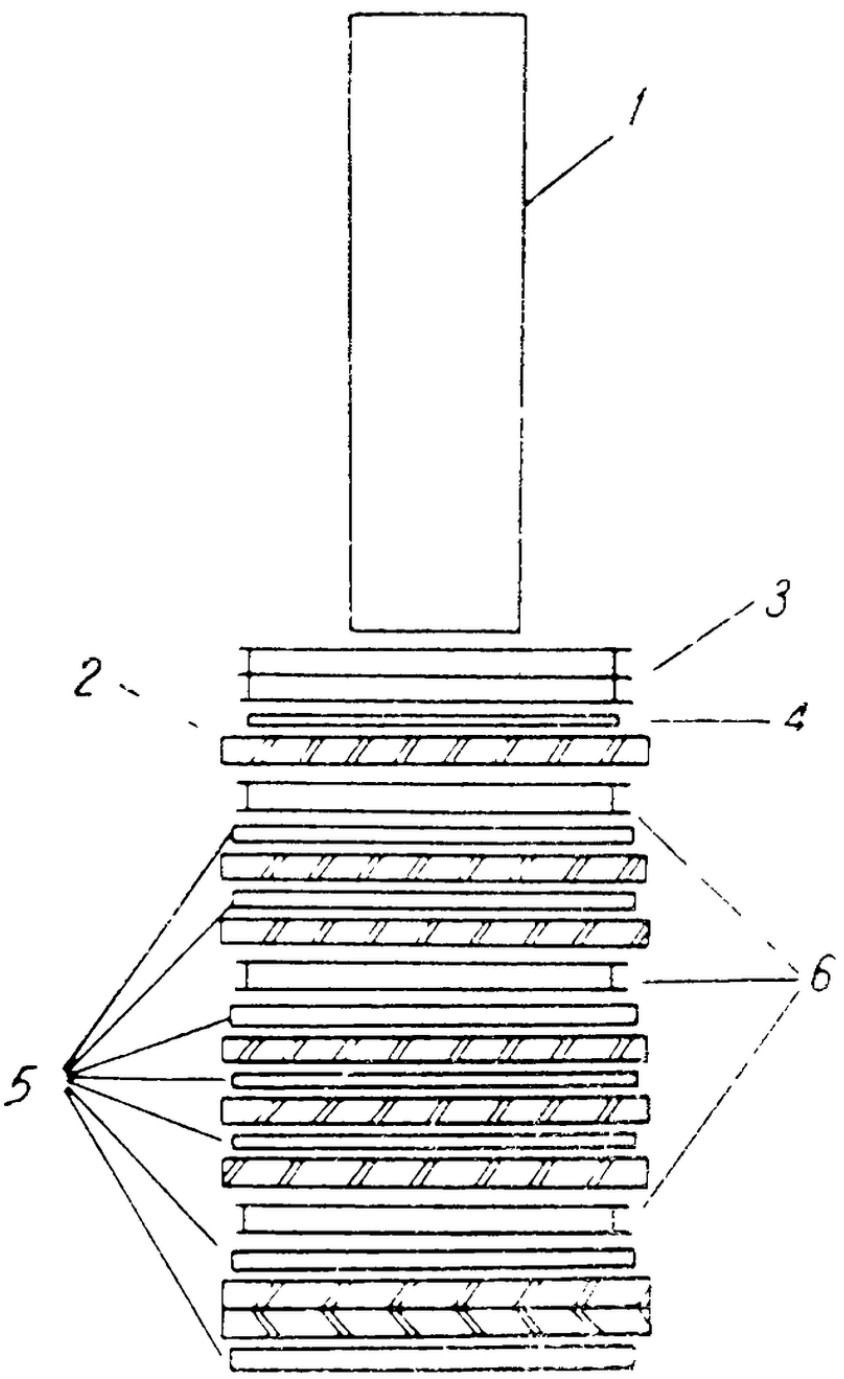

Fig. 20.7. Schematic of the University of Maryland facility [**75**.1, p. 2538]: 1—XTR detector, 2—iron absorbers, 3—streamer chamber for selecting “single” hadrons, 4—trigger scintillation counter, 5—scintillation counters, 6—streamer chambers.

## *C. Facility for the study of Cosmic Electrons*

In primary cosmic radiation, along with protons, there is a much weaker flux of electrons (by approximately 2–3 orders of magnitude). The study of cosmic electrons is associated with great difficulties due to the presence of a strong proton background, and therefore the data on the energy distribution (spectrum) of such electrons were rather inaccurate.

In 1974–1975, a group led by Müller (University of Chicago, USA) [**77**.1, **77**.9] carried out a balloon-borne experiment to measure the spectrum of primary cosmic electrons using a TRD to distinguish electrons from protons. The use of an XTR significantly reduced the proton background and thereby substantially improved the accuracy of the measurements.

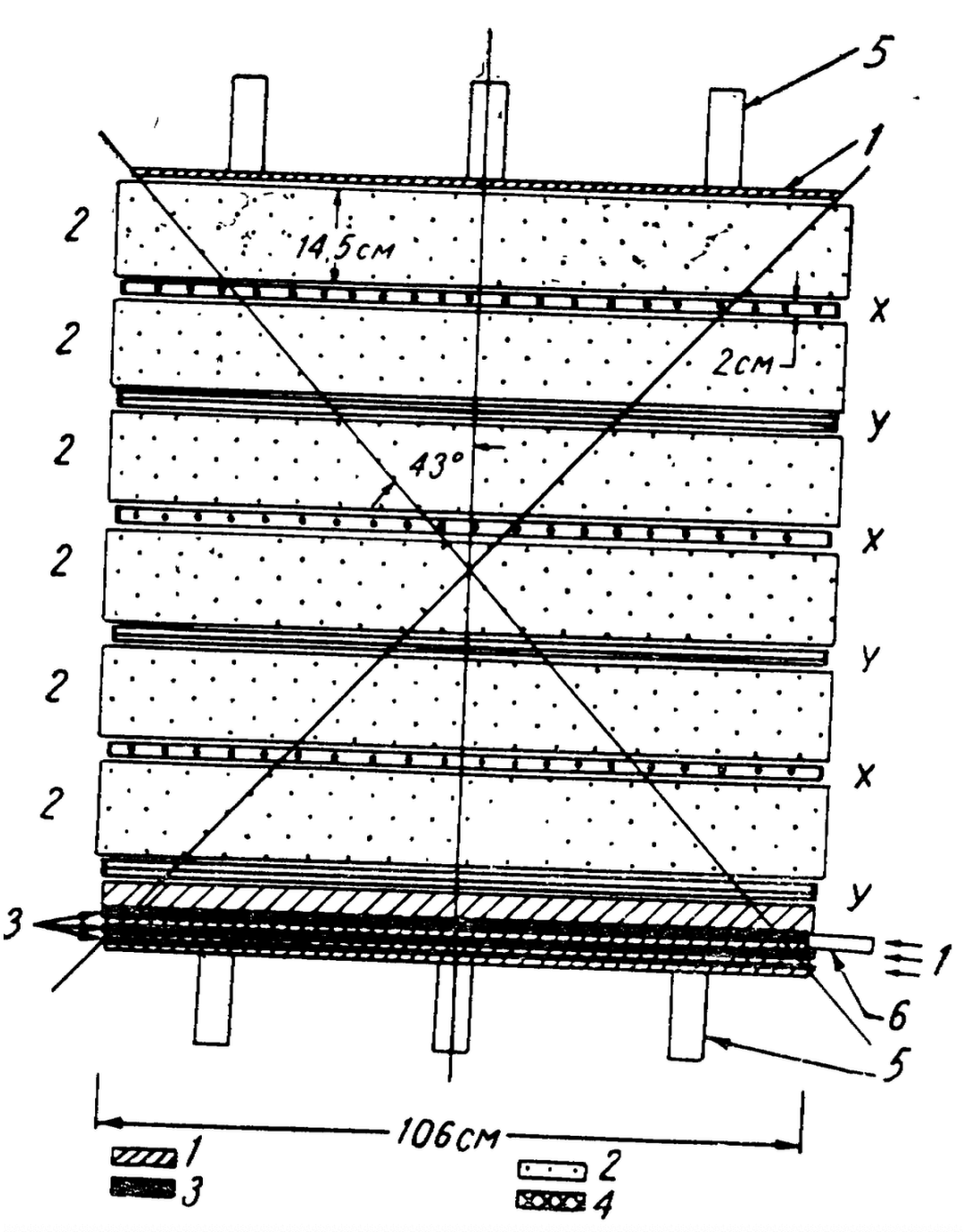

Fig. 20.7. Schematic diagram of the facility for studying cosmic electrons [**77**.9]: 1—organic scintillation counters, 2—polyethylene foam, 3—lead plates, 4—solid polyethylene, 5, 6—tubes, X, Y—multiwire proportional chambers with mutually perpendicular wire orientations

The schematic diagram of the facility used in this experiment is shown in Fig. 20.8. The facility consists of a scintillation telescope, a TRD, and a shower detector.

The scintillation telescope consisted of four scintillators and had a geometric factor of 0.48 $m^2 \cdot sr$. Time-of-flight analysis between the first and last scintillators made it possible to select particles moving from top to bottom.

The TRD was six-module. Its radiators were 14.5 cm thick layers of plastic foam, and the multiwire proportional chambers were 2 cm thick and filled with an 80% Xe + 20% $CO_2$ mixture. The chamber wires in neighboring modules were mutually perpendicular, so the chambers also operated as hodoscopes with a spatial resolving power of 5 cm.

The shower detector contained three lead plates with a total thickness of 8 radiation lengths. Showers were detected after each plate (i.e., at depths of 4, 6, and 8 radiation lengths, respectively) by scintillators included in the telescope.

Owing to the significantly higher measurement accuracy achieved through the use of the TRDs in the facility under consideration, it was established that the electron intensity falls with energy noticeably faster than previously thought, and that the power-law indices of the electron and proton spectra differ substantially. From such a result, if confirmed, certain astrophysical inferences can be drawn regarding the age of cosmic electrons.

### *D. Facility for the Study of J/ψ and Υ Particles*

Starting in 1977, on colliding proton-proton beams of the storage rings at CERN, the Willis group [**77**.12—**77**.14, **78**.4, **79**.3, **80**.4, **80**.5] carried out a series of experiments to study the production of electron-positron pairs with an invariant mass of ~2.5 GeV. To distinguish electrons and positrons from hadrons (mainly pions), a TRD was used in the facility.

The facility comprised four identical sections arranged around the interaction region of the colliding beams (Figs. 20.9, 20.10). Each section consisted of a two-module TRD, a scintillation hodoscope, and a lead and liquid-argon calorimeter. To reduce the absorption of soft XTR quanta in the radiators, the latter were made of layers of lithium foil ($N$=650–700, $a = 50$ μm, $b = 200$ μm). The multiwire proportional chambers of the TRDs had a thickness of 1.3 cm and were filled with a gas mixture of 80% Xe + 20% CO2. As special measurements showed, each such TRD module had a hadron rejection factor not higher than 6% at an electron (or positron) detection efficiency not lower than 90%. The liquid-argon calorimeter provided high accuracy in measuring particle energy and in determining the spatial position of the electromagnetic shower axis, and also further reduced the hadronic background.

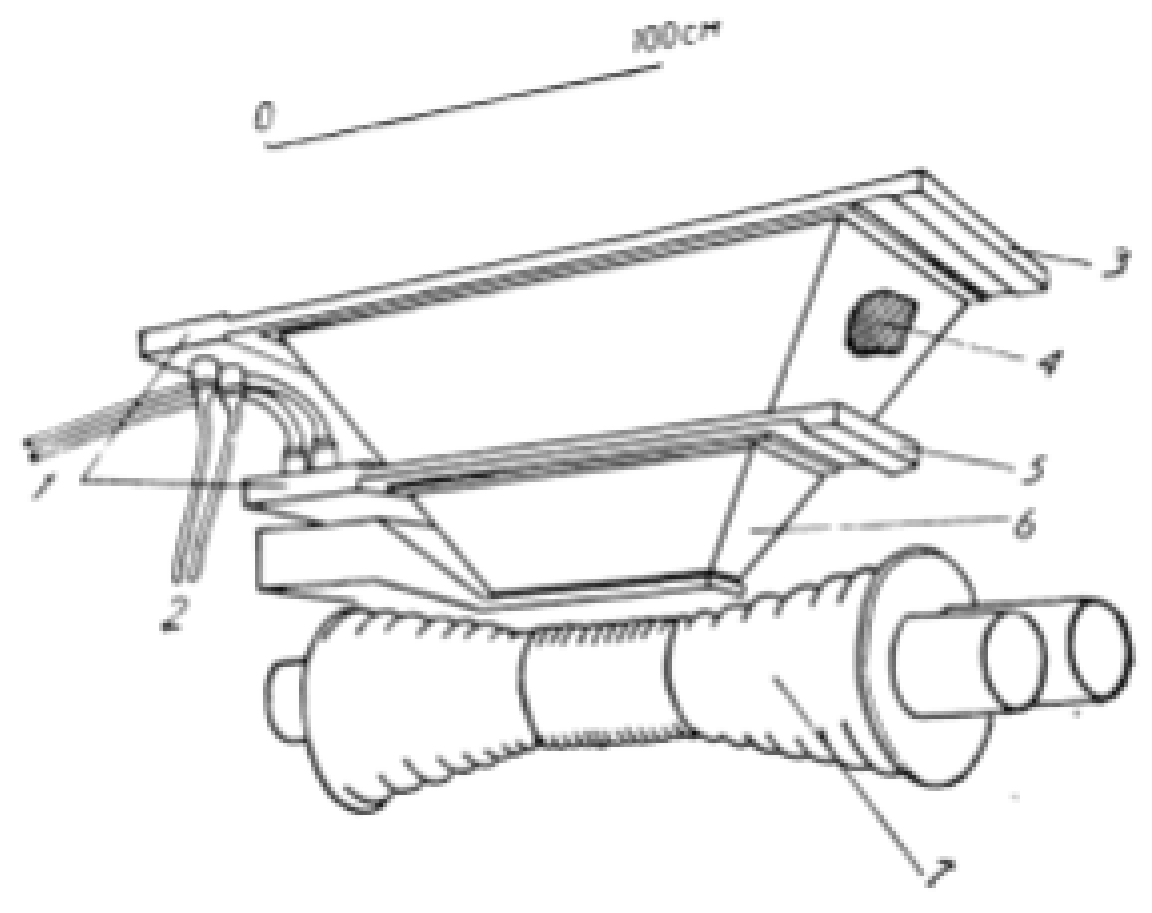

Fig. 20.9. Schematic representation of the interaction region of colliding proton beams and one section of the setup for studying the reaction $p + p \rightarrow e^+ + e^- ...$: 1—preamplifiers, 2—cables, 3 and 5—multiwire proportional chambers, 4 and 6—lithium XTR radiators, 7—the interaction region of the colliding beams [77.11].

In this facility a sufficiently large number of events was recorded, convincingly proving the existence of the *J/ψ* particle with a mass of about 3 GeV/$c^2$, as well as the ϒ particle with a mass of about 9.5 GeV/$c^2$. The TRD made a significant contribution to the event selection.

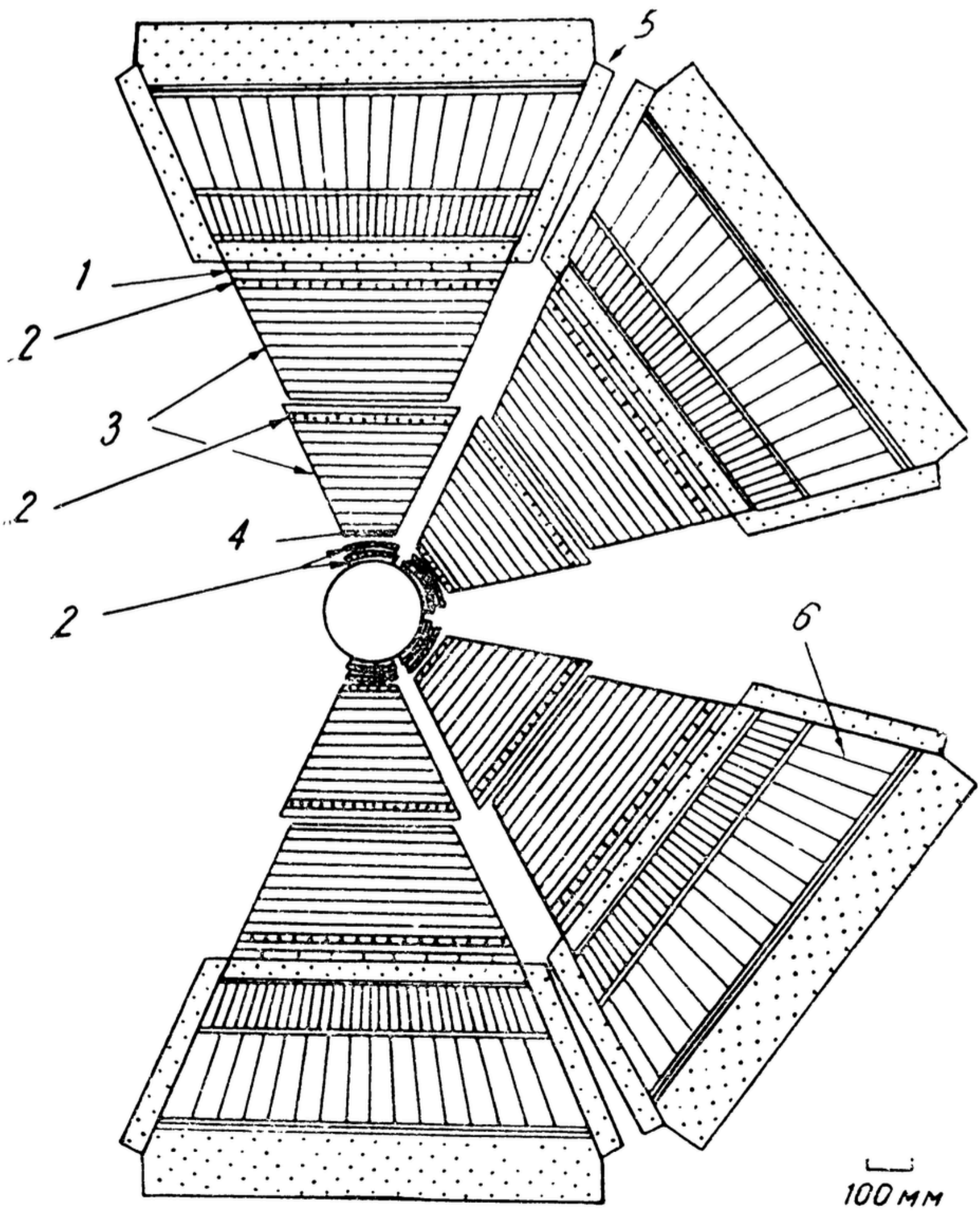

Fig. 20.10. Schematic diagram of all four sections of the facility for studying the reaction p + p → e+ + e–...: 1 and 4—scintillation counters,

2—proportional detectors, 3—lithium XTR radiators, 5—shielding, 6—calorimeters [**77**.12]

## E. Detector with Combined Radiator and Chamber

The group of Oganesyan, Atac, et al. (YerPhI and Fermilab) [**77**.15] developed an original TRD in which the radiator and the multiwire proportional chamber are combined in a single unit (Fig. 20.11). Such a design makes it possible to reduce the absorption of XTR quanta in the radiator material and to increase the relative contribution of XTR energy deposition in the chamber gas.

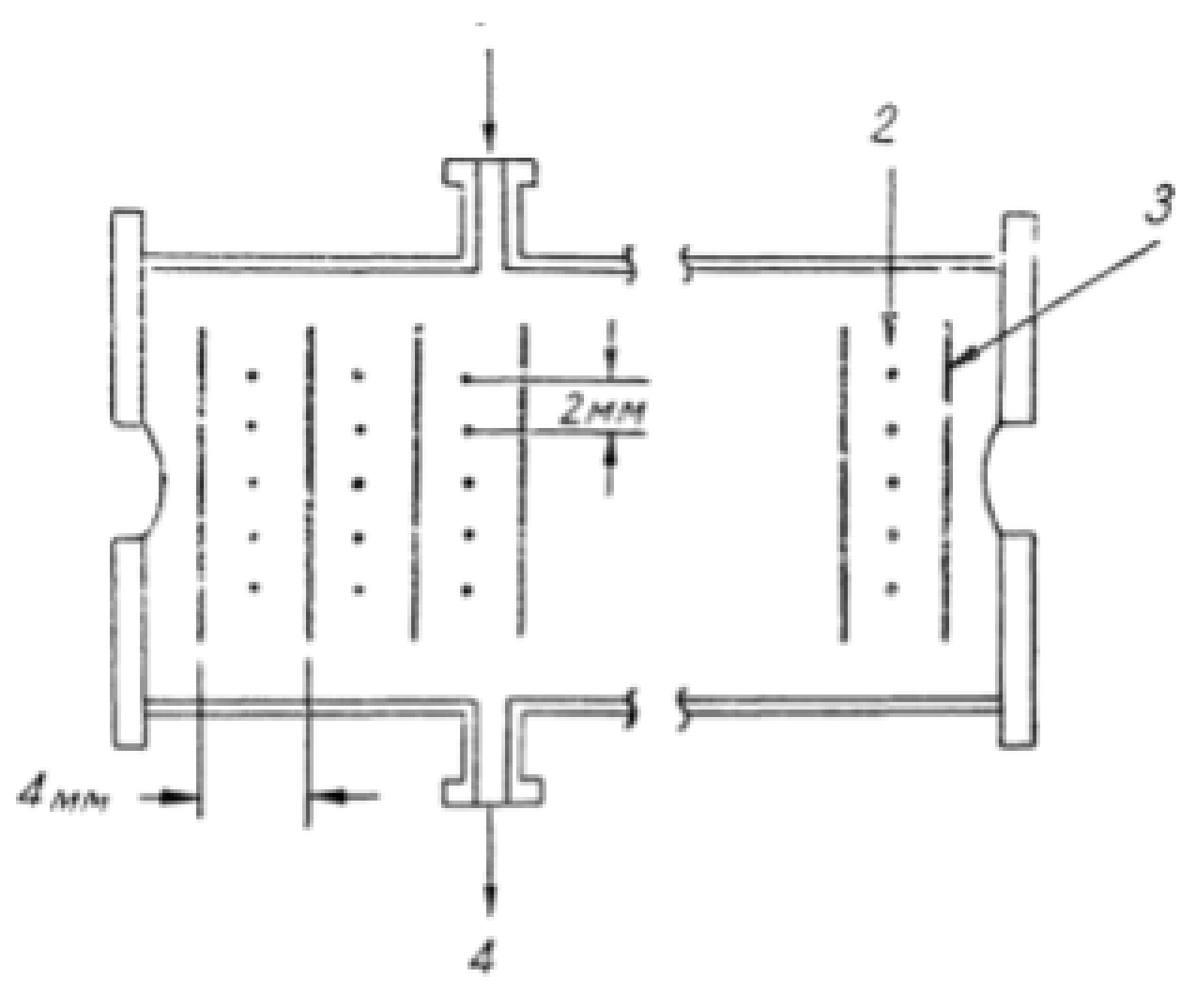

Fig. 20.11. Schematic of an XTR with combined radiator and multiwire proportional chamber [77.15]: 1—gas mixture, 2—signal electrodes, 3—aluminized Mylar, 4—pump

In the chamber, the signal electrodes (wires with a diameter of 25 μm) form 18 planes, which are separated from each other by 5 μm Mylar films. The films have 0.025 μm aluminum coatings on both sides and serve as the XTR radiator. The distance between the films is 4 mm. Gas mixture (80% Xe + 20% CO2) in the chamber is at a pressure on the order of 20–30 mmHg.

From the experimental results one can conclude that using a larger number of such modules in an XTR detector will make it possible to distinguish pions and protons with energies of about 100 GeV and higher with good accuracy.

On the basis of an TRD with an integrated radiator and chamber, an instrument [**82**.9] was created for studying the spectra and composition of multiply charged nuclei ($6 \le Z \le 30$) of primary cosmic radiation at energies up to 1000 GeV per nucleon.

The instrument's TRD made it possible to measure the Lorentz factor of nuclei up to values of order $10^3$. High-altitude balloon flights were carried out with this instrument on board. Processing of the obtained data provided information on the chemical composition of cosmic nuclei and their energy distribution [**82**.13].

## 20.3. Other TRDs

The previous section described TRDs already used in physical experiments. In recent years, efforts by specialists from various countries to develop new TRDs have not ceased.

Ispiryan et al. [**77**.1, p. 209] proposed a project for a multi-module TRD in which, after each radiator, there are several multiwire proportional chambers of small thickness. The essence of this project is as follows. The ionization energy losses of a particle traversing the proportional chambers are almost uniform along its entire trajectory, whereas the energy release due to the absorption of a XTR quantum is localized. Therefore, the smaller the thickness of each proportional chamber, the larger the ratio $W^{TR}/W^{ion}$, i.e., the higher the resolving power of the detector.

Lorikyan et al. [**80**.17, 81.15] proposed another method for reducing the signal due to the ionization energy losses of a traversing particle in a proportional chamber. This method is based on creating conditions to increase the spread of the collection time of ionization electrons. In this case, the signal from ionization losses, which are uniformly distributed along the particle's trajectory, is significantly reduced, while the signal due to the absorption of XTR quanta and having a localized character is not substantially changed. The authors preliminarily tested this method using radioactive photon and electron sources, as well as the electron accelerator of the Yerevan Physics Institute, and obtained encouraging results. However, the technique still requires further development and optimization.

Alongside these proposals, several other detectors with sufficiently good resolving power were tested.

### *A. Detector Based on Superheated Superconducting Granules*

Drukier, Yuan, et al. proposed and tested an original idea for creating a TRD based on superheated superconducting granules [**75**.18, **76**.6, **77**.1, pp. 344, 354].

It is known that a superconductor passes from the superconducting state to the normal state if its temperature exceeds a certain critical value $T_C$ or if an external magnetic field exceeds $H_C$. However, granules of type-I superconductors (In, Sn, Tl, Hg, etc.) can remain in the superconducting state if the temperature or magnetic field is not much greater than $T_C$ or $H_C$. Such a superheated state is metastable, and under the action of a traversing charged particle the granules can transition to the normal state.

For particle detection, Drukier, Yuan, and collaborators used a cylinder made of wax (or another light material) containing tin granules in the form of a colloidal suspension. The cylinder was placed in liquid helium and in a uniform magnetic field with a magnitude on the order of hundreds of oersteds. The average diameter of the granules was 5–7 μm, and the total volume of the granules was about 10% of the cylinder volume.

Several methods were developed to record the fact of the transition of granules from the superheated superconducting state to the normal one. One of the methods consisted in the following: due to the Meissner effect (expulsion of the magnetic field from a superconductor), the magnetic flux changes when the granules transition from the superconducting state to the normal state, which led to the appearance of an electromagnetic pulse in a measuring coil wound around a cylinder containing the granules. After amplification, this pulse was recorded in coincidence with the pulse from a particle traversing the granules. The method was sufficiently sensitive to record the transition in each individual granule occurring as a result of interaction with a single traversing ultrarelativistic particle.

The detector was tested at the DESY electron synchrotron (Hamburg, Federal Republic of Germany). A monotonic increase of the signal magnitude as a function of electron energy was obtained, close to a quadratic law in the energy range of 2–6 GeV. The authors of this experiment explain such a quadratic dependence by the influence of multiple scattering of electrons on XTR. However, in reality, quanta of the total radiation, which was the sum of bremsstrahlung and transition radiation, were generated and absorbed in the granules (see Chapter IV). The intensity (or number of quanta) of the total radiation does not have a quadratic dependence on the particle energy. Therefore, it should be considered that a theory of the experiment with superheated superconducting granules does not yet exist (for details, see Refs. [**76**.16, **77**.1, p. 374]).

### *B. Detector with Drift Chamber (Transverse Drift)*

An TRD using a drift chamber (Fig. 20.12) was developed by an international group of Dolgoshein, Willis, et al. (USSR, Switzerland, USA,

Federal Republic of Germany) [**81**.5, **82**.12, p. 165]. This detector was a further development of a similar detector by Alikhanian, Dolgoshein, et al. [**77**.1, p. 571, **79**.13, **82**.12, p. 158].

The operating principle of the detector was based on both the energy and spatial separation of XTR photons from the ionization products of the particle. When a particle and its accompanying XTR photons entered the drift chamber, some gas atoms in the chamber were ionized. In this case, the atoms ionized by the particle were located predominantly on the particle track, whereas the atoms ionized by the XTR photons were at some distance from the track because XTR was emitted at small but finite angles relative to the particle trajectory, and the chamber was at a certain distance from the radiator. The electrons knocked out of the atoms during their ionization began to drift in a direction transverse to the chamber anode. Since such electrons were produced at different distances from the anode, they arrived at the anode at different times (Fig. 20.13). Thus, in the presence of XTR photons the particle produced a signal with not only a larger total magnitude (the area of the shaded region in Fig. 20.13) but also a greater width. By performing a two-parameter analysis of both the total magnitude and the width of the signal, it was possible to significantly improve the resolving power of the detector. The experiment showed that when using a radiator made of lithium ($N$ = 1500, $a$ = 30 μm, $b$ = 200 μm) or Mylar ($N$ = 1000, $a$ = 5 μm, $b$ = 140 μm) plates, the described detector consisting of 15 modules is capable of satisfactorily distinguishing pions and kaons with energies on the order of 140 GeV.

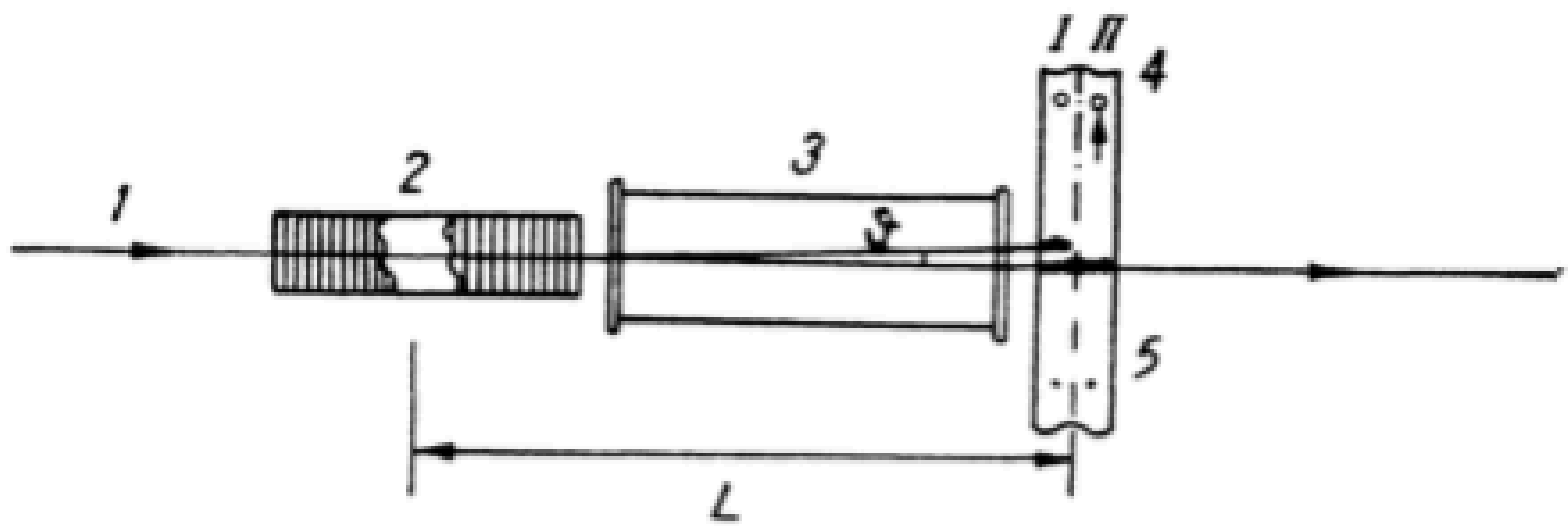

Fig. 20.12. Schematic diagram of a detector with a drift chamber (transverse drift) [**81**.5]: 1—beam, 2—XTR radiator, 3—helium beam pipe, 4—drift chambers, 5—anodes

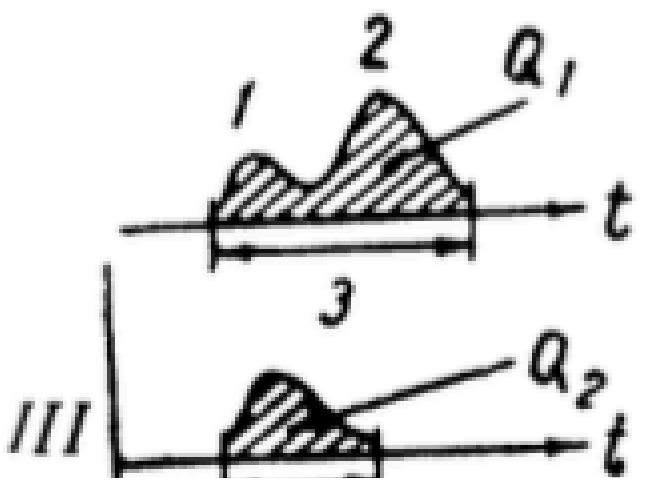

### *C. Cluster-Counting Detector (Longitudinal Drift)*

The use of the transverse-drift method considered in the previous section required some distance between the radiator and the drift chamber (no less than 0.5–1 m). This increased the size of the detector and reduced its light-gathering power, and therefore in some cases such a method was unacceptable. A group led by Dolgoshein, Willis, and others also developed a TRD with a drift chamber placed directly adjacent to the radiator. In this chamber, the drift of electrons knocked out of gas atoms by the charged particle and by XTR quanta occurred in the longitudinal direction. In doing so, the authors used the so-called cluster-counting method [**81**.6, **81**.7, **82**.12, pp. 173, 179], in a sense related to the method of studying XTR with a streamer chamber.

The essence of the cluster-counting method was as follows. XTR quanta were absorbed in the chamber gas and created clusters (localized regions of ionization) of limited size. For example, 10 keV photoelectrons produced by an XTR quantum in xenon gas at normal pressure created clusters of the order of 500 μm. Such a cluster size was mainly determined by the diffusion of the electrons produced by ionization in the gas. A traversing charged particle can also create clusters due to the production of δ-electrons, i.e., ionization electrons with sufficiently high energy $E_\delta$ (greater than, say, 1 keV). Such clusters led to background in the detection of clusters produced by XTR quanta. However, the number of clusters *N* produced by δ-electrons is relatively small (for example, at $E_\delta > 3$ keV for ultrarelativistic particles in xenon $N_\delta = 0.1$ cm$^{-1}$) and, what is very important, the fluctuation of the number $N_\delta$ is described by a Poisson distribution, i.e., it is substantially smaller than the fluctuation of the energy deposition due to ionization losses of the particle, which is described by the Landau distribution. These two circumstances were significant advantages of the cluster-counting method over the conventional energy-deposition method.

Moreover, the number of XTR quanta fluctuated less than the total XTR energy, albeit with a somewhat flatter γ-dependence. Nevertheless, the authors of the cited works note that there are cases in which the developed cluster-counting detector can be successfully used.

The optimized TRD with cluster counting consisted of 12 modules—radiators and proportional chambers (Fig. 20.14). Each proportional chamber had an area of 10×10 cm². The chambers were filled with a gas mixture of 50% Xe + 50% $CH_4$.

As radiators, stacks of lithium foils ($N \approx 100$, $a$=30 μm, $b$=160±80 μm; $N \approx 160$, $a$=35 μm, $b$=240±80 μm) or irregularly arranged fibers of pure carbon (with a thickness of 7 μm and a density of ) were used. The total length of the 12-module detector was 40–70 cm.

Current pulses from the proportional chambers, after amplification and shaping, were analyzed by appropriate electronic circuits.

Beam tests at CERN showed that, in distinguishing pions from electrons (at particle energies of about 15 GeV), the cluster-counting method yielded a pion rejection factor of the order of $10^{-3}$ at an electron detection efficiency of about 90%. When using the conventional energy deposition method with the same detector, the pion rejection factor was approximately an order of magnitude higher (up to a factor of 25).

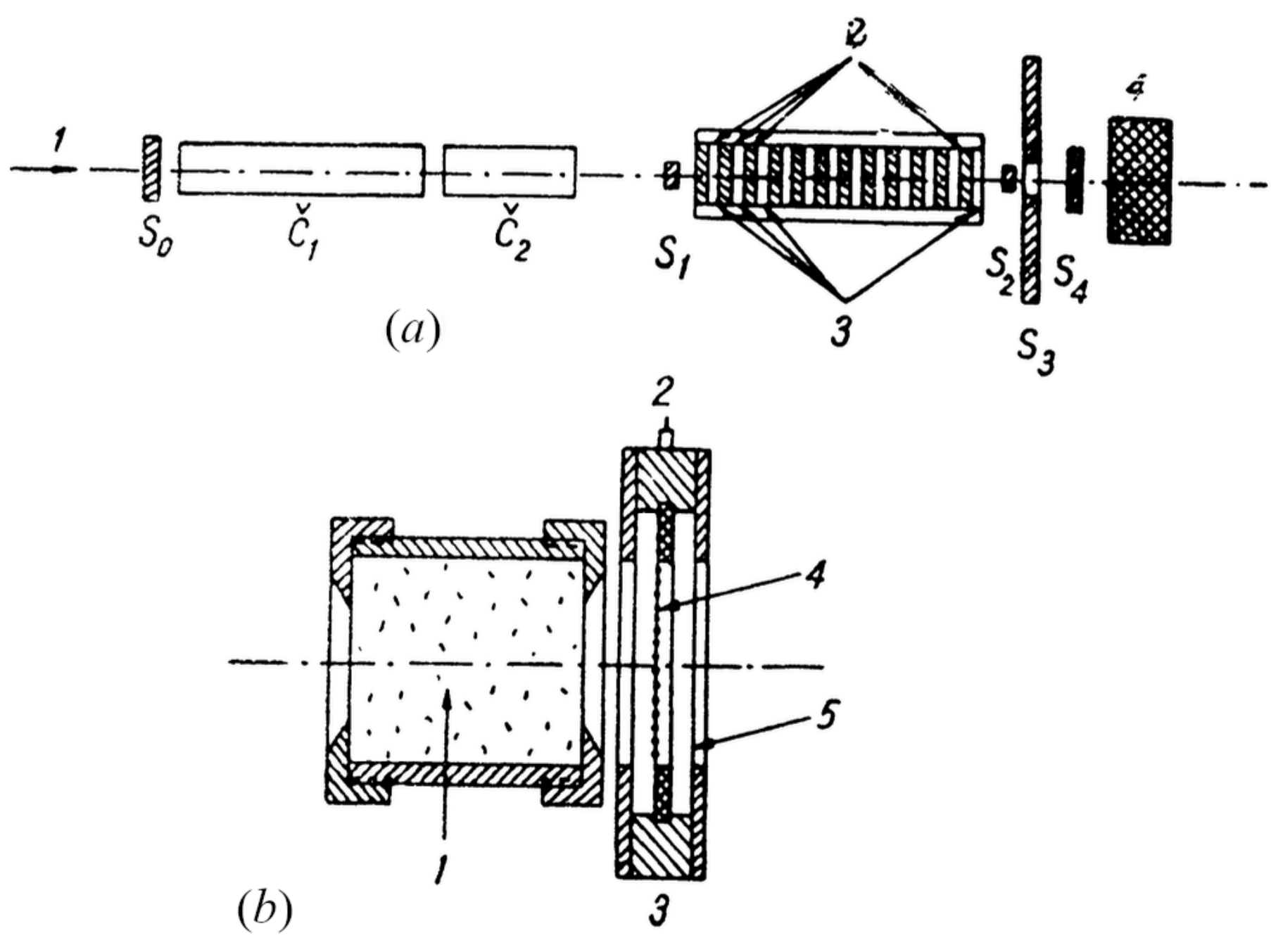

Fig. 20.14. Schematic diagram of a cluster-counting detector. [**81**.7]:
(a) Particle selection and identification system: $S_0$–$S_4$—scintillation counters, $Č_1$, $Č_2$—gas threshold Cherenkov detectors, 1—beam, 2—radiators, 3—multiwire proportional detectors, 4—shower detector consisting of lead and organic material;
(b) Diagram of one radiator module with chamber: 1—radiator, 2—high voltage, 3—multiwire proportional chamber, 4—anode wires, 5—cathode foil

An analogous detector with 24 modules (total length equal to 132 cm) made it possible to distinguish pions from kaons at energies of about 140 GeV with a pion rejection factor of about 10 at a kaon detection efficiency of about 90%. In this case the best results were obtained with a radiator made of irregular carbon fibers. In this case, the cluster-counting method also showed a significant advantage over the conventional energy deposition method.

## CONCLUSION

In summary, the following most important properties of X-ray transition radiation can be noted:

1. XTR is emitted mainly in the forward direction, at small angles relative to the particle trajectory. The maximum of the angular distribution (see Eq. (1.59)) of the radiation produced at a single interface between a medium and vacuum occurs at an angle of order the inverse of the particle Lorentz factor ($\gamma^{-1}$) and extends approximately up to an angle of order $(\gamma^2 + \omega_0^2/\omega^2)^{1/2}$, where $\omega^2$ is the plasma frequency of the medium. At the same time, the spectral intensity (see Eqs. (1.64) or (1.67)) of XTR drops sharply at frequencies exceeding the cutoff frequency $\omega_0\gamma$. The total XTR intensity (integrated over angle and over some frequency interval) increases substantially with increasing $\gamma$ if the upper frequency of the chosen range is of the order of the cutoff. In particular, when the upper frequency of the range exceeds the cutoff, the total intensity of XTR produced at a single interface grows in proportion to $\gamma$ (see Eq. (1.71)).

2. In the case of a plate, interference occurs between the radiations produced at different boundaries. When the thickness $a$ of the plate is much greater than the formation length of transition radiation $|z_{med}(\vartheta)|$ (see Eq. (2.25)), the interference maxima and minima in the spectral–angular distribution of the intensity are located very close to each other, so that after averaging over a small range of angles (or frequencies) there is an effectively independent addition of the intensities (numbers of quanta) of the radiations formed at the individual boundaries. When the thickness $a$ is much smaller than $|z_{med}(\vartheta)|$, the radiation intensity is significantly lower (compared with the intensity at large $a$, but at the same value of $|z_{med}(\vartheta)|$).

For relatively small values of the particle's Lorentz factor ($\gamma < a\omega_0/c$), the cutoff frequency of the XTR spectrum formed in a plate is of order $\omega_0\gamma$, as in the case of a single boundary. When $\gamma > a\omega_0/c$, the highest frequencies of the spectrum cease to increase with $\gamma$ and remain at values of order $a\omega_0^2/c$. In this case, the spectral intensity of XTR has maxima at following frequencies (see Eq. (2.58)):

$$\omega_n^{\rm pl} = \frac{a\,\omega_0^2}{2\pi c(2n+1)} \quad (n = 0, 1, 2, \ldots).$$

The dependence of the total radiation intensity on $\gamma$ for $\gamma < a\omega_0/c$ is rather strong (linear or stronger), and with further increase of $\gamma$ it becomes weaker and gradually changes over to logarithmic.

3. In the case of a regular stack of a large number of plates, at certain frequencies and angles (see condition in Eq. (3.27)), there is coherent superposition for the radiation generated in different plates of the stack (where, of course, absorption must be sufficiently small). The spectral–angular intensity of such resonant transition radiation is proportional to the square, $N_{eff}^2$, of the effective number of plates in the stack. The width (spectral or angular) of each such interference maximum is proportional to $N_{eff}^{-1}$, so that the intensity integrated over angle is proportional to $N_{eff}$ (see Eq. (3.31)).

When the distance $b$ between the plates of the stack (it is assumed that the plates are in vacuum) is much greater than the formation length $z_{\rm vac}(\vartheta)$ (see Eq. (2.26)), the above-mentioned interference maxima are located very close to each other, so that after averaging over a small range of angles (or frequencies) there is an independent addition of the intensities (numbers of quanta) of the radiation formed on individual plates. When $b$ is much less than $z_{\rm vac}(\vartheta)$ and the thickness $a$ of the plates of the stack is much less than $|z_{\rm med}(\vartheta)|$, the entire stack radiates as a single equivalent plate with thickness $a_{\rm eq} = N(a+b)$, plasma frequency $\omega_{0\,eq} = \sqrt{a/(a+b)}\omega_0$, and linear absorption coefficient $\eta_{eq} = a\eta/(a+b)$, where η is the linear absorption coefficient of the radiation in the material of the plates of the stack. Then, for $\gamma > N\sqrt{a(a+b)}\omega_0/c$ , additional maxima appear in the radiation spectrum at frequencies

$$\omega_n^{\mathrm{edge}} = \frac{N a\,\omega_0^2}{2\pi c(2n+1)}, \quad (n = 0, 1, 2, \ldots).$$

The intensities of these maxima increase with increasing γ.

The total radiation intensity for $\gamma < a\omega_0/c$ depends very strongly on γ (linearly or even more strongly); for $a\omega_0/c < \gamma < \sqrt{ab}\omega_o/c$ it depends somewhat more weakly; for $\sqrt{ab}\omega_0/c < \gamma < N\sqrt{a(a+b)}\omega_0/c$ it is almost independent of γ; and for $N\sqrt{a(a+b)}\omega_0/c < \gamma$ it again increases with γ.

4. In the case of an irregular stack of plates, there is a smoothing of the spectral–angular distribution of the transition radiation intensity. If the ratios of the root-mean-square deviations of the plate thicknesses and the spacings between plates to the corresponding squares of the formation lengths are much smaller than $N^{-1}$ (see Eq. (6.17)), then the spectral–angular distribution of the radiation intensity differs little from the corresponding distribution for a regular stack. If, however, the root-mean-square deviation of the spacings between plates (and of the plate thicknesses) is much larger than the squares of the corresponding formation lengths, then the intensities from $N$ plates (and correspondingly $2N$ interfaces) add incoherently, with due account, of course, taken of absorption of the radiation in the medium.

5. Multiple scattering of the particle in the medium leads to smoothing of the spectral–angular and spectral distributions of the XTR intensity and to the appearance of bremsstrahlung. In this case bremsstrahlung appears at frequencies higher than the cutoff frequency $\omega_0\gamma$ (the medium polarization effect). If one separates from the total radiation the part due to the presence of interfaces, then in its frequency spectrum at sufficiently large $\gamma$ there will be an enrichment with high frequencies, up to a frequency that depends quadratically on $\gamma$.

6. When a fast particle passes through a single crystal, along with the central XTR spot, side spots also appear (quasi-Cherenkov radiation, see Secs. 13, 14). This radiation is emitted at large (Bragg) angles with respect to the particle trajectory and has narrow directivity and good monochromaticity. The angular divergence of the radiation in each spot is of order $\gamma^{-1}$, and the spectral width of the radiation is tens of eV for a photon energy of order keV ($\Delta\omega/\omega\sim 10^{-2}$). The intensity of this radiation (integrated over frequency and angle for a given spot) is approximately $10^{-4}$–$10^{-5}$ quanta per particle.

Regarding the practical use of transition radiation, the following aspects can be noted at present.

1. Owing to the dependence of the XTR intensity on the particle Lorentz factor $\gamma$, TRDs (Sec. 20) have been constructed and are used for nondestructive detection and identification of high-energy charged particles with $\gamma$ of order $10^3$–$10^5$. Both regular and irregular layered media are used as XTR radiators. TRDs reliably distinguish pions from protons (not to mention particles with a larger mass ratio). There are preliminary experiments in which it is possible to distinguish kaons from protons.

2. Resonant transition radiation in a stack of plates can serve as a good radiation source, in particular, in the X-ray and microwave frequency ranges. The monochromaticity of such a source amounts to tens of percent, and the operating frequency range is easily tunable.

3. Quasi-Cherenkov radiation in crystals can be a source of narrowly directed and highly monochromatic ($\Delta\omega/\omega\sim 10^{-2}$) X-ray radiation.

4. Quasi-Cherenkov radiation in crystals and resonant microwave transition radiation in a layered medium can also be used for monitoring

and diagnostics of a beam of fast charged particles owing to the large emission angles of this radiation relative to the particle trajectories.

5. Transition radiation can also be used to study the properties of the medium itself in which it appears. For example, from the dynamical maxima of quasi-Cherenkov radiation one can investigate the structure of a crystal; from resonant XTR vacancy pores in solids can be studied; and from optical transition radiation the state of a body's surface can be investigated.

Some theoretical results await their experimental verification. Among them, in particular, is the formation of the Laue pattern of dynamical maxima of quasi-Cherenkov radiation by a charged particle in single crystals. At the same time, there are a number of experimental facts that have not yet received a theoretical explanation, such as the results of the experiment by Drukier, Yuan, et al. with superheated superconducting granules, anomalously intense optical radiation under grazing incidence of electrons on a metallic surface, and others.

* * *

As an ancient Eastern proverb goes, "Without a well-honed tool one cannot do quality work." In high-energy particle physics, the primary tools are particle detectors. It is well known that ionization and spark chambers and Cherenkov detectors have played—and continue to play—a tremendous role in the development of high-energy particle physics. Meanwhile, the creation and improvement of these instruments are inconceivable without a deep understanding of the physical processes that underlie them.

In recent years, fundamentally new detectors based on X-ray transition radiation—XTDs—have been developed and created. They were preceded by detailed theoretical and experimental studies of the properties of transition radiation carried out by physicists from many countries, above all in the Soviet Union. TRDs are still at an early stage of development. There is no doubt that they will continue to develop and improve, and this will be closely connected with further studies of XTR.

In addition, the phenomenon of transition radiation, including X-ray transition radiation, like any properly understood phenomenon of the material world, can find and is finding other, no less interesting applications, whose full value is difficult to foresee in advance. The authors of this book are confident that work in this area of physics will continue successfully.

# REFERENCES

The references are arranged in chronological order. The first two digits indicate the year of publication, while the number following the period denotes the sequential number within that year. At the end of each reference, after the □ symbol, a cross-reference is provided, indicating the location[25] in the book where the work is cited.

---

[25] The following abbreviations are used: Vv—the opening part of the Introduction; A, …, G—the corresponding sections of the Introduction; IV, V—the introductory sections of Chapters IV and V.